\documentclass{report}
\usepackage[utf8]{inputenc}
\usepackage{amsmath}
\usepackage{amsfonts}
\usepackage{amssymb}
\usepackage{stmaryrd} 
\usepackage{physics}
\usepackage{graphicx}
\usepackage[margin=1in]{geometry}
\usepackage{adjustbox}
\usepackage{appendix}
\usepackage{bm}
\usepackage{braket}
\usepackage{cite}
\usepackage{bbm}
\usepackage{tcolorbox}
\usepackage{subcaption}
\usepackage{mathtools}
\usepackage{xcolor}
\usepackage{hyperref}
\usepackage{stmaryrd}
\usepackage{epigraph}
\hypersetup{
    colorlinks=true,
    citecolor=blue,
    linkcolor=blue,
    filecolor=magenta,
    urlcolor=blue}

\usepackage[normalem]{ulem}

\newcommand{\eps}{\epsilon}
\newcommand{\lam}{\lambda}

\renewcommand{\th}{\theta}
\newcommand{\al}{\alpha}

\newcommand{\g}{\gamma}
\newcommand{\be}{\beta}
\newcommand{\co}{\nabla}
\newcommand{\pa}{\partial}

\newcommand{\si}{\sigma}
\newcommand{\sign}[1]{{\rm Sign}(#1)}
\newcommand{\ti}[1]{\Tilde{#1}}
\newcommand{\ffi}{\varphi}
\renewcommand{\cal}[1]{\mathcal{#1}}

\newenvironment{eqgroup}
{
\begin{equation}\begin{aligned}
}
{
\end{aligned}\end{equation}
}

\title{Thèse}
\author{Vassilis Papadopoulos}
\date{July 2022}

\begin{document}

\thispagestyle{empty}

\begin{center}

\Huge{Membranes, holography, and quantum information}

\vskip 1cm

\large{Vassilis Papadopoulos}

\vspace{1.0cm}

{\sl\small  Laboratoire de Physique de l'\'Ecole Normale Sup\'erieure, ENS, Universit\'e PSL}

\end{center}

\newpage

\tableofcontents

\chapter*{Acknowledgments}
While an essential part of any respectable thesis, the "acknowledgments" section feels more like a trap. This is because, while the people that are mentioned on here will be very happy with their inclusion, those that are not will be sorely disappointed. It does not help one bit that $99\%$ of people that will get their hands on this manuscript will read approximately $1\%$ of it, this $1\%$ being the "acknowledgments" section\footnote{N.B. : I don't blame them}. As literal hours\footnote{Not a lot of hours} separate me from the deadline as I write this final section, I am afraid that I will not be able to include the full list of people deserving to appear on this page. Thus, if you feel that your name should have been mentioned, know that it is not an oversight on my part just a lack of time, so you can redirect your complaints to the EDPIF for setting the deadline exactly on the day I am finishing the writing.

With this introduction out of the way, I would like to begin by thanking my parents, as it is their own love of mathematics and physics that was infused in me from a young age that ultimately lead me to this moment, submitting my PhD thesis (or at least its first draft). I should probably also thank them in advance, as they do not know it, but they will be the main contributors to my pot de thèse (they cook much better than I do). I should thanks also my brother and sister for being a great company ever since I can remember (which is however not reason enough not to go through the entire manuscript, so get to work).

I would like also to thank my girlfriend Pauline, which has had to endure me during these last three grueling PhD years. I know it must not have been easy to deal with my random and completely unconventional working hours, as well as my periodic obsessions with random subjects. She has always been present in the difficult and less difficult moments and has made the last three years much better than they would have been without her. Unfortunately, little does she know that my terrible working hours are not due to the fact I was preparing a Ph.D, but simply because I am unable to manage my time. I hope that this terrible revelation will not make her break up.

I should also thank my closest colleagues, with which I have shared about half of my waking time, at least when Covid wasn't in the way. On the one side, we have had myriad of interesting scientific discussions, and they certainly have elevated the quality of the work presented in this thesis. On the other, they have always been a great company, always filled with self-made memes, (less than) subtle jokes, and a few (too many) beers. It would take a full manuscript if I had to make a little clever joke about each one, so I will just list the names (randomly shuffled by use of a quantum random number generator) : Manuel, Augustin, Ludwig, David, Arthur, Hari, Zechuan, Gabriel, Farah, Gauthier, Marko, X. (please replace the X with your name if you have been unjustly forgotten. I swear this is not a judgment of your quality, I just forget most important things). I should also thank my fellow "PhD students in high energy", that I have met mainly during the Solvay school, but also throughout my PhD when Covid allowed it. They are way too many to cite, but each of them contributed to this thesis by way of (often heated) discussions.

My thanks also go to my friends from my previous studies, from EPFL and ENS. Although some have been lost to time and distance, all of them have contributed to making my studies more enjoyable and fun. I will cite them here in order, according to the probability that they read this paragraph of my thesis: Robin, Federico, Stéphane, Croquette, Simon, Adrien, Juliette, Arthur, Simon, André, Basile, Hortense, Antoine, Adrien, X.

Another subset of friends which I must cite here is my friends from school. This is because they firmly believe that it is thanks to the countless hours I have spent during our high-school years explaining to them our math lessons that have prepared me for this PhD. Thank you Javier, Felix, Matteo, Mattia, Eleonore, Raphael, Chloe, X.

Besides my friends, I should also thank the people that have contributed scientifically to my PhD, be it through discussion, e-mail exchange, or even anonymous review. Again, citing them all would require a whole new manuscript so I will abstain myself, but I would like to especially thank Pietro, Marco, and Julian for very stimulating discussions during my short visit in Geneva, as well as Zhongwu who was an amazing collaborator. Special thanks also to Mario, Frederic and Guilhem, with which it was a great pleasure to teach.

I must of course thank the jury, composed of Marco Meineri, Shira Chapman and Giuseppe Policastro and in particular the "rapporteurs", Marios Petropoulos and Johanna Erdmenger, which will have to go through this manuscript in detail. I may be wise to increase my chances to render this paragraph of thanks conditional to me being awarded the PhD.

Finally, I would of course like to thank my PhD advisor, Costa, which has been my main scientific reference and collaborator for the past three years (indeed the tradition in string theory is for the advisor to take only one PhD student at a time, a Padawan, one might say). Despite Covid preventing live discussions for a while, he managed to always be there (through the magic of Zoom) when I had questions and offered great insights which always managed to unblock me if I ever got stuck. I remember always leaving with dozens of possible directions to explore after every talk we had, although I must say that sometimes his expectations of my abilities were a bit too high, especially when it comes to numerical computation.

\chapter*{Abstract}
In this thesis, we study Interface Conformal Field Theories (ICFT) and their holographic dual, which is composed of two asymptotically Anti-de-Sitter (AdS) spaces glued through a thin gravitating membrane, or domain wall. We restrict our study to simple minimal models, which allow for analytic control while providing universally applicable results. Our analysis is set in 2D ICFT/3D gravity, but we expect much of the results to be generalizable to higher dimensions. 

We first consider this system at equilibrium and at finite temperature, in the canonical ensemble. By solving the equations of motion in the bulk, we find the allowable solution landscape, which is very rich compared to the same system without an interface. Classifying the different solutions among 3 thermodynamical phases, we draw the phase diagram outlining the nature of the various phase transitions. 

We then examine a simple out-of-equilibrium situation arising as we connect through an interface two spatially infinite CFT's at different temperatures. As we let them interact, a "Non-Equilibrium Steady State" (NESS) describes the growing region where the two sides have settled into a stationary phase. We determine the holographic dual of this region, composed of two spinning planar black holes conjoined through the membrane. We find an expression for the out-of-equilibrium event horizon, highly deformed by the membrane, becoming non-killing. This geometry suggests that the field theory interface acts as a perfect scrambler, a property that until now seemed unique to black hole horizons.

Finally, we study the entanglement structure of the aforementioned geometries by means of the Ryu-Takayanagi prescription. After reviewing a complete construction in the vacuum ICFT state, we present partial results for more general geometries and at finite temperature. For this purpose, it is necessary to introduce numerical algorithms to complete the computation. We outline the main difficulties in their application, and conclude by mentioning the Quantum Null Energy Condition (QNEC), an inequality that links entanglement entropy and energy, that can be used to test the consistency of the models.

\chapter*{Introduction}
Ever since the seminal paper of Coleman and De Luccia \cite{Coleman:1980aw,Coleman:1977py}, studies involving thin gravitating domain walls have appeared in a multitude of different contexts. For instance, they have been used in attempts to provide an alternative to compactification, by localising gravity on lower dimensional "brane-worlds"\cite{Randall:1999vf,Dvali:2000hr,Karch:2000ct}. They also appeared in efforts to embed inflation and de Sitter geometries in string theory, by studying inflating bubbles\cite{Freivogel:2005qh,Freivogel:2007fx,Barbon:2010gn,Banerjee:2018qey}, while they also enter in some of the swampland conjectures\cite{Ooguri:2020sua,Lanza:2020qmt,Bedroya:2020rac}.  More recently, they played an important role in toy models of black hole evaporation, in which an AdS black hole is connected to flat space to allow for its evaporation \cite{Rozali:2019day,Deng:2020ent,Geng:2020fxl,Bak:2020enw}. 

In this thesis we explore yet another facet of these gravitating walls, as holographic duals of conformal interfaces\cite{Karch:2000gx,Bachas:2001vj,DeWolfe:2001pq}. As such, they act as the border between two AdS space of possibly different radii. In the full UV complete version of the duality, the walls would presumably be smooth, interpolating continuously between the two different spacetimes. We will make the simplifying assumption that the transition region is "thin", meaning it cannot be resolved at the energy scales we will consider. In this "bottom-up" approach, we consider an effective model, and posit the duality on the grounds that there exists some UV theory from which this model descends. The pros of such a philosophy is that it allows for much more freedom on the model that we consider, the cons being of course that we are not assured that the holographic duality is applicable. Nonetheless, experience, the wealth of examples of holographic dualities, as well as independent checks of results seem to point to the usefulness of such bottom-up models.

The main interest of this thesis is focused on studying the holographic dual of a two-dimensional Interface CFT, which is composed of two asymptotically AdS spacetimes connected through a thin membrane/wall. We do not consider a particular realization of this duality, but rather focus only on universal quantities. This as the advantage of offering results that are applicable to any example of such a system, while keeping things sufficiently simple to have an analytical handle on them. The initial driving goal in considering such models was for their application in the Island constructions\cite{Anous:2022wqh,Chen:2020hmv}. However, they are also powerful playgrounds to study aspects of the holographic duality, such as the Ryu-Takayanagi (RT) conjecture\cite{Takayanagi:2011zk}. In addition, they offer deep and unexpected insights into the behavior of ICFT at large coupling, with potential applications in condensed matter physics. The richness of such seemingly simple models has been a great surprise in these 3 years of study, and certainly much is yet to be discovered about them.

I begin in chap.\ref{chap:basicsofholo} by reviewing the essential tools that will be needed to understand our work, as well as the motivations behind it. At the risk of being pedantic, I decided to start at the very basics, trying to have in mind the material that a student just starting in the field would need to apprehend the rest of the work. In secs. \ref{sec:generalitiesAdS}-\ref{sec:hawkingphasetranso} I review general facts about asymptotically Anti-de-Sitter spaces, which constitute the gravitational half of the holographic duality. I introduce black hole solutions, and explain their thermodynamics, emphasizing the three-dimensional case that will be of particular interest. Then, in sec.\ref{sec:CFT} I define and review the basic tools of Conformal Field Theory, after which I briefly describe in sec.\ref{sec:interfaces} the modifications that occur when one introduces an Interface. We focus the review on the universal properties of such models, which is what is needed to formulate the minimal models.

Having introduced both sides of the duality, in sec.\ref{sec:adscft} I formulate the AdS/CFT correspondence, only mentioning the most crucial results. The next section \ref{sec:bottomupapproach} presents the "bottom-up" approach to holography, where models "sur mesure" are considered as effective theories descending from a precise realization of the duality. I present the main ingredients of the "minimal" version of duality in the case of ICFT.

In sec.\ref{sec:entanglemententropy} we introduce entanglement entropy, and its role in QFT and CFT. I proceed with presenting the holographic way of computing it, by means of the Ryu-Takanayagi prescription and mention its quantum-corrected version. I finish in sec.\ref{sec:IslandsBHparadox} by outlining a direct application of these corrections culminating in the "Island formula", which allows the computation of the fine-grained entropy of Hawking's radiation. I sketch how this formula seems to resolve the black hole information paradox by recovering a unitary evaporation.

Having all the tools in hand, in chap.\ref{chap:phasesofinterfaces} I study the minimal ICFT model at finite temperature and at equilibrium through its holographic dual, which is composed of two AdS bulks (slices) connected on a membrane. I derive and solve analytically the Israel equations determining the shape of the gravitating membrane, which is dual to the interface. I classify the obtained solutions into 3 distinct thermodynamical phases, Hot, Warm and Cold, which are differentiated by the presence or absence of a black hole, and whether it intersects the membrane. A further phase structure is obtained by looking at the number of rest points for inertial observers (which acts as an order parameter), although I show they are not thermodynamic in nature by. We perform an analysis à la "Hawking-Page"\cite{Hawking:1982dh}, where the canonical parameters are the temperature as well as the relative size of the two CFTs on the boundary. The competing bulk solutions include gravitational avatars of the Faraday cage, black holes with negative specific heat, and an intriguing phenomenon of suspended vacuum bubbles corresponding to an exotic interface/anti-interface fusion. With the help of a numerical algorithm, I determine the dominant one at each point in phase space, displaying the phase diagram of the system for some chosen examples of the ICFT parameters.

In chap.\ref{chap:steadystatesofholo}, I consider the same ICFT model, but allow now for out-of-equilibrium solutions. I restrict to the tractable case of a non-equilibrium stationary state (NESS) which allows for an analytical resolution of the Israel equations, which we exhibit. Focusing first on the case of a single interface, the holographic dual contains a wall that necessarily falls into the flowing horizon. This restriction allows the recovery of the energy-transmission coefficients of the dual interface, which had already been obtained perturbatively \cite{Bachas:2020yxv}. By inspecting the dual horizon, we argue that by entangling outgoing excitations the interface produces entropy at a maximal rate, a surprising property that is usually exclusive to black hole horizons, but that could appear because of the strong coupling. Of great interest is also the far-from-equilibrium, non-killing event horizon in the bulk, which is highly deformed by the introduction of the wall, sitting behind the apparent horizon on the hotter side of the wall. We finish by looking at the thermal conductivity of a pair of interfaces, which jumps discontinuously when the wall exits the horizon, transitioning from a classical scattering behavior to a quantum regime in which heat flows unobstructed.

In the final chap.\ref{chap:entanglemententropyandholoint}, we present partial results on computations of (H)RT surfaces in the context of the geometries described in the previous chapters. For 3D bulks, (H)RT surfaces are simply spatial geodesics, thus we begin by showing how to compute them in any locally AdS geometry, by operating a coordinate change to the Poincaré patch. Following this train of thought, we review in detail the construction of RT surfaces in the vacuum ICFT state\cite{Anous:2022wqh}, and comment on their relation with the sweeping transition of chap.\ref{chap:phasesofinterfaces}, as well as the Island construction. We also show how to bootstrap it to compute the entanglement structure in more general states. We move to the application of the (H)RT prescription to the NESS state of chap.\ref{chap:steadystatesofholo}, concluding it requires the use of numerical methods and outlining some promising algorithms. We finish by introducing the QNEC and mentioning why it would be interesting to consider it in the ICFT geometries.

We end with a conclusion, reviewing the work from a broader perspective, and pointing out possible future research directions.

\chapter{Basics of Holography}
\label{chap:basicsofholo}
Holography, also known as "AdS/CFT correspondence" or "Gauge-gravity duality", was first discovered by Maldacena \cite{maldacena_large_1999} in 1997, and is one of the essential tools that has driven progress in Quantum Gravity in the 21st century. In a nutshell, holography describes an equivalence between a String theory in D-dimensions, and a quantum field theory in (D-1)-dimensions. We describe in this section the minimal ingredients needed to formulate this correspondence. 

We begin by describing Anti-de-sitter space. We then give a very brief  review of some concepts in Conformal Field Theory, and what happens we introduce an Interface. Equipped with the necessary concepts, we succinctly describe Maldacena's derivation and set the stage for the "minimal" version of the holographic correspondence that we will be using extensively. After that, we describe an application: how to compute entanglement entropies in CFT using the "Ryu-Takanayagi prescription", an holographic technique. Finally, we briefly present some recent progress toward the resolution of the black hole information paradox, which is based on the discovery of the "Island formula" for the entanglement entropy.

 \section{Anti-de-Sitter space, generalities}
 \label{sec:generalitiesAdS}
 Anti-de-Sitter (AdS) space can be efficiently described as the maximally symmetric Lorentzian manifold with (constant) negative curvature. Maximally symmetric Lorentzian manifolds of zero and positive curvatures are respectively Minkowski space and de-Sitter space. We will be mainly interested in the former, but efforts to develop some kind of holography in the other spacetimes are ongoing \cite{Strominger_2001,pasterski_lectures_2021}.

 Anti-de-Sitter space arises in gravity as a solution of the vacuum Einstein equations with a negative cosmological constant, usually denoted $\Lambda$. The equations can be derived from the Einstein-Hilbert action, in D-spacetime dimensions :
 \begin{eqgroup}
     S = \frac{1}{16\pi G_D}\int_{\cal{M}} d^Dx \sqrt{|g|}(R-2\Lambda)+\frac{1}{8\pi G_D}\int_{\pa\cal{M}}d^{D-1}y\sqrt{|h|}K\ .
     \label{EinsteinHilbertAction}
 \end{eqgroup}
 The integration is done on a Lorentzian manifold $\cal{M}$, with boundary $\pa\cal{M}$. We denote $g = {\rm det}g_{\mu\nu}$ the determinant of the metric, $R$ the Ricci-Tensor of $g_{\mu\nu}$ and $\Lambda$ the cosmological constant, which we will take to be negative. We include the counter-term that is necessary to make the variational problem well-defined, where $h_{\mu\nu}$ is the metric induced on $\pa{M}$, assumed to be timelike, and $K=K_{\mu\nu}h^{\mu\nu}$ is the trace of the extrinsic curvature\footnote{See Appendix \ref{sec:ExtrinsicAppendix} for more details on the definition of Extrinsic curvature}.
 
 Varying (\ref{EinsteinHilbertAction}), we recover Einstein's equations with a cosmological constant in the vacuum :
 \begin{eqgroup}
     G_{\mu\nu}= R_{\mu\nu}-\frac{1}{2}g_{\mu\nu}R+\Lambda g_{\mu\nu}=0\ .
     \label{EinsteinWithNegativeCosmo}
 \end{eqgroup}
 
 Contracting with $g^{\mu\nu}$ we can compute the value of the Ricci scalar. For the specific case of AdS (which is maximally symmetric), we can exploit this fact to recover the full Riemann tensor :
 \begin{eqgroup}
    R &= \frac{2\Lambda D}{D-2} \equiv -\frac{(D-1)D}{\ell^2}\ ,\\
    R_{\rho\mu\nu\si} &= -\frac{1}{\ell^2}(g_{\rho\nu}g_{\mu\si}-g_{\rho\si}g_{\mu\nu})\ ,
     \label{RiemannAdSValue}
 \end{eqgroup}
 where we introduced the "AdS radius" $\ell$, which is the characteristic length scale of the AdS spacetime.
 
 Another, more useful way of thinking about this spacetime is by embedding it in $D+1$ dimensions. Consider a flat spacetime of signature $(-,\underbrace{+,+,...}_{D-1},-)$. Denoting its coordinates by $X^M$, the metric reads :
 \begin{eqgroup}
     ds_{D+1}^2 = -(dX^0)^2-(dX^D)^2+\underbrace{(dX^1)^2+(dX^2)^2+...}_{D-1} \equiv dX^M dX^N g_{MN}\ .
     \label{embedmetric}
 \end{eqgroup}
 The isometry group of this spacetime is the Poincaré group in $D+1$ dimensions (with the appropriate signature). We now would like to find a spacelike hypersurface that preserves the $O(2,D-1)$ symmetry while breaking the translation. In that way, the induced metric will automatically have a $\frac{D(D+1)}{2}$ dimensional isometry group (the dimensionality of $O(2,D-1)$) and thus it will be maximally symmetric. 
 
 Taking this into account, it is easy to see that the sought-out surface is of the form :
 \begin{eqgroup}
     X^M X_M=-(X^0)^2 - (X^D)^2 + \underbrace{(X^1)^2+(X^2)^2+...}_{D-1} = -\ell^2\ .
     \label{AdSembedequation}
 \end{eqgroup}
 
 The advantage of working with this embedding is that it makes explicit many important coordinate systems, according to how we decide to parametrize the embedded surface. Note that the $\frac{D(D+1)}{2}$ Killing vectors of the Poincaré symmetry are simply :
 \begin{eqgroup}
     J_{MN} = X_M\frac{\pa}{\pa X^N}-X_N\frac{\pa}{\pa X^M}\mbox{  , } 0\leq M < N \leq D+1 \label{poincKillingVec}\ .
 \end{eqgroup}
Here we defined $X_M=X^K \eta_{KM}$, thus depending on the nature of $X^M$ and $X^N$ (spacelike or timelike), $J_{MN}$ will either induce boosts or rotations.

\subsection{Global coordinates}
 The first set of coordinates that will be of interest is the so-called "global" coordinate system. It parametrizes the hypersurface as :
 \begin{eqgroup}
     X^0 &= \ell \cosh{\rho}\cos{\tau},\mbox{    }X^D = \ell \cosh{\rho}\sin{\tau}\ ,\\
     X^i &= R\sinh(\rho) \Omega^i\;\;\mbox{ }(i\in \llbracket 1,D-1\rrbracket)\ ,
     \label{globalcoordinateembedding}
 \end{eqgroup}
 where $\Omega_i$ are coordinates describing the $(D-2)$-dimensional unit sphere, $\sum_i \Omega_i^2 = 1$. They can of course be explicitly parametrized by $D-2$ angles. One can easily verify that (\ref{globalcoordinateembedding}) is a solution to (\ref{AdSembedequation}), and that it covers the full hypersurface, explaining the name "global" for this coordinate system. The induced metric then takes the form : 
 \begin{eqgroup}
     ds^2 = \ell^2\left(-\cosh^2{\rho} d\tau^2 +d\rho^2+\sinh^2{\rho} d\Omega_{D-2}^2\right)\ ,
     \label{globalmetric}
 \end{eqgroup}
 where $d\Omega^2_{D-2}$ is the standard metric for the unit $(D-2)$-sphere.
 
We can immediately identify $\tau$ as the timelike coordinate. The natural range inherited from (\ref{globalcoordinateembedding}) is $\{\rho>0$, $0\leq\tau<2\pi\}$. This topology is problematic for a physical spacetime, as we can have closed timelike curves along the time-direction $\tau$. Thus, we will consider the same metric, with the range of $\tau$ uncompactified to take values $-\infty<\tau<\infty$. This is simply a topological change, that does not affect the local geometry, so this spacetime is still a solution to the Einstein equations.

In this coordinate system, not all of the isometries are manifest. We can identify the $SO(D-1)$ group of rotations of the $(D-2)$ sphere, along with the translation of $\tau$ which make up an $SO(2)$ before decompactification, and $\mathbb{R}$ after. We have of course $SO(2)\times SO(D-1) \subset SO(2,D-2)$. If one wants to recover the full isometry group, one simply has to project the Killing vectors (\ref{poincKillingVec}) on the hyperboloid, and re-express them in the new coordinate system. However, the isometries beyond the evident ones take very complicated form, and the Killing vectors often cannot be integrated to recover the associated finite symmetry.

One last important remark, in the context of AdS/CFT, is that the geometry of the boundary $\rho\rightarrow\infty$ is conformally equivalent to a cylinder $\mathbb{R}\times SO(D-1)$. This is simply seen by Weyl rescaling (\ref{globalmetric}) and taking $\rho=\rho_0\rightarrow \infty$. 

Another coordinate system that we will use extensively can be obtained by the change of coordinates $\ell\sinh{\rho}=r$, $\ell\tau = t$  and yields the following metric :
\begin{eqgroup}
    ds^2 = \left( -(1+\frac{r^2}{\ell^2})d\tau^2+\frac{dr^2}{1+\frac{r^2}{\ell^2}}+r^2d\Omega_{D-2}^2\right)\ .
    \label{radialglobalcoord}
\end{eqgroup}
While these coordinates are interchangeable with (\ref{globalmetric}), they are a little bit more intuitive since $r$ is essentially a radial coordinate. In addition to that, we will see that the coordinates describing Black Hole solutions will have a metric very similar to this one.

\subsection{Poincaré patch}
\begin{figure}[!h]
    \centering
    \includegraphics[width=0.3\linewidth]{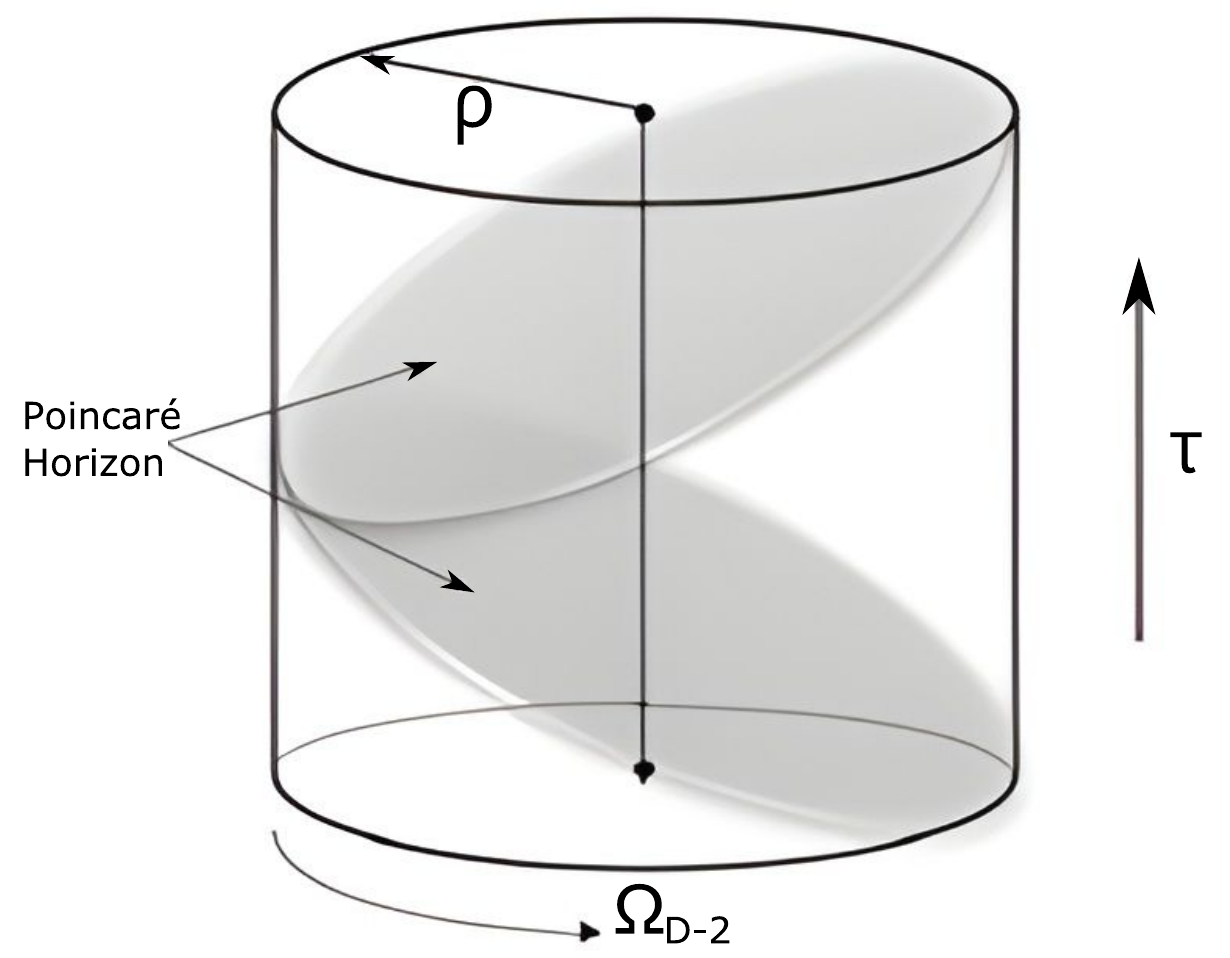}
    \caption{{\small Depiction of the AdS space in global coordinate as a cylinder (compactified $\rho$), for $0\leq\tau<2\pi$. Before unwrapping, the two ends of the cylinder are identified. The poincaré coordinates cover the volume in between the two grey wedges; the "poincaré horizons" correspond to the locations $z=\infty$. We can see we need two poincaré patches to cover the full AdS space, when $\tau$ is compact. }}
    \label{fig:poincarepatch}
\end{figure}

Another parametrization, which does not cover the entirety of the hyperboloid (\ref{AdSembedequation}), are the so-called "Poincaré coordinates" :
\begin{eqgroup}
X^0 &= \frac{1}{2z}\left(z^2+\ell^2+x^ix_i-t^2\right)\ ,\\
X^i &= \frac{\ell x^i}{z},\mbox{     }i\in\llbracket 1,D-2\rrbracket\ ,\\
X^{D-1} &=\frac{1}{2z}\left(z^2-\ell^2 +x^ix_i-t^2\right)\ ,\\
X^{D}&=\frac{\ell t}{z}\ ,
\label{poincarecoordinates}
\end{eqgroup}
where we denote $x^ix_i = \sum_{i=1}^{D-2}(x^i)^2$. This parametrization cannot cover the full hyperboloid, because we must either choose $z>0$ or $z<0$ as the point corresponding to $z=0$ is undefined. Thus, we have two charts for $z>0$ and $z<0$ which together cover the full hyperboloid. See fig.\ref{fig:poincarepatch} for a depiction of the region covered by one of the patches.

Note that after we unwrap the time-direction, we need an infinite number of Poincaré patches to cover the same manifold as the global coordinates (\ref{globalcoordinateembedding}). The convention is to pick the Poincaré patch with $z>0$.

In this coordinate system, the metric is conformally flat :
\begin{eqgroup}
    ds^2_{poinc}=\frac{R^2}{z^2}(-dt^2+dz^2+dx^i dx_i)\ ,
    \label{poincaremetric}
\end{eqgroup}
which is conformally flat. The boundary is located at $z=0$, and its topology is now $\mathbb{R}^{\{1,D-1\}}$, endowed with the flat metric.

The topology change of the boundary should not surprise us since this coordinate system covers only a
part of the full manifold and of its boundary. Of course, there exists a diffeomorphism that maps (\ref{poincaremetric}) to part of (\ref{globalmetric}). It can be worked out by comparing (\ref{globalcoordinateembedding}) and (\ref{poincarecoordinates}), but it will not be useful for us in what follows. Later we will see that the Holographic correspondence maps diffeomorphisms of AdS to conformal transformations of the boundary metric. As a teaser, we can already notice that the two boundary metrics can be related by a conformal transformation.

\section{Asymptotically Anti-de-Sitter space, 3D case}
The special case of three-dimensional Anti-de-Sitter space will be of particular interest for the purposes for this thesis. One might have thought that there is little qualitative difference between AdS in different dimensions. This is true for D $\geq$ 4, but below that threshold things change because there is no dynamical bulk gravity.

In D dimensions the Riemann tensor has $\frac{D^2(D^2-1)}{12}$ degrees of freedom. When $D=3$, that number is 6, the same as the Ricci tensor. Because of that it is possible generically to write the Riemann tensor as  :
\begin{eqgroup}
    R_{\mu\nu\rho\si} &= S_{\mu\rho}g_{\nu\si} - S_{\nu\rho}g_{\mu\si} + S_{\nu\si}g_{\mu\rho} - S_{\mu\si}g_{\nu\rho}\ ,\\
    S_{\mu\nu} &= R_{\mu\nu}-\frac{R}{4}g_{\mu\nu}\ ,
\end{eqgroup}
As a result, if $g_{\mu\nu}$ satisfies (\ref{EinsteinWithNegativeCosmo}), the full Riemann tensor is specified. Every solution of (\ref{EinsteinWithNegativeCosmo}) in 3D looks (locally) like AdS$_3$. Since the Riemann is completely fixed, there is no room for local degrees of freedom, like gravitational waves in higher dimensions. In fact, in 3-dimensions gravity can be re-formulated as a Chern-Simons theory\cite{Witten:1988hc3dgravityasanexactly,Witten:2007kt3drevisited,Carlip:1995zj}, which is a 3-dimensional Topological Quantum Field Theory. We will not delve in the details of this re-formulation. 

What is important for our purposes is that although the 3d theory does not have local degrees of freedom in the bulk, it does have degrees of freedom localized on the boundary of the spacetime. Although any two different solutions of (\ref{EinsteinWithNegativeCosmo}) can be connected by a gauge-transformation (i.e. a diffeomorphism), if the gauge transformation is non-vanishing on the boundary, then it will change the physical state on the boundary. Furthermore, two solutions will always look similar locally, but they may differ globally, the typical example being the "BTZ" black hole \cite{banados_geometry_1993} solutions we will discuss shortly.
  
\subsection{Symmetries of AdS$_3$}
\label{sec:symmetriesofAdS3}
Much of the discussion of Sec.\ref{sec:generalitiesAdS} still applies also in $D=3$ dimensions. However, there is an enhancement of the (asymptotic) symmetries of the spacetime. The isometries are the same as in the higher dimensional case, $SO(2,2)$ for $D=3$. In this low-dimensional case, it is sometimes convenient to rewrite the isometry group as $SL(2,\mathbb{R})\times SL(2,\mathbb{R})$. This is done by looking at the embedding space (\ref{embedmetric}) as $2\times2$ matrices as follows : 
\begin{eqgroup}
      (X_0,X_1,X_2,X_3)\mapsto g=\begin{pmatrix}
X_{1}-X_{0} & X_2-X_3 \\
X_2+X_3 & X_{1}+X_0 
\end{pmatrix}  \in {\rm Mat}_{2\times 2}\ .
\end{eqgroup}
In this representation, the hyperboloid describing the $AdS_3$ embedding is simply :
\begin{eqgroup}
    {\rm Det}(g)  = -\ell^2\ .
\label{ads3embedding}
\end{eqgroup}
Then, we can verify that there is a group action of $SL(2,\mathbb{R})\times SL(2,\mathbb{R})$ acting on  ${\rm Mat}_{2\times2}$, that preserves the metric (\ref{embedmetric}) :
\begin{eqgroup}
    (g_L,g_R)\in &SL(2,\mathbb{R})\times SL(2,\mathbb{R})\mapsto \rho(g_L,g_R)\\
    &\rho(g_L,g_R)g = g_L g g_R\ .
    \label{sl2rsymmetryads3}
\end{eqgroup}
The map $\rho$ is a homomorphism (double cover) of $SL(2,\mathbb{R})\times SL(2,\mathbb{R})\rightarrow O(2,2)$, which induces an isomorphism $SL(2,\mathbb{R})\times SL(2,\mathbb{R})/\mathbb{Z}_2\rightarrow O(2,2)$. As in this thesis we won't worry about discrete isometries of spacetime, we won't be very careful about which connected component we are considering. We will consider $SO^+(2,2)$ as the isometry group.

Finally, is easy to see that the map $\rho$ leaves invariant the equation (\ref{ads3embedding}). This more "group-theoretic" way of dealing with $AdS_3$ isometries is useful as it simplifies and clarifies some computations.

There is, as we hinted previously, an extended symmetry group of AdS$_3$ (and in fact also of asymptotically AdS$_3$ spacetimes ), which is harder to see geometrically. These extra "asymptotic symmetries"\cite{brown_central_1986} correspond to the infinite extension of the 2d conformal algebra. Going into depth into this subject would be outside the scope of this thesis, but we give a brief summary of the main ideas, as their discovery was a prelude to the holographic correspondence.

Naively, in a gauge theory one usually thinks of any two states that are related by a gauge transformation as the same state, described differently. In truth, one can show that this statement only holds for "small" gauge transformations, which vanish at the boundary of the manifold we consider\cite{Avery:2015rga}. This can also be seen by looking at the conserved currents, and the associated charges. For gauge transformations, the conserved current can be written as $j^\mu = S^\mu + \pa_\nu k^{\mu\nu}$, where $S^\mu$ vanishes on shell, and $k^{\mu\nu}=-k^{\mu\nu}$ is a two-form. Then the associated charge will be expressible as an integral on the boundary of the spacetime. Therefore, it will be vanishing for the "small" gauge transformations, thus showing they have no associated conserved charge. For "big" gauge transformations, the charge may be non-vanishing, which shows that they can act as bona-fide symmetry, rather than being simply a redundancy in parametrization.

Let us concentrate on General Relativity, where the gauge group are diffeomorphism. We begin by getting rid of the redundancy in the description by choosing some gauge-fixing conditions that picks out a single representative of the "gauge-orbit" of any given state. In this way, we get rid of the unphysical "small" gauge transformations. We denote as "residual gauge group" the gauge symmetries which remain un-fixed by this procedure, which will therefore be non-vanishing at the boundary.

Among the residual gauge symmetries, the transformations that alter the boundary conditions of the problem are discarded. The surviving diffeomorphisms then belong to the "asymptotic symmetry group" of spacetime, aptly named as it concerns gauge transformations acting on the (asymptotic) boundary. Of course, the asymptotic symmetry group will then depend on the boundary conditions we have chosen for the metrics. Picking a set of boundary conditions that allows for interesting solutions, while removing unphysical ones is still an ongoing problem \cite{Compere:2013bya}, and is mostly done by trial and error.

In this thesis, we will stick with the "Fefferman-Graham" \cite{fefferman_conformal_nodate} prescription for gauge-fixing and boundary conditions in AdS, which is the relevant prescription in the context of AdS/CFT. We denote in this prescription the coordinates as $x^\mu = (z,x^i)$. We set the range $z>0$, the boundary of AdS being located at $z=0$. Then the gauge-fixing reads \cite{reviewRuzzi} :

\begin{eqgroup}
    g_{zz} = \frac{\ell^2}{z^2}\mbox{ ,  }g_{z i}=0\ .
    \label{fgboundarycond}
\end{eqgroup}

As expected, we have three independent gauge fixing conditions, for the three independent parameters of the diffeomorphism gauge group. Thus, a gauge-fixed metric will take the form :

\begin{eqgroup}
    ds^2 = \frac{\ell^2}{z^2}dz^2+g_{ij}(z,x^i)dx^i dx^j\ .
    \label{FGmetric}
\end{eqgroup}

We can now proceed to compute the residual gauge symmetry. It is immediately clear that included in this residual symmetry group there will be general change of coordinates in $x^i$. The full equation that an infinitesimal diffeomorphism $x^\mu\rightarrow x^\mu + \xi^\mu$ must satisfy in order to preserve this gauge structure are simply :
\begin{eqgroup}
    \cal{L}_\xi g_{zz}=0\mbox{ ,  }\cal{L}_\xi g_{z i} = 0\ .
    \label{residualconditions}
\end{eqgroup}
This can be solved generally, yielding an infinitesimal description of the residual gauge group :
\begin{eqgroup}
    \xi^z = \si(x^i)z\mbox{ ,  }\xi^i = \xi^i_0(x^i)-\ell^2 \pa_k\si\int_0^z \frac{dz'}{z'}g^{ik}(z',x^l)\ ,
    \label{residualFGgroup}
\end{eqgroup}
where $\si(x^i)$ is an arbitrary function, as are the $\xi^i_0$.

To determine the asymptotic symmetry group, we must first describe the Fefferman-Graham boundary conditions :
\begin{eqgroup}
    g_{ij} &= \frac{\ell^2}{z^2}(g_{ij}^{(0)}+z^2 g_{ij}^{(2)}+O(z^4))\ ,\\
    g^{(0)}_{ij}dx^i dx^j &=  e^{2\phi}\eta_{ij}dx^i dx^j= e^{2\phi}(-dt^2+d\ffi^2)\ ,
    \label{boundarycondFG}
\end{eqgroup}
where we write $\{x^i\} = (t,\varphi)$. The coordinate $\varphi$ will be periodic of period $2\pi$. This choice is simply to conform to the global AdS coordinates (\ref{globalmetric}) asymptotic geometry. Since we are considering the conformal family of metrics, we can recover the "planar" case by the correct choice of $e^{2\phi}$ and a conformal transformation.

As we have already stated, it is a subtle matter to choose appropriate boundary conditions. To distill an interesting set of constraints, one usually looks at several solutions of the Einstein equations, and try to choose conditions that remove unwanted solutions without being too restrictive. In this case, the F-G boundary conditions are a good way to describe a spacetime that looks asymptotically like AdS$_3$\cite{fefferman_conformal_nodate}. Indeed, for the example $\phi=0$, the leading order metric in (\ref{boundarycondFG}) looks like AdS$_3$ in Poincaré coordinates (\ref{poincaremetric}).

We are now set to compute the asymptotic symmetry group. A generic residual gauge symmetry (\ref{residualFGgroup}), will preserve the F-G boundary conditions iff :
\begin{eqgroup}
    \cal{L}_\xi g_{ij} = \cal{O}(z^{-1})\ ,
    \Leftrightarrow &\cal{L}_\xi g_{ij}^{(0)}=0\ .
    \label{residualboundaryconditions}
\end{eqgroup}
Expanding the conditions we get the following set of equations for the killing vectors:
\begin{eqgroup}
    \si-\pa_i\phi\xi^i_0 &= \pa_t \xi^t_0=\pa_\ffi \xi^\ffi_0\ ,\\
    \pa_t \xi_0^\ffi &= \pa_\ffi\xi^t_0\ .
    \label{asymptoticKillingCondition}
\end{eqgroup}
By taking partial derivatives and combining the equations we derive the necessary condition :
\begin{eqgroup}
    &-\pa_t^2\xi^i_0+\pa_\ffi^2\xi^i_0=0\ ,\\
    &\Rightarrow \xi_0^i = f_-^i(\ffi-t)+f_+^i(\ffi+t)\ .
    \label{steponeconformalkilling}
\end{eqgroup}
Already, we recognize the equations of the conformal Killing vectors of $\eta_{ij}$ in 2D. Plugging back into (\ref{asymptoticKillingCondition}) forces the free functions to be the same up to a constant, and we find for the general solution :
\begin{eqgroup}
    \xi^t_0&=f_+(x+t)-f_-(x-t)\ ,\\
    \xi^x_0&=f_+(x+t)+f_-(x-t)\ ,\\
    \si&= f_+'(x+t)+f_-'(x-t)+\xi^t_0\pa_t\phi +\xi^\ffi_0\pa_\ffi\phi\ .
        \label{solutionAsymptoticKilling}
\end{eqgroup}
To study the algebra of the Killing vectors, it is convenient to consider instead the lightcone basis, $w^+=\ffi+t$, $w^-=\ffi-t$.
\begin{eqgroup}
    \xi^+_0&=f_+(w^+)\ ,\\
    \xi^-_0&=f_-(w^-)\ ,\\
    \si&= \frac{1}{2}\left(f_+'(w^+)+f_-'(w^-)\right)+f_+(w^+)\pa_+\phi + f_-(w^-)\pa_-\phi\ .
    \label{solutionAsymptoticKillinglightcone}
\end{eqgroup}
To identify a basis of Killing vectors, we expand in Fourier series the functions $f_-$ and $f_+$, by exploiting the periodicity of $\varphi$ :
\begin{eqgroup}
    f_{\pm}(w^\pm) = \sum_{n\in\mathbb{Z}}\al_n e^{in w^{\pm}}\ .
\end{eqgroup}
This allows us finally to define a basis of Killing vectors $\xi_n$ and $\ti{\xi}_n$ :
\begin{eqgroup}
    \xi_n &= i e^{inw^+}(\pa_{w^+}+z(in+\pa_+\phi) \pa_z)\ ,\\
    \ti{\xi}_n &=i e^{inw^-}(\pa_{w^-}+z(in+\pa_-\phi) \pa_z)\ .
    \label{basisAsymptoticKillingVector}
\end{eqgroup}

We can now go on to compute the algebra of this family of Killing vectors. It is important to note that since we are computing the asymptotic algebra, all computations should be done at the level of the first leading order in $z$. Indeed, the algebra will not be closed at higher orders.

We find the following Lie Brackets :
\begin{eqgroup}
    &[\xi_n,\ti{\xi}_m]=0\ ,\\
    &[\xi_n,\xi_m]=(n-m)\xi_{n+m}\ ,\\
    &[\ti{\xi}_n,\ti{\xi}_m]=(n-m)\ti{\xi}_{n+m}\ ,
    \label{liebrackasympt}
\end{eqgroup}
which are readily identified as two copies of the Witt algebra. In the special case of two dimensions, the asymptotic symmetry group is thus much bigger than the isometry group of the vacuum, AdS$_3$. Doing the same derivation in higher dimensions will show that the asymptotic symmetries of asymptotically AdS$_D$ spaces correspond to the isometries of AdS$_D$, that is $SO(D-1,2)$. As a foreshadowing of the Holographic correspondence, we can see that this matches the conformal group in $D-1$-dimensions, the symmetry group of a CFT$_{D-1}$.

The only thing missing from our derivation is the recovery of the central charge $c$, which appears in the quantization of CFT's, where the Witt Algebra is centrally extended to Virasoro. In the gravity perspective we consider here this can, remarkably, be obtained at the classical level by looking at the algebra of the conserved charges\cite{brown_central_1986} associated to the Killing vectors (\ref{basisAsymptoticKillingVector}). Carrying this computation yields the famous Brown-Henneaux formula, which expresses the central charge of the Virasoro algebra as a function of the Anti-de-Sitter radius :

\begin{eqgroup}
    c = \frac{3\ell}{2 G}\ .
    \label{brownhenneauxcentralcharge}
\end{eqgroup}

There is also a way to obtain this formula through our covariant formalism, by looking at the transformation of $g_{ij}^{(2)}$ (see (\ref{boundarycondFG})) under the asymptotic symmetries. It can be computed easily in when the metric is vacuum AdS, namely $g_{ij}=\frac{\ell^2}{z^2}\eta_{ij}$ in (\ref{FGmetric}). Under the asymptotic Killing vector (\ref{solutionAsymptoticKillinglightcone}) :
\begin{eqgroup}
    g_{++}^{(2)}{}'&=-\frac{1}{2}\left(\xi_0^+\right)'''\ ,\\
    g_{--}^{(2)}{}'&=-\frac{1}{2}\left(\xi_0^-\right)'''\ ,\\
    g_{+-}^{(2)}{}'&= 0\ ,
    \label{g2transformations}
\end{eqgroup}
and $g^{(0)}$ is left invariant (as it should). 

Now, through the arguments of sec.\ref{sec:minimalholoandrenorm}(or by computing the boundary Noether current associated to the asymptotic symmetry), one can identify $g^{(2)}$ with a "boundary" stress-energy tensor as $g^{(2)}=\frac{T_{ij}}{8\pi G \ell}$. Using (\ref{g2transformations}), we deduce that this stress-energy tensor does not transform covariantly, but has an additional contribution to its transformation law. Assuming it is the stress-tensor of a (dual) CFT consistency with (\ref{TTransform}) forces us to the identification (\ref{brownhenneauxcentralcharge}), recovering the Brown-Henneaux formula.

\subsection{Other vacuum solutions}
Despite the lack of local degrees of freedom of Gravity in 3D, we have seen that this does not mean that the solutions are completely frozen. Solutions may differ in their global structure, as well as in their behavior at the conformal boundary of spacetime. The best known non-trivial example is the celebrated BTZ black hole\cite{Ba_ados_1992}.

\subsubsection{The BTZ black hole}
For vanishing cosmological constant, the vacuum solutions of 3D Gravity are trivial and admit only the Minkowski vacuum. For a negative cosmological constant, the space of vacuum solution is reacher, as was discovered first by the authors of \cite{Ba_ados_1992}. The vacuum BTZ solution can be described by the following metric :

\begin{eqgroup}
     ds^2 &= -h(r)dt^2+\frac{\ell^2dr^2}{h(r)}+r^2(d\varphi-\frac{J\ell}{2r^2}dt)^2\ ,\\
     h(r) &= (r^2-M\ell^2+\frac{J^2\ell^2}{4r^2}) = \frac{(r^2-r_+^2)(r^2-r_-^2)}{r^2}\ ,\\
     r_\pm^2 & =\frac{1}{2}\left(M\ell^2\pm\sqrt{M^2\ell^4-J^2\ell^2}\right)\ .
     \label{banadosBTZmetric}
\end{eqgroup}
In this notation $r_+$ and $r_-$ are respectively the outer and inner horizon radii, and $M$ and $J$ can be identified (through the conserved charges of asymptotic symmetries) with the mass and spin of the black hole solution. Note that to avoid naked singularities, we need to satisfy $M\ell\geq |J|$, with the equality corresponding to an "extremal" black hole.

The crucial ingredient that distinguishes (\ref{banadosBTZmetric}) from a mere though non-trivial reparametrization of AdS$_3$ is that the angle coordinate $\varphi$ is periodic, $2\pi$. Without this identification, the solution becomes a "black string", and the event horizon disappears, as maximally extending the spacetime would reveal that the region $r<r_+$ is not causally disconnected. In other words, the apparent horizon of the black string solution is simply a coordinate artifact, and the geometry of the solution simply describes a portion of the regular AdS$_3$ spacetime. 

While this is true, note that the coordinate change linking these two geometries acts non-trivially on the boundary; therefore the boundary degrees of freedom of the two geometries are not equivalent.

Let us illustrate this claim by exhibiting a coordinate parametrization of the embedding of AdS$_3$, as in (\ref{ads3embedding}) :
\begin{eqgroup}
    X_0 &= \ell \sqrt{A(r)}\cosh\frac{\ti{\varphi}}{\ell}\ ,\\
    X_1 &= \ell \sqrt{B(r)}\sinh\frac{\ti{t}}{\ell}\ ,\\
    X_2 &= \ell \sqrt{A(r)}\sinh\frac{\ti{\varphi}}{\ell}\ ,\\
    X_3 &= \ell \sqrt{B(r)}\cosh\frac{\ti{t}}{\ell}\ ,\\
    \mbox { with }A(r)=\frac{r^2-r_-^2}{r^2_+-r^2_-}\mbox{ ,  }&B(r) = \frac{r^2-r_+^2}{r^2_+-r^2_-}\mbox{ ,  }\ti{t} = r_+ t-r_-\varphi\mbox{ ,  }\ti{\varphi} = -r_-t+r_+\varphi\ .
    \label{blackstringembedding}
\end{eqgroup}

The parametrization (\ref{blackstringembedding}) covers the exterior ($r>r_+$) region of the black string metric. To cover the interior as well, one needs alternative parametrization that however connects smoothly to (\ref{blackstringembedding}). Crucially, we see that in this parametrization $\varphi$ cannot be considered periodic, as it appears in hyperbolic functions. This shows that the black string solution is really vacuum AdS$_3$ in disguise. Despite this fact, it is still interesting from the point of view of the boundary degrees of freedom, that will differ between the two solutions as they are related by diffeomorphisms that do not vanish on the boundary!

The parametrization (\ref{blackstringembedding}) provides also additional insight on the geometrical construction of novel vacuum solutions, including the BTZ black hole. Following \cite{banados_geometry_1993}, consider a Killing vector $\xi$ of the hyperboloid (\ref{ads3embedding}). "Integrating" the infinitesimal coordinate change $x^\mu \rightarrow x^\mu+\xi^\mu$ yields a one-parameter subgroup of the isometries of AdS$_3$, whose elements we denote by $e^{t\xi}$. Following the discussion Sec.\ref{sec:symmetriesofAdS3}, we can see $e^{t\xi}$ as an element of $SL(2,\mathbb{R})\times SL(2,\mathbb{R})$. 

Let us now define the "identification subgroup", whose elements are :
\begin{eqgroup}
    \{e^{t\xi}\mbox{ s.t.  }t=2k\pi\mbox{ ,  }k\in \mathbb{Z}\}\ .
    \label{identificationsubgroup}
\end{eqgroup}

As the name implies, the new solution is then constructed by quotienting AdS$_3$ along the identification subgroup, meaning that points separated by the action of elements of (\ref{identificationsubgroup}) are identified. As this procedure does not modify the geometry locally, the quotiented spacetimes are automatically solutions of the Einstein vacuum equations.

The only caveat to this procedure is that the identification may generate causality paradoxes, for instance when causal curves become closed under the identification. A necessary condition to avoid this problem is to require that $\xi$ be spacelike, $\xi^\mu \xi^\nu \eta_{\mu\nu}>0$. Indeed, if this condition fails to be satisfied, then we would perform identifications of points lying on killing vector orbits which are causal, producing closed timelike curves.

The BTZ solution can be obtained through this process using :
\begin{eqgroup}
    \xi_{BTZ} = \frac{1}{\ell}\left(r_+J_{10}-r_-J_{23}\right)\ ,
    \label{BTZkillinggod}
\end{eqgroup}

where $J_{MN}$ is defined in (\ref{poincKillingVec}). Note that this is a tangent vector of the AdS$_3$ hyperboloid, and in BTZ coordinates it simply corresponds to $\pa_\varphi$. We see then that the identification along the orbits of this Killing vector is indeed realized by setting $\varphi$ to be periodic.

Computing $\xi_{BTZ}\cdot\xi_{BTZ}$ one realizes that it is not everywhere positive.
\begin{eqgroup}
    \xi_{BTZ}\cdot\xi_{BTZ} = \frac{r_+^2}{\ell^2}((X^0)^2-(X^2)^2)+\frac{r_-^2}{\ell^2}((X^3)^2-(X^2)^2)=  \frac{r_+^2-r_-^2}{\ell^2}((X^0)^2-(X^2)^2)+r_-^2\ .
\end{eqgroup}
We can check that plugging in the parametrization (\ref{blackstringembedding}) gives $\xi_{BTZ}\cdot \xi_{BTZ}=r^2$, which is indeed strictly positive.
  
We can then excise the regions $\xi_{BTZ}\cdot\xi_{BTZ}<0$ from AdS$_3$. This procedure seems unphysical, since it generates a geodesically incomplete spacetime. Indeed, geodesics crossing the region $\xi_{BTZ}\cdot\xi_{BTZ}=0$ are abruptly stopped. This problem is resolved because in the resulting geometry the region $\xi_{BTZ}\cdot\xi_{BTZ}=0$ becomes a singularity, whose nature is quite different from the higher dimensional black holes. It is a singularity in the causal structure, since beyond that point one encounters closed timelike curves. Contrary to the higher-dimensional counterparts, the curvature remains finite at the singularity since the solution is locally AdS$_3$.

\subsubsection{Spectrum of solutions and conical singularities}
Let us now restrict to the simpler case $J=0$. This yields a 1-parameter group of solutions of metric (\ref{nonspinningBTZmetric}):
\begin{eqgroup}
    ds^2 = -(r^2-M\ell^2)dt^2+\frac{\ell^2 dr^2}{r^2-M\ell^2}+r^2d\varphi^2\ .
    \label{nonspinningBTZmetric}
\end{eqgroup}
with $\varphi$ a $2\pi$ periodic coordinate as explained earlier. For $M>0$, the singularity is behind a horizon and except at $r=0$ it is a regular solution. For $M=0$, the solution is still regular except at $r=0$, but the horizon disappears. This solution is sometimes referred as the "zero-mass" black hole. What is interesting is that we do not recover Anti-de-sitter space when we send the mass $M$ to zero, something very different from what happens in higher dimensions. 

In 3D, the black hole spectrum is separated by a "gap" from the Anti-de-Sitter vacuum. Indeed, setting $M=-1$ in (\ref{nonspinningBTZmetric}) we recover AdS$_3$ in global coordinates, as in (\ref{globalmetric}). What about masses in the interval $-1<M<0$ ? Expanding them near the origin $r=0$ yields the metric :
\begin{eqgroup}
    ds^2  = M dt^2 -\frac{1}{M}dr^2+r^2d\varphi^2\ .
    \label{conical singu}
\end{eqgroup}
Redefining $\ti{r} = \frac{1}{\sqrt{-M}}r$ and $\ti{\varphi}=\sqrt{-M}\varphi$, we exhibit a conical singularity of deficit angle  $(1-\sqrt{-M})2\pi$. These solutions thus exhibit naked conical singularities, and are not considered in general as valid classical solutions. However they will be important in the quantization, as they still are valid saddle points of the Einstein-Hilbert action. This metric can be generated by placing an excitation on the AdS$_3$ vacuum, of energy $-1<M<0$. The conical singularity will presumably only appear away from the source, and be resolved as we get close to it.

Solutions with $M<-1$ are completely unphysical, one way to see it is by noting that a conical singularity is created by a string. If the string has positive tension, the deficit angle is positive, while if it has negative tension there is an angle excess. The latter is thus unphysical. From the boundary perspective we will present, $M =-1$ is just the Casimir energy of the vacuum CFT on the circle. Adding excitations can only increase the energy in a unitary theory.

\section{Hawking-Page phase transition and black hole thermodynamics}
\label{sec:hawkingphasetranso}
Having presented the black hole solutions in AdS$_3$, in this section we focus on the thermodynamical properties of the solutions, following the paper of Hawking and Page\cite{Hawking:1982dh}. We will work in the canonical ensemble, thus we consider equilibrium solutions at fixed temperature $T$. We consider for this analysis only non-spinning solutions,

\subsection{Adding temperature}
\label{sec:AddingTemperature}
Black hole solutions, like BTZ (\ref{nonspinningBTZmetric}) have a naturally associated temperature, which is the temperature of their Hawking radiation \cite{Hawking:1975vcx}. While this is a semi-classical effect, it is possible to recover this temperature in a much simpler manner using the following trick. Consider the Wick-rotated BTZ metric (\ref{BTZeuclideanmetric}), which essentially amounts to making the change of variables $\tau=it$, where we call $\tau$ the "Euclidean time". 
\begin{eqgroup}
    ds^2 = (r^2-M\ell^2)d\tau^2+\frac{\ell^2 dr^2}{r^2-M\ell^2}+r^2d\varphi^2\ .
    \label{BTZeuclideanmetric}
\end{eqgroup}
Notice that after Wick rotation the Signature becomes $(+++)$, hence the name "Euclidean".

Expanding close to the horizon at $r=\sqrt{M\ell^2}+\frac{\sqrt{M}}{2\ell}\rho^2$, the metric reads, at first order in $\rho$ :
\begin{eqgroup}
    ds^2 = d\rho^2+M\rho^2d\tau^2+M\ell^2 d\varphi^2\ .
    \label{euclideanBTZnearhorizon}
\end{eqgroup}

In the $(\rho,\tau)$ plane, the geometry looks locally like flat space, with $\tau$ playing the role of the angular coordinate. If we want to avoid conical singularities, then we must have the identification $\tau\rightarrow \tau+\frac{2\pi}{\sqrt{M}}$. As we know, considering a QFT on an Euclidean background with periodic time $\tau\rightarrow \tau+\beta$ is one way to compute the partition function at finite temperature $T=1/\beta$ (more details in sec.\ref{sec:entanglement}). For this reason, we interpret the period of the euclidean time $\tau$, enforced by (\ref{euclideanBTZnearhorizon}), as the temperature of the BTZ horizon. Although this derivation is heuristic at best, it can be checked that it corresponds to the semi-classical derivation. We find that the temperature of the BTZ black hole of mass $M$ is simply $T=\frac{\sqrt{M}}{2\pi}$.

Another saddle point of the Einstein-Hilbert action that contributes to the canonical ensemble is derived from the vacuum solution, pure AdS$_3$ (\ref{radialglobalcoord}). Giving a temperature to the vacuum solutions is again done formally by wick-rotation. Unlike the BTZ case, the wick-rotated AdS$_3$ metric (\ref{euclideanAdsMetric}) does not impose any constraint on the periodicity of $\tau$, as it is everywhere well behaved.
\begin{eqgroup}
    (r^2+\ell^2)d\tau^2+ \frac{dr^2}{1+r^2/\ell^2}+r^2 d\varphi^2\ .
    \label{euclideanAdsMetric}
\end{eqgroup}

We can thus arbitrarily choose the "temperature" of this space-time by setting the periodicity of $\tau$. The resulting spacetime is nicknamed "thermal AdS". In the Lorentzian version, it is not possible to "see" the temperature without adding some fields on the background. Indeed, since pure gravity does not have gravitons, there is no degrees of freedom that can have a temperature, and that is why we define it through the Wick-rotation. This procedure remains nonetheless correct even in the presence of bulk fields.

In the absence of additional fields, these two solutions are all that we need for the thermodynamic analysis. While there are other saddle points to the Euclidean action \cite{Maldacena:1998bwsl2zinstantons} it can be shown they are always subleading, and so we can discard them in a classical treatment.

\subsection{Euclidean action and Free Energy}
\label{sec:HawkingPage}
We have determined that the two competing solutions at temperature $T$ are respectively the non-spinning BTZ black hole and thermal AdS. To determine which one is dominant in the canonical ensemble, we must compute their respective Free Energies $F$. From a standard thermodynamical argument, the solution with lowest $F$ will be the dominant one.

To compute the free energy at temperature $\beta$, we consider the Wick rotated system with the compactified time coordinate, as explained in the previous section. Then, the Free Energy is defined as :
\begin{eqgroup}
    F = -T \ln(Z)\label{freeenergy}\ ,
\end{eqgroup}
where $Z$ is the partition function of the system at inverse temperature $\beta$. Using the path integral formulation, we have the following path integral identity (see sec.\ref{sec:entanglement} for more details):

\begin{eqgroup}
    Z = \langle e^{-\beta H}\rangle = \int \cal{D}g e^{-S_E(g)}\ ,
\end{eqgroup}
where $\cal{D}g$ denotes the path integral measure for metrics, and $S_E(g)$ the Euclidean action. We use the saddle point approximation to give it a value. The two competing saddle points are the BTZ black hole and thermal AdS. The goal then is to compute the Euclidean action for both of these solutions.

The full Einstein-Hilbert action (\ref{EinsteinHilbertAction}) will be in general divergent in asymptotically Anti-de-Sitter space. To compare Euclidean actions, we then need to introduce a cutoff at $r=r_{max}$. To get finite answers in the limit $r_{max}\rightarrow \infty$, we can add a counterterm to the action, which does not affect the equations of motions. The correct prescription is described in \cite{Hawking:1982dh} and results in the following Euclidean action\footnote{Note that performing the Wick-rotation in gravity is also a non-trivial operation, which might not be well-defined in general. To get to the Euclidean action (\ref{fullEinsteinHilbert}), we first linearize the theory around the AdS background, and then perform the Wick rotation for the linearized theory, before re-expressing everything in terms of the (now euclidean) Ricci scalar)} :
\begin{eqgroup}
    S_E=    -\frac{1}{16\pi G}&\bigg[ \int_{\cal{M}}d^3x \sqrt{g}(R-\frac{2}{\ell^2})+2\int_{r=r_{max}}\hspace{-0.7cm} d^2y\sqrt{h}(K-\frac{1}{\ell})\bigg]\ ,
    \label{fullEinsteinHilbert}
\end{eqgroup}
where the counterterm is simply the $-1/\ell$ substraction in the boundary term. 

The integrand over $\cal{M}$ will be the same for both geometries, which satisfy $R=-\frac{6}{\ell^2}$ and $g = \ell r^2$, see (\ref{RiemannAdSValue}). One difference will arise because of the range of the radial coordinate, which spans $0<r<r_{max}$ for thermal AdS and $r_{hole}=\sqrt{M\ell^2}<r<r_{max}$ for BTZ.

Starting with thermal AdS :
\begin{eqgroup}
   \int_{\cal{M}}d^3x \sqrt{g}(R+\frac{2}{\ell^2}) =- \int_{r=0}^{r_{max}}\int_{\varphi=0}^{2\pi}\int_{t=0}^{\beta}\ell rdrd\varphi dt\frac{4}{\ell^2}=-\frac{4\pi\beta r_{max}^2}{\ell}\ .
\end{eqgroup}

For the boundary term, the boundary surface is described by $r=r_{max}$, from which we deduce that the normal covector, taken to be outward facing is proportional to
\begin{eqgroup}
    n_\mu \propto (0,1,0)\ .
\end{eqgroup}

The normalized vector reads :
\begin{eqgroup}
    n^\mu =\frac{1}{\sqrt{1+\frac{r_{max}^2}{\ell^2}}}(0,1,0)\ .
\end{eqgroup}

Parametrizing the boundary metric in terms of the two remaining coordinates,
\begin{eqgroup}
h_{ij}dx^idx^j = (\ell^2+r_{max}^2)dt^2+r_{max}^2 d\varphi^2\ .
\end{eqgroup}

One can compute the extrinsic curvature $K$ following the prescription in the Appendix \ref{sec:ExtrinsicAppendix}. We find after some straightforward calculations :
\begin{eqgroup}
    \sqrt{h}K = \ell(1+2\frac{r_{max}^2}{\ell^2})\ .
\end{eqgroup}

Adding the contribution of the counterterm, and expanding in $r_{max}$ :
\begin{eqgroup}
    2\int_{r=r_{max}}\hspace{-0.7cm} d^2y\sqrt{h}(K-\frac{1}{\ell})&= 4\pi \beta\ell \left(1+2\frac{r_{max}^2}{\ell^2}-\frac{r_{max}^2}{\ell^2}(1+\frac{\ell^2}{2r_{max}^2})\right)+\cal{O}(r_{max}^{-2})\ ,\\
    &=4\pi\beta\ell\left(\frac{r_{max}^2}{\ell^2}+1/2\right)\ .
\end{eqgroup}

Adding all the terms we see that the diverging part indeed drops out, allowing us to safely take the $r_{max}\rightarrow \infty$ limit to obtain :
\begin{eqgroup}
    F_{AdS}=\frac{S_E^{AdS}}{\beta} =-\frac{\ell}{8G}\ .
\end{eqgroup}

The computation for BTZ proceeds in a similar way so we skip the details. For the bulk part we find :
\begin{eqgroup}
    \int_{\cal{M}}d^3x \sqrt{g}(R+\frac{2}{\ell^2}) = -\frac{4\pi\beta (r_{max}^2-M\ell^2)}{\ell}\ .
\end{eqgroup}

For the boundary part we have :
\begin{eqgroup}
    \sqrt{h}K=\ell(-M+\frac{2r_{max}^2}{\ell^2})\ .
\end{eqgroup}

Adding everything up as before we end up with :
\begin{eqgroup}
    F_{BTZ} = -\frac{M\ell}{8G}\ .
\end{eqgroup}

Comparing the free energies, we finally deduce that there is a phase transition at the critical value $M=1$, corresponding to a critial temperature $T=\frac{1}{2\pi}$. Below this value, the dominant saddle is thermal AdS, while above it the black hole geometry is favored. This is the so-called "Hawking-Page phase transition" \cite{Hawking:1982dh} and it will be a guiding thread throughout the first part of the thesis.

Just as a check, let us compute the energy and entropy of the states using thermodynamic identities. The energy $E$ of the states is given by :
\begin{eqgroup}
    \langle E\rangle=\frac{\pa}{\pa \beta}(\beta F) = \frac{M\ell}{8G}\label{energyofads}\ ,
\end{eqgroup}
where for thermal AdS $M=-1$. We can then compute the entropy of the solutions with $F=\langle E \rangle - \frac{S}{\beta}$. The entropy vanishes for thermal AdS as expected, while for the BTZ solution $S=\frac{\sqrt{M}\ell\beta}{4G}=\frac{2\pi r_+}{4G}$. It is equal to the Area of the horizon divided by $4G$, which is the Bekenstein-Hawking formula (\ref{bekensteinhawkingformula} that we will introduce later in the text.

Finally, let us point out that this phase transition is not restricted to the 3-dimensional case. In higher dimensions $D>3$, the temperature of the black hole as a function of the AdS radius and horizon radius $r_+$ can be easily shown to take the following form 
\begin{eqgroup}
    T = \frac{(D-1)r_+^2+(D-3)\ell^2}{4\pi r_+ \ell^2}\label{temperatureadsBHDbig3}\ .
\end{eqgroup}

From (\ref{temperatureadsBHDbig3}), there is a temperature below which there is no black hole solution. This critical temperature is realized for $r^+_{\rm min}{}^2=\frac{\ell(D-3)}{D-1}$. Furthermore, given any temperature above this threshold we will have two dinstinct black hole solutions called "small" and "big" black hole. The big one is the one that will be relevant for the Hawking Page transition. Indeed, the "small" black holes are always thermodynamically unstable. One can see this by computing the derivative $\frac{\pa T}{\pa r_+}$ which will be negative for $r<r_{min}$. As the small black hole radiates, its horizon will shrink, and its temperature increase, speeding up the radiation. Put it differently, in the canonical ensemble $\frac{\pa T}{\pa r_+}$ will be proportional to the specific heat of the solution, and negative specific heat signals an instability as described above. Small black holes are in this sense similar to the Schwarzschild black hole in flat spacetime.

\section{Conformal Field Theory}
\label{sec:CFT}
\textbf{C}onformal \textbf{F}ield \textbf{T}heory is the second essential ingredient necessary to the holographic correspondence. In this chapter we introduce the very basics that will be required to understand the bulk of the work. As our analysis is generally more focused on the gravity side, we won't dwell too much on details in this section. For a full study of the subject with emphasis on 2D CFT, see \cite{DiFrancesco:1997nk}.

\subsection{The conformal group}
The Coleman-Mandula theorem \cite{Coleman:1967ad} is a no-go theorem that applies to lorentzian Quantum Field Theories that have a mass gap, as well as some mild assumptions on scattering amplitudes. It states that the largest group of spacetime symmetries is the Poincaré group, and any internal symmetry must appear as a direct product with it (colloquially, spacetime and internal symmetries "don't mix"). The two famous "loopholes" of the theorem's assumptions are supersymmetry, where the extended symmetry algebra is a superalgebra, and conformal symmetry when the mass gap is zero.

If we remove this assumption, then we can have a spacetime symmetry group bigger than Poincaré, namely the Conformal group (we will denote it by CFT$_{D-1,1}$, to emphasize the signature of the spacetime). Conformal transformations are purely coordinate transformations that preserve angles. An example are dilatations, $x^\mu \rightarrow \lambda x^\mu$. To find the defining equations, consider first a euclidean metric (where the notion of angles is familiar, although the same derivation applies in Minkowski signature), and two vectors $v_1^\mu$ and $v_2^\mu$. The angle between them can be computed using the formula :
\begin{eqgroup}
    \cos(\th)=\frac{v_1^\mu v_2^\nu g_{\mu\nu}}{\sqrt{v_1^\mu v_1^\nu g_{\mu\nu} v_2^\mu v_2^\nu g_{\mu\nu}}}\label{angleeq}\ .
\end{eqgroup}
We see that the angle will be preserved iff the scalar product of vectors are rescaled, namely $v_1^\mu v_2^\nu g_{\mu\nu}\rightarrow e^{2\phi(x)}v_1^\mu v_2^\nu g_{\mu\nu}$. The peculiar parametrization of the scale parameter will come in handy later.

Under a change of coordinates $x\rightarrow x'$, the transformed vector fields components read $v'^{\nu}(x')=\frac{\pa x'^\nu}{\pa x^\mu}v^{\mu}(x)$. Plugging this in the "rescale" condition yields :
\begin{eqgroup}
    v_1'^\mu v_2'^\nu g_{\mu\nu}(x') &= v^{\al}(x)v^{\be}(x)\frac{\pa x'^\mu}{\pa x^\al}\frac{\pa x'^\nu}{\pa x^\be}g_{\mu\nu}(x')\stackrel{!}{=}v^{\mu}(x)v^\nu(x)e^{2\phi(x)}g_{\mu\nu}(x)\ ,\\
    \Leftrightarrow &\frac{\pa x'^\mu}{\pa x^\al}\frac{\pa x'^\nu}{\pa x^\be}g_{\mu\nu}(x')=e^{2\phi(x)}g_{\mu\nu}(x)\ ,
    \label{conditionconformal}
\end{eqgroup}
where in the second line we exploited the fact that the equality must hold for any two vector fields.

The condition on the metric (\ref{conditionconformal}) is the defining condition for conformal transformations. Even though this is given as a condition on the metric, one must keep in mind that conformal transformations act purely on the coordinates, leaving the metric unchanged. In that way, they are not to be confused with diffeomorphisms, which also preserve angles in a trivial way by also acting on the metric, nor with Weyl transformations which act solely on the metric, rescaling it locally.

Given an infinitesimal change of coordinates $x'^\mu = x^\mu+\xi^\mu$ (and thus an infinitesimal rescaling $\phi$), (\ref{conditionconformal}) reduces to :
\begin{eqgroup}
    \cal{L}_\xi g_{\mu\nu} = \xi^\rho \pa_\rho g_{\mu\nu}+g_{\mu\rho}\pa_\nu\xi^\rho+g_{\nu\rho}\pa_\mu\xi^\rho = 2\phi g_{\mu\nu}\ .
    \label{conformalkilling}
\end{eqgroup}
Vectors that satisfy (\ref{conformalkilling}) are aptly named conformal Killing vectors.

Let us now specify to the case of the Minkowski metric $\eta_{\mu\nu}$, as the conformal field theories that we will be interested in will live on a conformally flat background.
\begin{eqgroup}
    \pa_\mu\xi_\nu+\pa_\nu\xi_\mu = 2\phi \eta_{\mu\nu}\ .
    \label{flatconformalkilling}
\end{eqgroup}

In that case, tracing (\ref{flatconformalkilling}) with $\eta_{\mu\nu}$ immediately gives an expression for $\phi$ :
\begin{eqgroup}
    \phi = \frac{\pa_\rho \xi^\rho}{D}\equiv \frac{f}{D}\ .
\end{eqgroup}

Acting with $\pa^\nu$ on (\ref{flatconformalkilling}) yields yet another condition :
\begin{eqgroup}
    \square \xi_\mu = \frac{2-D}{D}\pa_\mu f \label{tracedflatconformalKilling}\ .
\end{eqgroup}

We immediately notice that the 2-dimensional case will be special; a fact that echoes the radically different asymptotic symmetry group of AdS in 3-dimensions, and is yet another hint of the holographic correspondence. We consider for now $D>2$ dimensions. Contracting (\ref{tracedflatconformalKilling}) with $\pa^\nu$ gives $\square f = 0$. Finally, applying the $\square$ operator on (\ref{flatconformalkilling}) gives us the simple equation :
\begin{eqgroup}
    \pa_\mu \pa_\nu f = 0\Rightarrow f=a+b_\mu x^\mu\ .
    \label{fintegrated}
\end{eqgroup}
To conclude we apply $\pa_\rho$ to (\ref{flatconformalkilling}) and choose a suitable linear combination of the equations obtained by permutations of $\mu,\nu,\rho$ to get :
\begin{eqgroup}
    2\pa_\mu \pa_\nu \xi_\rho=\eta_{\mu\rho}\pa_\nu f+\eta_{\nu\rho}\pa_\mu f-\eta_{\mu\nu}\pa_\rho f=\eta_{\mu\rho}b_\nu+\eta_{\nu\rho}b_\mu-\eta_{\mu\nu}b_\rho = c_{\rho\mu\nu}\ ,
\end{eqgroup}
where we used (\ref{fintegrated}).

Finally, a double integration gives us the form of the general conformal transformation :
\begin{eqgroup}
    \xi_\mu = a_\mu + b_{\mu\rho}x^\rho+c_{\mu\nu\rho}x^\nu x^\rho\ .
\end{eqgroup}
Plugging back into the original equation (\ref{flatconformalkilling}) yields an additional condition on $b_{\mu\nu}$ :
\begin{eqgroup}
    b_{\mu\nu} = \lam \eta_{\mu\nu}+\omega_{\mu\nu}\mbox{ ,  }\omega_{\mu\nu}=-\omega_{\nu\mu}\ .
\end{eqgroup}
Expanding everything in terms of the independent infinitesimal parameters :
\begin{eqgroup}
    \xi^\mu = a^\mu +\omega^\mu_{\;\nu} x^\nu+ \lam x^\mu +  (2(b_\nu x^\nu) x^\mu-x^2b^\mu)\ .
    \label{flatconformalkillingsolution}
\end{eqgroup}
In addition to the expected translations and rotation, we find the additional dilatation (associated to the $\lam$ parameter) and the so-called "special conformal transformations" (associated to the parameters $b_\nu$). Counting the number of free parameters, we obtain $\frac{(D+1)(D+2)}{2}$ which is then the dimensionality of the conformal group in $D$ dimensions.

Before specializing to the 2-dimensional case that will be of most interest to us, let us point some more facts in the higher dimensions. Consider the action of the conformal group on a scalar field $\varphi(x)$, where its transformation is induced by the change of the coordinates, $\varphi'(x')=\varphi(x)$ :
\begin{eqgroup}
    e^{-iw_aG^a} \varphi(x) =\varphi'(x)\ ,
    \label{basicrepresentationaction}
\end{eqgroup}
where $G^a$ are the generators of the conformal group, and $w_a$ an infinitesimal parameter, so that $x'^\mu = x^\mu+w_a f^{\mu\,a}(x)$. Taylor expanding (\ref{basicrepresentationaction}) furnishes a representation of the generators $G_a$, leading to the following expressions in the case of the conformal group:



\begin{eqgroup}
    \mbox{Translations : } &\;\;\; P_\mu = -i\pa_\mu\ ,\\
    \mbox{Rotations/Boosts : } &M_{\mu\nu}= i(x_\mu\pa_\nu-x_\nu\pa_\mu)\ ,\\
    \mbox{Dilatation : } &\;\;\;\;D=-ix^\mu \pa_\mu\ ,\\
    \mbox{Special conformal : } &\;\;\;K_\mu = i(x^2\pa_\mu-x_\mu x^\nu\pa_\nu)\ .
    \label{conformalgensinnormalrep}
\end{eqgroup}

The indices on the generators can of course be raised and lowered by the metric.

Computing the Lie bracket then reveals the commutation relations which \emph{define} the conformal algebra. For reference, we include the non-vanishing Lie brackets :
\begin{eqgroup}
        [M_{\mu\nu},M_{\rho\sigma}] &=  i(\eta_{\mu\rho}M_{\nu\sigma}-\eta_{\nu\rho}M_{\mu\si}+\eta_{\nu\si}M_{\mu\rho}-\eta_{\mu\si}M_{\nu\rho})\ ,\\
    [D,P_\mu] &= iP_\mu\;,\;\;\;[D,K_\mu] = -iK_\mu\ ,\\
    [K_\mu,P_\nu] &= 2i(\eta_{\mu\nu}D-J_{\mu\nu})\ ,\\
    [K_\mu,M_{\nu\rho}] &= i(\eta_{\mu\nu}K_\rho-\eta_{\mu\rho}K_\nu)\ ,\\
    [P_\mu,M_{\nu\rho}] &= i(\eta_{\mu\nu}P_\rho-\eta_{\mu\rho}P_\nu)\ .\\
    \label{commutatorsCFT}
\end{eqgroup}

Let us make a last comment on $CFT_{D-1,1}$ before specializing to $D=2$. One might notice that the group's dimensionality matches the dimension of the orthogonal group $SO(D+2)$. This is not a coincidence, as one can show that it is indeed isomorphic to $SO(D,2)$. Let $J_{AB},\; -1\geq A,B\geq D+1$ denote the generators of $SO(D,2)$, then one can verify that the map (\ref{isolorentz}) is a Lie Algebra isomorphism :
\begin{eqgroup}
    J_{\mu\nu}&=M_{\mu\nu}\ ,\\
    J_{-1,\mu}&=\frac{1}{2}(P_\mu-K_\mu)\ ,\\
    J_{(D+1),\mu}&=\frac{1}{2}(P_\mu+K_\mu)\ ,\\
    J_{-1,(D+1)}&=D\ ,
    \label{isolorentz}
\end{eqgroup}
where the $-1$ and $(D+1)$ indices denote the new timelike and spacelike coordinates respectively. It is through this isomorphism that we are able to identify the asymptotic symmetry group of $AdS_{D+1}$ with $CFT_{D-1,1}$.

\subsection{The conformal group in 2 dimensions}
Most treatments of two-dimensional CFTs consider a metric with Euclidean signature, as it allows the use of powerful complex analysis techniques. Therefore, in this section we will consider the following background metric :
\begin{eqgroup}
    ds^2 = d\tau^2+dx^2\ .
\end{eqgroup}

Lorentzian results can of course be recovered by the correct Wick rotation.
Let us now go back to (\ref{tracedflatconformalKilling}). For $D=2$, there are two independent equations (\ref{conformacond2D}) :
\begin{eqgroup}
    &\pa_\tau\xi^x=-\pa_x \xi^\tau\ ,\\
    &\pa_\tau \xi^\tau = \pa_x \xi^x\ .
    \label{conformacond2D}
\end{eqgroup}

If we consider then $\tau$, $x$ as the coordinates of the complex plane $z=x+i\tau$, (\ref{conformacond2D}) become exactly the Cauchy-Riemann equations for the complex function $\xi=\xi^x+i\xi^t$. Therefore, the general solution is given by 
\begin{eqgroup}
    \xi^x+i\xi^\tau=\xi(x+i\tau)\ .
    \label{solutionCFT2D}
\end{eqgroup}

To go back to real space, one replaces $\tau=it$ in (\ref{solutionCFT2D}). Then $\xi^\tau \pa_\tau = \xi^\tau(-i)\pa_t \equiv \xi^t \pa_t$, so that vector components are re-scaled with $i$ under the Wick rotation. Let us define, in real space, the lightcone coordinates $w^+=x+t$, $w^-=x-t$. By taking the real and imaginary part of (\ref{solutionCFT2D}), we conclude :
\begin{eqgroup}
    &\xi^x+\xi^t = \xi^+= \bar{\xi}(x+t) = \bar{\xi}(w^+)\ ,\\
    &\xi^x-\xi^t = \xi^-= \xi(x-t) = \xi(w^-)\ .
\end{eqgroup}

It follows that in Minkowski spacetime the solutions divide into right-moving and left-moving transformations, which can be chosen independently (in real space, $\xi$ and $\bar{\xi}$ are independent functions). Again, the comparison with the asymptotic symmetries of AdS$_3$ (\ref{solutionAsymptoticKillinglightcone}) is flagrant.
In fact, from the form of the metric in lightcone gauge $ds^2=dw^+dw^-$, one readily infers the finite form of the conformal transformations :
\begin{eqgroup}
    (w^+)' &= f^+(w^+)\ ,\\
    (w^-)' &= f^-(w^-)\ ,
    \label{lightcone2Dcftrealspace}
\end{eqgroup}
so that the scaling factor is $e^{2\phi} = f^+{}'(w^+)f^-{}'(w_-)$. This splitting into left and right-moving transformations will follow us throughout this chapter. All states and excitations will also split accordingly and we will mostly concentrate on one of the two sectors. This property of two-dimensional CFTs is sometimes referred to as "holomorphic factorization".

Let us introduce a last piece of machinery before finally entering into the field theory proper. We already saw that in the euclidean picture, conformal transformations can be seen as holomorphic functions. To simplify this notation and take the analogy even further, we can define the formal change of coordinates to the complex 2-plane :
\begin{eqgroup}
    z=x+i\tau\;,\;\; \bar{z}=x-i\tau\ .
    \label{complexificationchangeofcoord}
\end{eqgroup}

Although the notation $\bar{z}$ is suggestive, in this change of coordinate $\bar{z}$ should be considered as an independent complex variable to $z$. For this to make sense, we must consider the euclidean plane to be also complexified, namely $x,\tau \in \mathbb{C}$. This is unphysical, and at the end of the day we should impose reality on the original coordinates. This condition takes the form $\bar{z}=z^*$ where the ${}*$ operator is the bona fide complex conjugation.

This seems like a lot of trouble but it will simplify the notation greatly. In these complex coordinates, the metric is written as :
\begin{equation}
    ds^2 = dz d\bar{z}\ .
\end{equation}
In this form, passing to Lorentzian space is as simple as $z\rightarrow w^-$, $\bar{z}\rightarrow w^+$. Conformal transformations in complex space :
\begin{eqgroup}
    z'&=f(z)\ ,\\
    \bar{z}'&=\bar{f}(\bar{z})\ .
\end{eqgroup}

At the risk of hammering the point a bit too much, $f$ and $\bar{f}$ are considered independent functions until the end, where $\bar{f}$ must be identified with the complex conjugate of $f$. This coincides nicely with (\ref{lightcone2Dcftrealspace}).

\subsection{Primary fields}
In this section we get a first look at the restrictive power that the conformal symmetry will impose on the fields of the theory. We begin with a classical treatment.

The irreducible representations of CFT$_2$ will be constructed around the core concept of \emph{primary fields}. These fields will be labeled by two numbers $h$ and $\bar{h}$. To understand how they come about, consider first the transformation induced on a scalar field $\phi(z,z')$ by a conformal change of coordinates. For now we look only on the change brought by the coordinate change, thus we consider an otherwise invariant field $\phi'(z',\bar{z}')=\phi(z,\bar{z})$. Then with $z'=z+\xi(z)$ (and equivalently for the anti-holomorphic part)

\begin{eqgroup}
    \phi'(z,\bar{z}) &= \phi(z,\bar{z})-\xi(z)\pa\phi-\bar{\xi}(\bar{z})\bar{\pa}\phi\ ,\\
    \Leftrightarrow \delta\phi &= \sum_{n\in \mathbb{Z}}a_n l_n \phi(z,\bar{z})+\bar{a}_n \bar{l}_n \phi(z,\bar{z})\ ,
    \label{trivialfunctiontransfocft2d}
\end{eqgroup}
where we used the Laurent expansion of the parameter $\xi(z) = \sum_n a_n z^{n+1}$, which defines for us the generators of the conformal symmetry, $l_n = -z^{n+1}\pa_z$ and likewise for $\bar{l}_n$. The algebra is the Witt algebra described in (\ref{liebrackasympt}).

An important distinction is to be made here between the "global" and "local" conformal transformations. As it can be easily checked, the only generators that are well defined both at $z=0$ and $z=\infty$, and hence on the whole complex plane are $l_{-1}$, $l_{0}$ and $l_{1}$ (and likewise for the antiholomorphic part, so we will stop mentioning it from now). Together, they form the only non-trivial finite subgroup of the Witt algebra, which is $SL(2,\mathbb{C})$.  In Minkowski spacetime, the global subgroup $SL(2,\mathbb{C})\times SL(2,\mathbb{C})$ reduces to $SL(2,\mathbb{R})\times SL(2,\mathbb{R})$, which is the isometry group of AdS$_3$ (\ref{sl2rsymmetryads3}).

One useful parametrization of the finite global conformal transformations is :
\begin{eqgroup}
    z' = \frac{az+b}{cz+d}\;,\;\; ad-bc=1\ .
    \label{sl2ctranfofinite}
\end{eqgroup}

We would like now to expand the transformation rules (\ref{trivialfunctiontransfocft2d}), by introducing "internal" quantum numbers that will affect the transformation of the fields. Usually, the transformation rules of "primary fields" are simply defined right away as :
\begin{eqgroup}
    \phi'(z',\bar{z}') = \left(\frac{dz'}{dz}\right)^{-h}\left(\frac{d\bar{z}'}{d\bar{z}}\right)^{-\bar{h}}\phi(z,\bar{z})\ .
    \label{primaryfield}
\end{eqgroup}

Although this definition is natural (the field is rescaled with the local scaling factor, to the power of its "scaling dimension") we would like to provide a little more context as to the origin of this formula.

To do so, we consider something analogous to the "little group trick" to find the representations of the Poincaré group. For this purpose, we consider the subgroup of conformal transformations that leave the origin $z=0$ invariant. It is straightforward to see that it is generated by the $l_n$, $n\geq 0$. Let us denote the operators at $z=0$ by $\ti{l}_n$. We must now choose the action of the $\ti{l}_n$ on $\phi(0)$. Primary operators will act as the "highest weight" of the representations we will construct. Thus it is natural to define :
\begin{eqgroup}
    \ti{l}_0\phi &= -h\phi\ ,\\
    \ti{l}_n\phi &= 0 \ .
    \label{actionoflittlegrouponphi}
\end{eqgroup}

Indeed by the commutation relations $[\ti{l}_0,\ti{l}_n]=-n \ti{l}_n$, they act as lowering operators for the eigenvalue of $\ti{l}_0$, sending $-h\rightarrow -h-n$\footnote{Alternatively, we could call them raising operators of the scaling dimension, $h\rightarrow h+n$}.

To recover the action of the algebra on $\phi(x)$, all we need to do is translate the operators, exploiting the action of $l_{-1}$ which is the translation generator. Thus, we define the generic $l_n$, $n\geq 0$ at position $z$ as :
\begin{eqgroup}
    l_n = e^{-zl_{-1}}\ti{l}_ne^{zl_{-1}}\ .
    \label{lntranslation}
\end{eqgroup}

To compute (\ref{lntranslation}), we use the Hausdorff formula :
\begin{eqgroup}
    e^{-A}Be^{A} = B +\frac{1}{1!}[B,A]+\frac{1}{2!}[[B,A],A]+...\ .
    \label{hausdorff}
\end{eqgroup}

In our case $A=zl_{-1}$, so the $u$'th term in (\ref{hausdorff}) can be easily shown to give :
\begin{eqgroup}
    \frac{z^u}{u!}[...[\ti{l}_n,\underbrace{l_{-1}],l_{-1}],...],l_{-1}}_{u}]=\frac{z^u (n+1)!}{u!(n-u+1)!}\ti{l}_{n-u}\ .
\end{eqgroup}

As the action of $\ti{l}_{n},\;n>0$ on $\phi$ is trivial by definition, the only non-trivial term in the series of commutators will be the $u=n$ and the $u=n+1$. After that, the series terminates as we commute $l_{-1}$ with itself. Putting this together gives us the action of $l_n$ on $\phi$ :
\begin{eqgroup}
    l_n\phi(z) = \left((n+1)z^n \ti{l}_0 +z^{n+1}l_{-1}\right)\phi=-\left((n+1)z^n h + z^{n+1}\pa_z\right)\phi(z)\ .
    \label{actionofln}
\end{eqgroup}

It remains to find the action of the $l_{-n},\;n>0$. This can be done by noticing that under the transform $z=1/w$, $l_{-n}(z)=-z^{-n+1}\pa_z = w^{n-1}w^{2}\pa_w=w^{n+1}\pa_w$. Hence the point $w=0\Leftrightarrow z=\infty$ is fixed by the $l_{-n}$. The translation operator at $w=0$ is now given by $-l_1(z=\infty)=-\pa_w$, while the $l_0(\infty)$ eigenvalues remain the same, as can be seen by taking the $z=\infty$ limit in (\ref{actionofln}) for $n=0$.  

By going through the same procedure, we find that the expression (\ref{actionofln}) is valid also for $n<0$. Now, noticing the action of $l_n$ is associated to the infinitesimal transformation $z'=z+\eps z^{n+1}$ :
\begin{align}
    \phi'(z)-\phi(z)=&\delta\phi = \eps l_n\phi = -\eps(h\pa_z(z^{n+1})\phi+z^{n+1}\pa_z\phi)\;\forall\;n\in \mathbb{Z}\label{changeinducedbyconformal1}\ ,\\
    \Rightarrow &\delta\phi = -(h\phi\pa_z\xi(z)+\xi(z)\pa_z\phi)\;\forall\; \xi(z),\;z'=z+\xi(z)\ .
    \label{changeinducedbyconformal2}
\end{align}

Integrating (\ref{changeinducedbyconformal2}), we recover the formula (\ref{primaryfield}). To get (\ref{changeinducedbyconformal2}), we used that since (\ref{changeinducedbyconformal1}) is valid for all generators of the conformal transformations, it will be valid for a generic one.

The condition (\ref{primaryfield}) is very powerful, and it places very stringent constraints on the correlators of primary fields. Consider for instance a correlator of holomorphic fields :
\begin{eqgroup}
    \langle \phi_1(z_1)\phi_2(z_2)...\phi_n(z_n)\rangle= G(z_1...z_n)\ .
    \label{correlatorgeneric}
\end{eqgroup}
Denoting by $w(z)$ a global\footnote{Many thanks to Marco Meineri for graciously pointing out that we should underline the fact the transformation should be global... One can see that (\ref{transfocorrelator}) will fail for more generic conformal transformations, as they do not leave the vacuum invariant. Alternatively, we could keep this formula in the generic case, but the correlators should be computed in the state obtained by acting with the conformal transformation on the vacuum (which acts trivially in the case of a "global" conformal transformation).}conformal transformation of the coordinates, and then using (\ref{primaryfield}) with (\ref{correlatorgeneric}) yields the functional equation :
\begin{eqgroup}
    G(w_1,...,w_n) = \left(\frac{dw}{dz_1}\right)^{-h_1}\left(\frac{dw}{dz_2}\right)^{-h_2}\ldots\left(\frac{dw}{dz_n}\right)^{-h_n}G(z_1,...z_n)\ ,
    \label{transfocorrelator}
\end{eqgroup}
where we have used the conformal invariance of the theory through the following identity: $\langle\phi_1(z'_1)\ldots \phi_n(z'_n) \rangle=\langle\phi'_1(z'_1)\ldots \phi'_n(z'_n) \rangle$.

These functional equations fix the exact form of the $2$ and $3$-point functions, while there still remains some freedom for bigger correlators. We will only need the form of the two-point function :
\begin{eqgroup}
    \langle \phi_1(z_1)\phi_2(z_2) \rangle = \frac{C_{12}}{(z_1-z_2)^{2h}}\mbox{ if }h_1=h_2\mbox{,  0 otherwise} \ .
    \label{twopointfunction}
\end{eqgroup}

The constant $C_{12}$ is usually set to one by the freedom to re-normalize the fields.

Let us mention that for 3-point functions, conformal symmetry fixes them completely up to one constant, which will depend on the specifics of the theory. Intuitively, this is because conformal symmetry is able to map any three $(z_1,z_2,z_3)$ to $(1,0,\infty)$. This is no longer possible for more than three points, so 4-point correlators are fixed up to an undetermined function of a conformally invariant cross-ratio.

\subsection{The stress-energy Tensor}
As we know from Noether's theorem, to each symmetry corresponds a conserved current. For the conformal symmetry, it is embodied by the stress-energy tensor, denoted $T^{\mu\nu}$. This operator is central in the study of CFT. The first reason is that it is an universal operator, as any CFT will at least have a stress-energy tensor in its operator spectrum. The second reason is that it acts as the generator of the conformal transformations. 
To determine the stress-energy tensor, we consider first only ordinary translations, $x^\mu{}'=x^\mu+\xi^\mu$. The action $S$ of our CFT is then of course left invariant by this change of coordinate which we write as $\delta S$. This means that if we promote $\xi^\mu$ to depend on the coordinates $x^\rho$, we must have :
\begin{eqgroup}
    \delta S = \int d^2x T^\mu_\nu\pa_\mu \xi^\nu\ ,
\end{eqgroup}
such that it vanishes exactly when $\xi$ is constant. If we now consider ourselves to be on-shell, then any variation of the fields should make $\delta S$ vanish, by definition. Then an integration by parts shows conservation $\pa_\mu T^{\mu\nu}=0$, that should hold on shell.

Using this method, the resulting $T^{\mu\nu}$ isn't always symmetric, although there still remains some freedom to modify it by terms that have no physical effect (i.e. they do not modify the conservation law and conserved charges). The symmetrized tensor that can be obtained is called the "Belifante tensor". There is also a more direct technique to obtain the "nice" stress tensor directly from the variation, outlined in \cite{improvecnoether}. Here we opt for another trick that is more straightforward.

As we have stated before, in a CFT the metric is non-dynamical and fixed. Let us relax this condition just for a moment, and consider the same action $S$ which now also depends on a dynamical metric $g_{\mu\nu}$. By construction, such an action will now be invariant under the diffeomorphism induced by $x^\mu{}'= x^\mu+\xi^\mu(x^\rho)$. Then, writing the total change of the metric :
\begin{eqgroup}
    \delta S = 0 = \int d^2x \frac{\delta S}{\delta g_{\al\be}}\cal{L}_\xi g_{\al\be}+\underbrace{\delta S}_{\mbox{fixed metric}}\ ,
    \label{diffeoS}
\end{eqgroup}
where the $\delta S$ on the RHS is the variation of the action induced by other fields than the metric, as signified by the caption "fixed metric".

Thus if we take (\ref{diffeoS}) and evaluate it at $g_{\mu\nu}=\eta_{\mu\nu}$ we can determine the fixed metric variation as minus the change of the action when varying the metric. Using the expression of the Lie derivative (\ref{conformalkilling}) :
\begin{eqgroup}
    \underbrace{\delta S}_{\mbox{fixed metric}}=\int d^2x T^{\al\be}\pa_\al \xi_\be=-2 \int d^2x \frac{\delta S}{\delta g_{\al\be}}\pa_\al \xi_\be\ .
\end{eqgroup}

Thus, up to an arbitrary normalization factor :
\begin{eqgroup}
    T^{\mu\nu} = -\frac{2}{\sqrt{-g}}\frac{\delta S}{\delta g_{\mu\nu}}\ ,
\end{eqgroup}
which is automatically symmetric. Furthermore, if we choose $\xi_\be$ as parametrising a conformal transformation, we have $\cal{L}_\xi g_{\al\be}=\pa_\rho\xi^\rho g_{\al\be}$, as well as $\underbrace{\delta S}_{\mathclap{\mbox{fixed metric}}}=0$ assuming conformal invariance. Then by (\ref{diffeoS}):
\begin{eqgroup}
    0 = \int d^2x T^\al_\al \pa_\rho\xi^\rho\ .
\end{eqgroup}

While strictly speaking this does not force $T^\al_\al=0$ as $\pa_\rho \xi^\rho$ is not arbitrary, in the overwhelming majority of cases this will hold, so we will consider $T_{\mu\nu}$ to be traceless from now on. The converse is however true, namely that a traceless stress-tensor implies the theory is a CFT (classically)\cite{Jackiw:2011vz}.

In complex coordinates, these constraints are explicitly solved as :
\begin{eqgroup}
    T_{z\bar{z}}=0\;,\; T_{zz}=-\frac{T(z)}{2\pi}\;,\; T_{\bar{z}\bar{z}}=-\frac{\overline{T}(\bar{z})}{2\pi}\ ,
    \label{stressenergyconditions}
\end{eqgroup}
where the prefactors are introduced simply to obtain simpler expressions in what follows (and they are standard). The energy currents are thus also separated into holomorphic "right-moving" and anti-holomorphic "left-moving" currents.

Let us now finally move to the quantization of the CFT. Until now, most derivations and objects were defined assuming the theory could be formulated through an action principle. For CFTs, such a formulation is often lacking, and the fundamental objects are local operators which we will denote by $O(x)$. All the dynamics are then encoded in correlators of such operators as in (\ref{correlatorgeneric}). As such, from now on when we write an operator identity such as $O_1(z) = O_2(z)$ it should be taken to mean $\langle O_1(z)\ldots\rangle = \langle O_2(z)\ldots\rangle$, where "$\ldots$" is any insertions of operators away from $z$.

In trying to compute such objects, the \textbf{O}perator \textbf{P}roduct \textbf{E}xpansion (OPE) will be invaluable. It is an identity the describes what happens when we bring two operators to the same point :
\begin{eqgroup}
    O_i(z)O_j(w) = \sum_k C_{ij}^k(z-w)O_k(w)\ .
    \label{opegeneral}
\end{eqgroup}

Equation (\ref{opegeneral}) can be seen as arising simply from locality; as the two operators get too close to be distinguished, their combined action becomes local and it can be written as a single operator.

Returning to the stress-energy tensor, we would like to distill the quantum version of the conformal invariance constraints. This is done by simply deriving the Ward Identities associated to the conformal symmetry. The derivation is straightforward although lengthy, so we do not reproduce it here. We consider the Ward Identity for a conformal transformation $z'=z+\xi(z)$ localised around $z_1$, thus we assume $\xi(z_i)=0\;\forall i>1$ for $z_i$ in the correlator $\langle O_1(z_1)\ldots O_n(z_n)\rangle$. In other words, the conformal transformations only "hits" the first operator. The ensuing Ward identity can be expressed as an identity (\ref{wardidentity}) :
\begin{eqgroup}
    \delta_\xi O_1(z_1) = -\mbox{Res}_{z_1}[\xi(z) T(z)O_1(z_1)]\ ,
    \label{wardidentity}
\end{eqgroup}
where $\delta_\xi O_1(z_1)$ denotes the transformation of the operator under the conformal transformation, and $\mbox{Res}_{z_1}(f(z))$ denotes the residues of $f(z)$ in $z_1$. So if we know the OPE of the stress energy tensor with any operator, we also know how the operator transforms under any conformal transformation through (\ref{wardidentity}). This explains our earlier claim that the stress-energy tensor is the generator of the conformal transformations.

Conversely, if we know how $O_1$ transforms, we can deduce its OPE with $T$. Consider then a primary operator $\phi$, whose transformation law is given by (\ref{changeinducedbyconformal2}). We derive from it the OPE :
\begin{eqgroup}
    T(z)\phi(w) = \frac{h \phi(w)}{(z-w)^2}+\frac{\pa_w \phi(w)}{z-w}+\mbox{reg}\ ,
    \label{TOPE}
\end{eqgroup}
where "reg" denotes non-singular terms as $z\rightarrow w$. We will omit them from now on as they do not affect the physics in the limit $z\rightarrow w$ that we consider most of the time.

Operators that are not primary can also be assigned a scaling dimension with a similar derivation to (\ref{actionoflittlegrouponphi}). However, the transformation law (\ref{changeinducedbyconformal2}) will hold only for pure dilations, namely $\xi(z)=z$. Thus the OPE with $T(z)$ is only partially fixed, and we can state :
\begin{eqgroup}
    T(z)O(w)= \ldots+\frac{h O(w)}{(z-w)^2}+\frac{\pa_w O(w)}{z-w}\label{genericOPE}\ ,
\end{eqgroup}
where $\ldots$ denotes now terms that are more singular than $1/(z-w)^2$. An example of non-primary field can be obtained by differentiating a primary field. Indeed, applying $\pa_w$ to (\ref{TOPE}) we obtain :
\begin{eqgroup}
    T(z)\pa\phi(w) = \frac{2h\phi(w)}{(z-w)^3}+\frac{(h-1)\pa\phi}{(z-w)^2}+\frac{\pa(\pa\phi)}{(z-w)}\label{nonprimaryTOPE}\ ,
\end{eqgroup}
which tells us that $\pa\phi(w)$ has scaling dimension $(h-1)$. This was to be expected as applying a derivative is equivalent  to applying the operator $-l_{-1}$ on the primary field, which from (\ref{liebrackasympt}) acts as a lowering operator for $h$. Everything checks out !

The last piece of machinery we will need to introduce is the OPE of $T$ with itself. Under holomorphic dilations $z'=\lam z$, $T_{\mu\nu}$ being a two-tensor transforms as :
\begin{eqgroup}
    T'_{zz}=(\frac{dz'}{dz})^{-2}T_{zz}\ ,
\end{eqgroup}
from which we deduce its scaling dimension to be $h=2$ (likewise for antiholomorphic, $\bar{h}=2$). However, this is the "classical" dimension, which can also be obtained by dimensional analysis, but in general, this will not be the same in the quantum theory. One can however show that the dimension of a conserved current does not receive quantum corrections, and so this is true for the energy-momentum tensor.

From (\ref{genericOPE}), we know part of the $TT$ OPE. In general, we should allow all other possible singular terms. However the only operator guaranteed to exist in a CFT (other than $T$), is the identity or trivial operator, naturally of scaling dimension $(h,\bar{h})=(0,0)$. By dimensional analysis, the only term we can add to the OPE is $\frac{c/2}{(z-w)^4}$.

\begin{eqgroup}
    T(z)T(w)=\frac{c/2}{(z-w)^4}+\frac{2 T(w)}{(z-w)^2}+\frac{\pa_w T(w)}{z-w}\ ,
    \label{TTOPE}
\end{eqgroup}
and likewise for $\bar{T}$ with $\bar{c}$.

The constant $c$ is the so-called "central charge" of the CFT. By considering several examples and also from general theorems \cite{Zamolodchikov:1986gt,Cardy:2010fa}, it is apparent that this number somehow represents the number of degrees of freedom our theory has. This OPE also lets us derive the "quantum" version of conformal transformations, by expanding $T(z)$ into modes. The algebra that results is the Virasoro algebra, which is the Witt algebra of the "classical" conformal generators, with a central extension proportional to $c$. We will not go any deeper into the CFT machinery and refer the interested reader to one of the many excellent reviews \cite{DiFrancesco:1997nk,Tong:2009np,Cardy:2008jc}.

As for any operator, (\ref{TTOPE}) combined with (\ref{wardidentity}) allows us to find the infinitesimal transformation rules for $T(z)$. After integration, we expect to find a modified version of (\ref{primaryfield}) because of the $1/(z-w)^4$ term in the $TT$ OPE.

\begin{eqgroup}
    T'(w)&= \left(\frac{dw'}{dz}\right)^{-2}\left(T(z)-\frac{c}{12}\left\{w(z),z\right\}\right)\ ,\\
    \left\{w(z),z\right\}&=\frac{2w'''w'-3(w'')^2}{2(w')^2}\ ,
    \label{TTransform}
\end{eqgroup}
where the operation $\{w,z\}$ is called the "Schwarzian". 

As it turns out, the Schwarzian vanishes exactly under the $SL(2,\mathbb{C})$ subgroup of global conformal transformations. This makes sense because the vacuum of the theory will be defined as the state that vanishes under the action of $l_{-1}$, $l_{0}$ and $l_{1}$. Thus if we set $\langle T \rangle = 0$ for the vacuum state, it will remain unchanged under the global conformal transformation.

However, under more general transformations, the vacuum expectation value will change. Consider for example the following holomorphic map from the complex plane to the cylinder: $w(z) = \frac{L}{2\pi}\ln(z)$, where $e^{\frac{2\pi}{L} Re(w)}$ is the radial coordinate and $\frac{Im(w)}{L}$ the "angle". Computing the Schwarzian gives:
\begin{eqgroup}
    T_{cyl}(w) = \left(\frac{2\pi}{L}\right)^2\left( z^2 T_{pl}-\frac{c}{24}\right)\ .
    \label{planetocyl}
\end{eqgroup}

Assuming the expectation value in the plane vacuum state vanishes, $\langle T_{pl}\rangle = 0$ we obtain 
\begin{eqgroup}
    \langle T_{cyl}(w)\rangle = -\frac{\pi^2 c}{6 L^2}\ .
    \label{Casimircylinder}
\end{eqgroup}

This non-zero negative vacuum energy is to be interpreted as Casimir energy which appears because of the compactness of the cylinder. As we can see, it is proportional to $c$ which reinforces its interpretation as the number of degrees of freedom of the theory.

Let us compare this with the Energy of AdS found in (\ref{energyofads}). The energy density in real space is $\langle T_{tt}\rangle = \langle T_{++}\rangle+\langle T_{--}\rangle$. The energy of the state on the cylinder is then (\ref{Casimircylinder}) multiplied by $L$ (and the factor of $2\pi$ from (\ref{stressenergyconditions})). For the AdS geometry of (\ref{energyofads}), the boundary cylinder has period $L=2\pi$, then :
\begin{eqgroup}
    L\langle T_{tt}\rangle=E_{AdS} \Leftrightarrow -\frac{\pi c}{6 L}=-\frac{\ell}{8G}\Leftrightarrow c=\frac{3\ell}{2G} \label{energymatching}\ .
\end{eqgroup}
We see that we correctly recover the Brown-Henneaux formula (\ref{brownhenneauxcentralcharge}), that was obtained by looking at the asymptotic symmetry algebra! This is one of the simplest consistency checks of holographic duality.

\section{Interface CFT}
\label{sec:interfaces}
We would like to consider an extension to CFTs by introducing an interface that will allow us to bring to contact two distinct CFTs. The study of such a system through the holographic lens will be the main topic of this thesis.

Generic defects in a CFT are inhomogeneities localized on a lower dimensional hypersurface. Interfaces are special defects of codimension one, that separate spacetime in two parts. We will consider conformal interfaces which preserve a subset $SO(2,d-1)\subset SO(2,d)$ of the conformal symmetries. This will relax the conditions on correlators and allow for more general forms. For instance, scalar operators can acquire a vacuum expectation value.

\begin{figure}[!h]
    \centering
    \includegraphics[width=0.5\linewidth]{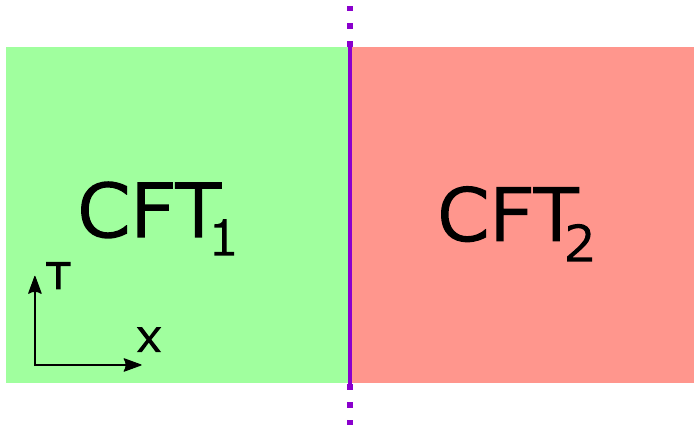}
    \caption{{\small Simplest timelike interface separating two different CFTs. The interface is depicted in purple, and is described by the equation $x=cst$. The two colors of the CFT underline the fact that they may be different theories living on each side.}}
    \label{fig:twocftinterface}
\end{figure}

The case that we will consider is an interface in two spacetime dimensions, as illustrated in fig.\ref{fig:twocftinterface}. The interface sits at position $x=0$ and is parametrized by the time $t=\tau$ ($z=i\tau$ in complex coordinates). The global conformal transformations that preserve the geometry of the interface form the group $SO(1,2)$ and include $\tau$ translations and scale transformations.

Consider the Action (\ref{2faceaction}) which has the generic form of the system depicted in fig.\ref{fig:twocftinterface}.

\begin{eqgroup}
    S = \int_{x<0} dxdt \cal{L}_1 + \int_{x>0}dxdt\cal{L}_2 + \int_{x=0} d\tau \cal{L}_{\rm int}\ ,
    \label{2faceaction}
\end{eqgroup}
where $\cal{L}_1$, $\cal{L}_2$ and $\cal{L}_{\rm int}$ respectively represent the lagrangians of CFT$_1$, CFT$_2$ and the interface degrees of freedom \footnote{While the derivation we outline here relies on the Lagrangian formulation of the CFT, it is by no means necessary, see \cite{Billo:2016cpy}. We decided to go the Lagrangian route to simplify the explanation.}. To derive the conditions imposed by the interface, let us first consider a generic change of coordinates, written as $x^\mu\rightarrow x^\mu + \eps^\mu$. We further assume that the systems on both sides are conformally invariant, which implies the following form for the variation :

\begin{eqgroup}
    \delta S &= \int_{x<0} dxdt T^{\mu\nu}_1\pa_\mu\eps_\nu + \int_{x>0}dxdt T^{\mu\nu}_2\pa_\mu\eps_\nu\ , + \int_{x=0}d\tau D^\mu \eps_\mu\ , 
\end{eqgroup}
where for now we didn't make any assumption about symmetry properties of the interface Lagrangian. In fact, there is some abuse of notation when writing this generic variation, as generically it will deform the interface. These deformations are also encapsulated in the quantity $D^\mu$ \footnote{To be a bit more careful, one should write the interface action as $\int dx d\tau \delta(x) \cal{L}_{\rm int}$. Then variations due to the deformation of the interface are accounted for in the variations of the Dirac delta.}.

Let us now assume we are making the variation around an on-shell configuration, s.t. $\delta S=0$. After integration by parts, we obtain :
\begin{eqgroup}
    0 &= -\int_{x<0}dxd\tau\pa_\mu T_{1}^{\mu\nu}\eps_\nu -\int_{x>0}dxd\tau\pa_\mu T_1^{\mu\nu} +\int_{x=0}d\tau (D^\mu+n_\nu(T_1^{\mu\nu}-T_2^{\mu\nu}))\eps_\mu \\
    &= -\int_{x<0}dxd\tau\pa_\mu T_{1}^{\mu\nu}\eps_\nu -\int_{x>0}dxd\tau\pa_\mu T_1^{\mu\nu}\eps_\nu +\int_{x=0}d\tau (D^\mu+(T_1^{\mu x}-T_2^{\mu x}))\eps_\mu\ ,
    \label{genericEpsTransl}
\end{eqgroup}
where $n_\mu$ is the normal vector to the interface pointing away from side 1.

Now, the volume and interface integrals should vanish independently, and by the arbitrariness of $\eps_\mu$ we thus conclude to the conservation of the stress-energy tensor in each bulk, $\pa_\mu T^{\mu\nu}_i=0$.

This also gives us a relation between $D^\mu$ and the values of the stress-tensors on the interface, but we would like to refine that using the fact that the interface preserves a subset of the symmetries. To this end, we specialize the $\eps_\mu$ parameter to first represent a $\tau$ translation ($\eps_\mu= \eps \delta_\mu^\tau$). In that case, the interface Lagrangian is invariant by assumption (i.e. $D^\mu \eps_\mu =0$ in this case) so that we obtain :
\begin{eqgroup}
    0 &= \int_{x=0} d\tau \left(T_1^{xt}-T_2^{xt}\right) \\
    &=\int d\tau \left((T^1_{++}-T^1_{--})-(T^2_{++}-T^2_{--})\right)\ .
    \label{conditiontranslation}
\end{eqgroup}

In the $x,t$ coordinates it can be seen that this condition amounts to the continuity of the time-averaged energy flow across the interface. We omitted it for ease of notation, but the stress tensors in the last integral are of course to be evaluated at $(x=0,t=\tau)$.

We can also do a similar procedure with scale transformations, which also leave the interface invariant, $\eps^\mu = \eps x^\mu$. On the interface at $x=0$, this becomes simply $\eps^\mu = \eps \delta^\mu_\tau \tau$, which yields :

\begin{eqgroup}
    0 &= \int_{x=0} d\tau \tau\left((T^1_{++}-T^1_{--})-(T^2_{++}-T^2_{--})\right)\ .
    \label{conditionscale}
\end{eqgroup}

That exhausts the group of transformations that leave the interface invariant. A general solution to the conditions (\ref{conditiontranslation}), (\ref{conditionscale}) can be written as :
\begin{eqgroup}
    \left(T^1_{++}-T^1_{--}-T^2_{++}+T^2_{--}\right)=\pa_\tau^2\theta(\tau)\ ,
\end{eqgroup}
where $\theta$ is an operator on the interface, vanishing as $\tau\rightarrow \pm \infty$. However, from dimensional analysis $\theta$ will have scaling dimension $0$, so from (\ref{twopointfunction}) we will have that $\langle \theta \theta \rangle = C$ and thus $\langle \pa_\tau \theta \pa_\tau \theta \rangle = 0$. Then from more sophisticated unitarity arguments \cite{Nakayama:2012ed} it can be shown that this implies $\pa_\tau \theta = 0$ as an operator equation. 

In the end, we will take that a conformal interface is such that :
\begin{eqgroup}
    &T^2_{++}(t)-T^2_{--}(-t)=T_{++}^1(t)-T_{--}^1(-t)\ ,\\
    \Leftrightarrow &T_2(i\tau)-\bar{T}_2(-i\tau)=T_1(i\tau)-\bar{T}_1(-i\tau)\ ,
    \label{conformalinterfacecondition}
\end{eqgroup}
where we include the Wick rotation for completeness.

Without the interface, the full symmetry group of the CFT is $\rm{Virasoro\times\overline{Virasoro}}$. After the joining through the interface, we obtain an additional constraint on the stress-tensor which relates the left-moving and right-moving modes. This restriction means that we only have half as many independent modes, and the symmetry group is reduced to just one copy of $\rm{Virasoro}$. This fact is of course also reflected in the asymptotic symmetry group of the dual \cite{Bachas:2001hpy}, which we will describe in more detail later.

There are many ways to satisfy (\ref{conformalinterfacecondition}). One extreme is to require independently that $T_i(i\tau)=\bar{T}_i(-i\tau)$ for each side. This is the case of a fully reflecting interface: an incoming right-moving mode is reflected and turned into a left-moving one as it hits the interface. The other extreme is a fully transparent interface, also called "topological", characterized by $T_1(i\tau)=T_2(i\tau)$ and $\bar{T}_1(-i\tau)=\bar{T}_2(-i\tau)$.

An important universal operator associated with the interface is the "Displacement operator", denoted D. This operator will arise in the Ward identities of the stress-energy tensor, when considering conformal transformations that are broken by the interface. In that sense, it is the generator of the deformations of the interface. An easy way to derive it is to go back to the formula (\ref{genericEpsTransl}), but now specialising $\eps^\mu$ as deformations in the $x$-direction. The variation $D^\mu\eps_\mu$ will no longer be vanishing, it will reduce to $D \eps_x$ in our case where the interface is one-dimensional (writing $D^x\equiv D$). Then, following similar steps to  (\ref{conditiontranslation}) :
\begin{eqgroup}
    0&=\int d\tau \left(T_1^{xx}-T_2^{xx}+D\right)\eps_x\ ,\\
    &=\int d\tau \left(T^1_{++}-T^2_{++}+T^1_{--}-T^2_{--}+D\right)\eps_x\ ,\\
    &\Leftrightarrow D = -2(T^1_{--}-T^2_{--})\ ,
    \label{displacementdefinition}
\end{eqgroup}
where in the last line we used that $\eps_x$ is arbitrary, as well as the conformal interface condition, (\ref{conformalinterfacecondition}). This derivation clearly shows that the Displacement operator is the generator of the coordinate transformations that deform the interface. 

Note that in (\ref{displacementdefinition}), the stress-tensor appearing has not been rescaled according to (\ref{stressenergyconditions}), so the equation differ by a sign w.r.t. \cite{Meineri:2019ycm}. Rescaling the stress-tensors by $-2\pi$, and the displacement operator by $2\pi$ recovers the agreement. In what follows, the stress-tensors are normalized in the usual CFT convention (\ref{stressenergyconditions}).

 The displacement operator, while it is a genuine operator of the interface, is determined by the stress tensors on either side. Its two point function depends on an undetermined constant, $\langle D(t_1) D(t_2) \rangle = \frac{C_D}{(t_1-t_2)^2}$, which will depend on the specifics of the interface.

Let us make a last comment on a different point of view for ICFT's, through what is called the "folding trick". By applying a parity transformation on CFT$_2$, we bring both CFT's on the same side. The full system is then described by the a tensor product CFT$_1$$\otimes$ CFT$_2$ living on $x<0$, where the two CFTs are completely decoupled except at the boundary $x=0$. Thus in this picture, we deal with a tensor product CFT in $x<0$, and the interface becomes a boundary. Both formulations are equivalent\cite{Oshikawa:1996dj}.

\subsection{Reflection and transmission of energy}
\label{sec:reflectiontransmisstioncoefficients}
One of the key quantities that will interest us in ICFT will be the transport of energy across the interface. In higher dimensions, this depends on the nature of excitations incident on the interface. But in two dimensions it turns out to be an universal quantity\cite{Meineri:2019ycm,Quella:2006de}, under some mild assumptions explained below.

To properly define the transmission and reflection coefficients, we must set up a scattering experiment on the interface, and measure the transmitted and reflected energy at infinity. While there are no proper asymptotic states in a CFT, it is possible to prepare a scattering experiment as explained in reference \cite{Meineri:2019ycm}, which we briefly outline here.

First, we must define the observable; the energy received at infinity. The energy density is of course given by $T^{tt}=T_{++}(w^+)+T_{--}(w^-)$. As excitations in the CFT propagate independently in the lightlike directions it is thus natural to integrate the energy in these direction. We define

\begin{eqgroup}
\cal{E} = \int_{-\infty}^\infty dw^- T_{--}(w^-),\;\;\overline{\cal{E}} = \int_{-\infty}^\infty dw^+ T_{++}(w^+)\ .
\label{totalEnergy}
\end{eqgroup}

\begin{figure}[!h]
    \centering
    \includegraphics[width=0.4\linewidth]{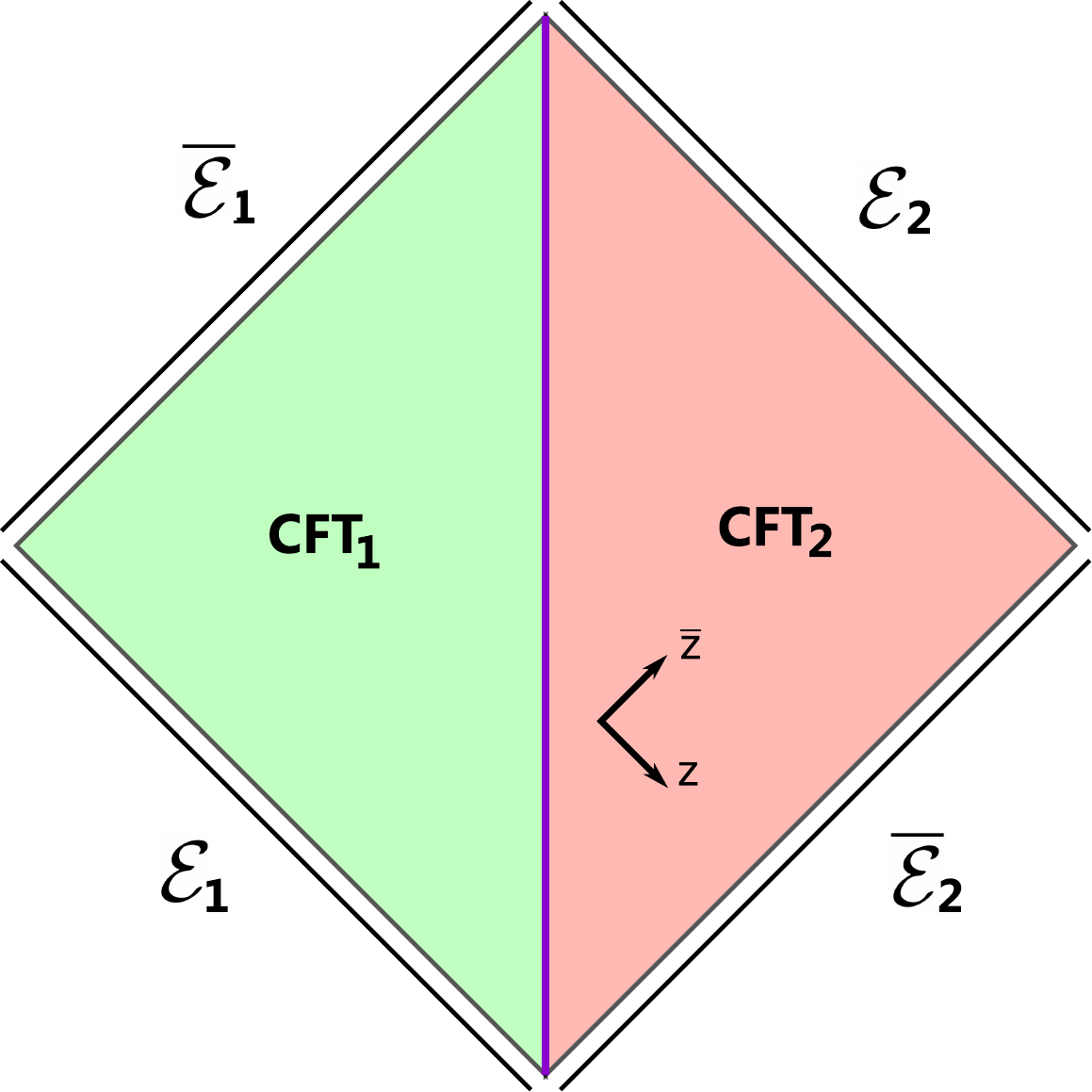}
    \caption{{\small Picture of the conformally compactified spacetime on which the ICFT lives, with the integration paths of the energies for each side. To avoid crossing the interface (in purple) the integrations of the energies must be taken to infinity. In the case where the interface is absent, the integration can be freely translated along $\bar{z}$($z$) for $\cal{E}$($\overline{\cal{E}}$). }}
    \label{fig:energiesinterface}
\end{figure}

The line over which the integration takes place is of course irrelevant for the integral, as $T_{--}$ (resp $T_{++}$) depend only on $w_-$ ($w_+$). Thus, these quantities indeed represent the total energy of the left-mover and right-movers respectively. In the interface picture, we will be able to define such operators for both CFTs, which we will label with $i=1,2$. The picture depicting the locations of the various integrations in this case is fig.\ref{fig:energiesinterface}.

We must now prepare the state that we will scatter on the interface. We would like to control its initial energy, but this is made harder by the presence of the interface. Indeed, stress-tensor interactions have only a power law decay in the conformal theory and thus the initial state will already be affected by the interface, which would make it hard to single out the scattering event to define the reflection and transmission coefficient.

The way to resolve this problem is to create the initial state infinitely away from the interface. Do to so we introduce a compactly supported kernel function $k(x)$ :
\begin{eqgroup}
    \int_{-\infty}^\infty |k(x)|^2 dx = 1,\;\; k(x)=0\;\mbox{if}\;|x|>r\ .
    \label{kernelfunc}
\end{eqgroup}

We prepare a generic state by applying any operator $O_1$ to the vacuum, and localize it using (\ref{kernelfunc}). To prepare a scattering experiment that will scatter at $x=0,t=0$, we prepare the state in the past by following back in time the lightrays intersecting this point. Instead of the lightray coordinates $w^+$, $w^-$, let us pass to the complex coordinates which make the notations more palatable (real space coordinates can be recovered by the "change of coordinates" $w^-=z$, $w^+=\bar{z}$). Our scattering state is :
\begin{eqgroup}
    |O_1,L\rangle_I = O_1^L|0\rangle_I=\int dzd\bar{z}k(z)k(\bar{z}+L) O_1(z,\bar{z})|0\rangle_I\ ,
    \label{scatteringstate}
\end{eqgroup}
where subscript $I$ on the vacuum states denotes that it is the vacuum in the presence of the interface.

The useful scattering state will then be obtained in the $L\rightarrow \infty$ limit of (\ref{scatteringstate}). In this limit, we can safely assume that the influence of the interface vanishes, and thus that the prepared state doesn't get contributions from the interface effects. In other words, in this limit we should be able to drop the subscript $I$, and consider the state as prepared in the true (i.e., no broken symmetries) vacuum of the CFT. We only need the "soft version" of this limit, meaning that we will allow ourselves to remove the index $I$ only in correlation functions. Crucially, this implies the identity $\lim_{L\rightarrow\infty}\langle O_1,L||O_1,L\rangle_I = \langle O_1,L||O_1,L\rangle$.

With that in mind, we are ready to define the reflection and transmission coefficients, which can now be defined very intuitively (especially with the help of (\ref{fig:energiesinterface})) :
\begin{eqgroup}
    \cal{T}_1 &= \lim_{L\rightarrow \infty}\frac{\langle O_1,L|\cal{E}_2|O_1,L\rangle_I}{\langle O_1,L|\cal{E}_1|O_1,L\rangle}\ ,\\
    \cal{R}_1 &= \lim_{L\rightarrow \infty}\frac{\langle O_1,L|\overline{\cal{E}}_1|O_1,L\rangle_I-\langle O_1,L|\overline{\cal{E}}_1|O_1,L\rangle}{\langle O_1,L|\cal{E}_1|O_1,L\rangle}\ ,\\
    \cal{T}_2 &= \lim_{L\rightarrow \infty}\frac{\langle O_2,L|\overline{\cal{E}}_1|O_2,L\rangle_I}{\langle O_2,L|\overline{\cal{E}}_2|O_2,L\rangle}\ ,\\
    \cal{R}_2 &= \lim_{L\rightarrow \infty}\frac{\langle O_2,L|\cal{E}_2|O_2,L\rangle_I-\langle O_2,L|\cal{E}_2|O_2,L\rangle}{\langle O_2,L|\overline{\cal{E}}_2|O_2,L\rangle}\ ,
    \label{transmissionreflectioncoeffdefinition}
\end{eqgroup}
where $\cal{T}_i$ are the transmission coefficients for excitations incident from side $i$, and $\cal{R}_i$ are the reflection coefficients. To obtain the coefficients for side $i$ the scattering state is prepared on side $i$.

From conservation of energy, we expect $\cal{T}_i+\cal{R}_i=1$, and this identity indeed holds thanks to (\ref{conformalinterfacecondition}). The proof is quite technical, and I present it here for completeness, but the reader may want to skip it since the result is intuitively obvious. We begin by defining the following correlator for ease of notation :
\begin{eqgroup}
    G_i(z) &= \langle O_1(z_1,\bar{z}_1) \ldots O_n(z_n,\bar{z_n}) T_i(z)\rangle_I\ ,\\
    \bar{G}_i(\bar{z}) &= \langle O_1(z_1,\bar{z}_1) \ldots O_n(z_n,\bar{z_n}) \bar{T}_i(\bar{z})\rangle_I\ ,
    \label{operatorG}
\end{eqgroup}
where the insertion of operators in $G_i$ are all on the $i$'th side. We will also need to re-express (\ref{conformalinterfacecondition}) for these operators :
\begin{eqgroup}
    G_1(i\tau)-\bar{G}_1(-i\tau)= G_2(i\tau)-\bar{G}_2(-i\tau)\ .
    \label{conformalconditionforG}
\end{eqgroup}

\begin{figure}[!h]
    \centering
    \includegraphics[width=0.4\linewidth]{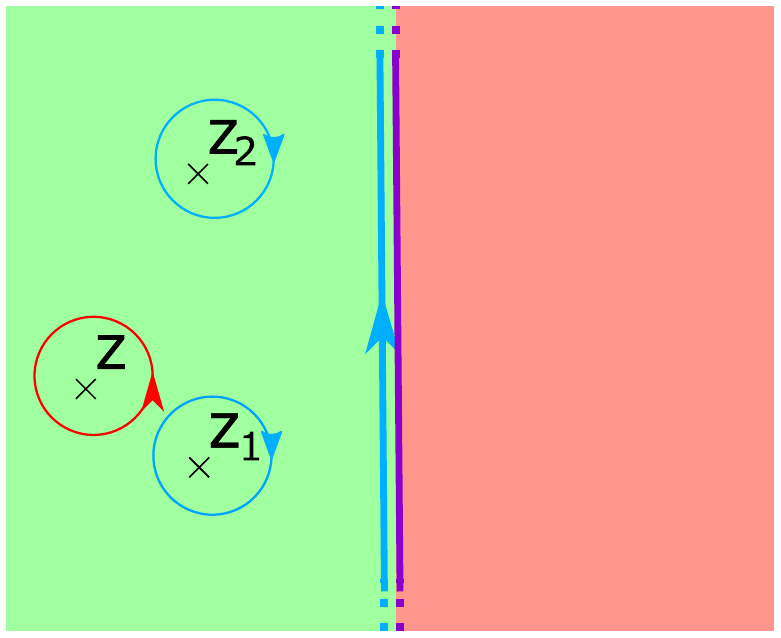}
    \caption{{\small In blue, the original contour given by Cauchy's formula for $G_1(Z)$. In red, the deformed contour of equal value. We depict here just two insertions of operators at $(z_1,z_2)$. The vertical integration is along the interface, we displaced it slightly to the right for clarity. Note that we used the absence of singularity at infinity, without which we would get an additional contribution.}}
    \label{fig:deformationcontour}
\end{figure}

By denoting $G^L_i$, the operator (\ref{operatorG}) in the case of only two insertions of scattering operators (\ref{scatteringstate}), we have :
\begin{eqgroup}
    \cal{T}_1 = \lim_{L\rightarrow \infty}\frac{\int dw G^L_2(w)}{\int dw \langle O_1,L|T_1(w)|O_1,L\rangle}\ ,
\end{eqgroup}
with similar identities for the rest of the formulas in (\ref{transmissionreflectioncoeffdefinition}). We would like to explicit the effect of the interface in the definition of (\ref{operatorG}). To do so, we will exploit the holomorphy to deform the contour in Cauchy's formula, as in fig.\ref{fig:deformationcontour}.

This yields the following formulas (let us specify to $i=1$ here):
\begin{eqgroup}
    G_1(z) &= \frac{1}{2\pi i}\oint dw \frac{G_1(w)}{w-z}=-\sum_{w=z_i} {\rm Res}\left(\frac{G_1(w)}{w-z}\right)+\frac{1}{2\pi i}\int_{w=-i\infty}^{i\infty}dw \frac{\bar{G}_1(-w)+G_2(w)-\bar{G}_2(-w)}{w-z}\ ,\\
    \bar{G}_1(\bar{z}) &= \frac{1}{2\pi i}\oint d\bar{w} \frac{G_1(\bar{w})}{\bar{w}-\bar{z}}=-\sum_{\bar{w}=\bar{z}_i} {\rm Res}\left(\frac{G_1(\bar{w})}{\bar{w}-\bar{z}}\right)+\frac{1}{2\pi i}\int_{\bar{w}=-i\infty}^{i\infty}d\bar{w} \frac{G_1(-\bar{w})-G_2(-\bar{w})+\bar{G}_2(\bar{w})}{\bar{w}-\bar{z}}\ .
    \label{masteridentity}
\end{eqgroup}

For the integral on the interface, we used the identity (\ref{conformalconditionforG}). To simplify further these integrals, we can again deform the contour for the integrands containing $G_2$. We deform the line to a circle in side $2$, since that is where $G_2$ is defined (the stress tensor $T_2$ is only defined on side $2$). However, since by assumption the insertions of the operators (\ref{operatorG}) are only on side $1$, the only singularity that can arise is from the denominator $w-z$. This yields the following identities :
\begin{eqgroup}
    &\frac{1}{2\pi i}\int_{w=-i\infty}^{i\infty}dw \frac{G_2(w)}{w-z}=0\ ,\\
    &\frac{1}{2\pi i}\int_{\bar{w}=-i\infty}^{i\infty}d\bar{w} \frac{\bar{G}_2(\bar{w})}{\bar{w}-\bar{z}}=0\ ,\\
    &\frac{1}{2\pi i}\int_{w=-i\infty}^{i\infty}dw \frac{\bar{G}_2(-w)}{w-z}=-\frac{1}{2\pi i}\int_{w=-i\infty}^{i\infty}dw \frac{\bar{G}_2(w)}{w+z}=\bar{G}_2(-z)\ ,\\
    &\frac{1}{2\pi i}\int_{\bar{w}=-i\infty}^{i\infty}d\bar{w} \frac{G_2(-\bar{w})}{\bar{w}-\bar{z}}=-\frac{1}{2\pi i}\int_{\bar{w}=-i\infty}^{i\infty}d\bar{w} \frac{G_2(\bar{w})}{\bar{w}+\bar{z}}= G_2(-\bar{z})\ ,
    \label{identityforG}
\end{eqgroup}
where (\ref{identityforG}) is easily obtained by the aforementioned assumptions of holomorphy of $G_2$($\bar{G}_2$), and by deforming the contour to a circle in side $2$. For the first two integrals, it can be shrunk to a point, while for the second we get a contribution from the pole at $-z$($-\bar{z}$). Recall that $\bar{z}$ and $z$ act as independent complex coordinates; therefore we could reparametrize an integral in $d\bar{w}$ as an integral in $dw$.

For the remaining contribution, we have to deform the contour on side $1$, and we will encounter potential singularities at every operator insertion :
\begin{eqgroup}
    \frac{1}{2\pi i}\int_{w=-i\infty}^{i\infty}dw \frac{\bar{G}_1(-w)}{w-z}=-\frac{1}{2\pi i}\int_{\bar{w}=-i\infty}^{i\infty}d\bar{w} \frac{\bar{G}_1(\bar{w})}{\bar{w}+z}=-\sum_{\bar{w}=\bar{z}_i} {\rm Res}_{\bar{z}_i}\left(\frac{\bar{G}_1(\bar{w})}{\bar{w}+z}\right)\ .
\end{eqgroup}

Plugging (\ref{identityforG}) in (\ref{masteridentity}) :
\begin{eqgroup}
    \bar{G}_1(\bar{z})+G_2(-\bar{z})&=-\sum_{\bar{w}=\bar{z}_i} {\rm Res}_{\bar{z}_i}\left(\frac{\bar{G}_1(\bar{w})}{\bar{w}-\bar{z}}\right)-\sum_{w=z_i} {\rm Res}_{z_i}\left(\frac{G_1(w)}{w+z}\right)\ ,\\
    G_1(z)+\bar{G}_2(-z)&=-\sum_{w=z_i} {\rm Res}_{z_i}\left(\frac{G_1(w)}{w-z}\right)-\sum_{\bar{w}=\bar{z}_i} {\rm Res}_{\bar{z}_i}\left(\frac{\bar{G}_1(\bar{w})}{\bar{w}+z}\right)\ .
    \label{resultofmastereq}
\end{eqgroup}

Notice the two equations are related through a complex conjugation, so we will keep only one of them from here on.
Let us restrict now to the case where in (\ref{operatorG}) there are only two insertions of the scattering operator. To obtain the residues needed in (\ref{resultofmastereq}), we employ the OPE of $O_1^L$ with the stress-tensor. This will give use some singularities in $w=z_i$, and what remains of $G_1(w)$ are correlators of the form $\langle O_1^L \pa^n O_1^L\rangle_I$. In the limit $L\rightarrow \infty$, those are far-removed from the interface we can ignore the subscript $I$. Walking backwards, with the same logic as (\ref{masteridentity}), but without the interface integral, we find :
\begin{eqgroup}
    \bar{G}_1(\bar{z})+G_2(-\bar{z})&=\int dz\langle O_1^L|\bar{T}_1(z)|O_1^L \rangle+\int dz \langle  O_1^L|T_1(-z)|O_1^L \rangle=\langle O_1^L|\cal{E}_1|O_1^L \rangle+\langle  O_1^L|\bar{\cal{E}}_1|O_1^L \rangle\ .
    \label{sum1finalstep}
\end{eqgroup}

Then, using (\ref{sum1finalstep}) it suffices to express $\cal{T}_1+\cal{R}_1$ in terms of the $G_i$, to obtain the conservation of energy identity :
\begin{equation}
    \cal{T}_1+\cal{R}_1=1\label{conservationenergy}\ .
\end{equation}

Naturally, an analogous relation exists for side 2, with an equivalent proof. That these identities hold is a consistency check of the definitions (\ref{transmissionreflectioncoeffdefinition}).

We turn now unto the proof of the universality of (\ref{transmissionreflectioncoeffdefinition}). We will only sketch it in a more restricted case, and the full proof can be found in \cite{Meineri:2019ycm}. From the conservation (\ref{conservationenergy}), we can simply focus on the transmission coefficient $\cal{T}_1$. Then, the main quantity we are interested in is the 3-point correlator of $T_2$ with scattering operators on side $1$. The easier case, to which we restrict, is that the operators $O_1^L$ are purely holomorphic. We would like to compute (\ref{computeTR}) :
\begin{eqgroup}
\langle O_1^L(z_1)T_2(z)O_1^L(z_2)\rangle_I\ .
    \label{computeTR}
\end{eqgroup}

We can do so by fusing the two defect operators using the OPE. As this is taken in the $L\rightarrow \infty$ limit, only operators from side $1$ may be produced. We will then obtain a linear combination of two-point function of the resulting operators with $T_2(z)$. However, for the two-point function of holomorphic operators, even in the presence of the interface, (\ref{twopointfunction}) still holds \cite{Meineri:2016jpp,Billo:2016cpy}. Thus only the operators of scaling dimensions $(h=2,\bar{h}=0)$ will contribute to the sought-out 3-point function. Assuming that there is no other spin 2 field than the stress-energy tensor, the only operator contributing to (\ref{computeTR}) is $T_1$.

Thus the two-point function that will determine the transmission coefficient is :
\begin{eqgroup}
    &\langle T_1(z)T_2(w)\rangle_I = \frac{c_{12}}{(z-w)^4}\ ,\\
    \Rightarrow & \langle O_1^L(z_1)T_2(z)O_1^L(z_2)\rangle_I = \alpha(z_1,z_2)\frac{c_{12}}{(z-z_1)^4}\ ,
    \label{t1t2correlator}
\end{eqgroup}
where the quantity $c_{12}$ is determined by the specifics of the interface, and $\alpha(z_1,z_2)$ comes out of the OPE fusion of the $O_1^L$. Doing the same thing for the three-point correlator with $T_1$, and taking the ratio we obtain :
\begin{eqgroup}
    \cal{T}_1 = \frac{c_{12}}{c_1}\ ,
    \label{transmissionresult}
\end{eqgroup}
where $c_1$ is the central charge of $CFT_1$. In (\ref{transmissionresult}), any specifics of the initial scattering operators are completely lost; hence this quantity is completely universal, and depends only on the properties of the interface through $c_{12}$. Of course here the proof is only for holomorphic operators, but it can be extended for arbitrary ones. The universality of the transport coefficients is the main take-away from this section.

We will limit our attention to non-chiral CFTs for which the central charge is the same for left and right movers. Repeating the procedure for a scattering coming from side $2$, we would obtain :
\begin{eqgroup}
    \cal{T}_2=\frac{c_{12}}{c_2}\ .
\end{eqgroup}

Comparing the formulas give :
\begin{eqgroup}\label{detailedbalance}
    c_1\cal{T}_1=c_2\cal{T}_2\ .
\end{eqgroup}

This important equation is known as the "detailed balance condition". It ensures that at equilibrium the net heat flow across the interface is zero, as we will see in our discussion later.

\section{The AdS/CFT correspondence}
\label{sec:adscft}
The first and most famous example of what is now known as "AdS/CFT correspondence" was discovered by Maldacena in his seminal paper \cite{maldacena_large_1999,Witten:1998zw,Gubser_1998}. It posits that "$\cal{N}=4$ $\rm{SU}(\rm{N})$ \textbf{S}uper-\textbf{Y}ang-\textbf{M}ills (SYM)" is dual to "Type IIB Strings on $AdS_5\times S^5$". The meaning of "duality" in this context is that the two theories describe exactly the same physics, with different mathematical formulations. In other words, any physical quantity that be can be computed from one theory, can also be obtained from the other one. The set of "rules" that dictate how to translate quantities from one theory to the other are usually referred to as the "holographic dictionary".  We will denote by $\al'$ the parameter governing the string tension $T=\frac{1}{2\pi\al'}$, $g_s$ the string coupling.

Let us sketch the way that this duality was first unearthed. We begin by considering the low-energy limit of $N$ stacked D3-branes ($\al'\rightarrow 0$). For the D-brane prescription to hold, we must assume that $g_s N = 4\pi \lam \ll 1$ (where $\lam$ is the 't Hooft coupling, see (\ref{holoconstantsdictio})), such that their back-reaction on the geometry is negligible. From the point of view of String theory in flat space, we can show that the only excitations that remain in this limit are the massless closed strings, which contribute to a $10$D supergravity sector, and the open strings ending on the $N$ branes, giving rise to the $SU(N)$ SYM sector.

Consider now the alternative point of view of strings on the black hole background generated by the stack of $N$ D3-branes, which is the picture that holds when the backreaction is large, $g_s N\gg 1$. In the same low energy limit (as observed by an asymptotic observer) we recover again a $10$D Supergravity sector (from the massless strings with low energy that don't see the black hole). There is also a second sector of strings propagating in the near-horizon region of the black hole, which has the geometry of $AdS_5\times S^5$. Because of the infinite redshift at the horizon, independently of their energy these excitations will look soft to the asymptotic observer at infinity. We thus conclude that this sector is the full String theory on an $AdS_5\times S^5$ background. 

Then, by "canceling" the 10D supergravity sectors of the two points of view, and since the SYM theory (living on the branes) should remain well-defined at any $\lam$, we can assume that this description still applies when $g_s N\gg1$, allowing us to identify it with String theory on AdS$_5\times S^5$.

Although this derivation heuristically establishes the equivalence of the two theories, we get only a small piece of the dictionary. We can see that the isometry group of the background, $SO(2,4)\times SO(6)$ is mapped one-to-one to the bosonic subgroup of the superconformal group; the $3+1$-dimensional conformal symmetry is isomorphic to $SO(2,4)$, and the $SU(4)$ R-symmetry is identified with the $SO(6)$. The more precise statement is about the equivalence of the asymptotic isometries $SO(6)$ of the 5-sphere. 

Still at the ground level, let us look at how the parameters of the two theories are related. From type IIB on AdS$_5\times S^5$ we have : the string coupling $g_s$, the string tension $\al'$ and the AdS radius $\ell$. From $\cal{N}=4$ $SU(N)$ : the size of the gauge-group $N$ and the Yang-Mills coupling $g_{YM}$. From Maldacena's derivation, those are found to be mapped as :
\begin{eqgroup}
    \sqrt{\lam} &= \sqrt{g_{YM}^2 N} = \frac{\ell^2}{\al'}\ ,\\
    4\pi g_s &= \frac{\lam}{N}\ ,\\
    \frac{\ell}{l_p} &= N^2\ ,
    \label{holoconstantsdictio}
\end{eqgroup}
where we introduced the 't Hooft coupling $\lam$, which is the relevant coupling in the limit of large $N$. As we will not enter in the details of the large N limit of field theories, the interested reader can look at \cite{tHooft:1973alw,Moshe_2003}.

The string-theory side of (\ref{holoconstantsdictio}), is well-understood only in the classical Supergravity limit, $g_s\ll 1$, $\ell^2\gg \alpha'$. This corresponds to the $\lam\gg 1$, $N\gg \lam$ limit of the field theory, which is a large $N$ strongly coupled limit. On the other hand, the weak 't Hooft coupling limit of the field theory, $\lam\ll 1$, corresponds to String theory on an AdS$_5$ background where the $\al'$ corrections are large. The duality thus interchanges weak with strong coupling on the two sides. 

To verify the duality, which is believed to hold for all ranges of parameters, one must thus be able to perform strong-coupling calculations on one of the sides. This is possible for some protected observables which do not receive quantum corrections thanks to supersymmetry. In the planar $N\rightarrow \infty$ limit, more progress could actually be achieved by the exact solution of the spectrum for all values of $\lam$\cite{Gromov:2009tv}, confirming the validity of the duality.

Conversely, assuming the hypothesis to be correct, the duality gives us a way to define string theory non-pertubatively on AdS spacetimes, a formulation that is still completely lacking in flat spacetime.


Although Maldacena's derivation gave a compelling argument for the duality, it did not outline how to use it. In other words, there was still no full "dictionary" that explained how to match quantities in both theories, besides the symmetry groups and parameters. This was explained in the papers \cite{Gubser_1998,wittenAdSHolo}. In a nutshell, the computation of correlation functions in the dual CFT is beautifully encapsulated by the formula (\ref{GKPWprescription})
\begin{eqgroup}
    \langle \exp(-\int d^Dx \phi_0(x)\cal{O}(x))\rangle_{CFT}=\cal{Z}_{\rm{string}}\left[\phi(\vec{x},z)\Bigr|_{z=0}=\phi_0(\vec{x})\right]\ .
    \label{GKPWprescription}
\end{eqgroup}
There is a lot to unpack in this formula. Let us first look at the LHS. This is the generating functional for correlation functions of the operator $\cal{O}$. By taking functional derivatives w.r.t. $\phi_0$, we can compute any correlation function containing $\cal{O}$. 
\begin{eqgroup}
    \langle O(x_1)O(x_2)\ldots\rangle_{CFT} = \frac{\delta}{\delta (-\phi_0(x_1))}\frac{\delta}{\delta (-\phi_0(x_2))}\ldots\left(\langle \exp(-\int d^Dx \phi_0(x)\cal{O}(x))\rangle_{CFT}\right)\Bigr|_{\phi_0=0}\ .
    \label{functionalderivativecomputecorrelator}
\end{eqgroup}
On the RHS, $\cal{Z}_{\rm string}$ denotes the partition function of the string theory on the asymptotically AdS background. It has a boundary condition on the field $\phi(x,z)$; it should match $\phi_0$ at the boundary of AdS which we assume to be located at $z=0$. Thus, for each field propagating in the String theory (examples are the dilaton, the graviton, or any other possible excitation of the string) there will be a corresponding operator in the dual CFT. 

Even defining $\cal{Z}_{\rm string}$ generically is a very difficult task. As we have already stated, computations on the string side are almost always done in the Supergravity limit. In this limit, we can use the saddle point approximation to find $\cal{Z}_{\rm string}$. Schematically, in the Euclidean picture :

\begin{eqgroup}
\cal{Z}_{\rm{string}}\left[\phi(\vec{x},z)\Bigr|_{z=0}=\phi_0(\vec{x})\right]=\int_{\phi(\vec{x},z)\Bigr|_{\mathclap{z=0}}=\phi_0(\vec{x})}\cal{D}\phi e^{-S_E}\approx e^{-S^c_E(\phi)}\int \cal{D}\phi e^{-\frac{\delta^2 S}{(\delta \phi)^2}(\delta\phi)^2+\cal{O}(\delta\phi^3)}
    \label{pathintegralholographic}\ ,
\end{eqgroup}
where $S^c_E(\phi)$ denotes the Euclidean action evaluated on the classical solution that satisfies the appropriate boundary conditions. Here we lumped all the fields into $\phi$, thus among them, there should always be the metric $g_{\mu\nu}$ as we are considering a string theory. This is the main reason the definition of the path integral (\ref{pathintegralholographic}) is hard. The saddle point integration gives the leading order contribution in the small $g_s$ (large $N$) expansion. The $(\delta\phi)^2$ term in the exponential accounts for the leading quantum corrections to the saddle point approximation.

Even at the classical level, computing this quantity is not straightforward as it is generally divergent as we take the limit $z\rightarrow 0$, and requires regularisation and the addition of counterterms to make sense of it.  This procedure is called "holographic renormalization"\cite{Skenderis_2002,Skenderis:1999nb}. Indeed, these IR divergences that arise in the computations of the action are the holographic manifestation of the UV divergences of the dual field theory. This highlights another key aspect of the AdS/CFT correspondence, the "UV/IR connection"\cite{susskinduvirconnection}, namely that UV effects in one theory are realized as IR effects in the other, and vice-versa.

One last interesting remark about (\ref{GKPWprescription}) is that it lends itself to the interpretation that the CFT leaves on the physical boundary of AdS. In other words, we view the gravity theory as living in the "bulk" or "volume" of spacetime, and the field theory as living on its boundary. Whether this interpretation has any real "physical" signification is up for debate. Nonetheless, this point of view is very useful for visualization, and we will use the associated jargon extensively.

Let us conclude by illustrating (\ref{GKPWprescription}) with the basic example of a scalar field. By solving the wave equation in pure AdS  (since what matters to us is the asymptotic behavior, where spacetime always looks like empty AdS), one finds that the propagating field $\phi$ has two independent modes with the following leading behaviours on the boundary (in units $8\pi G=1$):
\begin{eqgroup}
    \phi(x,z)= (A(x)+\cal{O}(z^2)) z^{\Delta}+(B(x)+\cal{O}(z^2)) z^{D-\Delta},\;\; \Delta = \frac{D}{2}+\sqrt{\frac{D^2}{4}+\ell^2 m^2}\ .
\end{eqgroup}
The subsequent $\cal{O}(z^2)$ corrections are algebraically determined from the equation of motion in terms of $A(x)$ and $B(x)$\footnote{There is an exception in the case where $\Delta$ is an integer. Then, there is a $\log{z^2}$ correction to $A(x)$. This coefficient is related to the conformal anomaly in the field theory, but we omit those details here.}.

We will assume $d<2\Delta$, which is the Breitenlohner-Freedman bound \cite{Breitenlohner:1982bm} necessary for the scalar field not to destabilize AdS. Note that this allows for tachyonic fields with $m^2<0$ which is not the case in flat space. Above this bound, the dominant solution at $z\rightarrow 0$ is the $B(x)$ mode, which is the one that will be matched to $\phi_0(x)$. To do so we must regularise, and we do so by putting a cut-off at $z=\eps$. We see then that for the boundary condition to be satisfiable, we should replace $\phi_0(x)\rightarrow \lim_{\eps\rightarrow 0}\eps^{D-\Delta}\phi_0(x)$. Then the boundary condition simply states $B(x)=\phi_0(x)$. Note that $A(x)$ is still undetermined. In the Euclidean theory it is determined by requiring regularity of the solution in the interior, while in the Lorentzian theory there could be additional freedom that corresponds to the choice of the boundary state. . 

If we go through all the steps to renormalize the action and compute $S^c_E(\phi)$, we can then compute the one-point function of the associated operator \cite{de_Haro_2001} with (\ref{functionalderivativecomputecorrelator}) and (\ref{pathintegralholographic}):
\begin{eqgroup}
    \langle O(x)\rangle = \frac{\delta S^c_E}{\delta \phi_0(x)}=-(2\Delta -D)A(x)+C(B(x))\label{1-pointexpectationvalueholo}\ ,
\end{eqgroup}
where $C(B(X))$ is a regularisation-scheme-dependent term. Note that we swept under the rug most of the computation, the point here being only to underline the fact that the second free coefficient $A(x)$ determines the expectation value of the operator that is dual to the scalar field.

\section{The bottom-up approach}\label{sec:bottomupapproach}
In the previous section, we presented a lightning overview of the AdS/CFT correspondence in a putative exact or "top-down" holographic setting. Here the correspondence is posited between UV complete gravity theory and a precise dual (S)CFT. While there are numerous such examples in string theory, these are often cumbersome to work with. Furthermore one would like to work in a more general framework, not restricted to a precise holographic setting. 

Such an approach is coined "bottom-up" holography. The basic idea is to take any AdS gravity model, with a chosen set of fields tailored to what we want to study. Then, we can identify with the holographic dictionary a putative CFT dual to our theory for which we can do computations in the gravity side. The putative dual CFT, if it exists, would most likely be strongly coupled. This is approach is useful in condensed matter physics\cite{Erdmenger:2020fqe,Zaanen:2015oix,Hartnoll:2009sz,Adams:2012th,Liu:2018crr,Sachdev:2010ch,Erdmenger:2018xqz}, where the microscopic UV Hamiltonian is rarely known. There is an alternative denomination in this case, "AdS/CMT"(AdS/Condensed Matter Theory).

The bottom-up holography relies on much flimsier ground than the precise correspondence that we outlined in the previous section. Indeed, we implicitly assume that our hand-picked model is realized as an effective theory of some UV complete quantum gravity with a holographic dual. In other words, it somehow assumes that given any CFT or AdS gravity theory, one could embed it into a precise correspondence of UV complete theories. More modestly one assumes that the bottom-up theory retains some qualitative features of strongly coupled CFTs that are hard to compute ab initio. An example are strange non-Fermi liquid phases of matter\cite{Chowdhury:2021qpy}.

\subsection{Minimal holography}
\label{sec:minimalholoandrenorm}
In the class of "bottom-up" approaches, one stands out from the others as the most general one. In this model, which we will call "minimal holography", we allow only the bare minimum needed to be able to formulate an AdS/CFT duality. While the resulting model is not very rich, it makes up for it in generality. Indeed the minimality of the model will also mean that results derived within it will be applicable universally in a broad class of holographic models. Furthermore, we will consider the appropriate limit so that computations in the bulk reduce to classical gravity, namely large $N$ (large central charge) and strong coupling for the CFT. Although we didn't review the large $N$ limit, we will need only the fact that it can be seen as a "classical" limit. By that we mean that correlators of reasonable\footnote{By "reasonable", we mainly mean operators which are not composed of a parametrically large number of fields. If this number scales parametrically with $N$, the large-N limit cannot be easily taken and the "classical" limit doesn't apply anymore.} operators will factorize, $\langle A B\rangle = \langle A \rangle \langle B \rangle +\cal{O}(1/N)$\cite{Yaffe:1981vf}. Thus all we will need to define our state completely will be the expectation value of the operators.

In fact we will restrict ourselves to a single operator whic is always present in any local CFT, the Stress tensor $T_{\mu\nu}$. In a theory living in flat space, the only choices left to us are the CFT central charge $c$ and its state defined by choosing the expectation values $\langle T_{\mu\nu}\rangle$. Of course, these must respect the constraints imposed by the conformal invariance. In the 2-dimensional theories in which we work, these constraints are most easily expressed in lightcone coordinates, see (\ref{stressenergyconditions}) :
\begin{eqgroup}
    \langle T_{+-}\rangle = 0,\;\; \langle T_{++}\rangle = \langle T_{++}\rangle(w_+),\;\; \langle T_{--}\rangle = \langle T_{--}\rangle(w_-)
    \label{lightconestressenergyconditions}\ .
\end{eqgroup}
As an example, the vacuum state will have a vanishing stress-tensor on the plane, or carry the Casimir energy (\ref{energymatching}) on the cylinder.

From the bulk point of view, the situation is equally simple. The only necessary field for a gravity theory to be defined is naturally the graviton. There will of course be a negative cosmological constant, and its value will be set to satisfy the Brown-Henneaux formula (\ref{brownhenneauxcentralcharge}). The action will be given by (\ref{fullEinsteinHilbert}) in Euclidean space, and by the Wick-rotated version in Lorentzian signature. As usual the metric will be the field dual to $T_{\mu\nu}$ which is the unique spin-two field in our simple theory.

To obtain the precise matching between the metric and the stress-tensor, we must follow the GKPW procedure. The first step is finding the general solution to the Einstein equations, with the constraint that the metric be Asymptotically AdS. We thus use the Fefferman-Graham gauge (\ref{FGmetric}) in which the constraints are explicit. In two dimensions, it can be shown that the generic solution to the equation of motions takes the form \cite{de_Haro_2001} :

\begin{eqgroup}
    g_{ij}(z,x) = \frac{\ell^2}{z^2}\left(g^{(0)}_{ij}+z^2 g^{(2)}_{ij}+\frac{z^4}{4}g^{(2)}_{ik}g_{(0)}^{kl}g^{(2)}_{lj}\right)\ .
    \label{generalsolutioneinsteineq}
\end{eqgroup}

$g_{ij}^{(0)}$ is immediately fixed to be equal to the metric of the manifold on which the CFT lives, flat in our case. Thus, as we expected from the general procedure, there remains one degree of freedom to fix in $g^{(2)}_{ij}$. Einstein's equations fix the general form (\ref{g2fixed}).

\begin{eqgroup}
g^{(2)}_{ij}&=\frac{1}{2}\left(R^{(0)}g^{(0)}_{ij}+ a_{ij}\right)\ ,\\
&\co^{(0)}_i a^{ij}=0,\;\;a_i^i=- R^{(0)}\ .
\label{g2fixed}
\end{eqgroup}

In our case, $R^{(0)}=0$ since the $g^{(0)}$ metric is flat. Going through the steps to compute (\ref{1-pointexpectationvalueholo})  we find, as should be expected by the suggesting constraints of (\ref{g2fixed}) :
\begin{eqgroup}
    a_{ij} = 2\frac{\langle T_{ij}\rangle}{\ell}\ .
    \label{g2value}
\end{eqgroup}
We see that the constraints on $a_{ij}$ are the same as those imposed on the stress-tensor by conformal invariance. In particular, we recover the conformal anomaly in the case where the CFT is put on a curved manifold :
\begin{eqgroup}
    \langle T_i^i \rangle &= -\frac{\ell}{2}R^{(0)}=-\frac{c}{24}{R^{(0)}}\ .
\end{eqgroup}

In the special case where $g^{(0)}$ is flat, we can then explicitly write the full bulk metric as follows :
\begin{eqgroup}
 ds^2 = \frac{\ell^2 dz^2}{z^2}+\frac{\ell^2}{z^2}\left(dx^-+z^2 \frac{\langle T_{++}\rangle}{\ell} dx^+\right)\left(dx^++z^2 \frac{\langle T_{--}\rangle}{\ell} dx^-\right)\ .
 \label{flatfeffermangraham}
\end{eqgroup}

\subsection{An extra interface}
\label{sec:MinimalICFT}
The previous section describes the minimal holographic model valid for the calculation of energy-momentum correlators in any homogeneous CFT. We would like now to slightly expand this model to include CFTs with an interface, as described in section \ref{sec:interfaces}. 

We now have to deal with two CFTs (denoted by numbers $1$, $2$) which are joined on an interface. For each CFT$_i$, the procedure from the previous section goes through unchanged, so we have two copies of each equation. As for the interface, in the minimal case it will be characterized by a single $\lambda$ (the tension of the dual brane, as we will see). This parameter will determine the value of the reflection and transmission coefficients\cite{Bachas:2020yxv}, defined in (\ref{transmissionreflectioncoeffdefinition}), as well as the entropy or g-factor\cite{Karch:2021qhd,Affleck:1991tk,Simidzija:2020ukv} of the interface. This is a special feature of our model, more generally the transport coefficients and entropy are independent quantities.

The displacement operator $D$, defined in (\ref{displacementdefinition}) is also a universal operator, which is however completely fixed from the choice of the state for the CFTs on either side.

For the holographic gravity dual of this minimal model, the derivation is a bit more heuristic. We know from (\ref{flatfeffermangraham}) that the metric close to the boundary will change as we cross the interface. Thus the metric dual to this configuration should be perturbed by the insertion of some object (dual to the interface) such that it modifies Einstein's equations to allow for the metric transition. In the UV-complete theory of gravity, this transition is expected to be smooth. But at low energies, when its internal structure cannot be resolved, the domain wall can be considered as infinitely thin.

We thus postulate the "thin membrane" approximation \cite{Coleman:1967ad,Coleman:1980aw}, in which the transition between the two metrics happens sharply along a membrane of codimension one, parametrized as $x^\mu(y^a)$. Naturally, this membrane should be anchored at the position of the CFT junction on the boundary, since we know from (\ref{flatfeffermangraham}) there is a sudden (although continuous) asymptotic metric change at that junction. To describe the dynamics of this object, we need to add a term to the Einstein-Hilbert action of the system. The minimal and most general way to do this is to simply imbue the membrane with a tension $\lam$. The term describing its dynamics is then :
\begin{eqgroup}
    S_{\rm{mem}} &= -\lam\int_{\rm Mem}d^{(D-1)}y\sqrt{-h}\ ,\\
    h_{ab} &= g_{\mu\nu}\frac{\pa x^\mu}{\pa y^a}\frac{\pa x^\nu}{\pa y^b}\ .
    \label{membraneaction}
\end{eqgroup}

So the number $\lam$ determining the minimal properties of the CFT interface acquires a much clearer interpretation in the bulk as the tension of a membrane. One might ask what is the dual of the displacement operator $\cal{D}$, and this is naturally encoded in the metric $h_{ab}$ of the wall. Since it is itself determined by $x^\mu_m(y^a)$ we can also see it as encoded in the shape of the membrane. As $\cal{D}$ was completely determined given the state of both CFTs, one should expect that to also be the case for the dual object, $x^\mu(y^a)$. As we will see in the main chapters, this expectation is true and the equations arising from (\ref{membraneaction}) added to the Einstein equations will completely fix the shape of the membrane. 

Let us finish by mentioning that one can augment the model, either by adding fields in the bulk and their dual operators, but also by adding fields restricted to the membrane, which should correspond to operators on the interface, from the CFT side (see \cite{Chen:2020uac,Chen:2020hmv} for an application of these more complicated walls).

\section{Entanglement entropy and the Ryu-Takanayagi prescription}
\label{sec:entanglement}
While presenting the holographic correspondence, we did not yet talks about a key ingredient that was the first clue to the holographic nature of gravity: entropy, and more precisely, its quantum counterpart, the fine-grained or "Von Neumann" entropy. 

By trying to save the second law of thermodynamics in their study of black holes, Hawking and Bekenstein \cite{bekensteinbound,bekensteinformula,Hawking:1976de} soon realised that the entropy of a black hole had to be curiously proportional to its area, instead of its volume as it usually is for ordinary objects. The precise Bekenstein-Hawking formula formula reads :
\begin{eqgroup}
    S_{\rm BH} = \frac{\cal{A}}{4G}\ ,
    \label{bekensteinhawkingformula}
\end{eqgroup}
where $\cal{A}$ is the black hole's horizon area.

This strongly suggests that the degrees of freedom of the black hole are localized on its surface. It is then natural to expect that this may hold more generally, since any system can be collapsed to a black hole if made compact enough.

If such a generalization of (\ref{bekensteinhawkingformula}) exists, it should come about naturally from AdS/CFT; this was indeed the ingenious conjecture of Ryu and Takayanagi.
\subsection{Entanglement entropy}
\label{sec:entanglemententropy}
Let us first recall the definition of von Neumann entropy. Consider a quantum system described by a density matrix $\rho$. For a "pure state", which can be described by a vector $|\psi\rangle$ in the hilbert space, the density matrix is $\rho = |\psi\rangle \langle \psi |$. However, it can also account for more general, "mixed" states, in which case the density matrix can be put in the general form :
\begin{equation}
    \rho = \sum_i p_i |\psi_i\rangle \langle \psi_i|,\mbox{ where }{\rm Tr}(\rho)=\sum_i p_i =1\ .
    \label{densitymatrixdef}
\end{equation} 
Such a probability mixture of quantum states arises because of a lack of knowledge on the state. For instance, for a thermal state, we would have :
\begin{eqgroup}
    \rho = \sum_{i}e^{-\beta E_i}|\psi_i\rangle \langle \psi_i|\ ,
    \label{thermaldensitymatrix}
\end{eqgroup}
where our ignorance comes from the fact we don't know the precise microscopic state of the system. Still, the knowledge of the macroscopic (inverse) temperature $\beta$ allows us to assign probabilities to the possible microstates, which is encoded in (\ref{thermaldensitymatrix}). We would like to define an "entropy", which quantifies our ignorance about the state of the system. Inspired from Shannon's entropy that does the same thing for probability distributions, we define the Von Neumann entropy, which we will also call "fine-grained entropy", and "entanglement entropy" when appropriate :
\begin{eqgroup}
    S = -{\rm Tr}(\rho \log \rho) \left( = -\sum_i p_i \ln(p_i)\right)\ .
    \label{entanglemententropy}
\end{eqgroup}

The way we introduced (\ref{entanglemententropy}) looks for now completely unrelated to entanglement. To make the connection clear, we introduce the purification of a density matrix. 

Consider a generic density matrix as in (\ref{densitymatrixdef}), constructed from states from the Hilbert space $H_A$. Then, introduce another arbitrary Hilbert space $H_B$ that we will use to "purify" $\rho$. For this purpose it is sufficient that $H_B$ be of the same dimensionality as $H_A$, but for any specific $\rho$ a much smaller Hilbert space might suffice. Then we define the pure density matrix in $H_A\otimes H_B$ :
\begin{eqgroup}
    \rho_{AB} = |\Psi_{AB}\rangle \langle \Psi_{AB}|,\mbox{ where }|\Psi_{AB}\rangle=\sum_i \sqrt{p_i}|\psi_i\rangle |b_i\rangle
    \label{purificationofrho}\ ,
\end{eqgroup}
where the $|b_i\rangle$ are vectors of $H_B$.

Then, $\rho_{AB}$ is called the purification of $\rho$. The state is obviously pure, and one can easily show that we recover the original density matrix by tracing on $H_B$, i.e. $\rho = {\rm Tr}_{H_B}\rho_{AB}$. Notice that $\rho_{AB}$ is not defined uniquely; in fact given any matrix $U$ s.t. $U^\dagger U =I$, the state $|\Psi_{AB}\rangle=\sum_i\sqrt{p_i}|\psi_i\rangle U|b_i\rangle$ is a valid purification. One can show that all purifications can be written in this form.

It is from here that we can justify the name "entanglement entropy" for $S$. Indeed, it can be seen as a measure of the amount of entanglement between the two systems $A$ and $B$ described by $\rho_{AB}$. When one traces out the degrees of freedom of $B$, one loses the information that was contained in the entanglement between the two systems, and we end up with a mixed density matrix of non-zero entanglement entropy.

As an easy check, for a pure matrix $\rho$ we get $S=0$, and for finite Hilbert spaces $S$ is maximized when $\rho$ is an uniform mixture of all states in $H_A$. The purification is then given by a state $|\Psi_{AB}\rangle$ which is maximally entangled.

Thus, we must be wary of what quantity does the Von Neumann entropy compute for us. If we consider a pure state, and compute the Von Neumann entropy of a subsystem, then that will in fact tell us about the amount of entanglement it had with the rest of the system. However, if we repeat the same procedure on an initially mixed state, there will be an additional contribution coming from our initial ignorance, unrelated to entanglement. While it is true that $S$ can always be viewed as measuring entanglement through a purification (\ref{purificationofrho}), one must keep in mind if the purifying system is physical, or merely a mathematical trick\footnote{There are ongoing efforts to find a quantity that can distinguish classical correlations from quantum entanglement, one example being the "negativity"\cite{Vidal:2002zz}}.

In practice, computing $S$ directly in QFT is often hopeless, in part because of the need to compute the logarithm of the density matrix, for which we need knowledge of the eigenvalues of $\rho$. An alternative way to compute it is as a limit of "Renyi entropies" $S^{(n)}$:
\begin{eqgroup}
     &S^{(n)}= \frac{1}{1-n}\ln\left({\rm Tr}(\rho^n)\right)\\
     &S = \lim_{n\rightarrow 1}S^{(n)}\ .
    \label{SaslimitofSn}
\end{eqgroup}

Of course, the limit in (\ref{SaslimitofSn}) is at best ill-defined, since $S^{(n)}$ is formally defined only for integer $n$. Underlying this procedure then there is a notion of analytic extension of $S^{(n)}$, whose legitimacy we will not attempt to justify. For functions specified on the integers, Carlson's theorem provides necessary conditions for the existence of a unique extension \cite{CarlsonSurUC}.

Before proceeding, let us mention that in QFT the entanglement entropy will always be a UV-divergent quantity. This is simply understood from the fact that there are degrees of freedom at every scale, and in particular at arbitrarily small scales. So, as we pick out a region $A$ of a QFT, there will be entangled pairs at arbitrarily small scales across the interface $\pa A$ of the region. We expect then a UV-divergent entanglement entropy proportional to the area of this interface, i.e. in a QFT in $D=d+1$ dimensions \cite{Eisert:2008ur} :
\begin{eqgroup}
 S_A = \frac{{\rm Area}(\pa A)}{\epsilon^{d-1}}+\ldots\ ,
\end{eqgroup}
where $\epsilon$ is a UV-cutoff. Thus when we speak about entanglement entropy of QFT and CFT's in what follows, we implicitly assume that they are regulated by a UV-cutoff. For $d=1$, the formula is ill-defined, and one finds a divergence in $\log(\eps)$.

The technical way in which $S^{(n)}$ is calculated in QFT is called the replica trick. Consider the Hilbert space $H$ within which $\rho$ is defined. The replica trick involves introducing an extended Hilbert space $H^{\otimes n}$, composed of $n$ "replicas" of $H$. We will show that $\rho^n$ can be obtained by computing a partition function in this extended Hilbert space.

To explain this method, we need first to understand how is the density matrix defined in QFT. The usual way to define a state in QFT is by preparing it with the help of an Euclidean path integral. Consider for simplicity a QFT with a single scalar field $\phi$. Then, a basis for the Hilbert space is $|\phi(x)\rangle$ where $\phi(x)$ is an arbitrary function of the spatial coordinates. One can define a state $|\Psi\rangle$ simply by specifying $\langle \phi(x)|\Psi\rangle$ for any $\phi(x)$. This can be done through the path integral by writing $|\Psi\rangle$ as an Euclidean evolution of a state $|\phi_1(x)\rangle$ over an arbitrary geometry  :
\begin{eqgroup}
     |\Psi\rangle = e^{-\beta H}|\phi_1(x)\rangle = \int^{\phi(\tau=\beta)=??}_{\phi(\tau=0)=\phi_1} \cal{D}\phi e^{-S_E(\phi)}
     \label{statedefinitioneuclidean}\ .
\end{eqgroup}

The "$??$" notation simply shows that the upper limit of the integral is unspecified. With (\ref{statedefinitioneuclidean}), overlaps can be computed naturally as :
\begin{eqgroup}
     \langle \phi_2|\Psi\rangle = \int^{\phi(\tau=\beta,x)=\phi_2}_{\phi(\tau=0,x)=\phi_1} \cal{D}\phi e^{-S_E(\phi)}
     \label{phicontractedwithPsi}\ .
\end{eqgroup}

In fact, (\ref{phicontractedwithPsi}) defines for us the wavefunction $\Psi(\phi_2)$.

\begin{figure}[!h]
    \centering
    \includegraphics[width=0.4\linewidth]{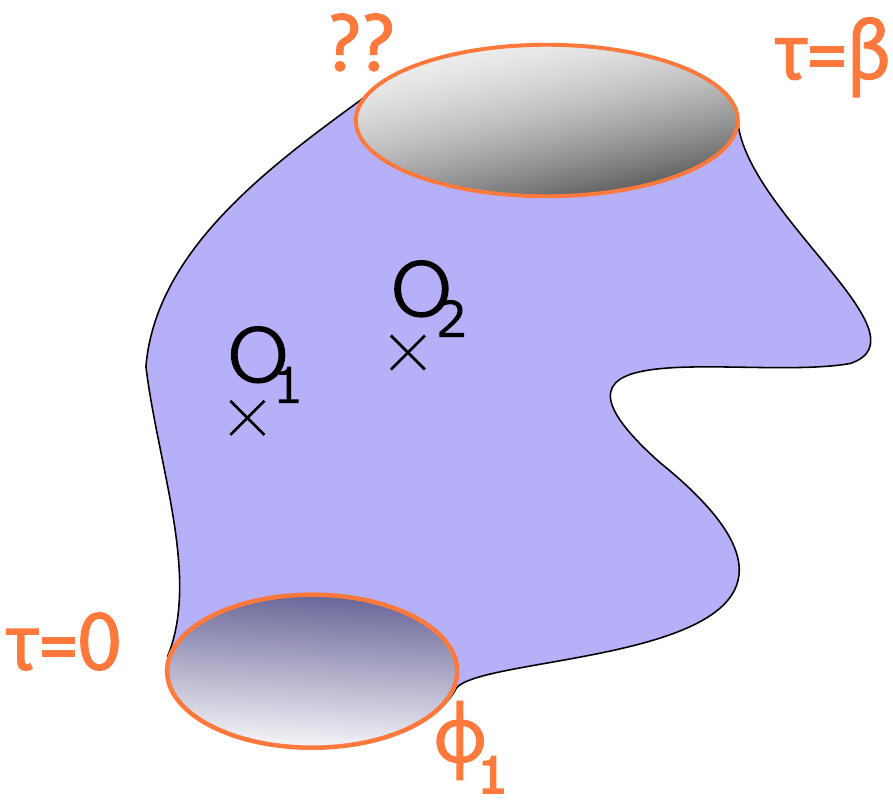}
    \caption{{\small Cartoon of an euclidean manifold which we may use to prepare a state. It contains two boundaries at $\tau=0,\beta$ (in orange) on which boundary conditions for the fields $\phi$ must be specified. The $\times$ denotes insertions of operators, which can be included to alter the prepared state.}}
    \label{fig:cartooneuclideanpathintegral}
\end{figure}

To choose a specific state, we can modify the geometry of the Euclidean manifold on which the path integral (\ref{statedefinitioneuclidean}) is computed, as well as change the boundary condition $\phi_1$. In addition to that, one may also include operator insertions on this manifold, i.e. one might add operators in the path integral (\ref{statedefinitioneuclidean}). A cartoon depicting this state preparation procedure is fig.\ref{fig:cartooneuclideanpathintegral}.

Although the procedure (\ref{statedefinitioneuclidean}) might seem strange, it turns out to be very useful to produce some of the widely used states.

One example is the vacuum state, which can be obtained by making an infinite time evolution from $\tau = -\infty$ to $\tau = 0$. As for the boundary condition at $\tau=-\infty$, it is irrelevant as long as we choose a state which has non-zero overlap with the vacuum state. Indeed, under Euclidean time evolution the coefficients of the energy eigenstates will be suppressed by $e^{-\tau E}$, and after an infinite time only the vacuum state will remain :
\begin{eqgroup}
    |0\rangle = \int^{\phi(\tau=0)=??}_{\phi(\tau=-\infty)=\phi_1} \cal{D}\phi e^{-S_E(\phi)}\ .
\end{eqgroup}

To pass from states to density matrices, there is but a step. Formally a density matrix will contract with a bra and ket, and spit out a number. Thus analogously to (\ref{statedefinitioneuclidean}), it can be defined by a euclidean path integral with two free boundary conditions on each side. Here, the most famous example is the thermal density matrix :
\begin{eqgroup}
    \rho=\frac{1}{Z}e^{-\beta H}=\frac{1}{Z}\int^{\phi(\tau=\beta/2)=??}_{\phi(\tau=-\beta/2)=??}\cal{D}\phi e^{-S_E(\phi)} 
    \label{thermaldensitypathintegral}\ .
\end{eqgroup}

From here on we will take (\ref{thermaldensitypathintegral}) as the prototypical example, but everything we will say applies for any density matrix defined with the Euclidean path integral method. $Z$ is such that the density matrix has unit trace, see (\ref{donutpartitionfunction}).

Product of density matrices are then computed straightforwardly. Pictorially, we glue the two manifolds at the free boundaries, combining them into one (bigger) path integral. For instance, $\rho^2$ would be the same integral as (\ref{thermaldensitypathintegral}), but on a time interval of $2\beta$. Finally, taking the trace is also relatively straightforward. Morally it is simply ${\rm Tr}\rho = \sum_i \langle \phi_i|\rho|\phi_i\rangle$, which from (\ref{thermaldensitypathintegral}) is the path integral with equal initial and final boundary condition, summed over all possible boundary conditions. This is equivalent to imposing periodic time on the euclidean manifold on which we do the path integral. Thus, the trace of $\rho$ can also be expressed as a path integral, on the compactified manifold in the $\tau$ direction.

As an example, for a theory defined on a spatial circle, the thermal partition function can be depicted as a path integral on a torus:
\begin{eqgroup}
    Z(\beta)={\rm Tr}e^{-\beta H}=\vcenter{\hbox{\includegraphics[width=0.2\linewidth]{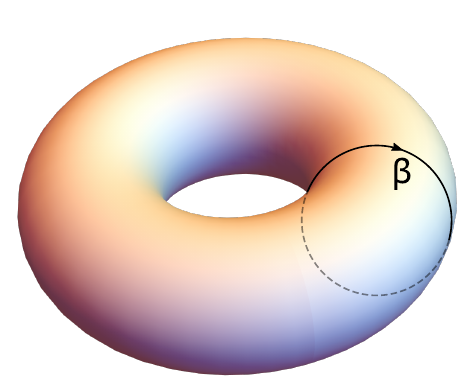}}}
    \label{donutpartitionfunction}\ ,
\end{eqgroup}
which justifies a posteriori the earlier claim we made that a QFT on an euclidean manifold with periodic time coordinate is at finite temperature. 

After this diversion we can go back to entanglement entropy. Consider now a QFT, whose state on a spatial slice is described by $\rho$, defined by the path integral method. We separate that spatial slice into two systems $A$ and $B$. We now make the intuitive (although rather subtle) assumption that the Hilbert space of a local QFT splits into two pieces $H=H_A\otimes H_B$.

As we have said, to compute the reduced density matrix on $A$, we need to trace out the degrees of freedom of $B$. This partial trace comes about again quite naturally from the diagrammatic picture; in the path integral (\ref{thermaldensitypathintegral}) we will impose periodic boundary conditions only on the spatial region $B$, while on the region $A$ we leave it unspecified. This will indeed give us the density matrix $\rho_A$, as now we need to feed in states defined only on $A$.

\begin{figure}[!h]
    \centering
    \includegraphics[width=0.47\linewidth]{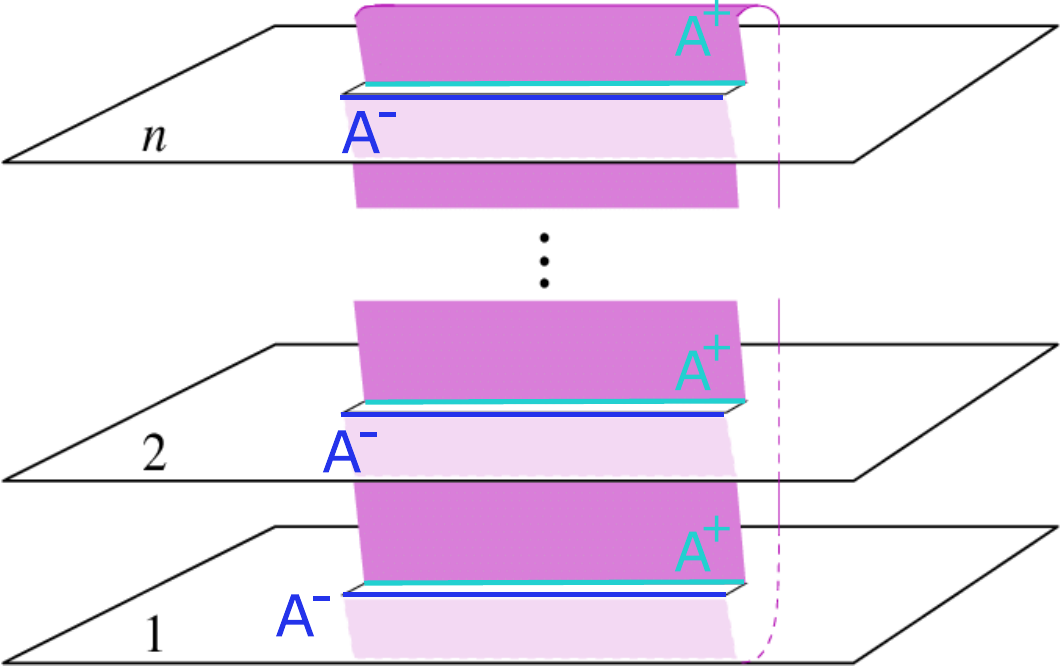}
    \caption{{\small  Riemann surface containing n sheets. The labeled segments $A^+$ and $A^-$ are the segment $A$ for which we seek the entanglement entropy, as approached either from above or below. As indicated, every $A^+$ segment is identified with the $A^-$ segment of the next sheet, and this is done cyclically. In the case of a thermal state, each sheet would instead be a torus.}}
    \label{fig:nsheetedriemanncomplex}
\end{figure}

Then, we need to do compute ${\rm Tr}(\rho_A)^n$. Once more, the simplest way to view this is diagrammatically. What we obtain is a path integral on a complex n-sheeted manifold (denoted $\cal{R}^{(n)}$), where the sheets are connected consecutively through the cut in the $A$ region, see fig.\ref{fig:nsheetedriemanncomplex}.

Notice that as circling around the cuts yields a deficit angle $2\pi(1-n)$. By this we mean that passing through all the sheets we make an angle $2\pi n$ before returning to the initial point. In this way the region $A$ for which we want to compute the Renyi entropy becomes a cut in the Euclidean geometry. 

While the path integral fig.\ref{fig:nsheetedriemanncomplex} can be done by considering fields living in the full Riemann surface, it is much more natural following the construction to consider also "replicated" fields $\phi_i(\tau,x)$, one copy living in each sheet. The gluing of the sheets along the cuts is then expressed as a cyclic boundary condition on the replicated fields, $\phi_i(\tau=0^+,x \in A)=\phi_{i+1}(\tau=0^-,x \in A)$. The action of the full system is given by a sum of the $n$ copies of the action for each of the $\phi_i$.

This picture induces a "replica symmetry", that is simply the cyclic permutation $\phi_i\rightarrow \phi_{i+1}$. Leveraging this symmetry, instead of considering the n-sheeted Riemann surface, we can consider a single sheet, with a cut spanning region $A$. Along this cut, we will have the aforementioned cyclic boundary conditions, which will link the fields $\phi_i$ that are otherwise uninteracting. In the 2d case, one can show that thanks to the replica symmetry, the shape of the cut is unimportant; only the two boundary points of $\pa A$ affect the computation of the partition function. This allows the interpretation of the cyclic boundary conditions as arising from the insertion of "twist" operators at $\pa A$. The partition function (\ref{fig:nsheetedriemanncomplex}) is then reformulated as the correlator of two twist operators located at the two points of $\pa A$. This correlator is formally computed in the theory on the complex plane, with $n$ uninteracting fields $\phi_i$. The cut and the associated boundary conditions are induced by the twist operator insertions.

Whatever the method, the name of the game then becomes the computation of the path integral (\ref{fig:nsheetedriemanncomplex}). This is impossibly complicated in a general QFT. Some results can be obtained in lower dimensions, especially in cases where conformal or other additional symmetries help to simplify the expression, see \cite{Calabrese_2009}. Here, we are content with the formal description, and will focus on the way this quantity can be computed holographically. Let us begin by the result, and we will briefly sketch a proof.

\subsection{The Ryu-Takayanagi prescription}
\label{sec:theRTprescription}
The \textbf{R}yu-\textbf{T}akayanagi (RT) prescription \cite{Ryu:2006bv} allows one to compute the entanglement entropy of a constant time slice $A$ of a holographic CFT in a static configuration. The static requirement is crucial here, as it allows to define the state with the Euclidean path integral prescription which in turns allows use (\ref{fig:nsheetedriemanncomplex}). A time-dependent state will in general require adding the time evolution on top of the Euclidean path integral that prepares the state. There is still a path integral representation of this process called the Schwinger-Keldysh path integral \cite{Keldysh:1964ud,SchwingerofKeldysh}, but we will not worry about it here.

\begin{figure}[!h]
    \centering
    \includegraphics[width=0.35\linewidth]{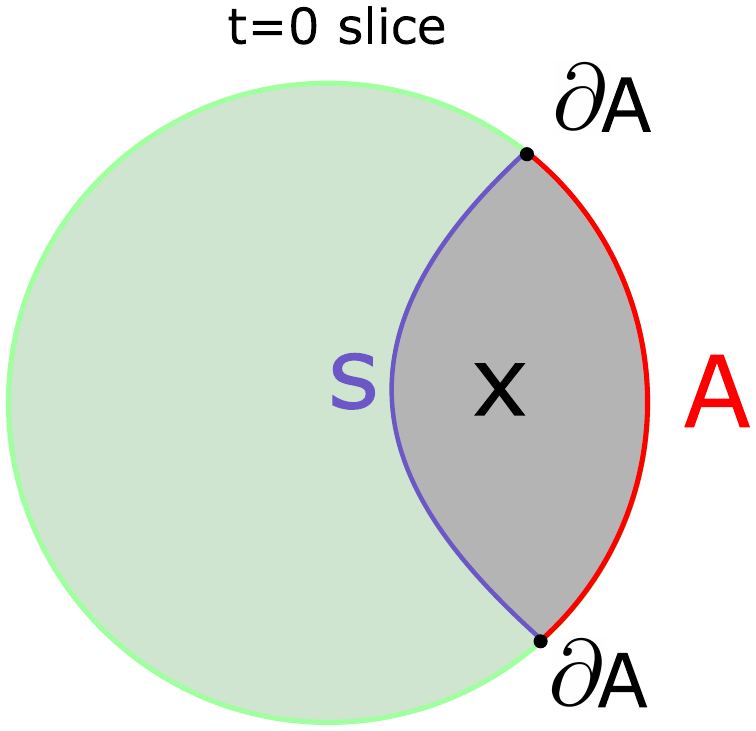}
    \caption{\small A sketch of the slice $t=0$ of an asymptotically AdS spacetime. The RT surface corresponding to the boundary interval $A$ is sketched in dark blue. The region $X$ is delimited by the RT surface and the boundary, as is called the "entanglement wedge" of $A$. It will become important later.}
    \label{fig:RTprescription}
\end{figure}

Consider the bulk gravity solution dual to our CFT state. The metric describing it will be static as well. Then, a natural foliation into spacelike slices is given by the surfaces orthogonal to the time killing vector $\pa_t$ which is guaranteed to exist by definition of the staticity. Denote the entanglement entropy of a sub-region $A$ of the spatial slice $t=0$ in the CFT as $S_A$, and the associated spatial slice in the bulk as $\cal{M}_{D-1}$. Then the RT prescription states :
\begin{eqgroup}
    S_A &= \frac{\cal{A}}{4G}\\
    \cal{A}&= {\rm Min}_{S\in \cal{S}}({\rm Area}(S))\ ,
    \label{RTprescription}
\end{eqgroup}
where $\cal{S} = \{S\subseteq\cal{M}_{D-1}\mbox{ codimension 2 surface homotopic to }A\mbox{ s.t. }\pa S = \pa A\}$. See fig.\ref{fig:RTprescription} for a visualization of $S$ in the case of $D=3$-dimensional bulk.

The prescription (\ref{RTprescription}) is very powerful, since the computation reduces to a minimization problem in classical geometry. It is in any case much simpler than (\ref{fig:nsheetedriemanncomplex}), and indeed in higher dimensions it is often the only way for computing entanglement entropies, provided the theory is holographic. We should however caution that this formula is only valid when the bulk dual reduces to classical (super)-gravity, hence in the large N and strong 't Hooft coupling limit. Both $1/N$ and stringy corrections are not included in this prescription.

In chap.\ref{chap:entanglemententropyandholoint} we provide several examples where the validity of this formula is confirmed. Impressively, we can do better; there exists a "proof" of (\ref{RTprescription})\cite{Fursaev:2006ih,Lewkowycz:2013nqa, Faulkner:2013yia} leveraging the Euclidean construction fig.\ref{fig:cartooneuclideanpathintegral}. For the holographic derivation of the Renyi entropy, we would like to compute ${\rm Tr}(\rho^n)$. According to the dictionary, we should find a dual bulk for which the boundary metric approaches that of the n-sheeted Riemann surface $\cal{R}^{(n)}$. Calling this bulk manifold $\cal{M}_n$ :
\begin{eqgroup}
    {\rm Tr}(\rho_A)^n = \frac{Z^{CFT}_{\cal{R}^{(n)}}}{(Z^{CFT}_{\cal{R}})^n}\hspace{0.3cm}\underbrace{=}_{\mathclap{holography}}\hspace{0.3cm}\frac{\cal{Z}^{gravity}_{\cal{M}_n}}{(\cal{Z}^{gravity}_{\cal{M}})^n}\ .
    \label{holotocomputerenyi}
\end{eqgroup}

Let us remember that in the saddle approximation, computing $\cal{Z}^{gravity}$ simply amounts to evaluating the Einstein-Hilbert action for a solution of the equations of motion.

\begin{figure}[!h]
    \centering
    \includegraphics[width=0.7\linewidth]{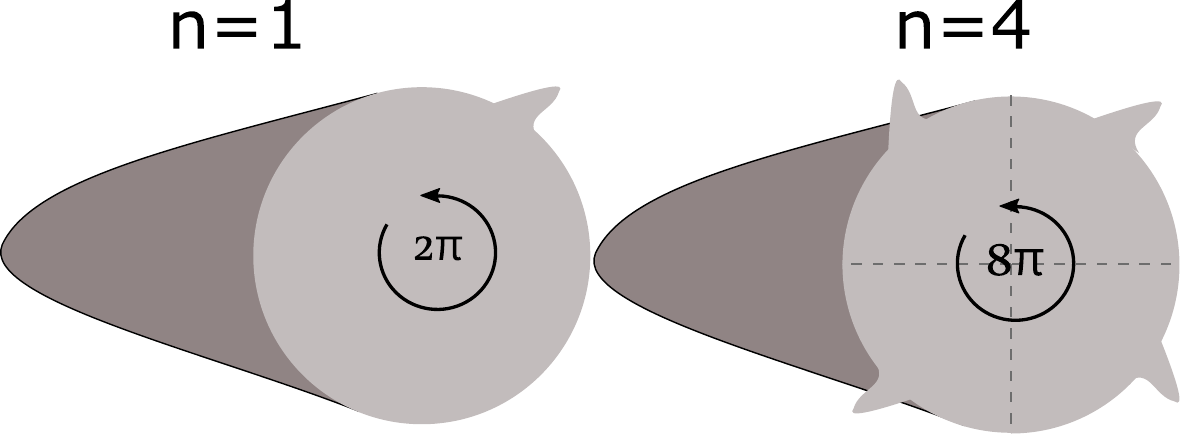}
    \caption{{\small One the left, one copy of the cigar which is one possible dual geometry to the thermal CFT state. On the right, the geometry replicated 4 times. We included a "kink" in the boundary, to make clear the replicating procedure. We glue 4 times the same boundary, and the goal is finding a spacetime geometry with the correct boundary, which in this case is depicted again by a cigar.}}
    \label{fig:replicageometry}
\end{figure}

As an example, consider the simpler to visualize case where $A$ is the full spatial slice, and the state is thermal. The replica geometry in question is an asymptotically AdS space where the time coordinate has an extended periodicity of $2\pi n$ instead of the usual $2\pi$ in the simple case. This geometry is sketched in fig.\ref{fig:replicageometry}.

Let us get a feeling for the replica bulk geometry which we denote by $\cal{M}_n$. Looking at the metric close to the boundary, as we go through the cut on a $\tau$ circle we expect the deficit angle to carry into the bulk. This naturally implies the existence of a codimension 2 surface extending the branch points into the bulk. As we cross it, we go through the  bulk geometry replicas creating a deficit angle in the bulk.

To restrict our search space, we will assume that the replica symmetry of the n-sheeted boundary geometry extends into the bulk. This is a reasonable assumption to find the dominant saddle, as it usually respects the maximum amount of symmetry. Replica symmetry breaking will in general contribute to higher $1/N$ corrections to the formula \cite{Almheiri:2019qdq,Penington:2019npb}. Proceeding as in the field theory construction, one can consider the quotiented geometry $\ti{\cal{M}}_n$, on which we consider $n$ copies of the action. The branch cut resulting in the geometry should carry a deficit angle of $\frac{2\pi}{n}$, which will enforce the cyclic boundary conditions for the $n$ copies of the fields. In the same vein of the "twist fields" construction, this branch cut can be naturally generated by introducing codimension 2 object with tension $T^{(n)}=\frac{n-1}{n}$, which will backreact the geometry appropriately\cite{Rangamani:2016dms}.

The result of this construction is that one has to consider the Euclidean action (\ref{effectiveorbifoldactionfornreplicas}) in order to compute $\cal{Z}^{gravity}_{\cal{M}_n}$ :
\begin{eqgroup}
 \ti{I}^n = I^{\ti{M}_n}_{EH}+\frac{T^{(n)}}{4G}\int_{\cal{N}} d^{D-2}x\sqrt{h}\ ,
    \label{effectiveorbifoldactionfornreplicas}
\end{eqgroup}
where $\cal{N}$ is a codimension-two spacelike surface, $h_{ij}$ its induced metric and $I_{EH}$ is the Einstein Hilbert action including the Gibbons-Hawking term as well as counterterms. According to the holographic prescription, we must find solutions of (\ref{effectiveorbifoldactionfornreplicas}) with boundary conditions set by the CFT state on the boundary. In particular, $\cal{N}$ should be anchored to $\pa A$, as we mentioned.

The equations of motion for the metric are the usual ones away from $\cal{N}$. Because of the staticity, we can immediately restrict $\cal{N}$ to the $\tau=0$ spatial slice. On this slice, the equations of motions from (\ref{effectiveorbifoldactionfornreplicas}) then tell us that the surface should be of minimal area.

The final step lies in the evaluation of $\cal{Z}^n_{grav}= e^{-I^n} = e^{-n \ti{I}^n}$, and its subsequent differentiation. Consider $\ti{I}^n$ evaluated on shell. The differentiation $\pa_n \ti{I}^{n}$ then can be seen as a variation of all the fields including the metric. Since we are on-shell, $\delta \ti{I}^{n}$ will vanish for a generic variation of the fields, up to boundary terms. Here we have two boundaries; the asymptotic boundary and the location of the bulk branch cut. The first one is canceled by the Gibbons-Hawking term and does not contribute. The location of the makes however a non-vanishing contribution.

To compute its value, we excise a codimension 1 tubular region of radius $\epsilon$ around the cut, denoted $\cal{N}^\eps$. This region is an additional boundary of our spacetime, thus we know that the variation of the E-H action will produce a boundary term of the Gibbons-Hawking-Type :
\begin{eqgroup}
 \pa_n \ti{I}^n = - \frac{1}{8\pi G}\pa_n \int_{\cal{N}^\eps} d^{D-1}x \sqrt{h_{\eps}}K_{\eps}\ ,
 \label{tubularaction}
\end{eqgroup}
where quantities indexed by $\eps$ refer to the aforementioned tubular neighborhood of the branch cut.

One can compute this contribution at leading order in $\eps$ by considering the branch cut induced metric $h_{ij}$ and expanding it locally. By this method on shows that $K_\eps\approx\frac{1}{n\eps}$ thus :
\begin{eqgroup}
 \pa_n \ti{I}^n = \frac{{\rm Area}(\cal{N})}{4 G n^2}\ ,
\end{eqgroup}
where being on shell, $\cal{N}$ satisfies precisely the conditions explained in (\ref{RTprescription}) (minus the homology constraint, which is included a posteriori\cite{Headrick:2007km}). With $\cal{Z}^{grav}_{\cal{M}_n}=\exp^{-n \ti{I}^n}$, and expanding (\ref{holotocomputerenyi}) for $n\approx 1$ :
\begin{eqgroup}
 S^{(n)}&=\frac{1}{1-n}\ln({\rm Tr}(\rho_A)^n)=-\frac{1}{1-n}\left(n\ti{I}^n-n I^0\right)\\
 &=-\frac{1}{1-n}\left(n(n-1)\pa_n\ti{I}^n\Bigr|_{n=1}+o((n-1)^2)\right)=n\frac{{\rm Area}(\cal{N})}{4G}+\frac{o(n-1)}{n-1}\ .
\end{eqgroup}

Taking the limit $n\rightarrow 1$ gives us finally the Von Neumann entropy according to (\ref{SaslimitofSn}), and demonstrates the validity of the RT conjecture.

\subsection{Non-static spacetimes and HRT prescription}
With the RT formula in hand, we have a powerful geometric method that computes entanglement entropies of holographic CFTs, provided that they are in a static configuration. The natural extension that one wants to consider is more general, non-equilibrium situations.

There are a few obstacles that appear in principle to the naive generalization of the RT prescription. In the RT formula (\ref{RTprescription}), one can seek a surface of minimal area, because the search is restricted to a spatial slice of the spacetime. In a non-equilibrium situation, given a spacelike slice on the boundary, there is no preferred way to extend it into the bulk. So how can we choose which slice to search the minimal surface on? One might suggest to not restrict our search to a predefined spacelike slice; instead, we could search for the minimal codimension two spacelike surface anchored at $\pa A$ on the boundary. However, this problem is ill-posed because of the timelike direction to which we now have access in the bulk. Indeed, given any spacelike surface, we can deform it so that it "zigzags" along nearly lightlike directions; in this way, one can bring the area arbitrarily close to zero.

The correct extension was found by \textbf{H}ubeny, \textbf{R}angamani, \textbf{T}kayanagi and is referred to as the HRT prescription. As explained in \cite{Hubeny:2007xt}, there are several equivalent ways to formulate it. The key takeaway is that the important property that must be preserved from the RT prescription is not the minimality of the area, but rather, its extremality. For any codimension one spacelike region $A$ of the boundary (and for any CFT state), the HRT formula is the same as (\ref{RTprescription}), with $\cal{S}$ now being the set of codimension 2 spacelike surfaces which extremise the action $\int d^{D-2}y \sqrt{h}$. As in general there might be several extremal candidates, we are instructed to pick the minimum of this set. Notice this trivially reduces to the RT prescription when the state is static, as by time-invariance and time-reversal symmetry one can restrict the extremal surface to lie in the constant time slice.

In the case of the 3-dimensional bulk that will be of interest here, an extremal codimension 2 spacelike surface is simply a spacelike geodesic. Thus in order to compute HRT surfaces we will need to find spacelike geodesics anchored on the two boundary points, and compute their lengths. 

\subsection{1-loop correction and the Quantum Extremal Surface}
As already stated, the result that we derived holds at leading order in $\frac{1}{N}$. To find the next-to-leading order contribution, we need to consider the quadratic fluctuations around the saddle point. As usual, this can be done in the field theory by considering the path integral expression of the replicated partition function, and computing the quantum corrections to the saddle point. Through the AdS/CFT correspondence, these one-loop corrections should be also computable on the gravity side, by appropriately amending the RT prescription (\ref{RTprescription}).

This expectation is correct, and we can again use the dictionary to translate the corrections of the field theory path integral to bulk quantities \cite{Faulkner:2013ana,Solodukhin:1994yz,Fursaev:1994ea,Barrella:2013wja}. This time, we skip the justification and go straight to the result :
\begin{eqgroup}
 S_A = \frac{\cal{A}}{4G}+S_{bulk}(X)+\cal{O}(1/N^2)\ ,
 \label{quantumcorrectiontoRT}
\end{eqgroup}
where $\cal{A}$ is defined in the same way as for (\ref{RTprescription}), and $X$ is the bulk region contained between the RT surface and the boundary region $A$, see fig.\ref{fig:RTprescription}.

$S_{bulk}$ then denotes the entanglement entropy of bulk fields as computed by tracing their density matrix over the complement of region $X$. For instance, if the gravity theory has a propagating scalar field, it will contribute to $S_{bulk}$ with a contribution of the form $-\rho_X \ln(\rho_X)$, where $\rho_X$ is the density matrix of the scalar field restricted on $X$. 

The important thing to realize is that in (\ref{quantumcorrectiontoRT}), we begin first by finding the extremal RT surface in the same way as before. After that, we compute the additional $S_{bulk}$ contribution. Note that the derivation of \cite{Faulkner:2013ana} is applied in the static case, but we will assume it generalizes to arbitrary states.

This leading-order prescription for quantum corrections admits an elegant generalization to all orders called the Quantum Extremal Surface(QES) prescription\cite{Engelhardt:2014gca}. The proposal is to extremize the quantity :
\begin{eqgroup}
 S^q(X) = \frac{{\rm Area}(\pa X)}{4G}+S_{bulk}(X)\ ,
 \label{extremalizableentropy}
\end{eqgroup}
where $X$ is again the region enclosed by the candidate surface $\pa X$ and the region $A$, as in (\ref{fig:RTprescription}). Note that in the case of the HRT prescription, the region $X$ is found by identifying a Cauchy slice which contains the entirety of $\pa X$.

The key difference from (\ref{quantumcorrectiontoRT}) is that instead of finding the $\pa X$ that extremalizes the Area functional, we search for $\pa X$ that extremalizes $S^q(X)$. If $X$ satisfies this condition, it is called "quantum extremal". Call the set of all quantum extremal regions $\cal{X}$. Then the entropy $S_A$ is :
\begin{eqgroup}
 S_A = {\rm Min}_{X\in \cal{X}} \left( S^q(X)\right)\ .
 \label{quantumextremalprescription}
\end{eqgroup}

The claim of \cite{Engelhardt:2014gca} is that (\ref{quantumextremalprescription}) can be used to compute the entropy of $A$ at all orders, so long as $S^q(X)$ is also computed at all orders. 

At first leading order in $N$, we recover the RT prescription. as at this order $S_{bulk}(X)$ does not contribute to $S^q(X)$. It is also possible to show that at the next leading order, we recover the quantum correction (\ref{quantumcorrectiontoRT}). This is not as straightforward however, because the quantum extremal surface and the classically extremal one of (\ref{quantumcorrectiontoRT}) will not coincide in general. However, it can be shown\cite{Engelhardt:2014gca} that this difference contributes only to the $O(1/N^2)$ corrections, confirming that the two prescriptions coincide up to $O(1/N^2)$ corrections.

While this formula is satisfying in its simplicity, using it to compute quantum corrections in the gravity theory is difficult since it entails calculations of QFT entanglement entropies in $S_{\rm bulk}(X)$. Nonetheless, we will see in the next and last section of this chapter that it is a crucial ingredient in one of the groundbreaking advances in the black hole information problem.

Let us finish by touching on the issue of "bulk reconstruction"\cite{Harlow:2018fse,Almheiri:2014lwa}. This is the question of how is the bulk encoded in the boundary CFT, and what portion can be reconstructed if we consider only part of the boundary system. The subject is extremely rich, but in essence, the prevailing theory is that bulk reconstruction and entanglement entropy of the dual CFT are intimately related. It is conjectured\cite{Dong:2016eik,Headrick:2014cta,Wall:2012uf} that given a boundary region $A$ (as in fig.\ref{fig:RTprescription}), the reconstructible region of the bulk (also called the entanglement wedge of A) is precisely $X$. More generally, if one considers also the time direction, the reconstructible region is the region which lies in the Causal diamond of $X$; namely the points which are causally connected only to points in $X$.

\section{The Black hole information paradox and the Island resolution}
\label{sec:IslandsBHparadox}
This section is a lightning review of the Island formula, a new prescription to compute the entanglement entropy of the Hawking radiation.  It was introduced in \cite{Penington:2019npb,Almheiri:2019psf} where it was argued that it could help resolve on aspect of the black hole information paradox.
\subsection{The information paradox}
Let us begin by very quickly reviewing the aforementioned paradox. For in-depth reviews, see \cite{Chakraborty:2017pmn,Mathur:2009hf,Raju:2020smc}. The black hole information paradox was first brought forth by Hawking \cite{Hawking:1976ra}, who noticed that the thermal evaporation process evolved pure states to mixed ones, thereby violating unitarity.

Naively, we would tend to explain away this problem by appealing to the fact that we do not know the underlying quantum gravity theory, and that this problem may go away in this framework. However, the paradox can be stated in a controlled manner, by considering a "nice" Cauchy slicing of the evaporating black hole spacetime, on which the curvatures and energies are much lower than the planck length $\ell_p$. Such a slicing can be found for most of the black hole's lifetime, when it is big enough such that the curvatures away from the singularity are not extreme. It will break down when the black hole becomes Planckian, but we will be able to formulate the paradox long before that time.

Consider a black hole that is formed by collapse of a pure state, $|\psi\rangle_{matter}$. We let it evaporate, and collect the Hawking radiation that is produced. The evaporation process can be seen as generated by successive emission of pairs of Hawking quanta, which are produced at the horizon. For our purpose, we can assume the state of the two Hawking pairs to be maximally entangled :
\begin{eqgroup}\label{hawkingpair}
\left|\psi\right\rangle_{pair}=\frac{1}{\sqrt{2}}\left(|0\rangle_{in} |0\rangle_{out}+|1\rangle_{in} |1\rangle_{out}\right)\ .
\end{eqgroup}

In practice, the Hawking pair state will be more complicated (see \cite{Hawking:1976ra,Giddings:1992ff}) but the essential fact that we wish to capture is that the state has an amount of entanglement of order unity. Then, the quanta labeled "out" makes it out to infinity and it is what constitutes the Hawking radiation, while the quanta labeled "in" falls into the horizon and towards the singularity.

One important fact that distinguishes this process of evaporation from something more mundane like a piece of burning wood, is that the particle creation happens mostly near the horizon, which is located very far from the matter constituting the black hole for the overwhelming majority of its lifetime. One can make this precise by looking at the geometry of a collapsing shell of matter, and the accompanying "nice" Cauchy slicing, but we will not get into such detail here. This picture and the crucial assumption of locality suggest that the total state of the Hawking pair and the black hole matter factorizes :
\begin{eqgroup}\label{totalPsi}
|\Psi\rangle_{\rm total}=|\psi\rangle_{\rm matter}\otimes \left|\psi\right\rangle_{pair}\ .
\end{eqgroup}

Then after $n$ timesteps, the total state will be of in the form (\ref{totalPsin}):
\begin{eqgroup}
|\Psi\rangle_{\rm total}=|\psi\rangle_{\rm matter}\otimes\left|\psi\right\rangle_{pair}^{\otimes n}\ ,
\label{totalPsin}
\end{eqgroup}
where $\left|\psi\right\rangle_{pair}^{\otimes n}$ stands for the emitted $n$ quanta. This is a rough argument, but it has been argued that corrections, e.g. due to the fact that two consecutive pairs can be created "near" one another, will not alter the final conclusion \cite{Mathur:2009hf}.

To obtain the state of the Hawking radiation, one has to trace out the black hole degrees of freedom ($|\psi\rangle_{\rm matter}$) as well as the infalling Hawking modes. Tracing out $|\psi\rangle_{\rm matter}$ does not produce any entanglement, but for each partially traced hawking pair, we increase the entanglement by $\ln(2)$. This shows that as the black hole evaporates, the entanglement entropy of the collected radiation increases. 

This is the crux of the paradox. Letting the black hole evaporate up until the point where it becomes Planck-sized (and our approximation ceases to apply) leads to a mixed density matrix $\rho_{\rm rad}$ for the collected Hawking radiation. It contains an arbitrarily high amount of entanglement, determined essentially by the size of the initial black hole. Assuming that the evaporation continues until the black hole's disappearance, we reach a contradiction with unitarity: the evaporation turned a pure state $|\psi\rangle_{\rm matter}$ into a highly mixed state $\rho_{\rm rad}$. Of course, in our explanation we swiped under the rug all the details that render this derivation truly convincing, but they can be found in the review \cite{Mathur:2009hf}.

One possible alternative is that the black hole does not disappear, but some remnant (whose precise description depends on the theory of quantum gravity)\cite{Chen:2014jwq} is left behind. Although this is a logical possibility, such remnants would have unbounded degeneracy (due to the fact that they should be able to have arbitrary amounts of entanglement), while having bounded energy and size, which would make them extremely exotic compared to usual matter.

\begin{figure}[!h]
    \centering
    \includegraphics[width=0.5\linewidth]{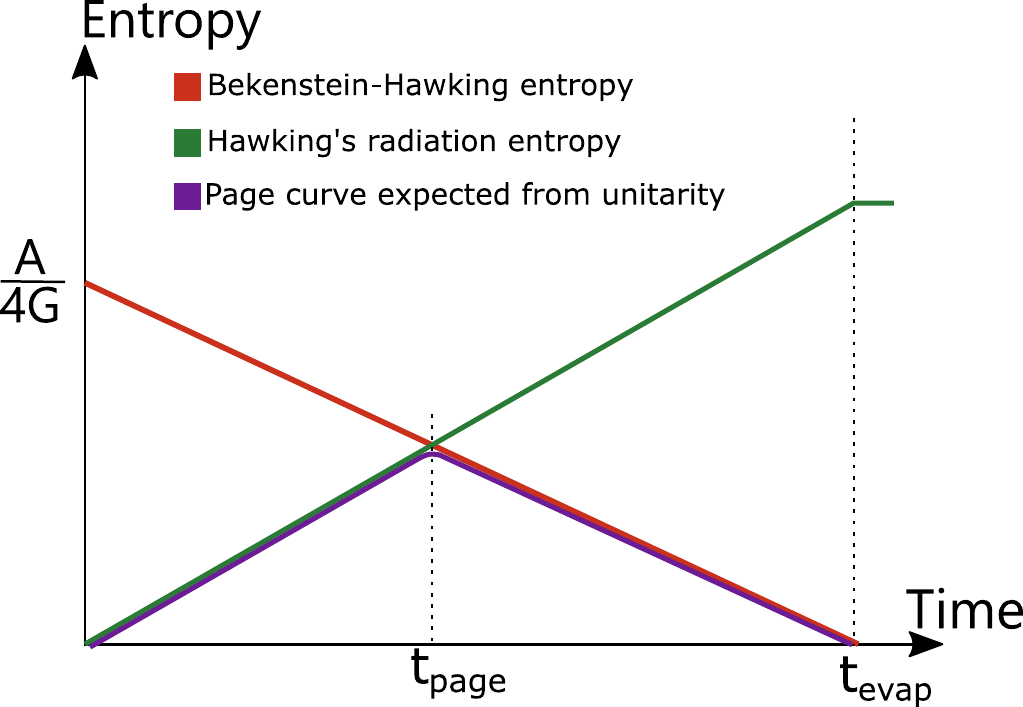}
    \caption{{\small Sketch of the different curves of interest for the information paradox. In red is the thermodynamic entropy of the black hole, proportional to its horizon area, decreasing as the black hole evaporates. In green, the entanglement entropy of radiation according to Hawking. At the page time, we reach an inconsistency as the green curve should be bounded by the red one if the combined system was pure initially. In purple is the Page curve expected from a unitary evaporation, we see it follows roughly the minimum of the red and green curve. Note that the curves are drawn linearly only for simplicity; their precise shape is not important.}}
    \label{fig:pagecurve}
\end{figure}

To make things worse, if one believes this argument and the Bekenstein-Hawking formula, the paradox appears much before the black hole approaches planckian sizes. Indeed, the thermodynamic entropy of the black hole (given by (\ref{bekensteinhawkingformula})) gives us an upper bound on its Von Neumann entropy, as the thermodynamic version can be seen as a "coarse grained" value of the entanglement entropy\footnote{The thermodynamic entropy of a state can be defined as the maximal possible entanglement entropy of the microscopic states allowable, given the macroscopic constraints on the state.}. Therefore, somewhere after the half-point of the evaporation, the entanglement entropy of $\rho_{rad}$ will exceed the thermodynamic entropy of the Black hole. But when a pure state is partitioned into two, the two subsystems have the same entanglement entropy. Thus we have a clash already at the mid-point of evaporation with the usual laws of quantum mechanics. This was made precise by Page \cite{Page:1993wv}, who modeled the Black hole dynamics as applying random unitaries on its Hilbert space, emitting random quanta as time passes. With this model, he recovered the so-called "Page curve" for the entanglement entropy of the radiation, depicted in fig.\ref{fig:pagecurve}. This is the curve that the entanglement entropy of the radiation should follow assuming the black hole has random unitary dynamics.

The goal of any potential resolution of Hawking's paradox should be to reproduce the curve fig.\ref{fig:pagecurve} for the entanglement entropy of the radiation. This seems an impossible task without abandoning at least one of the following seemingly well-tested concepts :
\begin{itemize}
    \item Locality of physics
    \item The equivalence principle
    \item Unitarity
\end{itemize}

For instance, proposals like fuzzballs\cite{Skenderis:2008qn,Lunin:2001jy,Lunin:2002qf,Mathur:2002ie} and firewalls \cite{Almheiri:2012rt,Almheiri:2013hfa} do away with the equivalence principle, while preserving locality and unitarity. The Island proposal that we will present shortly finds an interesting loophole that allows it to recover unitary evaporation while preserving all 3 of the aforementioned principles. It does so by modifying the procedure by which one computes $S_{rad}$, the entanglement entropy of the Hawking radiation. What this implies is that in a theory with gravity, (unexplained) gravitational effects render incorrect the entropy computation of the radiation in the naive way we have outlined above. While the Island prescription tells us the "correct" way to compute this entropy, it does not explain the full story of how the radiation is purified. It is nevertheless striking that a semiclassical calculation manages to reproduce the Page curve which one expected to be only computable in quantum gravity.

\subsection{The Island prescription}
\begin{figure}[!h]
    \centering
    \includegraphics[width=0.7\linewidth]{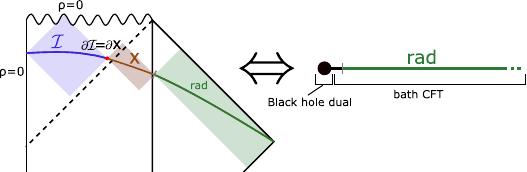}
    \caption{{\small Sketch of the two dual systems that we consider. On the left, a portion of the Penrose diagram of the gravitating system containing the evaporating black hole, connected at the asymptotic boundary to a flat "bath" system. We depict a Cauchy slice of the geometry, and depict the various portions of interest. In green, the radiation system, namely the portion of the Cauchy slice that lies in the bath. The red dot represents the HRT surface $\pa X$. It defines the Island region $\cal{I}$ in blue, delimits the region X, in brown (see also fig. \ref{fig:RTprescription}). The shaded colored regions denote entanglement wedges. On the left, a spatial slice of the dual system in which the gravitating system is dualized to a D-1 dimensional CFT system, which is coupled to the $D$-dimensional radiation bath. We indicate again in green the radiation region for which we wish to compute the entanglement entropy.}}
\label{fig:whatisIslandandX}
\end{figure}

The Island prescription for the entanglement entropy of the radiation is best stated in the context of AdS/CFT, where it was first derived. The setup is that of an asymptotically AdS spacetime coupled to a flat "bath" at its boundary. The coupling at this boundary is such that it is transparent to any outgoing excitations; in particular, Hawking radiation reaching the boundary passes through and is collected at future lightlike infinity.  The reason for including this auxiliary system is that otherwise the black hole would reach equilibrium with its radiation reflected at the AdS boundary\footnote{Strictly speaking, this is only true when the dimension of the bulk is $\leq 3$, see (\ref{temperatureadsBHDbig3}). In higher dimensions, one could in principle consider "small" black holes which do evaporate even in AdS. This unstable case is however much less understood from the holographic perspective, and furthermore one cannot easily separate the radiation from the black hole Hilbert space. Thus all applications of the Island formula are up until now in setups that include a bath system to collect the radiation.}. We skip over the details of how to actually confection such a setup (see\footnote{In these works, they start with an eternal black hole geometry and couple it to the vacuum bath at a finite time, letting the black hole evaporate. Alternatively, one could start with the coupled system, and prepare a shell collapsing to a black hole.} \cite{Almheiri:2019psf,Almheiri:2019hni}), and show in fig.\ref{fig:whatisIslandandX} the portion of the Penrose diagram that will be of interest to us.

If we consider the holographic dual of an AdS$_D$ black hole spacetime, we obtain a (D-1)-dimensional CFT in a thermal state (which we will call "DBH" for "Dual Black Hole") coupled to D-dimensional bath system (which we will call "bath"). We assume that the initial state is a tensor product of the two, and is pure. For convenience, we take the bath to also be a CFT, in the vacuum state. We then expect DBH to radiate into the bath, until it cools down to zero temperature. Our goal is to compute the entanglement entropy of the radiation we collect in the bath, which on the gravity side is precisely the entanglement entropy of Hawking radiation. To do so, we will apply the prescription (\ref{quantumextremalprescription}), the region $A$ being the full DBH system. This computes the entanglement entropy of the DBH system, but since the state of the full system is pure, it is also equal to the entanglement entropy of the radiation.

With this reasoning, we obtain the "Island formula" \cite{Penington:2019npb,Almheiri:2020cfm}:
\begin{eqgroup}\label{islandFormula}
S_{rad} = {\rm Min}\left\{{\rm ext}_{\cal{I}}\left(\frac{{\rm Area(\pa \cal{I})}}{4G}+S_{\rm matter}({\rm rad}\cup \cal{I})\right)\right\}\ ,
\end{eqgroup}
see fig. \ref{fig:whatisIslandandX} for a clarification of the different quantities entering (\ref{islandFormula}), and for an explanation of the term "Island" in the name.

Let us stress that this is simply a rewriting of the QES formula (\ref{quantumextremalprescription}), from the perspective of the complementary system. The two prescriptions are one and the same, simply viewed from two different perspectives.  

Armed with (\ref{islandFormula}) and the setup in fig.\ref{fig:whatisIslandandX}, we can perform the explicit computation of the entanglement entropy in simple AdS$_2$ toy models based on Jackiw-Teitelbom gravity\cite{Penington:2019npb,Almheiri:2019hni,Almheiri:2019yqk,Almheiri:2019psf}. The main reason for considering such simple 2D models is that they allow the computation of $S_{matter}$ which is otherwise impossible, as it boils down to a computation of entanglement entropy in QFT. We will not reproduce the full computations here, but let us indicate qualitatively how it reproduces the Page curve. We will make the argument both by considering (\ref{extremalizableentropy}), and by using the Island formula (\ref{islandFormula}), in order to show that they are indeed two facets of the same coin.

Consider first early times, just after the black hole formation and before it had time to evaporate much. The only candidate quantum extremal (H)RT surface is a vanishing $\pa X$ sitting at $r=0$ (we will call this the "trivial" surface). Then, the region $X$ spans the full gravitational Cauchy slice, which contains a few outgoing Hawking quanta on their way to the bath, and most importantly all the ingoing hawking quanta on their way to the singularity. In this way, $S_{bulk}(X)$ roughly grows by $\ln(2)$ for each emitted Hawking quanta. By invoking the purity of the state on the full Cauchy slice including the bath part, we recover Hawking's result for the radiation entropy, in the early times of evaporation.

Applying the Island formula, the arguments are similar. In the early times, the extremizer of (\ref{islandFormula}) is the empty set Island, $\cal{I}=\varnothing$. Thus we find that the "true" entanglement entropy of the radiation is simply given by the usual Hawking's result, in which we compute $S_{matter}({\rm rad})$ in the usual way by the Von Neumann formula (\ref{entanglemententropy}). This gives of course a steadily growing entropy because of the state's entanglement with the infalling modes.

\begin{figure}[!h]
    \centering
    \includegraphics[width=0.5\linewidth]{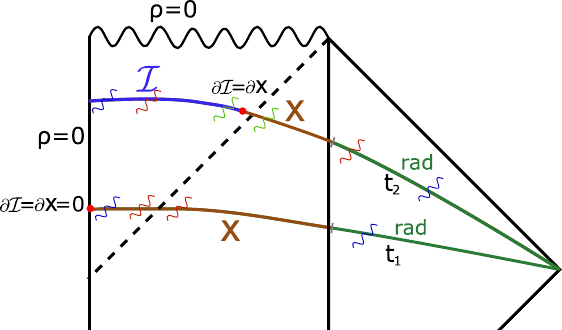}
    \caption{{\small Portion of the Penrose of the evaporating black hole coupled to the bath. We depict the two competing QES. For the earlier Cauchy slice at time $t_1$, the vanishing RT surface is depicted in red, or equivalently, the vanishing Island. Hawking pairs are drawn as squiggly lines of matching colors. As depicted, for the vanishing Island surface, Hawking pairs created earlier (in blue) have reached the bath in green. Therefore, they will contribute to $S_{\rm matter}(rad \cup \cal{I})$ as there is no Island. A small subset of recently created pairs (in red) will not contribute, but these effects will be negligible. The Island surface is depicted in a later Cauchy slice at time $t_2$. This time, the area of the dot is non-vanishing and will give a contribution. As a counterpart, pairs created earlier (blue and red) will lie in the union $\cal{I}\cup{\rm Rad}$, purifying the state and giving a vanishing contribution. There will be again small corrections for the pairs created recently (green), but that will be negligible.}}
    \label{fig:penrosediagramwithIsland}
\end{figure}

At later times, when a non-negligible amount of radiation has been emitted, another QES appears, which is characterized by $\pa X$ sitting close to the horizon\footnote{According to the toy model, this surface could be either slightly inside or slightly outside the horizon\cite{Almheiri:2019yqk}, but this does not alter the conclusion.}, or alternatively by an island $\cal{I}$ spanning the black hole interior, see fig. \ref{fig:penrosediagramwithIsland}.

Let us explain this new extremal surface (called "Island" surface) from the perspective of (\ref{extremalizableentropy}). In the case depicted in fig. \ref{fig:penrosediagramwithIsland}, ${\rm Area}(\pa X)$ gives a contribution akin to the Bekenstein hawking entropy $\approx \frac{\cal{A}}{4G}$ where $\cal{A}$ is the horizon area. This is a trade-off for a substantial reduction in $S_{\rm bulk}(X)$; indeed, in this way we excise most of the ingoing Hawking pairs, removing most of the contribution due to $S_{\rm bulk}(X)$ in (\ref{extremalizableentropy}). The main contribution to the trivial surface is thus the entanglement entropy of the ingoing Hawking pairs, while the main contribution to the Island surface is the Area of the horizon. At the page time, these will be equal, and the minimal surface will transition from the trivial to the Island one.

The computation using the formula (\ref{islandFormula}) proceeds likewise. This time, we introduce the Island region $\cal{I}$ as a means to reduce the contribution $S_{matter}(\rm{Rad}\cup\cal{I})$. Indeed, by including ingoing Hawking quanta together with the radiation, we include both pairs of (\ref{hawkingpair}), purifying the state and removing their contribution to the entanglement. Then, the main contribution comes from ${\rm Area}(\pa \cal{I})$, which is roughly the Bekenstein Hawking entropy when $\cal{I}$ is extremal.

With these facts established, we can easily see that we recover the page curve for $S_{rad}$. Before the Page time, the entanglement entropy of the radiation follows Hawking's result, and steadily increases as we collect more. However, at the Page time, the Island surface takes over as the dominant contribution to $S_{rad}$, and the entanglement entropy of the radiation goes down with the horizon area as the black hole continues to evaporate, recovering exactly the Page curve fig.\ref{fig:pagecurve}.

This is a satisfying result, especially because the QES prescription (\ref{quantumextremalprescription}) (equivalently the Island formula (\ref{islandFormula})) can be "proved" by an Euclidean path integral construction akin to the one showcased in sec.\ref{sec:theRTprescription}. In one sentence, the QES prescription arises when one considers wormhole geometries connecting different replicas\cite{Penington:2019kki}. Such saddle points were not considered in the derivation of the RT surface, and should be allowable in the replicated geometry since they satisfy the boundary condition, despite the topology change.

One final interesting remark concerns the entanglement wedge of the radiation in this evaporation model. Before the Page time, this entanglement wedge comprises only the radiation system, as we would expect. However, after the Page time, the entanglement wedge of the radiation contains also a portion of the black hole interior, namely the entanglement wedge of the Island region (see fig. \ref{fig:whatisIslandandX}). This seems to imply at first glance that some non-local effects are taking place; but at no point in the derivation did we abandon locality. We must conclude that the degrees of freedom of the black hole interior are somehow encoded in the Hawking radiation, which is not really surprising since we expect the information inside the black hole to be accessible from the radiated quanta. However, this is exactly what the Island prescription does not tell us, how precisely the information escapes from the black hole. In this sense, the black hole information paradox is still somehow an open question, as we have shown the information comes out, but not how.

Let us finish by mentioning that one does not necessarily need to consider an evaporating black hole geometry to study Hawking's information paradox. A version of this paradox can be obtained in the maximally extended eternal black hole geometry roughly as follows\cite{Almheiri:2019yqk}. The full geometry contains two boundaries, which house the dual system composed of two CFTs in a thermofield double state\cite{Maldacena:2013xja}. This is a purification of the thermal state of each individual CFT.

\begin{figure}[!h]
    \centering
    \includegraphics[width=0.5\linewidth]{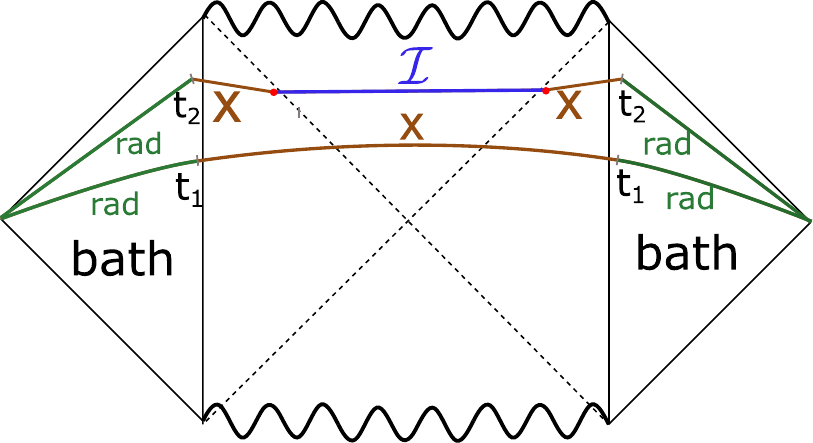}
    \caption{{\small Penrose diagram of the eternal black hole in Ads, with two flat baths glued at the two asymptotic boundaries. We depict the Island surfaces that compute the entanglement entropy of two Cauchy slices of the radiation at time $t_1$ and $t_2$. The same discussion from the evaporating black hole applies also here, except every contribution is double because of the two sides. The red dots are the RT surface, or equivalently the Island's boundary. This setup provides a much easier framework to study Hawking's paradox, as the geometry is static.}}
    \label{fig:maximallyextendedwithtwobathspenrose}
\end{figure}

To set up the information paradox, we couple the geometry to two CFT baths (see fig.\ref{fig:maximallyextendedwithtwobathspenrose}), which are in thermal equilibrium with the black holes. Because of this equilibrium, the eternal black hole geometry is not perturbed. Crucially, however, there is radiation exchange between the bath and black hole. We continuously collect Hawking radiation in the bath, and the energy lost in this way is compensated by ingoing radiation due to the thermal state of the bath.

What we do next is consider the Entanglement entropy of the two copies of the CFT baths as a function of time (two relevant Cauchy slices are sketched in fig.\ref{fig:maximallyextendedwithtwobathspenrose}). According to Hawking's computation, this entropy will increase indefinitely as we collect more and more radiation from the black hole. This is a paradox, because the thermodynamic entropy of both black holes is $2\frac{\cal{A}_{BH}}{4G}$, where $\cal{A}_{BH}$ is the area of one horizon. Therefore we should expect by unitarity that the entanglement entropy of the outgoing radiation does not exceed this value.

Again, the Island prescription saves the day. At late times, an Island surface spanning the interior of the double-sided black hole will purify the outgoing Hawking's radiation, while contributing a factor of $2\frac{\cal{A}_{BH}}{4G}$ to $S_{\rm rad}$ from the $\frac{{\rm Area}(\pa \cal{I})}{4G}$ term of (\ref{islandFormula}). This successfully caps the entropy of the radiation at just the right value to preserve unitarity.

\subsection{Doubly holographic models}
One major inconvenience of (\ref{islandFormula}) is that it requires the computation of entanglement entropies of quantum fields on a curved background. The cases where this can be done are few and far between, which is the reason for using the simple JT gravity model. As explained previously, in two and three dimensions gravity is quite different, and so the extrapolation of the computations to higher dimensions is far from obvious, but see \cite{Uhlemann:2021nhu,Demulder:2022aij,Chen:2020hmv} for some interesting work in this direction.

As it turns out, a case where the computation of the entanglement entropy is easier to handle is the case of a holographic system. The idea introduced in \cite{Almheiri:2019hni,Chen:2020uac,Chen:2020hmv} is to consider the DBH$+$bath system as a holographic BCFT model. Then, the bulk dual of such a system is an asymptotically AdS spacetime, capped off by an \textbf{E}nd-\textbf{O}f-the-\textbf{W}orld (EOW) brane, which is dual to the boundary of the CFT. 

\begin{figure}[!h]
    \centering
    \includegraphics[width=0.6\linewidth]{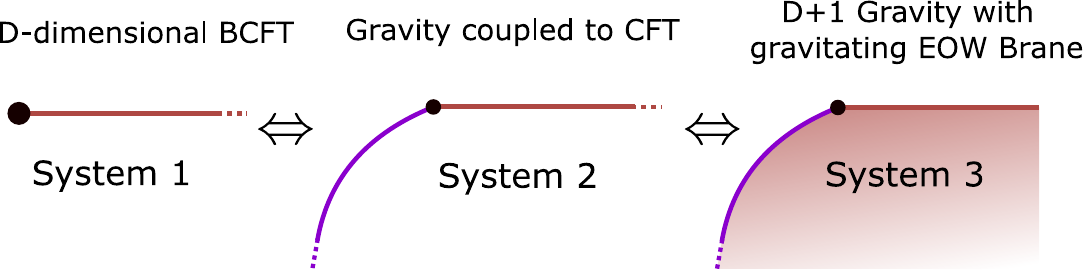}
    \caption{{\small Three systems which are presumably dual to each other through two applications of holography. In system 2, we dualize the boundary of system 1 into a gravitating system, depicted in purple. By applying the holographic correspondence to the full BCFT of system 1, we obtain system 3, which is composed by a higher dimensional gravitating bulk, along with an End Of the World gravitating brane depicted in purple. Integrating out the bulk of system 3 yields an effective theory on the purple brane, recovering system 2.}}
    \label{fig:doublyholographictriple}
\end{figure}

The clever trick lies in a third representation of the system, in which, roughly speaking, we "unfold" holographically only the CFT boundary. In this picture, we have the CFT bath joined through the interface to a gravitating system of the same dimension (the "brane-bath" picture). Choosing the boundary state accordingly, we could generate a black hole on the brane, and thus setup an evaporation experiment. The three dual systems are depicted in fig.\ref{fig:doublyholographictriple}.

System 1 and 2 already appeared in the previous section. From the holographic dictionary for BCFT's, systems 1 and 3 in fig.\ref{fig:doublyholographictriple} are related by a duality. The intermediate system 2 is not as obvious, as it is not clear we can consider the gravity dual of the theory living on the boundary while leaving the connected bath CFT untouched. In \cite{Chen:2020hmv}, system 2 is obtained by starting with system 3, and integrating out the bulk degrees of freedom, which produce an effective action on the EOW brane. By tuning the brane's tension and Lagrangian, one can setup a Randall-Sundrum \cite{Randall:1999ee,Randall:1999vf,Karch:2000ct} type scenario, where the brane's worldvolume effectively acquires a dynamical graviton.

If one accepts the legitimacy of this "doubly holographic" system, this is the perfect playground to study black hole evaporation, as the computation of the entanglement entropy of the radiation is completely geometrized through the (H)RT prescription in system 3. We skip over the details, but one needs to amend the RT prescription as we are in presence of a gravitating EOW brane in the geometry. It turns out\cite{Chen:2020hmv} that the RT surfaces are allowed to end on the EOW brane, and that this produces an additional contribution of $\frac{\cal{A}(\si_{int})}{4 G_{\rm brane}}$, where $G_{\rm brane}$ is the Newton's constant for the brane's effective gravity and $\cal{A}(\si_{\rm int})$ is the area of the region $\si_{\rm int}$ intersected by the RT surface.

\begin{figure}[!h]
    \centering
    \includegraphics[width=0.4\linewidth]{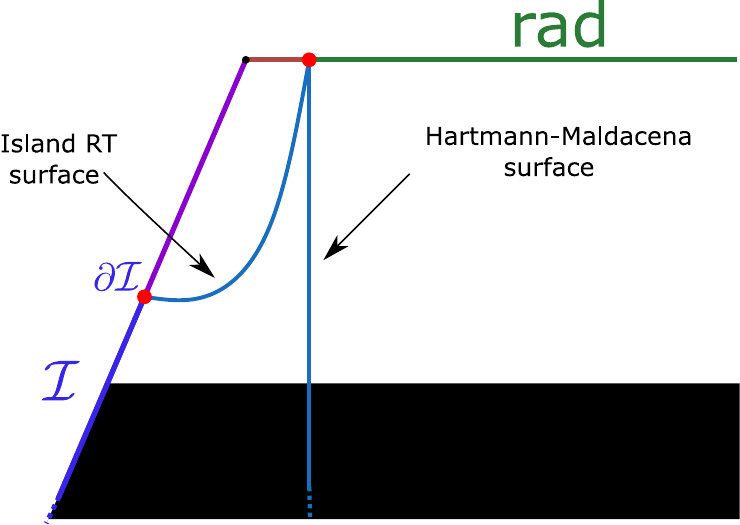}
    \caption{{\small One side of the eternal black hole geometry cutoff by the EOW brane. We depict the two candidate RT surfaces in light blue. The Hartmann-Maldacena surface crosses the wormhole and connects the two disconnected boundaries. The other curves back to end on the EOW brane. From the point of view of system 2, this generates an Island region, disconnected from the radiation bath.}}
    \label{fig:thetwoextremalsurfaces}
\end{figure}

We can now see from these models the simple origin of the Island prescription qualitatively as follows. We want to add a black hole in the gravity part of system 2 to recreate the eternal black hole setup of the previous section. We thus begin with two copies of system 1, in the thermofield double state. The bulk black hole of system 3 intersects the brane, which generates the black hole that lives on it (see fig.\ref{fig:thetwoextremalsurfaces}). We then use the RT prescription in system 3 to compute the entanglement entropy of both bath states.

There are two extremal surfaces (see fig.\ref{fig:thetwoextremalsurfaces}). The first one (called the Hartman-Maldacena surface \cite{Hartman:2013qma}) passes through the wormhole in the black hole interior to connect the two baths directly. This contribution will yield the initial growth of the entanglement entropy of the bath. Indeed, as when we evolve forward in time the wormhole throat is stretched out, increasing the RT surface's area accordingly.

The other extremal surface has a constant contribution as we evolve forward in time. Depicted in fig. \ref{fig:thetwoextremalsurfaces}, it curves and terminates on the EOW brane. The full RT surface is composed of two disconnected copies of the one depicted in fig. \ref{fig:thetwoextremalsurfaces}. They will have a contribution of approximately $2\frac{\cal{A}_{BH}}{4G_{\rm brane}}$ coming mainly from the anchor point which will be close to the brane horizon. Thus in the doubly unfolded system 3, the appearance of the Island region $\cal{I}$ on the brane is not so surprising anymore. In the bulk, this happens when the entanglement wedge of the two bath CFTs includes a portion of the EOW brane. The results obtained are the same as if we used the Island prescription in system 2. The advantage of the doubly holographic model is that we can perform the computation in arbitrary dimensions, as the computation is purely geometrical.

Let us conclude this section by mentioning that while the Island prescription does successfully recover the Page curve, it is still being debated whether we are indeed solving the information paradox in doing so. One criticism already mentioned is that in low dimensions where gravity is "non-standard"\cite{Raju:2021lwh}, behaving very differently thant $D\geq 4$. This problem is formally addressed with the use of doubly holographic models, but as was noticed in \cite{Geng:2021hlu,Geng:2020qvw} the coupling of the gravity to the CFT bath in system 2 gives a mass to the graviton due to the non-conservation of the gravity stress-energy tensor induced by the energy seeping out in the bath. It was furthermore argued in these works that attempting to make the graviton massless would make the Island disappear. Another possible issue in the doubly holographic models is simply the intermediate point of view of system 2. This is an effective description that involves integrating out degrees of freedom in the bulk, so it is not evident how robust it is.

\chapter{Phases of holographic interfaces}
\label{chap:phasesofinterfaces}
\epigraph{This chapter is based on 2101.12529}{}
This chapter is dedicated to the study of minimal holographic ICFT models that were described in sec.\ref{sec:MinimalICFT}, for the restricted set of equilibrium states at finite temperature. The analysis that we will perform is very similar in spirit to the Hawking-Page analysis of AdS at finite temperature described in sec.(\ref{sec:HawkingPage}). The presence of the interface that might seem innocuous at first will on the contrary enrich the system quite considerably, and produce a plethora of different equilibrium solutions that we will describe in detail. Our analysis is mainly focused on the gravity side in the limit where it is classical, and as such it should be seen as describing phenomena at strong coupling in the dual field theory. 

The initial motivation to consider such a model was for potential application to the Black Hole information paradox. As we have seen in sec.\ref{sec:IslandsBHparadox} the setup in which the Island construction is applied involves connecting two spacetimes, one containing the black hole and the other acting as a bath to collect the radiation. Even more directly, the doubly holographic constructions necessitate the introduction of EOW branes, which are a special case of the domain walls we consider here. Initially, the hope was to produce natural setups in which one could let a black hole on one side of the interface evaporate, and somehow collect the radiation on the other side. Although we found solutions on which this could be realized, the separation provided by the membrane is not sufficient to be able to repeat the arguments of sec.\ref{sec:IslandsBHparadox}. Later, \cite{Anous:2022wqh} managed to exploit the model in order to setup an Island computation in the doubly holographic picture, in which the two CFTs constitute a "double bath" for the radiation.

Of course, the study of this model at finite temperature is also interesting on its own, if only because it is yet another probe into the physics of ICFT. In that context, it can also provide an interesting model for some condensed matter systems. Indeed, two quantum wires joined at an impurity would be modeled by an ICFT at criticality \cite{Bernard:2014qia,Bernard:2016nci}. As an added bonus, entanglement entropy computations which have become an important tool are extremely facilitated when one has access to the holographic dual\cite{Ryu:2006bv}. The study of those is delegated to chapter \ref{chap:entanglemententropyandholoint}.

The main results of this chapter are the fully analytic description of dual geometries of the equilibrium ICFT state. The phase space and the nature of phase transitions is analyzed in detail, and we finish by the (numerical) computation of the phase diagram for selected values of the parameters. We comment on the interpretation of the different phases in the holographic context. 

To lighten notations, throughout this chapter we use units in which $8\pi G =1$.

\section{ICFT model and topology of slices}
\label{sec:icftmodelandtopology}
\begin{figure}[!h]
    \centering
    \includegraphics[width=0.7\linewidth]{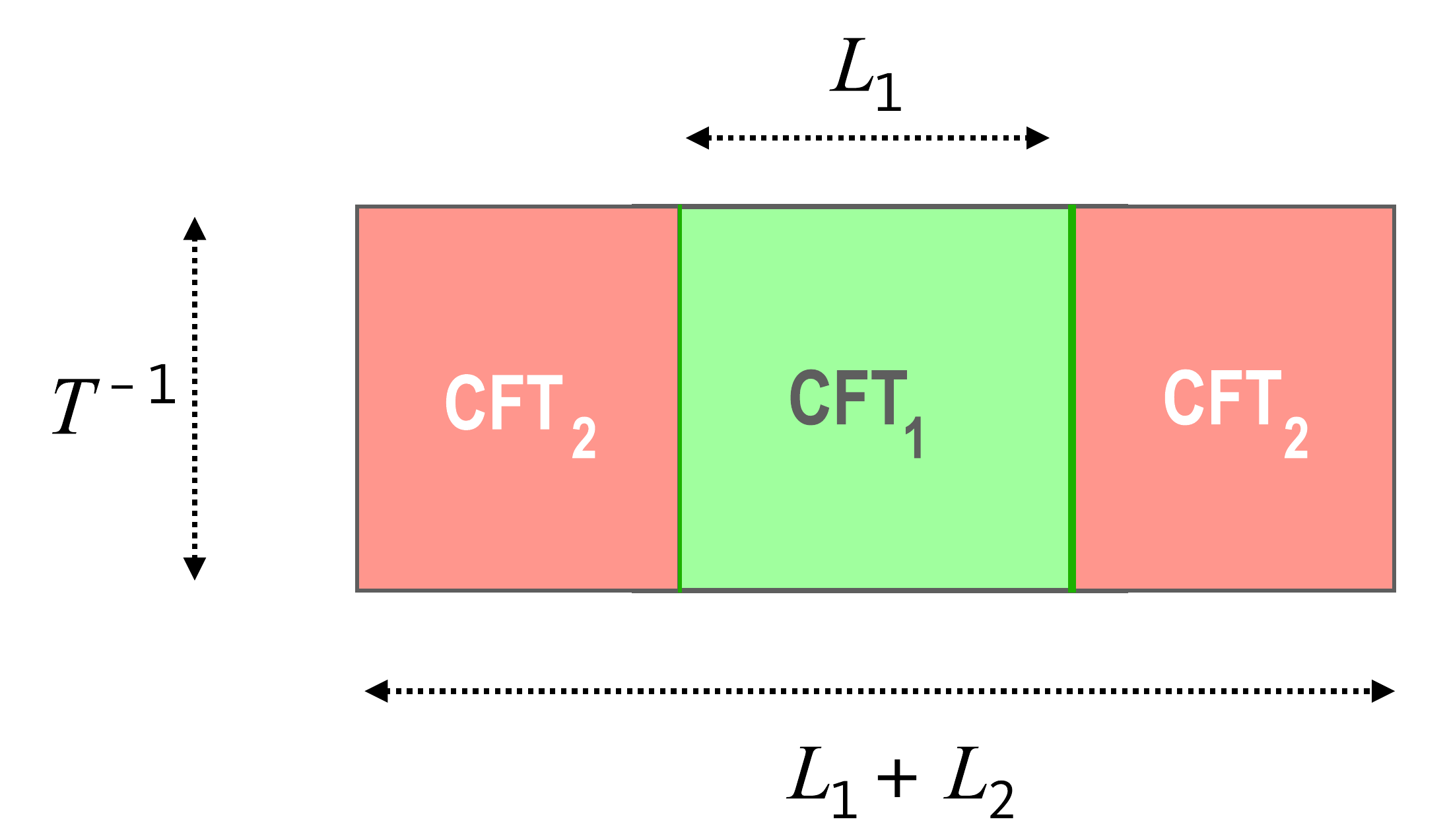}
    \caption{\small Finite temperature ICFT. The horizontal and vertical sides are identified, so that the topology is a torus.}
    \label{fig:stripedtorus}
\end{figure}

Let us define precisely the model we will be studying starting from the field theory description. We consider CFT$_1$ and CFT$_2$, two CFTs coexisting at thermal equilibrium on a circle. The manifold on which they live can thus be seen as a "striped torus", see fig.\ref{fig:stripedtorus}.
We will be working in the minimal holography bottom-up model described in sec.\ref{sec:MinimalICFT}. As such, the available parameters are the respective central charges $c_1$ and $c_2$, as well as the states on each side, determined by $\langle T^i_{\mu\nu}\rangle$ in the large-N limit. With the further assumption that we are in an equilibrium state, the only non-vanishing components are $\langle T_{++}^i\rangle = \langle T_{--}^i\rangle$, namely there is no net flux of energy. We have additional parameters related to the Manifold geometry; the temperature $T$ and the size of each CFT, $L_1$ and $L_2$. Only dimensionless parameters will be conformally invariant, so we can consider $\tau_1 = TL_1$ and $\tau_2 = TL_2$ as the two physically meaningful quantities. As for the interface, it is minimally characterized by one number $\lam$ as explained in sec.\ref{sec:MinimalICFT}. Without loss of generality, we henceforth assume $c_1\leq c_2$. We will refer to either theory and their associated bulk as being on "side 1" or "side 2".

The gravity dual will then be composed of two asymptotically Anti-de-Sitter spacetimes, as well as a gravitating membrane of tension $\lam$ (also referred to as "string" or "wall" in the text), anchored at the interface location on the boundary. Using the minimal holographic dictionary, more specifically the Fefferman-Graham prescription, we can determine the metric of the spacetime as a function of $\langle T_{++}^i\rangle$ (we will opt for the reverse process in this case to work with parameters adapted to the gravity theory). In a suitable coordinate system, the corresponding Euclidean metric for each side can be written in the form (\ref{EquilibriumAdSmetric}) :
\begin{eqgroup}
 ds^2= (r^2-M\ell^2)d\tau^2+\frac{\ell^2 dr^2}{r^2-M\ell^2}+r^2dx^2\ .
 \label{EquilibriumAdSmetric}
\end{eqgroup}

According to the sign of $M$, this metric either describes thermal AdS $(M<0)$ or the BTZ black hole $(M>0)$ with an horizon at $r=\sqrt{M}\ell$. To respect the boundary geometry, we have to require $\tau\sim \tau+\frac{1}{T}$ and $x\sim x+L$. Additionally to that, there are some constraints on the parameter $M$ in order to avoid conical singularities as explained in sec.\ref{sec:HawkingPage}. These conditions impose :
\begin{eqgroup}
 &M=(2\pi T)^2,\mbox{ if }M>0\ ,\\
 &M=-\left(\frac{2\pi}{L}\right)^2,\mbox{ if }M<0\ .
 \label{regularityconditions}
\end{eqgroup}

Finally, the AdS radius $\ell_i$ for each side will be of course determined by the Brown-Henneaux formula (\ref{brownhenneauxcentralcharge}), which in our units system reads $c =12 \pi \ell$. Borrowing nomenclature from Coleman and De Luccia \cite{Coleman:1980aw} in their study of vacuum decay, we will also refer to side 1 as the "true vacuum" and side 2 as the "false vacuum". This nomenclature is explained by considering the cosmological constant as arising from the vacuum expectation value of a minimally coupled scalar field, in which case lower values of $\ell_i$ correspond to a lower minimum for the scalar field potential.

The parameter $M$ is determined by the associated CFT state. Through the dictionary (see (\ref{flatfeffermangraham})) we find the correspondence $\langle T_{--}\rangle =\langle T_{++}\rangle =\frac{1}{4}M\ell$, which results in an energy density of $\frac{1}{2}M\ell$. From (\ref{regularityconditions}), when $M>0$ this energy depends on the temperature and is to be interpreted as a thermal energy density. When $M<0$, this energy is negative and scales as $\frac{1}{L}$, it is a Casimir energy as explained in (\ref{Casimircylinder}). 

\begin{figure}[!h]
    \centering
    \includegraphics[width=0.25\linewidth]{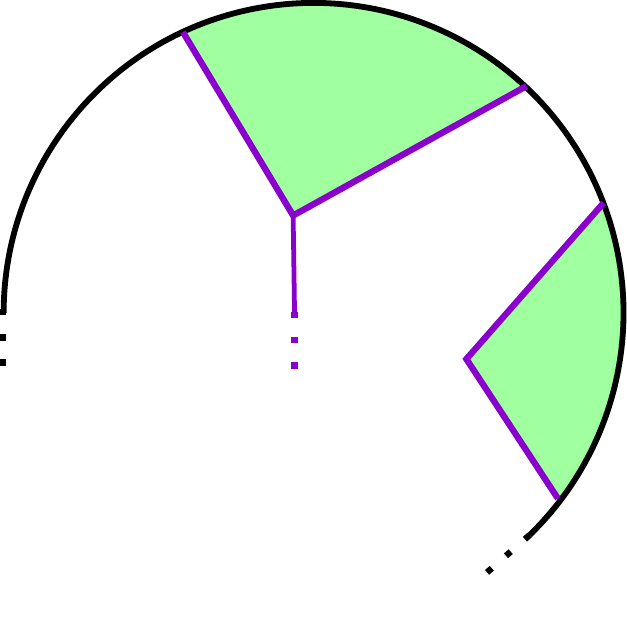}
    \caption{{\small Two examples of wall fusion which is not considered in our model. In the case on the right, we would need to add some other interaction to the picture, as the tension of the wall would tend to smooth out the kink. In the case of the triple fusion depicted on the top, no extra interaction would be needed beside the Hayward term (\ref{fullEuclideanAction}). However, one would need to understand how to glue two spacetimes in this configuration.}}
    \label{fig:nonsmoothjoining}
\end{figure}

The minimal ICFT dual involves a membrane of tension $\lam$ anchored at the CFT interface location on the boundary. One should specify its shape to complete the description of the bulk state. This is non-trivial in the general case, and one of the goals of this paper. We can however begin by specifying the different allowable topologies of each spacetime slice. Before doing so we will have to make a few additional restricting assumptions about the gravity dual model. While we know from the dictionary that there should be two membrane pieces attached on the boundary interfaces, there is no a priori requirement that they join smoothly in the bulk. Two membranes could in principle fuse at an angle instead of joining smoothly, as depicted in fig.\ref{fig:nonsmoothjoining}.

Although such wall junctions should enter in the full holographic picture of the model, we will restrict ourselves to consider only smooth walls. 

\begin{figure}[!h]
\centering
\includegraphics[width=.7\textwidth]{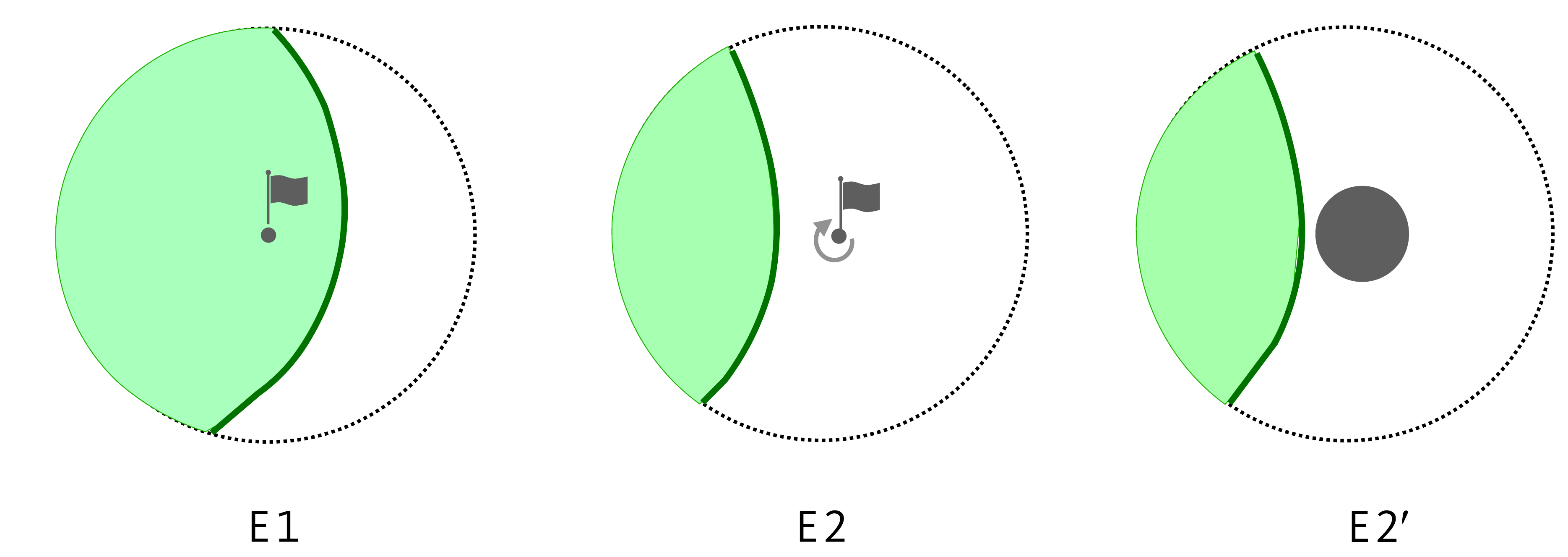}
 \includegraphics[width=.45\textwidth]{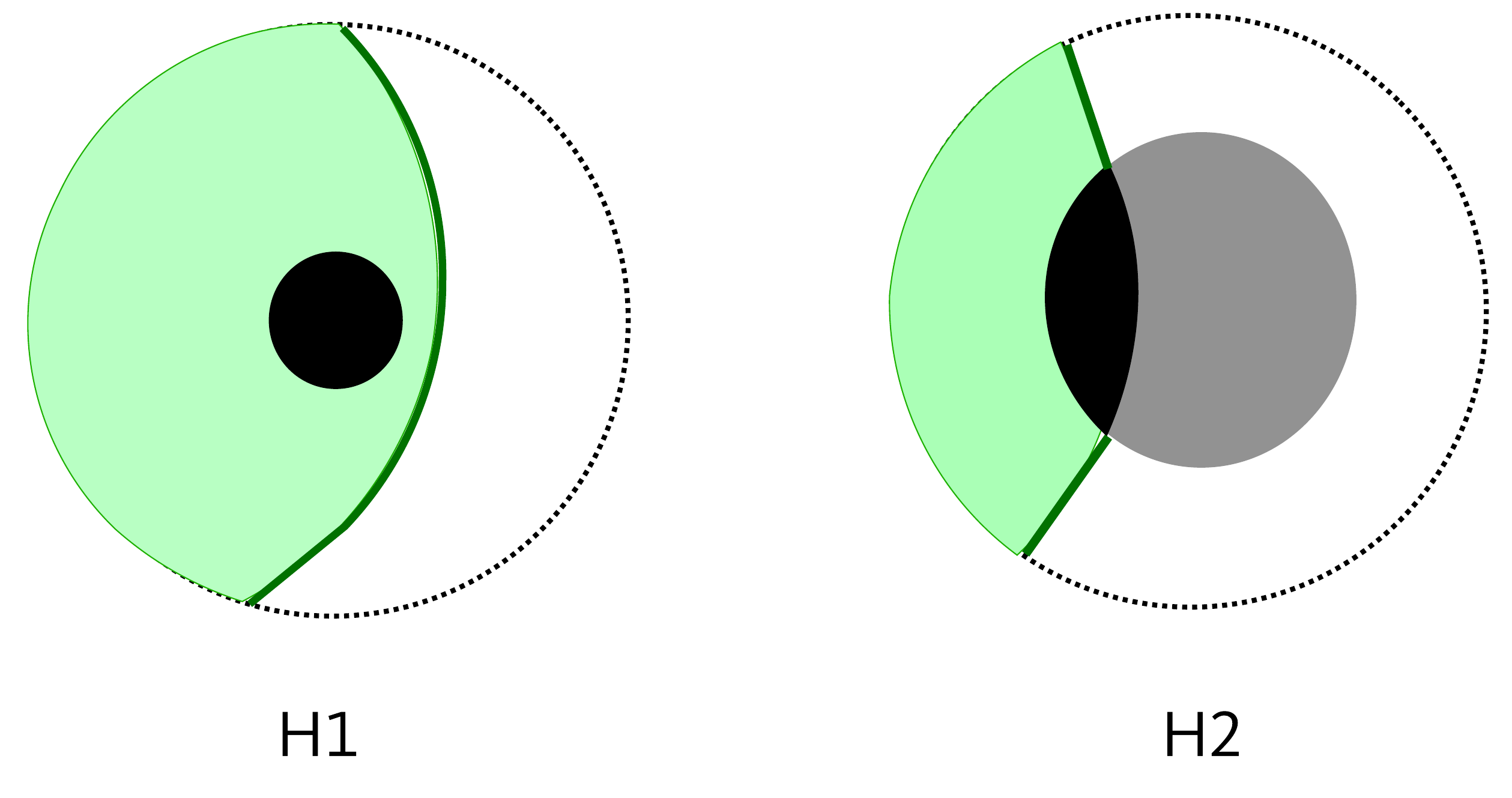}
   \caption{\small  The different types of   space-time slices.  The actual slice is colored in green, the  complementary region  is excised. 
     The letters `E' and `H' stand for `empty' and `horizon', and the grey flag  denotes the  rest point 
     of an inertial observer.  Note that since this is excised in E2, a conical singularity in its place is permitted. The centerful slice  E1  can act as a gravitational Faraday cage.}
\label{fig:slices}
 \end{figure}

With that in mind, we can identify 5 types of "half-spaces", or "slices", which are delimited by the membrane, see fig.\ref{fig:slices}.

The metric (\ref{EquilibriumAdSmetric}) of slices labeled by {\small E1},{\small E2} has $M<0$, while {\small E2}',{\small H1},{\small H2} have $M>0$. The different slices in (\ref{fig:slices}) are distinguished topologically according to which cycle ($\tau$ or $x$) is contractible in the bulk, if any. More simply put, they are distinguished by whether the center of AdS, or the black hole are excised from the geometry. In this sense, {\small E2} and {\small E2}' are of the same "topological type", and that is why we distinguish them only by a prime. On the contrary, {\small E1} and {\small E2} are of different topological type because the center of AdS is not excised in {\small E1}, while it is in {\small E2}.

Note that the regularity conditions (\ref{regularityconditions}) are applicable only when the relevant cycle is contractible in the excised slice. In other words, for {\small E2} and {\small E2}', the choice of the parameter $M$ is unconstrained, even if it would produce a conical singularity in the full spacetime, since such a singularity is excised in the final geometry. On the other hand, for {\small E1}, {\small H1} and {\small H2}, (\ref{regularityconditions}) must be satisfied.

\begin{figure}[!h]
    \centering
    \includegraphics[width=0.8\linewidth]{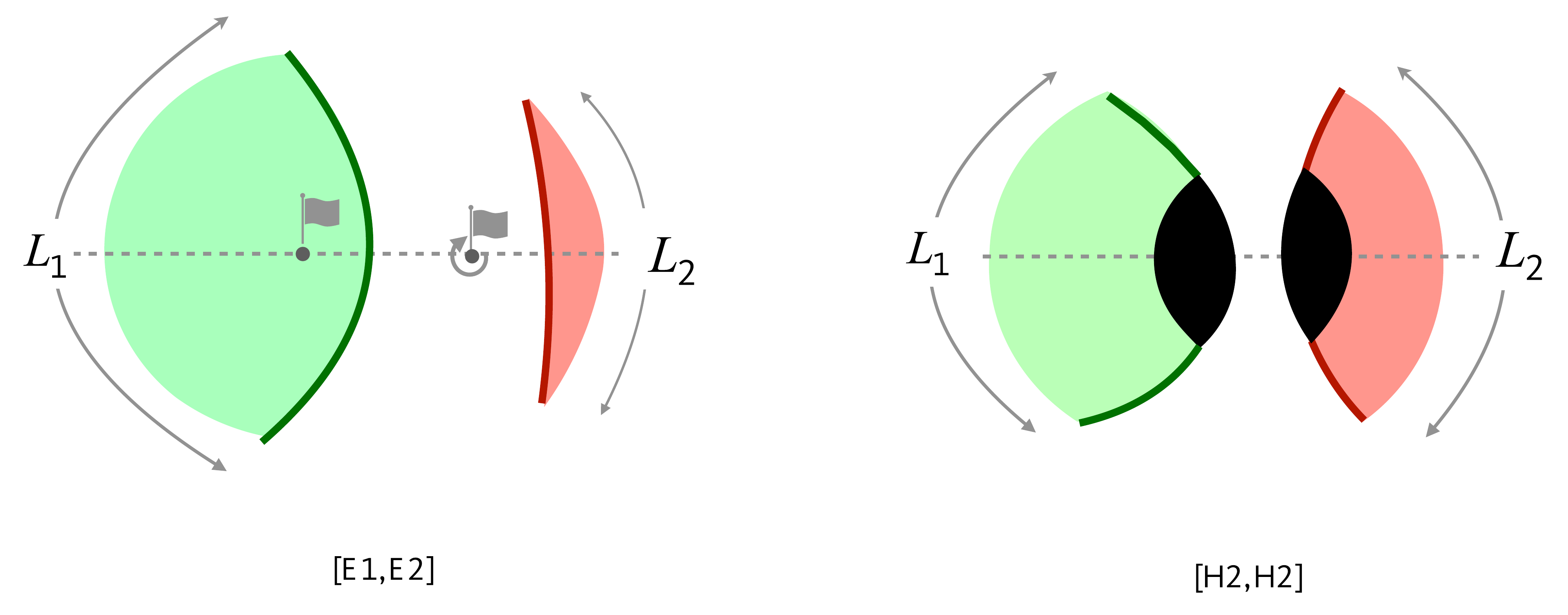}
    \caption{\small Given the CFT data described at the start of the section, one picks the appropriate slices and glue them together along the membrane.}
    \label{fig:gluingexample}
\end{figure}

To construct a generic solution dual to an equilibrium ICFT state, one should pick two slices from fig.\ref{fig:slices}, and glue them together along the membrane. This generates a bulk that has the required asymptotic metric as per the holographic dictionary. This is pictorially depicted in \ref{fig:gluingexample}.
Of course, the gluing of the slices must satisfy some conditions, which are a result of Einstein's equations coupled to the membrane of tension $\lam$. We will see that as a result, not all slice pairings will be allowed. These constraints, coming from the gravity theory, tell us something about the strongly coupled physics of the field theory dual. Determining what exactly is that is far from obvious, and will be explored in the main text.

\section{The gluing equation}
\label{sec:thegluingequation}
In this section, we re-derive the "matching conditions", i.e. the equations that determine how the gluing of the two spacetimes should be done. Of course, this study is not the first time these equations appear, and they are attributed to Israel \cite{Israel:1966rt} (hence the name "Israel matching conditions"). However, we provide here a derivation directly from the action principle of relativity. Usually, the equations are derived directly from the equations of motion, by integrating on a thin shell around the gravitating wall. Although the derivation from the action principle is also probably not original, we think it might be useful, especially in the case one would like to extend the equations to more general cases, such as interface junctions.

The first step is determining what is the correct action to vary. The full system is comprised of two manifolds with gravity, and one membrane of constant tension $\lam$. For the membrane, the action is simply proportional to the area of the surface, the proportionality value being the tension. For the two bulks, the action will simply be the Einstein-Hilbert one. Since the membrane constitutes a boundary of spacetime, we also need to include the Gibbons-Hawking boundary contributions to make the variational problem well-posed. Considering for simplicity there is no other boundary other than the membrane :
\begin{eqgroup}
S_{\rm gr} =    \frac{1}{2}&\int_{{\mathbb  S}_1}d^Dx \sqrt{-g_1}(R_1+\cal{L}^1_{mat})  +\frac{1}{2}
 \int_{{\mathbb  S}_2}d^Dx\sqrt{-g_2}(R_2+\cal{L}_{mat}^2) \\&-   \lam \int_{\mathbb{M}} d^{D-1}s
     \sqrt{-h_m} 
     + \int_{\pa\mathbb{S}_1} d^{D-1}s \sqrt{-h_1}K_1 
     + \int_{\partial {\mathbb  S}_2} d^{D-1}s \sqrt{- g_2}K_2   
     \label{Fullactionwithdomainwall}\ .
\end{eqgroup}

Some precisions are in order. In (\ref{Fullactionwithdomainwall}) we consider essentially two separate manifolds $\mathbb{S}_i$, which are formally identified at their boundary, $\pa \mathbb{S}_i$. This contains in particular the less general point of view in which the $\mathbb{S}_i$ are simply two sections of a bigger manifold $\mathbb{S}$. The manifold $\mathbb{M}$ can thus equivalently be seen as equal to either $\pa \mathbb{S}_i$.

$\cal{L}^i_{mat}$ are the Lagrangians for the matter content on each side. The $h_i$ are the induced metric on the boundaries, and the $K_i$ are their extrinsic curvatures. Note that with the sign conventions we have used in (\ref{Fullactionwithdomainwall}), the normal vectors with which we compute $K_i$ should be pointed outward from the bulk $\mathbb S_i$. The D-1 coordinates "$s$" parametrize the boundaries. 

Because the matter content on each side can be completely different, there is no reason to believe that the solutions to the Einstein's equation $g_1$ and $g_2$ should be smoothly related across the separating membrane. How they are connected will be given by the variation of (\ref{Fullactionwithdomainwall}). However, in writing this action, we implicitly made an assumption for it to be well defined. Indeed, we must assume $h_1=h_2$, since the boundaries are identified, $\pa \mathbb{S}_1=\pa \mathbb{S}_2=\mathbb{M}$, there is only one metric for this manifold, and as such we must have $h_1=h_2=h_m$ for consistency.

This first consistency condition is called the "metric matching condition". Parametrise the surface by coordinates $s^a$ as $x_i^\mu(s^a)$ on each side $i$. The identification of $\pa\mathbb{S}_i$ is done by identifying $x_1^\mu(s^a)\equiv x_2^\mu(s^a)$. Then, the metric matching condition reads :
\begin{eqgroup}
h^1_{ab}(s)=g^1_{\mu\nu}\frac{\pa x_1^\mu}{\pa s^a}\frac{\pa x_1^\nu}{\pa s^b}=g^2_{\mu\nu}\frac{\pa x_2^\mu}{\pa s^a}\frac{\pa x_2^\nu}{\pa s^b}=h^2_{ab}(s)\ .
\label{metricmatching}
\end{eqgroup}
This condition is a constraint on the equations of motion for the membrane, it is non-dynamical.

With that out of the way, we proceed to vary $S_{gr}$ w.r.t. to the metrics. Since the problem is completely symmetric on both sides, we will consider only the variation of $g_1$. We ignore the bulk matter fields as they simply give a stress-tensor contribution to the Einstein equations. Thus consider :
\begin{eqgroup}
 S_{gr} = \frac{1}{2}\int_{\mathbb{S}}d^Dx \sqrt{-g} R+\int_{\pa \mathbb S}d^{D-1}s \sqrt{-h}(K-\lam)\ .
 \label{epuratedAction}
\end{eqgroup}

The variation of the bulk term is well-known, we have :
\begin{eqgroup}
    &\delta \sqrt{-g}=\frac{1}{2}\sqrt{-g} g^{\mu\nu}\delta g_{\mu\nu}\ ,\\
    &\delta R = -R^{\mu\nu}\delta g_{\mu\nu}+\co^\mu\left(\co^\nu \delta g_{\mu\nu}-g^{\nu\lam}\co_\mu \delta g_{\nu\lam}\right)\ .
    \label{EHvariation}
\end{eqgroup}

To obtain (\ref{EHvariation}), one identity is useful :
\begin{eqgroup}
\delta \Gamma^\rho_{\mu\nu} =\frac{g^{\rho\si}}{2}\left(\co_\mu \delta g_{\nu\si}+\co_\nu\delta g_{\mu\si}-\co_\si \delta g_{\mu\nu}\right)\ .
 \label{christovariation}
\end{eqgroup}

Thus the variation of the bulk term yields :
\begin{eqgroup}
 \frac{1}{2}\int_{\mathbb S}d^Dx \sqrt{-g}\left(\frac{R}{2}g^{\mu\nu}-R^{\mu\nu}\right)\delta g_{\mu\nu}+\frac{1}{2}\int_{\pa \mathbb S}d^{D-1}s\sqrt{-h}\;n^\mu\left(\co^\nu \delta g_{\mu\nu}-g^{\nu\lam}\co_\mu \delta g_{\nu\lam}\right)\ .
 \label{EHvariationplugged}
\end{eqgroup}

The first term gives us the Einstein equations in the bulk, while we also get a contribution on the boundary. For the variation of the boundary term in (\ref{epuratedAction}), we will use the formula (\ref{Extrinsic1}) for the Extrinsic curvature. 

The variation of the boundary metric $h^\mu_\nu$ and the normal covector $n_\mu$ can be obtained from (\ref{deltanormal}):
\begin{eqgroup}
 \delta n_\mu &= \frac{1}{2}n_\mu n^\rho n^\si \delta g_{\rho\si}\ ,\\
 \delta h^\mu_\nu &= \delta(-n_\nu n_\al g^{\al \mu})=n_\nu n^\rho h^{\si\mu}\delta g_{\rho\si}\ .
 \label{deltanormal}
\end{eqgroup}
See (\ref{metricisprojection}) for the formula of $h_{\mu\nu}$. Notice also the identity $h^\al_\nu \co_\be n_\al = \co_\al n_\nu$. 

Then :
\begin{eqgroup}
 \delta{\sqrt{-h}}&= \frac{1}{2}\sqrt{-h}h^{\rho\si} \delta g_{\rho\si}\ ,\\
 \delta(K) =\delta (K_{\mu\nu}g^{\mu\nu})&=-K^{\mu\nu}\delta g_{\mu\nu}+\delta K_{\mu\nu} g^{\mu\nu}\ .
 \label{startingpointofKvariation}
\end{eqgroup}

The main difficulty lies in the computation of $\delta K_{\mu\nu}$. We break it down into three pieces using (\ref{Extrinsic1}), after some simplifications :
\begin{eqgroup}
 \delta K_{\mu\nu} =\underbrace{\delta h_\mu^\al \co_\al n_\nu}_A + \underbrace{h^\al_\mu h_\nu^\be  \co_\al \delta n_\be+h^\al_\mu \delta h^\be_\nu \co_\al n_\be}_B- \underbrace{h^\al_\mu \delta \Gamma^\rho_{\al\nu}n_\rho}_C\ .
 \label{deltaKmunuABC}
\end{eqgroup}

We find :
\begin{eqgroup}
 A&= K_\nu^\si n_\mu n^\rho \delta g_{\rho\si}\ ,\\
 B&= \frac12 K_{\mu\nu}n^\rho n^\si \delta g_{\rho\si}+n_\nu K_\mu^\rho n^\si \delta g_{\rho\si}\ .
 \label{AandB}
\end{eqgroup}
Combining this with the identity $n_\nu K^\nu_\rho =0$ :
\begin{eqgroup}
g^{\mu\nu}\delta K_{\mu\nu}&=g^{\mu\nu}\left(\frac12 K_{\mu\nu}n^\rho n^\si \delta g_{\rho\si}+(n_\nu K_\mu^\rho+n_\mu K_\nu^\rho) n^\si \delta g_{\rho\si}-h^\si_\nu h^\al_\mu n_\rho \delta \Gamma^\rho_{\si \al}\right)\ ,\\
&=\frac{1}{2} K h^{\rho\si} \delta g_{\rho\si}-h^{\al\si}n_\rho \delta \Gamma^\rho_{\si\al}\ .
\label{gmunudeltaKmunu}
\end{eqgroup}

We can now combine all the expressions we have collected to recover :
\begin{eqgroup}
 \delta K= \left[\frac{1}{2}K h^{\rho\si}-K^{\rho\si}\right]\delta g_{\rho\si}-h^{\al\si}n_\rho \delta \Gamma^\rho_{\si\al}
 \label{variationofboundaryterm}\ .
\end{eqgroup}

To be able to extract equations of motions, the combined sum of the variation must take the form $A^{\rho\si}\delta g_{\rho\si} + \cal{D}^\mu c_\mu$, where $\cal{D}^\mu$ denotes the covariant derivative that can be partially integrated w.r.t. the measure $\sqrt{-h}d^{D-1}s$. Indeed, after a bit of guesswork, one can re-arrange (\ref{variationofboundaryterm}) to find :
\begin{eqgroup}
 \delta K = -\frac{1}{2}K^{\rho\si}\delta g_{\rho\si}&-\frac{1}{2}n^\mu\left(\co^\nu \delta g_{\mu\nu}-g^{\nu\lam}\co_\mu \delta g_{\nu\lam}\right)-\frac{1}{2}h^{\al\si}\co_\al(h^{\mu\si}n^\nu \delta g_{\nu\si})\ ,\\
  \Rightarrow& \cal{D}_\mu c^\mu \equiv -\frac{1}{2}h^{\al\si}\co_\al(h^{\mu\si}n^\nu \delta g_{\nu\si})\ .
  \label{finaldeltaK}
\end{eqgroup}

From which we infer $c^\mu = \frac{1}{2}h^{\mu\si}n^\nu \delta g_{\nu\si}$. Conveniently, a big portion of (\ref{finaldeltaK}) cancels the boundary term of (\ref{EHvariationplugged}). That is of course to be expected, it is precisely what we require of the Gibbons-Hawking term to do. Combining all the formulas together, we finally get the full boundary variation as :
\begin{eqgroup}
 \int_{\pa \mathbb S}d^{D-1}s \sqrt{-h}\left[\frac{1}{2}\left(Kh^{\rho\si}-K^{\rho\si}-\lam h^{\rho\si}\right)\delta g_{\rho\si}+\cal{D}_\mu c^\mu \right]\ .
 \label{finalvariation}
\end{eqgroup}

Let us make a few comments before concluding. In the case where the surface $\pa \mathbb S$ is the asymptotic boundary of spacetime, one usually imposes asymptotic "Dirichlet" boundary conditions for the metric. In other words, the surface metric is fixed, and thus the tangential variation vanishes, $h^{\rho\si}\delta g_{\rho\si}=0$. When this holds, the expression (\ref{finalvariation}) vanishes, so that the boundary term is cancelled exactly by the variation of the Gibbons-Hawking term, as it should. 

In our case, the boundary $\pa \mathbb{S}$ will itself have no boundary. Hence the divergence $\cal{D}_\mu c^\mu$ can be integrated away. In models where this condition doesn't hold (for example if we have two boundaries joining at a kink, or if we have a "dangling" membrane), the partial integration of this term will yield further conditions. We leave these cases for later study. 
Thus the additional equation of motion that we obtain once we consider both side is :
\begin{eqgroup}
 K^1_{\rho\si}+K^2_{\rho\si}-(K^1+K^2)h_{\rho\si}=-\lam h_{\rho\si}\ .
 \label{IsraelMatching}
\end{eqgroup}
Equation (\ref{IsraelMatching}), together with (\ref{metricmatching}) are the so-called Israel matching conditions. Note that more generally $\lam h_{\rho\si}$ can be replaced with $T^{mem}_{\rho\si}$ where $T^{mem}_{\rho\si}$ is a stress-energy tensor for the surface; it should be symmetric for consistency, although it need not be conserved as one can check that the covariant derivative of (\ref{IsraelMatching}) need not vanish. In general, we will contract (\ref{IsraelMatching}) with the tangent vectors $t^\mu_a$, to obtain an equation expressed directly with surface tensors (instead of ambient space tensors).

It may be useful to trace-reverse (\ref{IsraelMatching}):
\begin{eqgroup}
 K^1_{ab}+K^2_{ab}=\frac{\lam}{D-2}h_{ab}\ .
 \label{tracereverseIsrael}
\end{eqgroup}

\section{Solving the wall equations}
\label{sec:solvingthewall}
In this section, we find the general solution of the domain wall equations in terms of the mass parameters $M_j$ and the AdS radii $\ell_j$, as well as the tension of the wall $\lam$. Part of this analysis is related to that of ref.\cite{Simidzija:2020ukv} by double Wick rotation; namely by exchanging the roles of $t$ and $x$. While in Euclidean space the distinction between time and space is formal, when considering the Lorentzian interpretation of the result, the conclusions are completely different.
We would like to solve the Israel matching conditions in our ICFT model. The full action is given by :
\subsection{The wall equations}
\begin{eqgroup}
S_{\rm gr} &=    \frac{1}{2} \int_{{\mathbb  S}_1}d^3x 
\sqrt{-g_1}(R_1+\frac{2}{\ell_1^2})  +\frac{1}{2}
 \int_{{\mathbb  S}_2}d^3x\sqrt{-g_2}(R_2+\frac{2}{\ell_2^2}) +\lam \int_{\mathbb   M} d^2s
     \sqrt{-h}\\
    & + \int_{\partial {\mathbb  S}_1} d^2s \sqrt{-h_1} K_1 
     + \int_{\partial {\mathbb  S}_2} d^2s \sqrt{-h_2} K_2 
     +\frac{1}{\ell_1}\int_{\mathbb{B}_1} \sqrt{-h_1}  
   +  \frac{1}{\ell_2}\int_{\mathbb{B}_2} \sqrt{-h_2}\ ,
\end{eqgroup}
where we have included the counterterms that will be needed to get well-defined on-shell actions. They contribute only to the asymptotic boundaries, which we denoted by $\mathbb{B}_i$, whereas $\pa \mathbb{S}_i$ will include also the boundary constituted by the membrane. We denoted by a blanket notation $h_i$ the induced metrics on the various boundaries.

The bulk equations are already solved by picking the bulk metrics to be in the form of (\ref{EquilibriumAdSmetric}). In general, the coordinates charts $(\tau_j,r_j,x_j)$ may be discontinuous at the interface, so we label them by their side. However, the assumptions of staticity mean the time-coordinate will be globally defined, so we write $\tau_1=\tau_2=\tau$. The membrane will be parametrized by two parameters, $\ti{\tau}$ and $\si$. From staticity, a generic parametrisation is written $\tau=\ti{\tau}$, $x_j = x_j(\si)$ and $r_j = r_j(\si)$. By abuse of notation, we will drop the $\ti{\tau}$ parameter and identify it with $\tau$. The two (Euclidean) metrics thus can be written as :
\begin{eqgroup}
 ds^2_j = (r_j^2-M_j\ell_j^2)d\tau^2+\frac{\ell_j^2dr_j^2}{r_j^2-M_j\ell_j^2}+r_j^2 dx_j^2\mbox{ with }(r_j,x_j)\in \Omega_j\ .
 \label{labeledmetricsstatic}
\end{eqgroup}

The range of the coordinates $\Omega_j$ will depend on the specific geometry, and is delimited as follows :
\begin{itemize}
 
\item by the embeddings of the
static wall in the two coordinate systems, 
   $\{x_j(\sigma), r_j(\sigma) \}$; 

\item by the horizon whenever the slice contains one, i.e. in cases {\small  H1}
   and {\small  H2}; 

\item by the cutoff surface $r_j  \approx  1/\epsilon\to    \infty$. 
\end{itemize}

The metric matching conditions then give us two (non-trivial) equations :
\begin{align}
  r_1^2 -M_1 \ell_1^2   = r_2^2 -M_2 \ell_2^2&\equiv f(\sigma)
  \label{matching1}\ ,\\
f^{-1} {\ell_1^2    r_1^{\prime\, 2}   } +  r_1^2  x_1^{\prime\, 2} =
f^{-1} {\ell_2^2\,   r_2^{\prime\, 2} } +  r_2^2  x_2^{\prime\, 2}  &\equiv g(\sigma)  \ ,
\label{matching2}
\end{align}
where we defined $f(\si)$ and $g(\si)$ as the coefficients of the surface metric, 
\begin{eqgroup}
 ds_{\mathbb{M}}^2 = f(\si)d\tau^2+g(\si)d\si^2\ .
 \label{membranemetric}
\end{eqgroup} 

In writing the equations in this manner, we still have a gauge freedom corresponding to reparametrizations of the membrane coordinates. We will use it to fix $f(\si)=\si$ henceforth. We name this the "blue-shift" or "red-shift" parametrization, since $\sqrt{\si}=\sqrt{g_tt^i}$ gives the blue-shift factor.

To write down the other surface equation (\ref{tracereverseIsrael}), we need first to compute the extrinsic curvature. The normal (co)vector is given by :
\begin{eqgroup}
 n^j_\mu &= \frac{1}{\cal{N}_j}\left(0,-x_j^{\prime}(\si),r^\prime_j(\si)\right)\ ,\\
 \cal{N}_j &= -\frac{\ell_j^2 r^{\prime\,2}_j}{\si}+r_j^2 x^{\prime 2}_j\ ,
 \label{normalvector}
\end{eqgroup}
where when we write $r_j$ or $x_j$ they are to be read as functions of $\si$.

Notice that in our convention (\ref{normalvector}) should be seen as being pointing outwards from the spacetime. Thus, in the pictures (\ref{fig:slices}), the normal vector should point away from the green region. In particular, since $r^\prime>0$, the part of the spacetime that is kept is the one lying on the smaller x side of the membrane. The computation of the extrinsic curvature tensor is relegated to the appendix (\ref{app:appendixextrinsiccomputationstatic}). Once done, one notices that (\ref{tracereverseIsrael}) furnishes only one independent equation :
\begin{eqgroup}
\frac{r_1^2 x_1^\prime}{\ell_1}+\frac{r_2^2 x_2^\prime}{\ell_2}=-\lam \sqrt{\si g(\si)}\ .
\label{extrinsicIsraelequationstatic}
\end{eqgroup}

In total, having fixed the parametrization, we are left with 3 independent equations to solve, for three unknown functions $x_j^\prime$ and $g(\si)$. Furthermore, the equations only involve first derivatives of $x_j$, so the integration constants are irrelevant choices of the origin of the $x_j$ axes. For a given $\ell_j$ and $\lam$, the wall embedding functions $x_j(r_j)$ are thus uniquely determined by the parameters $M_j$. However, different choices of $(M_1,M_2)$ may correspond to the same boundary data $(L_1,L_2,T)$. Two such solutions are then competing phases of the system.

\subsection{Near boundary solution}
\label{sec:nearboundarysol}
\begin{figure}[!h]
    \centering
    \includegraphics[width=0.5\linewidth]{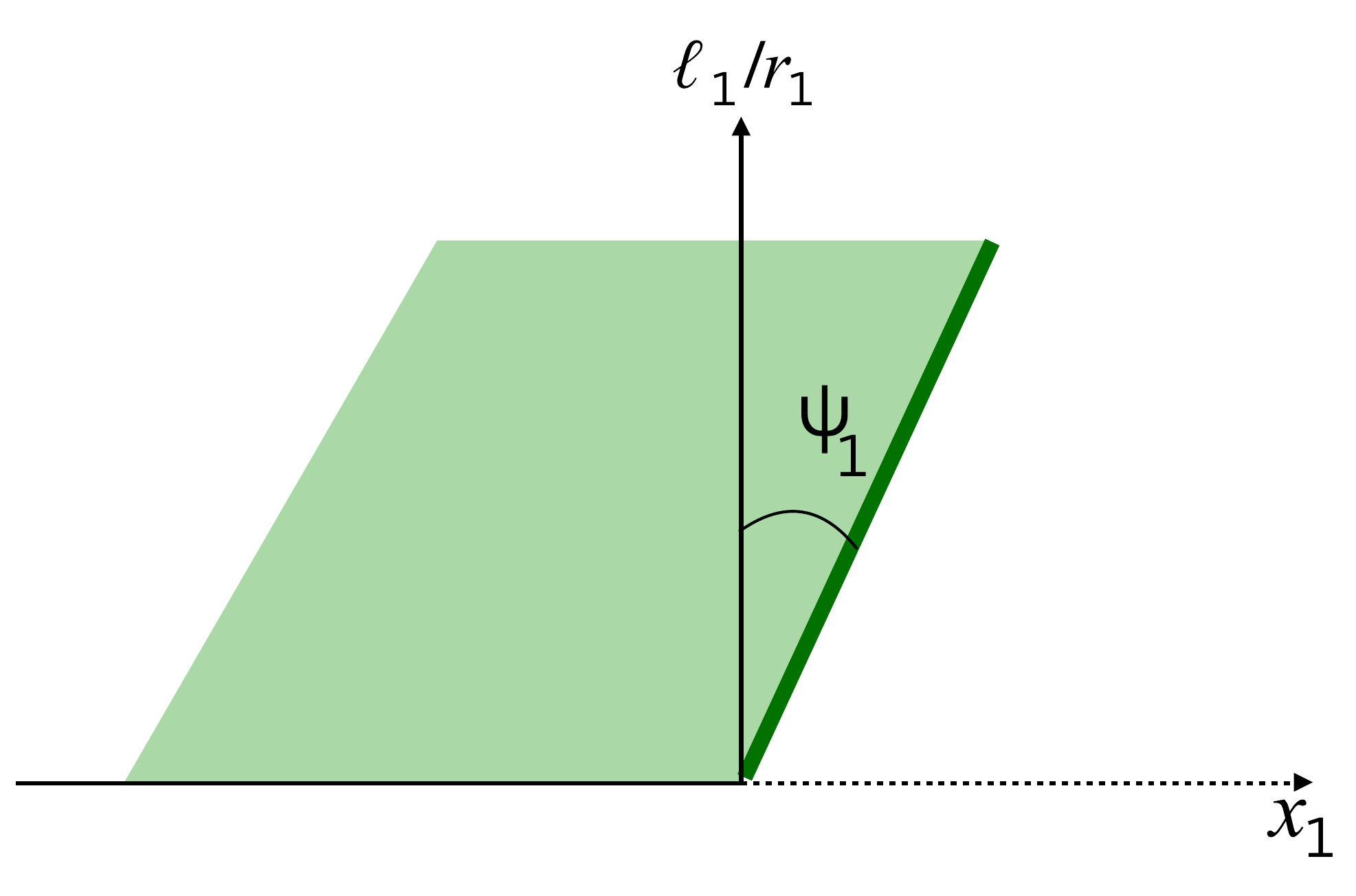}
    \caption{\small Near the AdS boundary in the $(x_j,\frac{l_j}{r_j})$ plane the string is a straight line subtending an angle $\psi_j$ with the normal.}
    \label{fig:nearboundarymem}
\end{figure}

It is instructive to look at the limiting behavior of the wall near the asymptotic boundary. Indeed, as $r_j\rightarrow \infty$, the parameters $M_j$ can be neglected and the metrics approach AdS$_3$ in Poincaré coordinates. In this limit, the solution to the gluing equation simplifies and reads \cite{Bachas:2001hpy} :
\begin{eqgroup}
 r_1\approx r_2 (\approx \si),\;\;\; x_j \approx \ell_j\frac{(\tan(\psi_j))}{r_j}\ ,
 \label{solutionbigr}
\end{eqgroup}
where $\psi_j$ is the angle in the $(x_j,\frac{\ell_j}{r_j})$ between the boundary and the interface, see figure \ref{fig:nearboundarymem}.

Naturally, the angles $\psi_j$ are completely fixed by the Israel conditions :
\begin{eqgroup}
 \frac{\ell_1}{\cos\psi_1}=\frac{\ell_2}{\cos\psi_2}\equiv\ell_m\;\;\; \mbox{and} \;\;\; \tan\psi_1+\tan\psi_2=\lam \ell_m\ ,
 \label{constraint on angles}
\end{eqgroup}
where one can identify $\ell_m$ with the radius of the membrane worldsheet, which will be AdS$_2$ in this limit. We have $-\pi/2<\psi_j<\pi/2$. 

In the assumption $c_1\leq c_2$ that we mentioned earlier, we have $\ell_1\leq \ell_2$ and the equations (\ref{constraint on angles}) impose $|\tan(\psi_1)|\geq|\tan(\psi_2)|$, as well as $\psi_1>0$ provided the tension $\lam$ is positive, which is the physical assumption. The sign of $\psi_2$ depends on the precise value of $\lam$. Combining equations (\ref{constraint on angles}) yields an equation for $\ell_m$ :
\begin{eqgroup}
 \sqrt{\frac{1}{\ell_1^2}-\frac{1}{\ell_m^2}}+{\rm Sign}(\psi_2) \sqrt{\frac{1}{\ell_2^2}-\frac{1}{\ell_m^2}}=\lam\ .
 \label{equationforlm}
\end{eqgroup}

For each value of the worldsheet radius $\ell_w$, there corresponds two tensions $\lam$, according to ${\rm Sign}(\psi_2)$. We obtain some bounds for the tension:
\begin{eqgroup}
 \lam_{\rm min}<\lam<\lam_0\;\;\mbox{for }\psi_2<0\mbox{ and } \lam_0<\lam<\lam_{\rm max}\;\;\mbox{for }\psi_2>0\ ,
 \label{tensionbounds}
\end{eqgroup}
where the three critical tensions read :
\begin{eqgroup}
   \lambda_{\rm min}  =   {\frac{1}{\ell_1}} - {\frac{1}{ \ell_2}} ,\; 
 \lambda_{\rm max} = \frac{1}{\ell_1}+  \frac{1}{\ell_2}\ ,\;
 \lambda_0  = \sqrt{\lambda_{\rm max}\lambda_{\rm min}}\ .
 \label{criticaltensions}
\end{eqgroup}

Let us briefly pause to discuss the significance of these critical tensions.

\section{Critical Tensions}
\label{sec:criticaltensions}
The  meaning of the  critical tensions $\lambda_{\rm min}$ and $\lambda_{\rm max}$   has been    understood in  the work of Coleman-De Lucia  \cite{Coleman:1980aw} and Randall-Sundrum \cite{Randall:1999vf}. Below  $\lambda_{\rm min}$   the false vacuum is unstable to  nucleation of true-vacuum bubbles, so  the two phases cannot coexist in equilibrium.\footnote{Ref.\,\cite{Coleman:1980aw}  actually computes the critical tension for  a  domain wall separating  Minkowski from  AdS spacetime. Their result can  be  compared to $\lambda_{\rm min}$   in the limit   $\ell_2\to \infty$.} The holographic description of such  nucleating  bubbles  raises   fascinating questions  in its own right, see e.g. refs.\cite{Freivogel:2005qh,Freivogel:2007fx,Barbon:2010gn}. It has been also advocated that expanding true-vacuum bubbles  could realize accelerating cosmologies  in string theory \cite{Banerjee:2018qey}. Since  our  focus here is on equilibrium  configurations, we will not  discuss  these interesting issues any further. 

The maximal tension $\lambda_{\rm max}$ is a stability bound of a different kind.\footnote{Both the  $\lambda = \lambda_{\rm max}$ and the  $\lambda = \lambda_{\rm min}$ walls can arise as flat BPS walls in supergravity theories coupled to scalars \cite{Cvetic:1992bf, Cvetic:1996vr}. These two  extreme types of flat wall,  called  type II and type III in \cite{Cvetic:1996vr}, differ  by  the fact that the   superpotential   avoids, respectively passes through zero as  fields extrapolate  between the AdS vacua \cite{Ceresole:2006iq}.} For $\lambda >   \lambda_{\rm max}$ the two phases can  coexist, but the   large  tension of the wall forces this latter   to inflate \cite{Karch:2000ct}. The phenomenon  is familiar  for gravitating domain walls in asymptotically-flat spacetime \cite{Vilenkin:2000jqa}, i.e. in the limit  $\ell_1, \ell_2\to \infty$. Again, such tensions are irrelevant in our search of equilibrium configurations.

The  meaning of $\lambda_0$  is  less clear,  its role will partially emerge  later. For now, note  that it is the turning point at which the worldsheet radius $\ell_{w}(\lambda)$  reaches its minimal value $\ell_2$. From (\ref{constraint on angles}), it also the point where the wall becomes perpendicular to the boundary, in the false vacuum side. The range $\lambda_{\rm min}< \lambda<\lambda_0$ only exists for non-degenerate AdS vacua, that is  when  $\ell_1$  is strictly smaller than $\ell_2$. 

    As explained in (\ref{sec:MinimalICFT}), in this model the interface is described only by one parameter, so all its CFT data will be determined by it. These include the energy transmission-reflection coefficients (\ref{transmissionreflectioncoeffdefinition}) as well as the interface entropy $S_{\rm int}$, or "g-factor", which is related with the number of localized degrees of freedom. Both of these quantities can be computed exactly. The entropy was computed for instance in \cite{Simidzija:2020ukv,Fu:2019oyc} and reads :
\begin{eqgroup}
 S_{\rm int} = 2\pi \ell_1 \ell_2 \left[\lam_{\rm max} \tanh^{-1}\left(\frac{\lam}{\lam{\rm max}}\right)-\lam_{\rm min}\tanh^{-1}\left(\frac{\lam_{\rm min}}{\lam}\right)\right]\ .
 \label{gfactor}
\end{eqgroup}

It varies monotonically between $-\infty$ and $\infty$ as $\lam$ varies inside its allowed range.

Using holographic techniques, the energy transmission coefficients were computed in (\cite{Bachas:2020yxv}) with the result :
\begin{eqgroup}
 \cal{T}_{1\rightarrow 2}=\frac{\lam_{\rm max}+\lam_{\rm min}}{\lam_{\rm max}+\lam},\;\; \cal{T}_{2\rightarrow 1}=\frac{\lam_{\rm max}-\lam_{\rm min}}{\lam_{\rm max}+\lam}\ .
 \label{transmissioncoeffintermsoflambda}
\end{eqgroup}

Using the identities $\lam_{\rm max}+\lam_{\rm min}=2/\ell_1$ and $\lam_{\rm max}-\lam_{\rm min}=2/\ell_2$ on can check that these coefficients obey the "detailed-balance" condition $c_1 \cal{T}_{1\rightarrow 2}= c_2 \cal{T}_{2\rightarrow 1}$ (see (\ref{detailedbalance}) for its microscopic origin).

The larger of the two transmission coefficients reaches the unitarity bound when $\lam = \lam_{\rm min}$, and both coefficients attain their minimum when $\lam = \lam_{\rm max}$. Total reflection (from the false-vacuum to the true-vacuum side) is only possible if $\ell_1/\ell_2 \rightarrow 0$, i.e. when the “true-vacuum” CFT$_1$ is almost entirely depleted of degrees of freedom relative to CFT$_2$, which confirms the universal results found in \cite{Meineri:2019ycm}.

\subsection{Turning point and horizon}
We will now derive  the general solution of the   equations (\ref{matching1},\ref{matching2},\ref{extrinsicIsraelequationstatic}), and then relate the geometric parameters $M_j$ to the data $(T, L_j)$ of the boundary torus shown in fig. \ref{fig:stripedtorus}.   

We use the parametrization of the membrane in terms of the  blueshift factor  of the worldsheet metric (\ref{membranemetric}). Let $\sigma_+$ correspond to the minimal value of the blueshift attained on the membrane, which is either zero or positive. If $\sigma_+ =0$ the string enters the horizon. While it may be interesting to understand what happens to the string after entering the black hole (and we discuss this in Chapter 3), in the Euclidean picture in which we focus our analysis, spacetime is capped at the horizon. Hence, for the purposes of this Chapter, we will not look behind the horizon.  

On the other hand, if $ \sigma_+ >0$ then, as we will confirm shortly,  this is  the  turning point  of the membrane where   both  $x_1^\prime$ and $x_2^\prime$ diverge, while remaining integrable.  

\begin{figure}[!h]
\centering
\includegraphics[width=.7\textwidth]{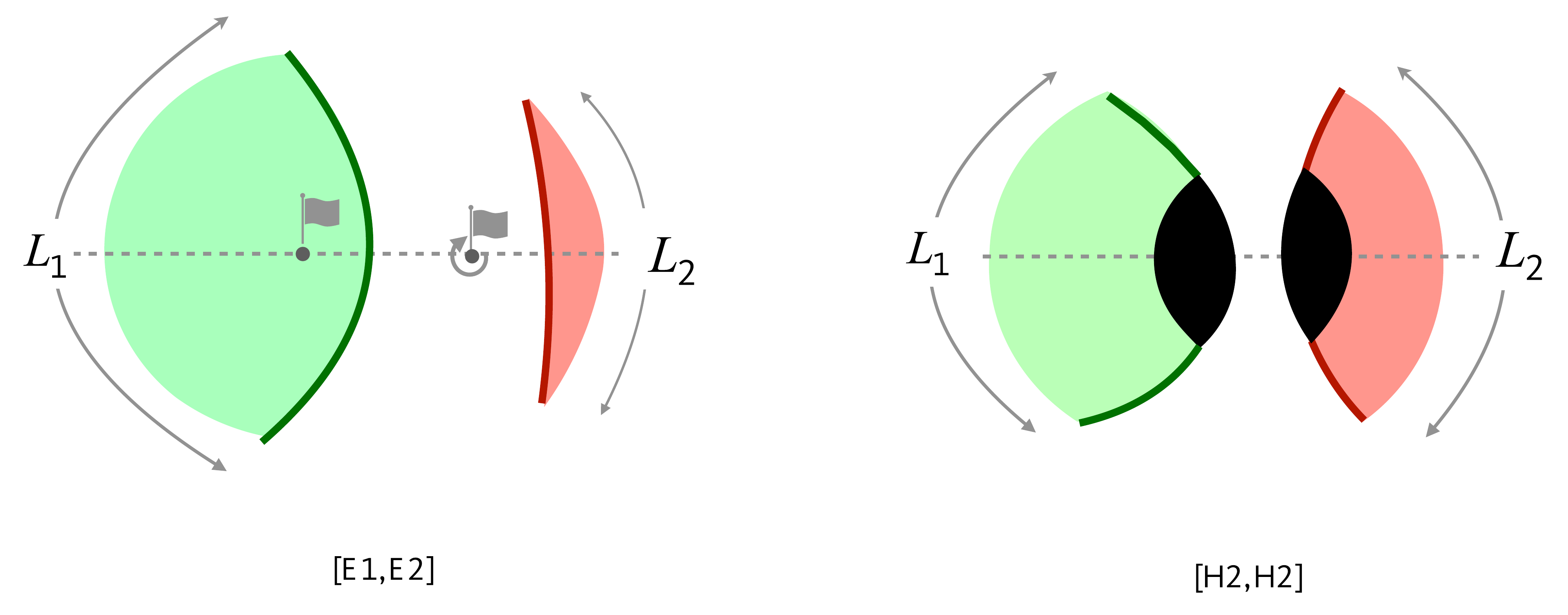}
   \caption{\small   Schematic drawing  of  a low-temperature  and a high-temperature  solution, corresponding to pairs of type [E1,E2] and [H2,H2]. The broken line is the axis of reflection symmetry. The   blueshift parameter  $\vert\sigma\vert $ decreases  monotonically until the string reaches either the turning point or the black-hole horizon.}
   \label{fig:z2symmetrygluing}
 \end{figure}

A  static string  has (at most)  one  turning point, and is symmetric under reflection in   the axis that passes through the centers  of the  boundary  arcs\footnote{In ref.\cite{Simidzija:2020ukv} this corresponds to the time-reflection symmetry of the  instanton solutions.}, as  illustrated  in  figure  \ref{fig:z2symmetrygluing}.  It follows that the blue-shift parametrization is one-to-two. Henceforth we focus only on one half string, corresponding to the boundary solution (\ref{solutionbigr}). The other half string is obtained by $x_j \to -x_j$\footnote{Seemingly, the sign-inverted solution solves (\ref{extrinsicIsraelequationstatic}) for negative tension. However, this is just the equations way of telling us that the conventions we chose (outer normal) are inverted. So in this case, the part of spacetime that is kept, is also sign-inverted. Thus by combining the two $\mathbb{Z}_2$ solutions, we can enclose a portion of spacetime as depicted in fig.(\ref{fig:z2symmetrygluing})}.

Eqs.(\ref{matching1}) imply that $2 r_j  r_j^\prime =1$. Inserting in eq.(\ref{matching2}) gives
\begin{eqgroup}
\label{xprime}
  (x_j^\prime)^2 =  r_j^{-2} \left( g(\sigma) - \frac{\ell_j^2}{4\sigma r_j^2} \right)\ .
 \end{eqgroup}
 
Squaring now twice eq.(\ref{extrinsicIsraelequationstatic}) and replacing $(x_j^\prime)^2$ from the above expressions leads to a quadratic   equation for $g(\sigma)$, the ${\tiny \sigma\sigma}$ component of the worldsheet metric. This equation has a singular solution, $g=0$, and a non-trivial one :
 \begin{equation}
    g(\sigma) =   \lambda^2 \left[  \left(\frac{2r_1  r_2 }{\ell_1 \ell_2 }\right)^2  - 
     \left(\frac{r_1^2}{\ell_1^{\, 2}}  + \frac{r_2^2}{\ell_2^{\, 2}} -   \lambda^2 \sigma\right)^2\, 
    \right]^{-1} 
    =  \frac{\lambda^2}{ A \sigma^2 + 2B\sigma + C}\ .
    \label{gsolution}
\end{equation}
where in the second equality we used  eq.(\ref{matching1}), and  
\begin{eqgroup}
   A &= (\lambda_{\rm max}^2 - \lambda^2 )  (\lambda^2 - \lambda_{\rm min}^2) \ ,\\
 B &=    \lambda^2 (M_1+M_2)  -  \lambda_0 ^2(M_1-M_2)\ ,\\
C &= - (M_1 - M_2)^2\ .
\label{abc}
\end{eqgroup} 

We  expressed   the    quadratic polynomial appearing in the denominator of \eqref{gsolution} in terms of  $M_j, \lambda$ and the critical tensions\eqref{criticaltensions}, in order to render manifest the fact that for $ \lambda$ in the allowed range, $ \lambda_{\rm min}< \lambda  < \lambda_{\rm max}$, the coefficient $A$ is positive. This is required for $g(\sigma) $ to be positive near the boundary where   $\sigma\to\infty$. In addition,    $ AC  \leq   0$ which ensures that the two roots of the denominator in \eqref{gsolution}  
\begin{eqgroup}
\sigma_\pm   =\frac{ -B \pm  (B^2 - AC)^{1/2}}{A} \ ,
\label{rootsofg}
\end{eqgroup}
are real, and that the larger root $\sigma_+$  is non-negative.  Inserting  \eqref{gsolution} in \eqref{xprime} and fixing the sign of the square root near the conformal boundary gives after a little algebra
\begin{eqgroup}
\frac{x_1'}{\ell_1}   = -  \frac{\sigma(\lambda ^2 + \lambda _{0}^2)+M_1-M_2}
{{2(\sigma + M_1\ell_1^{ 2})}\sqrt{A \sigma ( \sigma- \sigma_+)(\sigma-\sigma_-)}} \ ,\\ 
\frac{x_2'}{\ell_2}  =    - \frac{\sigma(\lambda ^2 -  \lambda _{0}^2)+M_2-M_1}
{{2(\sigma + M_2\ell_2^{ 2})}\sqrt{A \sigma ( \sigma- \sigma_+)(\sigma-\sigma_-)}}\ .
\label{fullsolutionstatic}
  \end{eqgroup}
  
We may  now confirm our  earlier claim that  if $\sigma_+ > 0$  then both $x_1^\prime \propto dx_1/dr_1$  and $x_2^\prime \propto dx_2/dr_2$ diverge at this point, with an integrable divergence. Thus the two $\mathbb{Z}_2$ symmetric solutions can be joined smoothly at this point.

Furthermore, since $\sigma + M_j\ell_j^{\, 2}  = r_j^2$ is  positive,\footnote{Except for the measure-zero set of solutions in which the string passes through the center of global AdS$_3$.} the $x_j^\prime$ are   finite at all $\sigma >\sigma_+$. Thus $\sigma_+$ is the unique turning point of the string, as advertised. 
 
Eqs.\eqref{matching1} and \eqref{fullsolutionstatic} give the general solution  of the string equations for arbitrary mass  parameters $M_1, M_2$ of the green and pink slices. These must be related to the torus parameters by interior regularity, and by the Dirichlet  conditions at the conformal boundary. Explicitly,  the  boundary  conditions for the  different slice types  of figure \ref{fig:slices} read: 
 \begin{align}
L_j &=     2  \int_{\sigma_+}^\infty d\sigma   x_j^\prime  \qquad
\qquad   \mbox{for {\small E2, E2}}^{\prime} \label{Dira}\ ,\\
L_j &=     n  P_j  +    2  \int_{\sigma_+}^\infty d\sigma   x_j^\prime   \qquad 
    {\mbox{for {\small E1, H1}}} \label{Dirb}\ ,\\
L_j  &=      \Delta x_j\bigl\vert_{\rm Hor} +   2  \int_{\sigma_+}^\infty d\sigma  x_j^\prime
     \qquad\quad{\mbox{for {\small H2}}} \label{Dirc}   \ .
\end{align}

The  integrals in these equations are the opening arcs, $\Delta x_j$, between the two endpoints of a half string. They can be expressed as complete elliptic integrals of the first, second and third kind,  see appendix \ref{app:2}.  For the   slices  {\small E1, H1} where $x_j$ is a periodic coordinate, we have  denoted by   $P_j >0$ its period, and by $n$ the string winding number. Finally, for strings entering the horizon we denote by $\Delta x_j\vert_{\rm Hor}$ the opening arc  between the two  horizon-entry points  in the  $j$th coordinate chart. 
  
Possible phases of the ICFT for given torus parameters  $T, L_j$ must be solutions to one pair of conditions. Apart from interior regularity, we will also require  that the string does not self-intersect. In principle, two string  bits intersecting at an angle $\not= \pi$ could   join into  another  string.   Such string junctions would be  the gravitational counterparts of interface fusion  \cite{Bachas:2007td},  and allowing them  would  make  the holographic model much richer.\footnote{Generically  the intersection point  in one slice will correspond to two points that must be identified in the other slice; this  may  impose further conditions.} To keep, however, our discussion  simple we will only allow a single type of  domain wall in this work. It would be particularly interesting to understand what membrane intersections means from the point of view of the field theory. 

The reader can easily convince him/herself that to avoid string intersections we must have $P_j > L_j$ and $n=1$  in \eqref{Dirb}, and $\Delta x_j\bigl\vert_{\rm Hor} >0$ in \eqref{Dirc}.

\section{Phases: cold, hot $\&$ warm}
\label{sec:phasescolshotwarm}        
               
Among  the five slice types of figure \ref{fig:slices}, {\small H2} stands apart because it can only pair  with itself.  This is because  a  horizon  is a closed surface, so it cannot end on the domain wall.\footnote{Except possibly in the limiting  case where the wall is the boundary of space.} An equivalent way to see that is that if we were to pair any other slice with {\small H2}, an observer could escape the black hole by traversing to the other side through the membrane.  We will now show that the  matching equations    actually  rule out  several other pairs   among   the  remaining slice  types.  
        
One pair that is easy  to  exclude is [{\small H1,H1}], i.e. solutions that describe two black holes  sitting on either side of the wall. Interior regularity would require  in this case  $M_1 = M_2 = (2\pi T)^2$. But eqs.\eqref{abc} and \eqref{rootsofg} then imply that $\sigma_+=0$, so the wall cannot avoid the horizon leading to a contradiction. 
  
This gives our first no-go lemma:
 \begin{center}  
 \begin{tcolorbox}[width=13.5cm]
     {Two black holes} 
     on either side of a static domain wall are ruled out. 
\end{tcolorbox}
\end{center}    
Note by  contrast that superheavy domain walls  ($\lambda > \lambda_{\rm max}$) inflate and  could   thus  prevent  the black holes from  coalescing.\footnote{Asymptotically-flat domain walls, which have been studied a lot in the context of Grand Unification \cite{Vilenkin:2000jqa}, are automatically in this range.} 

A second class of pairs one  can exclude are the `centerless geometries' [{\small E2,E2}], [{\small E2,E2$^\prime$}], [{\small E2$^\prime$,E2}] and    [{\small E2$^\prime$,E2$^\prime$}]. We use the word  `centerless' for geometries that contain  neither a  center of global AdS, nor a black hole in its place (see  fig.\,\ref{fig:interfacefluxes}). If such solutions existed, all inertial observers would necessarily hit the domain wall since there would be neither a center where  to rest, nor  a  horizon where to escape.\footnote{In  the double-Wick rotated context  of   Simidzija and Van Raamsdonk the   [{E2,E2}] geometries  give   traversable wormholes  \cite{Simidzija:2020ukv}. 
}

The argument   excluding  such  solutions is based on a simple  observation: What distinguishes the centerless slices {\small E2} and {\small E2$^\prime$} from those with an AdS center ({\small E1}) or a black hole  ({\small H1}) is the sign of   $x_j^\prime$  at the turning point,   
\begin{eqgroup}
\label{signs5}
{\rm sign} \left(x_j^\prime\bigl\vert_{\sigma\approx \sigma_+} \right) 
           = \begin{cases}  +  {\rm for \ {\small E2}, {\small E2}}^\prime\,, \\
           -  {\rm for \ {\small E1}, {\small H1}}\ . 
           \end{cases}
\end{eqgroup}
This is not a deep result, rather simply a consequence of the conventions that we picked. Since we keep the slice of spacetime lying to the left of the string\footnote{More precisely, we keep the slice that lies in $x<x(\si)$, where $x(\si)$ is the string in the $(x,\si)$ plane.}, the sign at the turning point instructs us whether to insert the $\mathbb{Z}_2$ symmetric part on the left or the right of our first wall, in order for them to join properly. Then (\ref{signs5}) follows from this discussion.

Now from eqs.\eqref{fullsolutionstatic} one has
\begin{eqgroup}
        (\sigma + M_1 \ell_1^{\,2}) \frac{x_1'}{\ell_1}  + 
         (\sigma + M_2 \ell_2^{\,2}) \frac{x_2'}{\ell_2}  <  0\ ,
\end{eqgroup}
 so  both  $x_j^\prime$  cannot be  simultaneously positive. This holds for  all $\sigma$, and hence also  near the turning point. This is our second no-go lemma:
   \begin{center}  
   \begin{tcolorbox}[width=13.2cm]
  '{Centerless}'static spacetimes in which  all
   inertial observers  would inevitably hit the domain wall are ruled out.
  \end{tcolorbox}
\end{center}

We can    actually   exploit this   argument further. As is clear from \eqref{fullsolutionstatic},  if $M_1>M_2$ then    $x^\prime_1$ is manifestly  negative, i.e. the green slice  is  of type {\small E1} or {\small H1}. The  pairs [{\small E2$^\prime$,E1}] and [{\small E2$^\prime$,E2}] for which the above inequality is automatic are thus ruled out. 

One can also show that $x^\prime_2\vert_{\sigma\approx \sigma_+}$  is  negative if  $M_2 > 0 > M_1$. This is obvious from \eqref{fullsolutionstatic} in the range $\lambda > \lambda_0$, and less obvious but also true for $\lambda < \lambda_0$, as can be checked by explicit calculation.\footnote{The tedious algebra is  straightforward and not particularly instructive, so we chose not to present it here. The inequalities where proven to hold by simplification using Mathematica.} 

The pairs  [{\small E1,E2$^\prime$}] and [{\small E2,E2$^\prime$}] for which the above mass inequality is  automatic, are thus also excluded.
Recall that the energy density of the $j$th CFT reads $\langle T_{tt}\rangle = \frac{1}{2}\ell_j M_j$. Ruling out  all pairs  of  {\small E2}$^\prime$ with    {\small E1} or {\small E2} implies  therefore   that  in the ground state the energy density must be everywhere negative. When one $L_j$ is much smaller than the other, the Casimir energy scales like $E_0 \sim \#/L_j$. The fact that the coefficient $\#$ is negative means that the Casimir force is attractive, in agreement with general theorems \cite{Kenneth:2006vr, Bachas:2006ti}. This is the third no-go lemma: 
\begin{center}  
\begin{tcolorbox}[width=13.7cm]
   A slice of global AdS$_3$  cannot be paired with a horizonless BTZ slice.  
This implies that in  the  ground state  of the putative dual ICFT
the  energy density is everywhere negative. 
 \end{tcolorbox}
\end{center}

We have collected for  convenience  all these conclusions in fig.\ref{fig:gluingexample}. The table shows the eligible slice pairs, or the allowed topologies of static-domain-wall spacetimes.  It also defines a color code for phase diagrams. 

\begin{figure}[!h]
\centering
 \includegraphics[width=.7\textwidth]{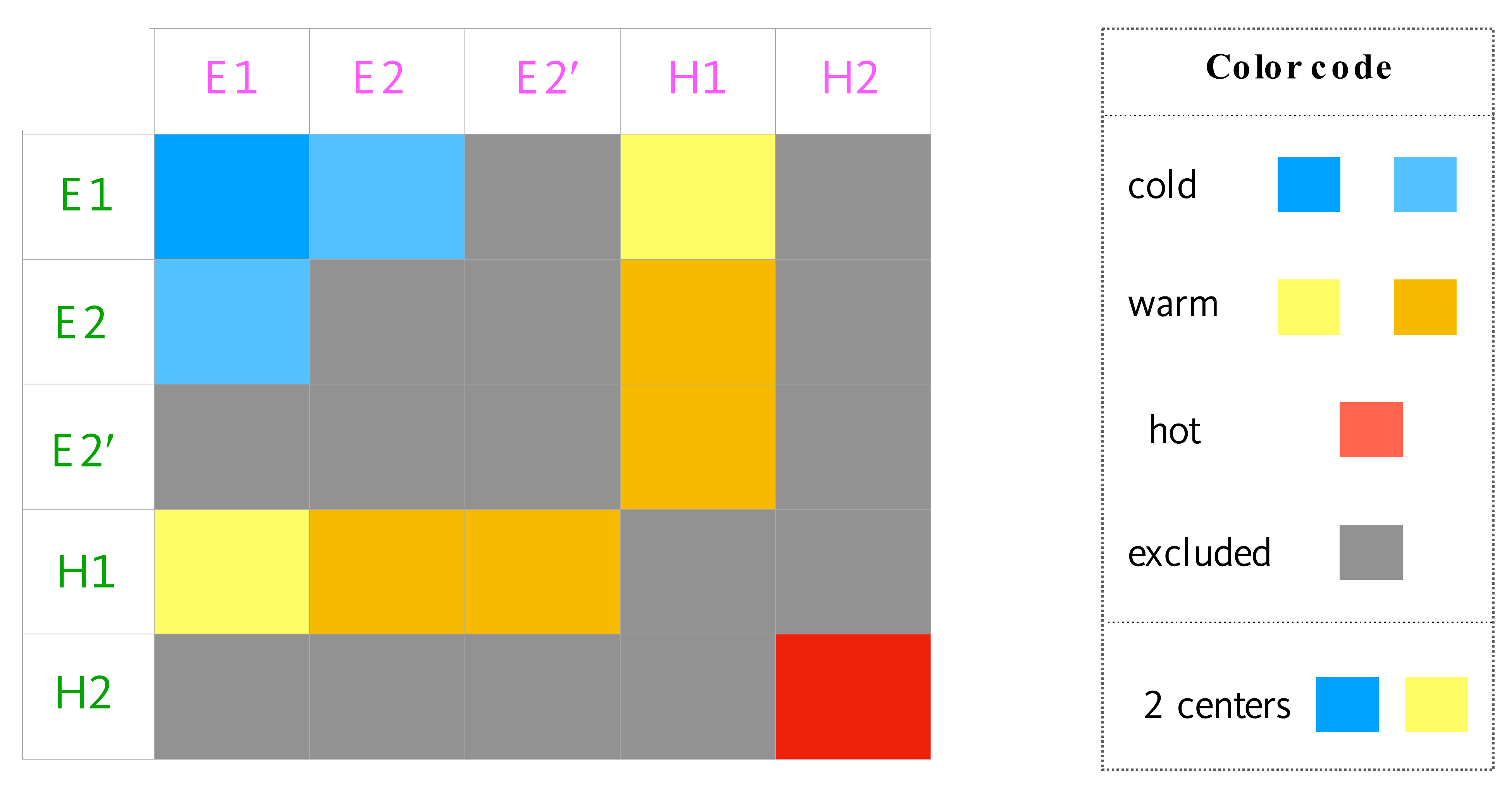}
    \caption{\small  
  Phases of the domain-wall spacetime.  The   type of the green  slice labels  the rows  of the table, and that  of the pink  slice  the columns. In the hot (red) phase the wall   enters  the black-hole  horizon,  while in the warm   (yellow)  phases  it avoids it. The cold (blue) phases have  no black hole. Geometries in which an inertial observer   is attracted to two  different centers are indicated by a  different shade (light yellow or darker blue).}
    \label{fig:gluingtable}
    \end{figure}

The light yellow  phases  that feature  a wall  between the  black hole and an AdS restpoint  are  the gravitational avatars of the Faraday cage. Indeed, an 
inertial observer lying on the thermal AdS slice can be "shielded" from the horizon, and rest happily in the other slice. In the full BTZ spacetime, an inertial observer will inevitably fall in the black hole. Such solutions  are easier to construct for  larger $\lambda$. Domain walls lighter than $\lambda_0$, in particular, can never shield from   a  black hole in the   `true-vacuum'  side. Indeed, as follows easily from  \eqref{fullsolutionstatic},  for $\lambda <  \lambda_0$ and $M_1>0> M_2$, the sign of $x_2^\prime\vert_{\sigma\approx\sigma_+}$  is  positive, so   geometries  of type [{\small H1,E1}] are excluded.

\section{Equations of state}\label{eqstate}
The   different colors in fig.\ref{fig:gluingtable} describe  different  phases of the system, since the corresponding geometries are topologically distinct. They differ in how  the wall, the horizon (if one exists) and inertial observers intersect or avoid each other.
       
Let us now think thermodynamics. For fixed Lagrangian parameters $\lambda, \ell_j$, the canonical variables that determine the state of the system are the temperature $T$ and the volumes $L_1, L_2$. As previously stated, because of scale invariance only  two  dimensionless ratios matter:
\begin{eqgroup}\label{variablenames}
\tau_1 := TL_1\ , \quad \tau_2:= TL_2\qquad {\rm or}\quad 
 \quad \gamma  := \frac{L_1}{L_2} = \frac{\tau_1}{\tau_2}\ .
\end{eqgroup}

On the other hand, natural variables for the interior geometry are the mass parameters $M_j$, and the size of the horizon $\Delta x_j\vert_{\rm Hor}$, if any exists. When several phases coexist, the dominant one is the one with the lowest  free energy  $F = \sum_j (E_j - TS_j)$.  

The free energy $F$ is determined from the renormalized on-shell gravitational action, as in sec. \ref{sec:HawkingPage}. The detailed computation is relegated to the appendix, app.\ref{app:renormonshellactoin}. From this formula, and since the energy density is given by $\langle T_{tt} \rangle$, we find :
\begin{eqgroup}
\label{othervars}
 \frac{E_j}{L_j} = \frac{\ell_j}{2}  M_j\qquad {\rm and}\qquad
S_j =  \frac{r^{\rm H}_j \Delta x_j\vert_{\rm Hor}}{4 G} =  2\pi \ell_j \sqrt{M_j} \Delta x_j\vert_{\rm Hor}\ ,
 \end{eqgroup}
where $E_j$ denotes the total energy of slice $j$, while $S_j$ denotes the total entropy. Note that in deriving the entropy we recover the Bekenstein-Hawking formula, as $r^{\rm H}_j \Delta x_j\vert_{\rm Hor}$ is the "area" of the horizon.

Thus the microcanical parameters (Energy, Entropy) are those naturally related to the interior geometry.  The entropies are scale invariant. The other key dimensionless variable is the mass ratio, viz.  the ratio of energy densities per degree of freedom in the two  CFTs
\begin{eqgroup}
\mu  := \frac{M_2}{M_1}\ .
\end{eqgroup}

The Dirichlet  conditions, (\ref{Dira}-\ref{Dirc}),  give for each type of geometry two relations among the above variables  that  play the role of equations of state.\footnote{In homogeneous systems there is a single equation of state. Here we have one equation for each subsystem.} They relate the natural interior parameters $S_j$ and $\mu$ to the variables $\tau_j$ and $\gamma$ of the boundary torus. Note that in each phase of the system there will remain one free interior parameter per slice, since  for horizonless slices $S_j=0$ and for slices with horizon $M_j = (2\pi T)^2$.  In computing the phase diagram we will have to invert these equations of state.


 \subsection{High-$T$ phase}
 \label{sec:highTphase}
For fixed $L_j$ and very high temperature, we expect the black hole to grow  so large that it eats away a piece of the domain wall and the  AdS rest points. The dominant solution is thus of type {\small [H2,H2]} and regularity fixes  the  mass parameters in  both  slices, $M_1=M_2= (2\pi T)^2$. 
The boundary  conditions \eqref{Dirc} reduce in this case to simple equations for the opening  horizon arcs $\Delta x_j\vert_{\rm Hor}$. Performing explicitly the integrals (see app. \ref{app:2})) gives :

\begin{eqgroup}\label{HighTarcs}
  L_1 - \Delta x_1\bigl\vert_{\rm Hor}  &=  - \frac{1}{\pi T}{\rm  tanh}^{-1}\left( \frac{\ell_1  (\lambda^2 + \lambda^2_0)
  }{2\lambda}\right)\ ,\\
  L_2 - \Delta x_2\bigl\vert_{\rm Hor}  &=   - \frac{1}{\pi T} {\rm  tanh}^{-1}\left( \frac{\ell_2 (\lambda^2 -  \lambda^2_0)
  }{2\lambda}\right)\ . 
\end{eqgroup}  

For consistency we must have $\Delta x_j\vert_{\rm Hor}>0$, which is automatic if  $\lambda > \lambda_0$. If   $\lambda < \lambda_0$, on the other hand,   positivity of $\Delta x_2\vert_{\rm Hor}$ puts a lower bound on $\tau_2$, 

\begin{eqgroup}
\label{tau2star}
 \tau_2   \geq  
    \frac{1}{ \pi  } {\rm tanh}^{-1}\left( \frac{\ell_2 (\lambda_0^2 -  \lambda^2) }{2\lambda}\right):= \tau_2^*\ .
\end{eqgroup}

We see here a first interpretation of the critical tension $\lambda_0$ encountered in section \ref{sec:criticaltensions}. For walls lighter than $\lambda_0$ there is a region of parameter space where the hot solution  ceases
to exist, even as a metastable phase. It is worth noting however, that the consistency condition is only valid if we ignore wall fusion. If we were to repeat the analysis including those more general spacetimes, $\tau_2^*$ could become a critical temperature signifying the transition to another phase.
The total energy and  entropy in the  high-$T$  phase read
\begin{align}
  E_{\rm [hot]}&=   \frac{1}{2}(\ell_1 L_1M_1 + \ell_2 L_2M_2 ) = 2\pi^2  T^2 \,(  \ell_1 L_1 +  \ell_2 L_2) \label{hotEnergy}\ ,\\
  S_{\rm [hot]} &=  4\pi^2  T \bigl(  \ell_1 \Delta x_1\bigl\vert_{\rm Hor} + 
  \ell_2 \Delta x_2\bigl\vert_{\rm Hor}\bigr)   =   
  4\pi^2  T^2 (  \ell_1 L_1 +  \ell_2 L_2)  + 2 \log g_I\label{hotEntropy}  \ ,
\end{align}
where $\log g$ is given by \eqref{gfactor} and the rightmost expression of the entropy follows from \eqref{HighTarcs} and  a  straightforward reshuffling of the arctangent functions. This is a satisfying result. Indeed, the first term in the right-hand side of \eqref{hotEntropy} is the thermal   entropy  of the two CFTs (being extensive these entropies  cannot depend on the ratio $L_1/L_2$), while the second term is the entropy of the two interfaces on the circle (this justifies a posteriori why we called $log(g)$ an entropy in \eqref{gfactor}). The Bekenstein-Hawking formula captures nicely both contributions.

Eqs.\eqref{HighTarcs}  and \eqref{hotEntropy}  show that  shifting  the $L_j$ at fixed $T$ does not change the entropy  if  and only if $\ell_1\,\delta L_1 = -\ell_2\, \delta L_2$.  Moving in particular  a defect (for which $\ell_1=\ell_2$) without changing the volume $L_1+L_2$   is an adiabatic process while moving a more general  interface generates/absorbs  entropy by  modifying the density of degrees of freedom.


\subsection{Low-$T$ phase(s)}
\label{sec:lowTphase} 
Consider next  the ground state   of the system, at $T=0$. The only geometries which exist at zero temperature are of the horizonless types: the double-center geometry  {\small [E1, E1]}, or  the single-center ones  {\small [E1, E2]} and {\small [E2, E1]}  (see fig.\ref{fig:gluingtable}).  
Here the entropies $S_j=0$, and the only relevant dimensionless variables are the volume and energy-density ratios,  $\gamma$ and $\mu$.  Note that they are both  positive since  $L_j>0$ and $M_j<0$ for both $j$. 

  The   Dirichlet  conditions for  horizonless geometries  read
\begin{eqgroup}
\label{dirichletHorizonless}
 \sqrt{\vert M_1\vert}  L_1 = 2\pi  \delta_{{\mathbb S}_1, {\rm E1}}- f_1(\mu)\ , \qquad  
 \sqrt{\vert M_1\vert}  L_2 = \frac{2\pi }{ \sqrt{\mu}}\delta_{{\mathbb S}_2, {\rm E1}} - f_2(\mu)\ ,
\end{eqgroup}   
where  $\delta_{{\mathbb S}_j, {\rm E1}} = 1$ if the $j$th slice is of type {\small E1} and $\delta_{{\mathbb S}_j, {\rm E1}}=0$  otherwise, and 

\begin{align}
\label{mua}
  f_1(\mu) &=   
   \frac{\ell_1}{\sqrt{A}} 
    \int_{s_+}^\infty   ds  \frac{
     s(\lambda^2+\lambda_0^2)  - 1 + \mu }{
   (s  -  \ell_1^{\,2}) \sqrt{s(s-s_+)(s-s_-) } }\ ,\\
\label{mub}
  f_2(\mu) &=   
   \frac{\ell_2}{ \sqrt{A} } 
    \int_{s_+}^\infty    ds  \frac{
     s(\lambda^2 - \lambda_0^2)  + 1  -  \mu}{
   (s -  \mu \ell_2^{2} ) \sqrt{s(s-s_+)(s-s_-) } }\ ,
 \end{align}
  with  
 \begin{eqgroup}
 \label{s+-cold}
  A s_\pm = \lambda^2 (1+\mu ) - \lambda_0^2 (1- \mu ) \pm 2\lambda 
  \sqrt{\frac{1-\mu }{\ell_2^2} + \frac{\mu^2 -\mu  }{\ell_1^2}  + \mu \lambda^2 }\ .
 \end{eqgroup}
 The dummy integration variable $s$ is the  appropriately rescaled blueshift factor of the string worldsheet, $s= \sigma/\vert M_1\vert$\,.

Dividing the two sides of  eqs.\,\eqref{dirichletHorizonless} gives $\gamma$ as  a function of $\mu$  for each of the three possible topologies.\footnote{The functions  $f_{j}(\mu)$ are combinations of   complete elliptic integrals of the first, second and third kind, see app. \ref{app:2}. The value $\mu=1$ gives $\gamma = 1$, corresponding  to the scale-invariant AdS$_2$ string worldsheet. The known  supersymmetric top-down solutions  live at this special point in phase  space.}    
If the  ground state of the putative dual quantum-mechanical system was unique, we  should find a single slice-pair type and value of $\mu$ for  each value of $\gamma$. Numerical plots  show that this is indeed the case. Specifically, we   found that $\gamma(\mu)$ is a monotonically-increasing function of $\mu$ for any given  slice pair, and that it changes continuously from one  type of pair to another. We will return  to these  branch-changing  `sweeping transitions'    in   section \ref{sec:sweepingtransition}. Let us stress that the  uniqueness of the cold solution  did not have to be automatic in classical gravity,  nor in the dual large-$N$ quantum mechanics.

For  most of the $(\ell_j, \lambda)$ parameter space, as $\gamma$ ranges   in $(0, \infty)$   the mass ratio $\mu$ covers also  the entire range $(0, \infty)$. However, we found that if  $\ell_1 <  \ell_2$ (strict inequality)  and  for sufficiently light domain walls, $\gamma$ vanishes at  some  positive $\mu= \mu_0(\lambda, \ell_j)$.   Below this critical value   $\gamma$   becomes
negative signaling that the wall self-intersects and the solution must be discarded. This leads to a  striking  phenomenon  that we discuss   in section \ref{sec:Faraday}.


\subsection{Warm  phases}

The last set  of solutions  of the model are the yellow- or orange-colored  ones in   fig.\ref{fig:gluingtable}. Here the string avoids the horizon, so the slice pair  is of type {\small [H1,X]} or {\small [X,H1]} with {\small X} one of the horizonless types: {\small E1, E2} or {\small E2}$^\prime$. 
 
Assume first  that the black hole is on the green side of the wall, so that $M_1 = (2\pi T)^2$.  In terms of  $\mu$  the  Dirichlet conditions  (\ref{Dira}, \ref{Dirb})  read: 
\begin{equation}
    \label{tildeDir}
    2\pi T \Delta x_1\bigl\vert_{{\rm Hor}}   - \, 2\pi \tau_1 = \tilde f_1(\mu)\ ,   \qquad
   2\pi \tau_2  =  \frac{2\pi}{\sqrt{-\mu}} \,\delta_{{\mathbb S}_2, {\rm E1}} - \tilde f_2(\mu) \ , 
\end{equation}
where
\begin{align}
\label{tildemua}
  \tilde f_1(\mu) &=   
   \frac{\ell_1}{\sqrt{A} } 
    \int_{\tilde s_+}^\infty   ds  \frac{
     s(\lambda^2+\lambda_0^2) +1  -\mu  }{
   (s  +   \ell_1^{\,2}) \sqrt{s(s- \tilde s_+)(s- \tilde s_-) } }\ ,\\
\label{tildemub}
  \tilde f_2(\mu) &=   
   \frac{\ell_2}{\sqrt{A} } 
    \int_{\tilde s_+}^\infty    ds  \frac{
     s(\lambda^2 - \lambda_0^2)  - 1  +  \mu  }{
   (s  +  \mu \ell_2^{\,2} ) \sqrt{s(s-\tilde s_+)(s-\tilde s_-) } }\ ,
\end{align}
 and the  roots  $\tilde s_\pm = \sigma_\pm/M_1$ inside  the square root are  given by   
\begin{eqgroup}
\label{s+-greensidebh}
  A \tilde s_\pm = - \lambda^2 (1+\mu ) + \lambda_0^2 (1- \mu ) \pm 2\lambda 
  \sqrt{\frac{1-\mu}{\ell_2^2} + \frac{\mu^2 -\mu}{\ell_1^2}  + \mu \lambda^2 }\ .
\end{eqgroup}
In the first condition \eqref{tildeDir} we have used the fact that  the period of the  green  slice that contains the horizon  is $P_1 =  \Delta x_1\vert_{{\rm Hor}}$. This comes simply from the fact that if the full black hole is included in the slice, we can compute its area if we know the x-periodicity.

If the black hole is on the  pink  side of the wall,  the  conditions  take a similar form in terms of the inverse
mass ratio  $\hat\mu = \mu^{-1} = M_1/M_2$, 
\begin{eqgroup}\label{hatDir}
  2\pi T \Delta x_2\bigl\vert_{{\rm Hor}}   -   2\pi \tau_2 = {\hat  f}_2(\hat \mu) ,   \quad
   2\pi \tau_1  =  \frac{2\pi }{ \sqrt{-\hat \mu}} \delta_{{\mathbb S}_1, {\rm E1}} - \hat f_1(\hat \mu)\ ,
\end{eqgroup}
where here
\begin{align}
\label{hatmua}
  \hat f_1(\hat \mu) &=  
   \frac{\ell_1}{\sqrt{A} } 
    \int_{\hat s_+}^\infty   ds  \frac{
     s(\lambda^2+\lambda_0^2) +  \hat\mu   - 1 }{
   (s  +  \hat\mu  \ell_1^{\,2}) \sqrt{s(s- \hat s_+)(s- \hat s_-) } }\ ,\\
\label{hatmub}
  \hat f_2(\mu) &=\,    
   \frac{\ell_2}{\sqrt{A} } 
    \int_{\hat  s_+}^\infty    ds  \frac{
     s(\lambda^2 - \lambda_0^2)  - \hat \mu   +  1}{
   (s  +    \ell_2^{\,2} ) \sqrt{s(s-\hat s_+)(s-\hat  s_-) } }\ ,
\end{align}
and the  roots  $\hat  s_\pm = \sigma_\pm/M_2$ inside  the square root are given by   
\begin{eqgroup}\label{s+-pinksidebh}
  A\, \hat  s_\pm = - \lambda^2 (\hat\mu + 1) + \lambda_0^2 (\hat\mu - 1 ) \pm 2\lambda 
  \sqrt{\frac{\hat\mu^2- \hat \mu}{\ell_2^2} + \frac{1 -\hat \mu}{\ell_1^2}  + \hat \mu \lambda^2 }\ .
\end{eqgroup}

The  functions $\tilde f_j$ and $\hat f_j$,  as well as   the $f_j$ of the cold phase,  derive from the same basic  formulae  \eqref{fullsolutionstatic}  and differ only by  a few signs.  We chose to write them out separately because these signs are important. Note also that while in  cold solutions  $\mu$ is always positive, here $\mu$ and its inverse $\hat\mu$  can have either sign.

All  the values of $\mu$ and $\hat\mu$  do not,  however,  correspond to admissible   solutions. For a pair of type {\small [H1,X]} we must demand  (i) that  the right-hand sides in \eqref{tildeDir} be positive -- the non-intersection requirement, and (ii)  that  $x_1^\prime\vert_{\sigma\approx \sigma_+}$  be negative -- the  turning point condition  \eqref{signs5}.  Likewise for solutions   of type {\small [X, H1]} we must demand that the right-hand sides in \eqref{hatDir} be positive and that $x_2^\prime\vert_{\sigma\approx \sigma_+}$  be negative. 

The turning-point requirement  is easy to implement. In the  {\small [H1,X]} case, $x_1^\prime\vert_{\sigma\approx \sigma_+}$ is negative when the numerator of the integrand in  \eqref{tildemua}, evaluated at at $s=\tilde s_+$,    is positive. Likewise for the  {\small [X,H1]} pairs,  $x_2^\prime\vert_{\sigma\approx \sigma_+}$ is negative when the numerator of the integrand in \eqref{hatmub}, evaluated at at $s=\hat s_+$, is positive.  After  a little algebra these conditions take  a simple form
\begin{eqgroup}
\label{conditionswarmextra}
{\rm for}{\rm [H1,X]} \quad  \mu\in (-\infty, 1] ; 
\qquad 
{\rm for} {\rm  [X,H1]}   \quad  \hat \mu = \mu^{-1}\in (-\infty, 1]  \ .
\end{eqgroup}

Recalling that $\mu = \hat\mu^{-1} = M_2/M_1$, we conclude that in all the cases the energy density per degree of freedom  in the horizonless slice is lower than the  corresponding density  in the black hole slice. 

This agrees with physical intuition:  the energy density per degree of freedom in the cooler  CFT   is less than the thermal density $\pi  T^2/6 $  --  the interfaces did not let the theory  thermalize. When $\mu\to 1$ or $\hat\mu \to  1$, the   wall  enters  the horizon and the energy is equipartitioned.

This completes our  discussion of the equations of state. To summarize, these equations relate the parameters of the interior geometry ($\mu, S_j$) to   those of the conformal boundary  ($\gamma , \tau_j$).  The relation involves  elementary functions  in  the hot phase, and was reduced   to a single function $\gamma(\mu)$,  that can be readily plotted, in  the cold phases. Furthermore  {\it {at  any given point in parameter space  the hot and cold solutions,
when they exist,  are unique.}} The excluded regions are   $\tau_2< \tau_2^*(\lambda, \ell_j)$  for the hot solutions,  and  $\mu > \mu_0(\lambda, \ell_j)$ for the cold solution with  $\mu_0$   the point where
$\gamma =0$. 
 
In  warm phases  the story  is richer  since more than one  solutions typically coexist at any given value of $(\gamma, \tau_j)$.  Some   solutions  have  negative specific heat, as we will discuss later. To find the parameter regions where different solutions exist requires inverting the relation between ($\gamma, \tau_j$) and ($\mu, S_j$). We will do this analytically in some limiting cases, and numerically  to  compute   the full phase diagram in section \ref{sec:phasediagrams}. 
 
 \section{Phase transitions}
 \label{sec:PhaseTransitions}
 In the previous section we have outlined three phases classified by the presence of a horizon, and according to whether the strings enter the horizon. Among these, we must differentiate phases that have a different number of "centers". We can then identify three types of transitions between the different phases :
    
\begin{itemize}
\item \underline{Hawking-Page} transitions describing  the formation of a  black hole. These  transitions  from the cold to the hot or warm phases of fig.\ref{fig:gluingtable} are always first order;
         
\item \underline{Warm-to-hot}   transitions  during which  part of   the   wall  is captured by the   horizon. We will show that   these   transitions are also  first-order;  
 
\item  \underline{Sweeping}   transitions where the   wall sweeps away   a  center  of global AdS, i.e.\,\,a rest point for  inertial observers. These are continuous transitions  between the one and two-center phases of fig.\ref{fig:gluingtable}. 
\end{itemize}

\begin{figure}[!h]
\centering
\includegraphics[width=.7\textwidth]{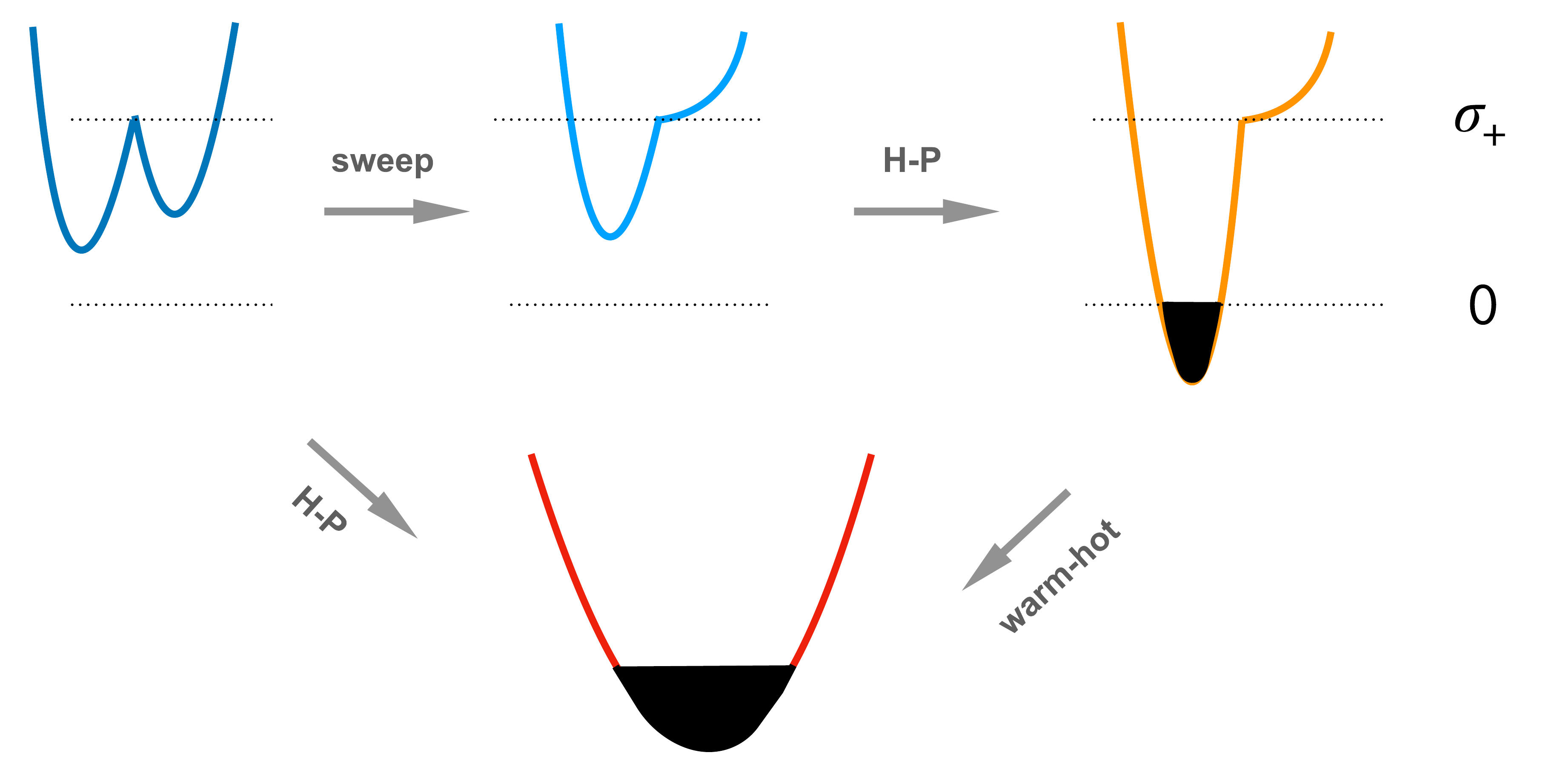}
      \caption{\small  Curves of the blueshift  factor $g_{tt}$ as one traverses space  along the $Z_2$ symmetry axis. The  color code is the same as in fig.\ref{fig:gluingtable}. The wall is located at the turning point  $g_{tt}=\sigma_+$ where the curve presents a kink. The grey arrows indicate possible transitions. The blackened parts of the curves are  regions behind the horizon.}
    \label{fig:phasetransitionsketch}    
  \end{figure}

It is  instructive    to picture these  transitions  by plotting the metric factor $g_{tt}$  while  traversing space along  the axis of reflection symmetry, see fig.\ref{fig:phasetransitionsketch}.
      
Before embarking in numerical plots,   we will first do  the following   things:(i) Comment on the ICFT interpretation of these  transitions; (ii) Compute the  sweeping transitions analytically;  and (iii) Prove that the warm-to-hot transitions are  first order, i.e. that one cannot lower the wall to   the horizon continuously by varying the boundary data.


\subsection{ICFT  interpretation}
\label{puzzles}
When   a holographic  dual exists,  Witten has argued that the appearance of a  black hole  at the Hawking-Page (HP) transition signals  deconfinement in  the gauge theory \cite{Witten:1998zw}. Within this interpretation \footnote{There is an extensive literature on the subject including  \cite{Sundborg:1999ue,Aharony:2003sx}, studies specific to two dimensions \cite{Aharony:2005ew,Keller:2011xi}, and recent discussions in relation with  the superconformal index  in  $N=4$ super Yang Mills \cite{Cabo-Bizet:2018ehj,Choi:2018hmj,Choi:2018vbz,Copetti:2020dil}. For an introductory  review see \cite{marsanophd}} leads to the conclusion that in warm phases a confined theory coexists with a deconfined one.  We will see below that such coexistence is easier when the confined theory is CFT$_2$, i.e. the theory  with the  larger central charge.\footnote{Even though for homogeneous 2-dimensional CFTs  the critical temperature,  $\tau_{\rm HP}= 1$,  does not depend on  the central charge by virtue of modular invariance.} 

This is natural from the gravitational perspective. Solutions of type {\small [H1, X]} are more likely than solutions of type {\small [X, H1]} because a  black hole forms more readily on the 'true-vacuum' side of the wall. We will actually provide some evidence later that if $c_2 > 3c_1$ there are no equilibrium  phases at all  in  which   CFT$_2$ is deconfined while CFT$_1$ stays  confined. 
 
The question that jumps to one's mind is what happens for thick walls, where one  expects a warm-to-hot crossover rather than a sharp transition. One possibility is that the coexistence of confined and deconfined phases is impossible in microscopic holographic models.  Alternatively, an  appropriately defined Polyakov loop \cite{Witten:1998zw} could provide  a sharp order parameter for this transition.

For sweeping transitions the puzzle is  the other way around.  Here  a  sharp order parameter exists in classical gravity --  it is the number of rest points for inertial observers. This can be defined both for  thin-  and  for thick-wall geometries. The interpretation on the field theory side is however unclear.  The transitions could be related to properties of the low-lying spectrum at infinite $N$, or to the entanglement structure of the ground state. 

More particularly, one hypothesis was that the entanglement wedge of the subsystem containing CFT$_2$ would include part of the side 1 bulk only for $\lam<\lam_0$. This hypothesis was motivated by looking at the near-boundary behavior of the RT surfaces. In that case RT surfaces are semi-circles anchored at the boundary, and thus they will necessarily intersect the membrane when $\lam<\lam_0$, and will avoid it when $\lam>\lam_0$. As promising as this initial lead was, it turned out not to be correct, as one can find crossing RT surfaces even for $\lam>\lam_0$. So the meaning of this sweeping transition in the field theory remains to be found. We will make a deeper dive into RT surfaces for ICFT in chap.\ref{chap:entanglemententropyandholoint}.


\subsection{Sweeping transitions}
\label{sec:sweepingtransition}
Sweeping transitions are continuous transitions that happen at fixed values of the mass ratio $\mu$.  We will prove these statements here. 

Assume for now continuity, and let the $j$th slice go from type {\small E1} to type  {\small E2}. The transition occurs when the string turning point and  the center of  the $j$th AdS   slice coincide, i.e. when  
\begin{eqgroup}
\label{pointofsweeping}
r_j(\sigma_+)  = \sqrt{ \sigma_+ + M_j\ell_j^2}  = 0\ .
\end{eqgroup}
Clearly, this has a solution only if $M_j<0$. Inserting  in \eqref{pointofsweeping} the expressions \eqref{rootsofg},\eqref{abc} for $\sigma_+$ two equations for the critical values of $\mu$ with the following solutions

\begin{eqgroup}
\label{mucritsweeping}
\mu_1^* &=    \frac{ 1 - \ell_2^{\,2} \lambda^2}{ \ell_2^2/ \ell_1^2    }  \ ,\\ 
\mu_2^* &=     \frac{\ell_1^2/ \ell_2^2}{1 - \ell_1^{\,2} \lambda^2} \ .
 \end{eqgroup}

In the low-$T$  phases  both $M_j$ are negative and $\mu$ is positive. Furthermore, a little  algebra  shows   that  for all $\lambda\in (\lambda_{\rm min},\, \lambda_{\rm max})$  the following is true
\begin{eqgroup}
x_1^\prime\bigl\vert_{\sigma\approx\sigma_+} <0&   \quad  {\rm at}\ \  \mu\gg 1\ ,\\
x_2^\prime\bigl\vert_{\sigma\approx\sigma_+} <0&  \quad  {\rm at}\ \ \mu\ll 1\ .
\end{eqgroup}

This means that for  $\mu\gg 1$ the green slice is of type {\small E1}, and for 
$\mu\ll 1$ the pink  slice is of type {\small E1}. A sweeping transition can occur if the critical mass ratios \eqref{mucritsweeping}  are  in the allowed range. We   distinguish three regimes of  $\lambda$: 

\begin{itemize}
\item \,  {  Heavy}\,{($\lambda > 1/\ell_1$):} 
None of the $\mu_j^*$  is positive, so the  solution  is of type {\small [E1,E1]} for all $\mu$, i.e. cold  solutions are  always double-center;

 \item  { Intermediate} {(${1/ \ell_1}> \lambda > 1/\ell_2$):} 
Only $\mu_2^*\,$ is positive. If  this   is  inside the  range  of  non-intersecting walls, the solution  goes   from {\small [E1,E2]} at large $\mu$,  to    {\small [E1,E1]} at small $\mu$. Otherwise the geometry  is  always of the single-center type  {\small [E1,E2]}; 
 
 \item \,{  Light}\,{($\lambda < 1/\ell_2$):}   Both  $\mu_1^*\,$ and $\mu_2^*\,$ are  positive, so there is   the possibility of two sweeping transitions: from   {\small [E2,E1]}  at small $\mu$   to {\small [E1,E2]}  at  large $\mu$ passing  through the double-center type {\small [E1,E1]}\,. Note that since $\lambda_{\rm min} = 1/\ell_1 - 1/\ell_2$,  this range of $\lambda$ only exists if  $\ell_2< 2\ell_1$,  i.e.\,\,when  CFT$_2$  has no more than   twice the number of   degrees of freedom of  the more depleted   CFT$_1$.  
\end{itemize}           

We can now confirm that sweeping transitions are continuous, not only in terms of the mass ratio  $\mu$ but also in terms of the ratio of volumes $\gamma$. To this end, we expand the relations \eqref{dirichletHorizonless} around the above critical points  and show that the $L_j$ vary indeed continuously across the transition. The calculations can be found in the appendix \ref{app:3}. In fact, numerically we can see that it is not only continuous but completely smooth. Thus, the phase transition in question is not to be understood in the thermodynamical sense, which is what we mean by labeling it as "continuous".
   
For  the warm phases, we proceed along  similar lines. One of the two $M_j$ is now equal to $(2\pi T)^2 >0$, so sweeping transitions may  only occur for negative  $\mu$.  Consider first  warm solutions of type {\small [H1,X]} with the  black hole in  the `true vacuum'  side.  A little calculation shows that $  x_2^\prime\vert_{\sigma\approx\sigma_+}$  is  negative,i.e.  {\small X=E1},\, if and only if 
\begin{eqgroup}
 \lambda > \frac{1}{\ell_1}\qquad{\rm and}\qquad  \mu <  \mu_2^* 
<0  \ .
\end{eqgroup}

Recall that when {\small X=E1}  some inertial observers can be shielded from the black hole by taking refuge at the  restpoint of the pink slice. We see that this is only possible for  heavy walls ($\lambda >1/\ell_1$)  and for $\mu <\mu_2^*$.   A sweeping transition  {\small [H1,E1]} $\to$ {\small [H1,E2]} takes place  at  $\mu= \mu_2^*$.

Consider  finally a black hole  in the `false vacuum' side, namely warm solutions  of  type  {\small [X,H1]}. Here   $x_1^\prime\vert_{\sigma\approx\sigma_+}$ is negative, i.e. $X$ has a rest point, if and only if the following conditions are satisfied
\begin{eqgroup}\label{BHsweep}
 \lambda > \frac{1}{\ell_2}\qquad {\rm and}\qquad \hat  \mu := \mu^{-1}  <  (\mu_1^*)^{-1} <0\ .
 \end{eqgroup}
 
Shielding from the black hole looks here easier,  both heavy and intermediate-tension walls can do it.  In reality,  however,  we have found  that solutions with the black hole in the `false vacuum' side are rare, and that the above inequality pushes  $\hat \mu$ outside the admissible range. Even rarer are the cases where such a solution is dominant, in terms of free-energy. 

The  general trend emerging from the analysis is that  the heavier the wall the  more likely are  the two-center geometries.  A suggestive  calculation actually  shows that
\begin{eqgroup}
\frac{\partial \sigma_+}{\partial\lambda}\Bigl\vert_{M_j {\rm fixed}} \ \ {\rm is} 
\begin{cases} {\textrm {  positive\ \  for \ two-center\ solutions}}\\ \,
  {\textrm {negative\ \  for \ single-center\ solutions}}
\end{cases}
\end{eqgroup}
where the word `center' here includes  both  an  AdS restpoint and  a black hole. At fixed energy densities, a single center is therefore pulled closer to a heavier wall, while two centers are instead pushed away. It might be interesting to also compute ${\partial \sigma_+ /\partial\lambda}$ and  
${\partial V/\partial\lambda}$ at $L_j$ fixed,  where $V$ is the regularized  volume of  the interior space. In the special case of the vacuum solution with 
an AdS$_2$ wall,  the volume  (and the associated complexity  \cite{Chapman:2018bqj}) can be seen to grow with the tension $\lambda$. 
 


\subsection{Warm-to-hot  transitions}
\label{sec:warmtohottransition}
In warm-to-hot transitions the thin  domain wall enters the black-hole horizon. One may have expected this to happen continuously,i.e.  to be able to lower   the wall to the horizon  smoothly, by  slowly varying the  boundary data   $L_j, T$. We will now show that, if  the tension $\lambda$ is  fixed, the transition is actually always  first order.

Note first that in warm solutions   the  slice  that  contains the black hole has   $M_j= (2\pi T)^2$. If  the  string  turning point approaches continuously the horizon, then $\sigma_+ \to 0$. From eqs.\,(\ref{abc}, \ref{rootsofg}) we see that this  can happen if and only if $(M_1-M_2)\to 0$,  which implies in passing that  the solution must necessarily  be of type {\small [H1,E2$^\prime$]} or  {\small [E2$^\prime$,H1]}. Expanding   around  this putative point where the wall touches the horizon we set 
\begin{eqgroup}
    \frac{M_1 - M_2}{M_1+M_2} := \delta    \quad {\rm with}\quad  \vert \delta\vert \ll 1      \Longrightarrow 
    \sigma_+ \approx  \biggl( \frac{2\pi T}{\lambda}\biggr)^2 \delta^2 \ .
\end{eqgroup}
Recalling that the horizonless slice has the smaller $M_j$ we see that for  positive $\delta$  the black hole must be  in the green slice and $\mu = 1 - 2\delta + {\cal O}(\delta^2)$, while for negative $\delta$  the black hole is  in the pink  slice and  $\hat \mu = 1+  2\delta + {\cal O}(\delta^2)$. 

The second option  can be  immediately ruled out  since   it is impossible to satisfy the boundary conditions \eqref{hatDir}.  Indeed, $\hat f_1(\hat\mu \approx 1)$ is manifetsly positive, as is clear  from \eqref{hatmua}, and  we have assumed that  $\mathbb{S}_1$  is  of type  {\small E2$^\prime$}. Thus the second condition \eqref{hatDir} cannot be obeyed. By the same reasoning we see that for  $\delta$ positive, and since now $\mathbb{S}_2$  is  of type  {\small E2$^\prime$}, we need that $\tilde  f_2(\mu \approx 1)$ be negative. As is clear  from the expression \eqref{tildemub} this  implies that  $\lambda <\lambda_0$. 

The upshot of the discussion  is that a warm solution  arbitrarily close to the hot solution may exist only if $\lambda <\lambda_0$ {\it and} if the black hole is on the true-vacuum side. 

It is easy to see that under these conditions the  two branches of solution indeed  meet at  $\mu =1$, $\Delta x_2\vert_{{\rm Hor}}=0$ and  hence from \eqref{HighTarcs}
\begin{eqgroup}
\label{tau2starr}
\tau_2 = \frac{1}{\pi} {\rm tanh}^{-1}\left( \frac{\ell_2 (\lambda_0^2 -  \lambda^2 )
  }{2\lambda}\right) :=     \tau_2^*\ .
\end{eqgroup}

Recall from section \ref{sec:highTphase} that this is the limiting value for the existence of the hot solution -- the solution ceases to exist at $\tau_2<  \tau_2^* $. The nearby warm solution could in principle  take over in this forbidden range, provided that $\tau_2(\delta)$ decreases as  $\delta$ moves away from zero. However, it actually turns out that $\tau_2(\delta)$  initially  increases  for small $\delta$, so  this last possibility  for a continuous warm-to-hot transition is also ruled out.

To see why this is so, expand \eqref{tildeDir} and \eqref{tildemub} around $\mu=1$, 
\begin{eqgroup}
 \tilde s_+ = \frac{\delta^2}{\lambda^2} + {\cal O}(\delta^3)\,,\quad
 \tilde s_-  =   - \frac{4\lambda^2}{A}\Bigl(1-\delta (1+ \frac{\lambda_0^2}{ \lambda^2})\Bigr) 
   + {\cal O}(\delta^2)\ ,\quad
\end{eqgroup}
and shift  the  integration variable $s := y+\tilde s_+$ so that \eqref{tildemub} reads
\begin{eqgroup}\label{delta}
2\pi\tau_2(\delta)  =  \frac{\ell_2}{\sqrt{A}}\int_0^\infty  dy \left[  \frac{ y(\lambda_0^2 - \lambda^2) + 2\delta}{
(y+ \mu\ell_2^2) \sqrt{y(y+  \tilde s_+) (y - \tilde s_-) } } + {\cal O}(\delta^2)
\right] \ .
\end{eqgroup}
We  neglected in the integrand all contributions  of ${\cal O}(\delta^2)$ except for the  $ \tilde s_+$   in the denominator
that regulates the logarithmic divergence of the ${\cal O}(\delta \log\delta)$ correction. Now use the inequalities (\ref{inequalitiesforwarm}):
\begin{eqgroup}\label{inequalitiesforwarm}
\frac{ y(\lambda_0^2 - \lambda^2) + 2\delta}{\sqrt{(y+  \tilde s_+) (y - \tilde s_-) } } >   \frac{ y(\lambda_0^2 - \lambda^2) + 2\delta}{
 \sqrt{ (y+  \delta^2/\lambda^2) (y +4\lambda^2/A ) } } >
\frac{\sqrt{ y} (\lambda_0^2 - \lambda^2)}{\sqrt{  (y +4\lambda^2/A ) } }\ ,
\end{eqgroup}
where the second one  is equivalent to $2\delta > (\lambda_0^2/\lambda^2 - 1)\delta^2$, which is true  for small enough $\delta$. Plugging in \eqref{delta} shows that $\tau_2(\delta)  > \tau_2(0)$ at the leading order in $\delta$, 
proving our claim. 

\begin{figure}[!h]
\centering
\includegraphics[width=0.6\linewidth]{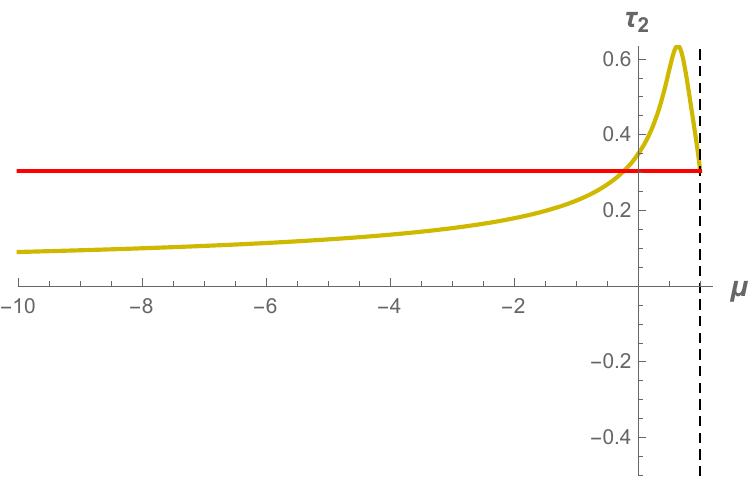}
       \caption{\small  The function $\tau_2(\mu) $  in the {\small[H1,E2]} and  {\small[H1,E2$^\prime$]} branches of solutions, for $2\ell_2 = 3\ell_1$ and $\lambda = {3/5\ell_2} < \lambda_0 = {\sqrt{5}/ 2\ell_2}$. The red line indicates the bound  $\tau_2^*$ below which the hot solution ceases to exist. }
    \label{fig:hotceaseexist}    
  \end{figure}
A typical   $\tau_2(\mu)$  in the  {\small[H1,E2]} and  {\small[H1,E2$^\prime$]}
branch of solutions, and for $\lambda <\lambda_0$, is plotted in figure \ref{fig:hotceaseexist}. The function grows initially as $\mu$ moves away from 1, reaches a maximum value and then turns around and goes to zero as $\mu \to -\infty$. The red line indicates the limiting value  $\tau_2^*$ below which there is no hot solution. For $\tau_2$ slightly above  $\tau_2^*$ we see that there are three coexisting black holes, the hot and two warm ones. For $\tau_2 < \tau_2^*$, on the other hand,  only one warm solution survives, but it describes a  wall at a finite distance from the horizon. Whether this is the dominant solution or not,  the transition is therefore necessarily first order.

 \section{Exotic fusion and bubbles}
\label{sec:Faraday}
Before proceeding to  the  phase diagram,  we pause here to discuss  the peculiar phenomenon announced earlier, in section \ref{sec:lowTphase}. This arises in the limits $\gamma= L_1/L_2\to 0$ or $\gamma \to \infty$,  with $L_1+L_2$ and $T$ kept fixed. In these limits, the conformal boundary of one slice   shrinks  to a point.

Consider for definiteness  the limit $L_1 \to 0$. In the language of the   dual field theory the interface and anti-interface fuse in this limit  into a defect of CFT$_2$.  The naive expectation, based on free-field calculations  \cite{Bachas:2007td,Bachas:2012bj,Bachas:2013ora}, is that this is the trivial (or identity) defect. Accordingly, the green  interior  slice should  recede to   the conformal boundary, leaving as the only remnant a (divergent)  Casimir energy.  

We have found that this expectation is not always borne out as we will now explain. Suppose first that the surviving CFT$_2$ is in its ground state, and that the result of the interface-antiinterface  fusion is the expected trivial defect. The geometry should in this case  approach  global AdS$_3$ of radius $\ell_2$,   with  $M_2$  tending to  $ -(2\pi/L_2)^2$. Furthermore,  $\sigma_+$  should   go to infinity in order for the green slice to  shrink towards the ultraviolet region.

As seen  from eqs.\,(\ref{rootsofg},\,\ref{abc})  this requires $M_1 \to -\infty$, so that  $\mu$ should vanish together with  $\gamma$. This is indeed  what happens in  much of the $(\lambda, \ell_1, \ell_2)$ parameter space. One finds $\mu \sim \gamma^{2}\to 0$, a scaling compatible   with  the expected Casimir energy $\sim \#/L_1$.   

\begin{figure}[!h]
\centering
\includegraphics[width=.7\textwidth]{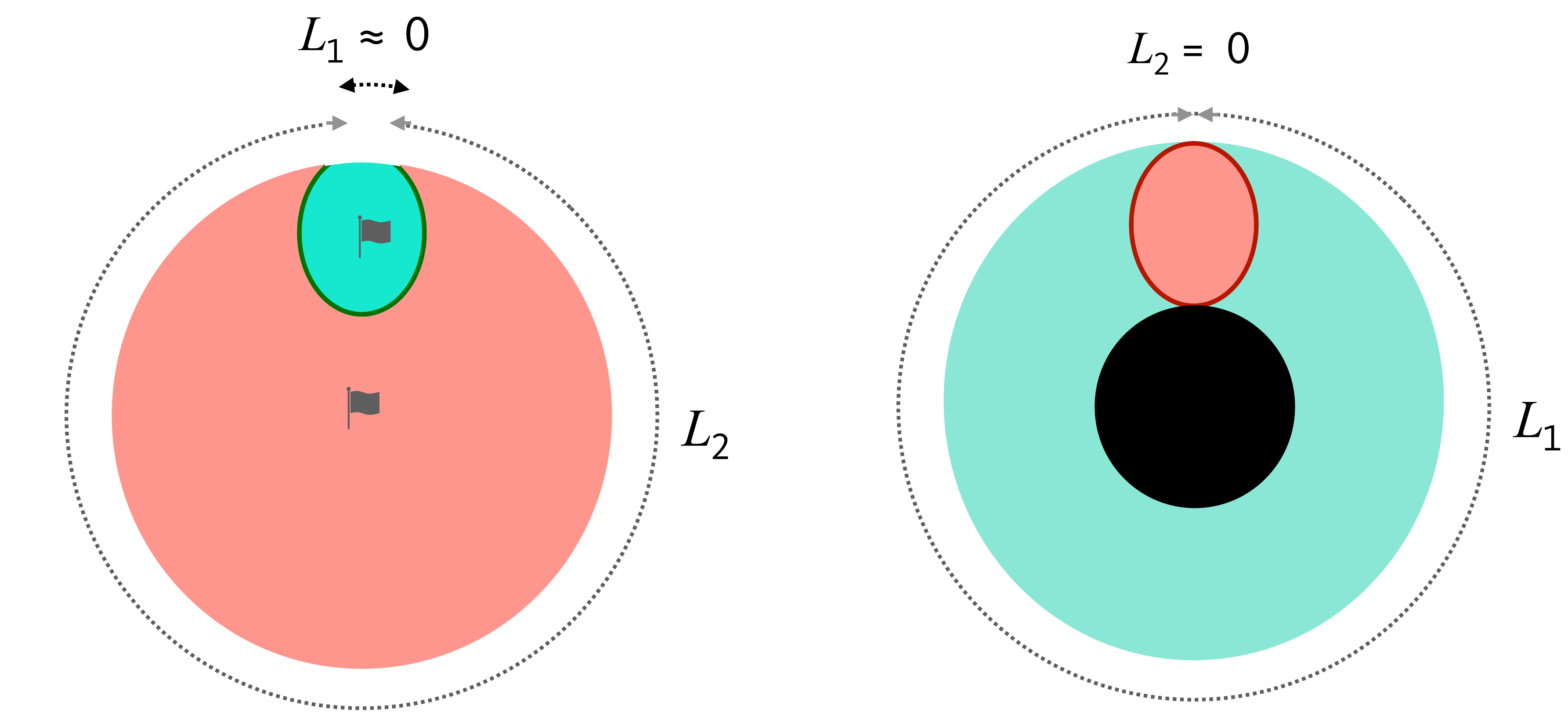}
\caption{\small  Left: A bubble of true vacuum  that survives inside the false vacuum despite  the fact that its conformal  boundary  shrinks to a point.   Right: A bubble of false vacuum with $\lambda =\lambda_0$
inscribed  between the boundary and the horizon of a black hole.}
    \label{fig:bubblesolution}    
  \end{figure}

Nevertheless,  sometimes $\gamma$ vanishes at finite $\mu_0$. In such cases, as $\mu\to\mu_0$  the green slice does not disappear even though its conformal boundary has shrunk to a point.  This is illustrated by the left  figure \ref{fig:bubblesolution}, which shows  a  static bubble of `true vacuum'   suspended from  a point on the  boundary of the `false vacuum'.\footnote {These are static solutions,  not to be confused with `bags of gold' which are  cosmologies  glued onto the backside of a Schwarzschild-AdS spacetime, see e.g.\cite{Marolf:2008tx,Fu:2019oyc}. The phenomenon is reminiscent of  spacetimes  that realize `wedge'  or codimension-2 holography, like those in \cite{Bachas:2017rch,Akal:2020wfl,Miao:2020oey}. } 

To convince ourselves that the peculiar phenomenon is real, we give an analytic proof in appendix \ref{app:bubbles} of the existence of such suspended  bubbles in at least one  region of parameters  ($\ell_2 > \ell_1$ and $\lambda\approx \lambda_{\rm min}>0$). Furthermore, since the vacuum solution for a given $\gamma$ is unique, there is no other competing solution. In the example of app. \ref{app:bubbles}, in particular,  $\gamma$ is finite and negative at $\mu=0$.

\noindent  

In the language  of field theory, this is a striking phenomenon. It implies that interface and anti-interface do not annihilate, but fuse into an exotic defect  generating  spontaneously a new scale in the process. This is the blueshift at the tip of the bubble, $\sigma_+(\mu_0, L_2)$, or better the corresponding frequency scale $r_2(\sigma_+)$  in  the D(efect)CFT.  

The   phenomenon is not symmetric under the exchange $1\leftrightarrow 2$. Static bubbles of the  false vacuum (pink)  spacetime inside  the  true (green) vacuum   do not seem to exist. We  proved this analytically for $\lambda < \lambda_0$, and  numerically for all other  values of the tension. We have also found that the suspended  green bubble  can be of type {\small E1}, i.e. have a center. The redshift factor $g_{tt}$ inside the bubble can even be lower than in the surrounding  space,  so that the bubble hosts the excitations of lowest energy. We did not show  this  analytically, but the numerical evidence is compelling. 

\begin{figure}[h!] 
 \centering
\begin{subfigure}{.45\textwidth}
  \centering
  \includegraphics[width=1\linewidth]{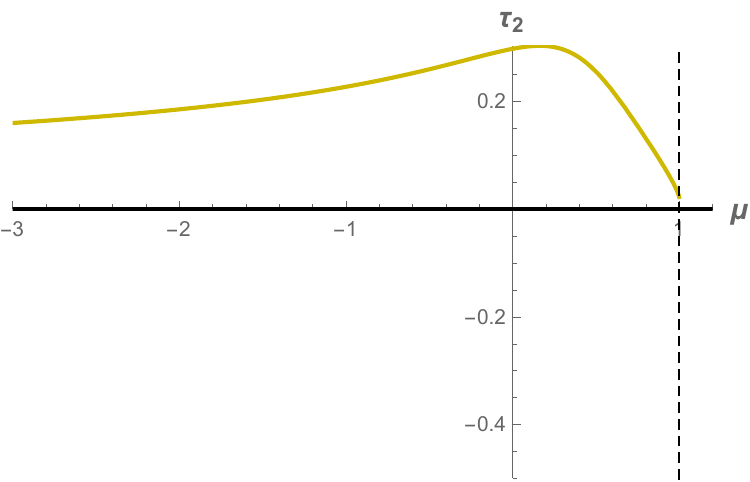}
 \end{subfigure}%
\hspace{10mm}
\begin{subfigure}{.45\textwidth}
  \centering
  \includegraphics[width=1\linewidth]{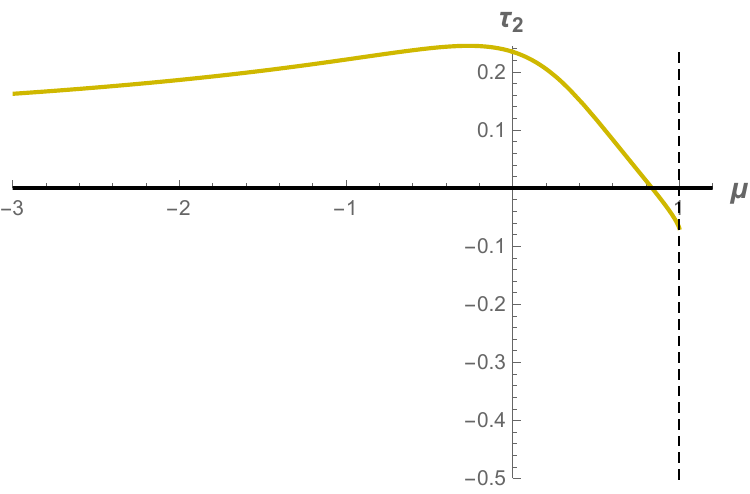}
 \end{subfigure}
 \caption{\small   Plots of the  function $\tau_2(\mu)$ in the [H1,E2] or  [H1,E2$^\prime$] branch of solutions for $\ell_2/\ell_1= 3/2$. The critical tension  is 
$\lambda_0 \ell_2 \approx 1.12$.  The curve on the left is for $ \lambda_0 \ell_2 = 1.05$, and the
curve on the right for $\lambda_0 \ell_2 = 1.35$. 
 }
\label{fig:continuouswarmgraph}
\end{figure}

Do  suspended  bubbles  also exist when the   surrounding  spacetime contains  a   black hole\,? The answer is affirmative as one can show semi-analytically by focussing on the region $\lambda \approx \lambda_0$. We have seen in the previous  section that near this critical tension there exist warm solutions of type {\small[H1,E2$^\prime$]} with the wall arbitrarily close the horizon. Let us consider the function $\tau_2(\mu, \lambda)$ given in this branch of solutions by \eqref{tildemub} and \eqref{tildeDir}  (with $\mathbb{S}_2 \not=${\small E1}). This is a continuous function in both arguments, so as $\lambda$ increases past $\lambda_0$, $\tau_2(1)$  goes from positive to negative with the overall shape of the function varying smoothly. This is illustrated in figure \ref{fig:continuouswarmgraph}, where we plot $\tau_2(\mu)$ for $\lambda$ slightly below and slightly above $\lambda_0$. It  should be  clear from these plots that  for  $\lambda > \lambda_0$ (the plot on the right) $\tau_2$ vanishes at  a finite  $\mu\approx 1$. This is a  warm bubble solution, as  advertised. 
 
We have found more generally that warm  bubbles  can  be also  of type {\small E1}, thus acting as a suspended Faraday cage that protects  inertial observers from falling towards the horizon of the black hole. Contrary,  however,  to what happened for  the ground state, warm bubble solutions are  not unique. There is always a  competing solution at $\mu\to -\infty$, and it is the dominant one  by virtue of its divergent negative Casimir energy. A stability analysis would show if warm  bubble solutions can be metastable and long-lived, but this is  beyond our present  scope.
 
As for  warm bubbles of type {\small[X, H1]},  that is with the black hole in the false-vacuum slice, these also exist but only if  $\ell_2<3\ell_1$.   Indeed, as we will see in a moment, when  $\ell_2> 3\ell_1$ the wall cannot  avoid a  horizon  located on the false-vacuum side.

Finally, simple inspection of  fig.\ref{fig:continuouswarmgraph} shows that by varying the tension,  the   bubble solutions for $\lambda > \lambda_0$  go over smoothly to the hot solution at  $\lambda=\lambda_0$.  At  this critical tension the bubble is inscribed between the horizon and the conformal boundary, as in  figure \ref{fig:bubblesolution}.  This gives another possible meaning to   $\lambda_0$:   Only walls with  this  tension  may  touch  the horizon without falling inside.

\section{Phase diagrams} 
\label{sec:phasediagrams}
In this last section of the paper we present numerical plots of the phase diagram of the model. We work in the canonical ensemble, so the  variables are the temperature and volumes, or by scale invariance  two of the dimensionless ratios  defined in \eqref{variablenames}.We choose these to be  $\tau  = \tau_1+\tau_2 = T(L_1+L_2)$ and $\gamma = L_1/L_2$. The color code is as in  fig.\ref{fig:gluingtable}. 

We plot  the  phase diagram   for different   values of  the action parameters $\ell_1, \ell_2, \lambda$. Since our analysis is classical in gravity, Newton's constant $G$ plays no role. Only two dimensionless ratios   matter,\footnote{Dimensionless in gravity, not in the dual  ICFT. } for instance
\begin{equation}\label{kappa}
  b := \frac{\ell_2}{\ell_1} = \frac{c_2}{c_1} \geq 1 \qquad {\rm and}\quad  \kappa := \lambda \ell_2 \in (b-1, b+1)\ .
\end{equation}

The value  $b=1$ corresponds to a defect  CFT, while  $b\gg 1$ is the opposite  ``near void" limit in which the  degrees of freedom of  CFT$_2$ overwhelm  those of CFT$_1$. The true vacuum approaches in this limit the infinite-radius AdS, and/or  the  false  vacuum approaches flat spacetime. The critical tension $\lambda_0$ corresponds to $\kappa_0 = \sqrt{b^2-1}$.   

\subsection{Inversion algorithm}
As mentioned before, to be able to plot the phase diagrams one needs to invert the equations (\ref{Dira}-\ref{Dirc}), to find which pairs of $M_1$, $M_2$ and $\Delta x|_{\rm Hor}$ yield the desired canonical parameters $\tau$ and $\gamma$. These equations are not invertible in general, since the integrals on the RHS of (\ref{Dira}-\ref{Dirc}) yield Elliptic functions, as shown in the appendix \ref{app:2}.

We must thus resort to numerical methods. Given $\tau$ and $\gamma$, we want to find all bulk parameters which satisfy the conditions (\ref{Dira}-\ref{Dirc}). To do so, we need to compartmentalize the search, treating possible bulks [{\small E},{\small E}], [{\small H},{\small E}], [{\small E},{\small H}] and [{\small E},{\small E}] separately. This is because according to the nature of the slice (hot, warm, cold) the equations that one needs to solve are different, see (\ref{Dira}-\ref{Dirc}).

Let us begin with the simpler case $[H,H]$. For this one, all we have to do is determine $\Delta x_i|_{\rm Hor}$ on both sides, which is immediately given by (\ref{HighTarcs}). One removes the solution if the necessary condition (\ref{tau2star}) is not satisfied, and we are done.

The cases [{\small H1},{\small E}] and [{\small E},{\small H1}] are symmetric, let us specify to the case where the black hole is on side 1. One gets equation (\ref{Dirb}) for side 1 and either of (\ref{Dira},\ref{Dirb}) for side two, depending on if it is {\small E1} or {\small E2}. The free parameters are $P_1$ and $M_2$, $M_1$ being set by the temperature. The tricky part here the possible jump in the relevant equation as we move $M_2$ through a sweeping transition. By implementing the condition (\ref{mucritsweeping}), we can immediately know which equation to consider, and as the two are connected continuously due to smoothness of the sweeping transition, it will not throw off the numerical algorithm. What we do then is to solve the equation for side 2, which depends only on $M_2$. We do this numerically, using a secant algorithm. From our analysis of the warm solution (see fig. \ref{fig:hotceaseexist}), we expect to find at most two solutions of this type. The main difficulty here is that the numerical algorithm will find a single solution, given an initial value. In addition, it might leave the allowed range which is $M_2<M_1=(2\pi T)^2$, necessary for the validity of the warm solution. To remedy this problem, we repeat the application of the algorithm for several different starting points, and stop it as soon as it leaves the allowed range. We find that this is sufficient to yield both solutions consistently, as long as the $\tau$ and $\gamma$ remain reasonable.

The case [{\small E},{\small E}] is comparatively simpler. Again, we know by our analysis sec.\ref{sec:lowTphase} that we should expect exactly one solution. We have two coupled equations (\ref{Dira},\ref{Dirb}), for the two free parameters $M_i<0$. By taking the ratio, we find an equation which depend only on $\mu=\frac{M_1}{M_2}$, reducing the problem to a single equation. Again, we solve it by means of a secant algorithm, stopping it when it finds a solution, or exits the allowed range $M_i<0$. We again apply the algorithm for evenly distributed initial conditions, to ensure that we find the solution. Similarly to the warm case, we have found that we always find the solution provided the inputs $\tau$ and $\gamma$ are not extreme. Once the solution is found, we should verify that $\frac{2\pi}{\sqrt{-M_j}}>L_j$, which is the "no self-intersection" condition.

This provides us with a set of candidate solutions in the form of quadruplets $(M_1,M_2,\Delta x_1|_{\rm Hor},\Delta x_2|_{\rm Hor})$. All is left to do is to find the dominant solution by plugging the quadruplet in (\ref{finalI}) and selecting the one with the lowest free energy. To plot the phase diagram, we discretize a chosen $(\tau,\gamma)$ region, and for each point perform the above computation. This leads to the "pixelization" that can be noticed in the phase diagrams of the next section. 

\subsection{Phase diagrams plots examples}
As explained in the introduction, although the interpretation is different,  our  diagrams are related to the ones of Simidzija and Van Raamsdonk \cite{Simidzija:2020ukv} by double-Wick rotation (special to 2+1 dimensions). Since time in  this reference is non-compact, only the boundaries of our phase diagrams, at $\gamma = 0$ or $\gamma= \infty$, can be compared. The roles of thermal AdS and BTZ are also exchanged

\subsubsection{Defect CFT}    
\begin{figure}[h!]
 \centering
\begin{subfigure}{.40\textwidth}
  \centering
  \includegraphics[width=1\linewidth]{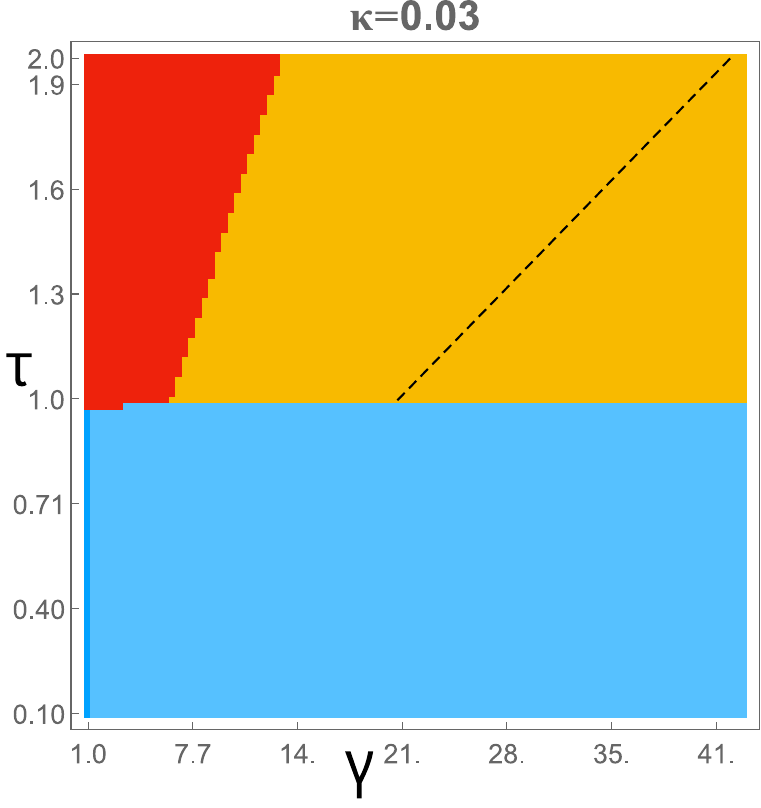}
 \end{subfigure}%
\hspace{19mm}
\begin{subfigure}{.4\textwidth}
  \centering
  \includegraphics[width=1\linewidth]{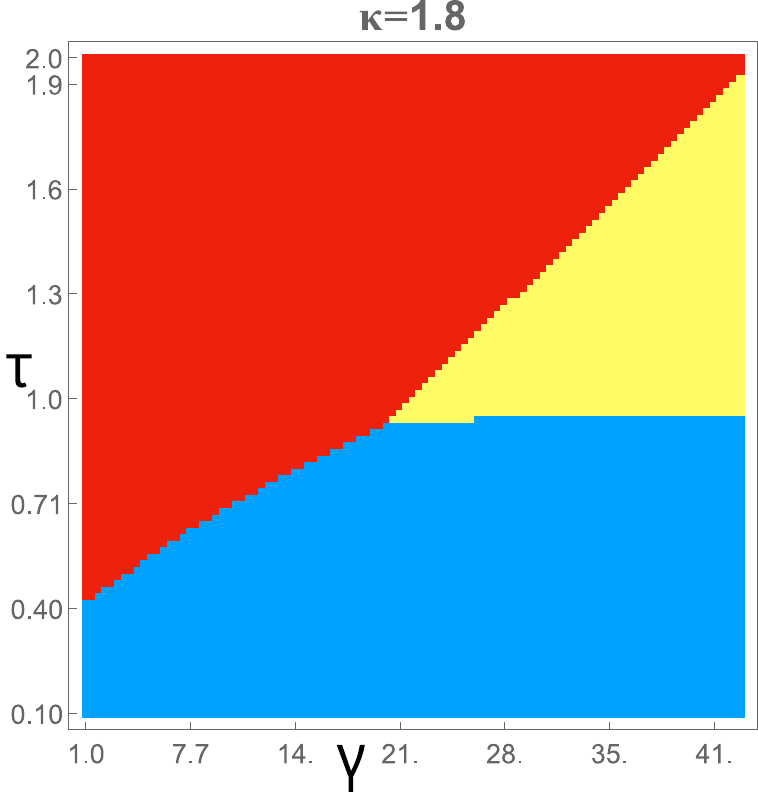}
 \end{subfigure}
 \caption{\small   Phase diagrams  of  a very light (left)  and a very heavy (right) domain wall between   degenerate  vacua ($b=1$). The horizontal and vertical axes are $\gamma$ and $\tau$. The broken  line in the left diagram separates  solutions of type [H1, E2$^\prime$] and  [H1,E2] that only differ in the sign of the energy of  the horizonless slice. The color code is as in fig.\,\ref{fig:gluingtable}.
 }
 \label{fig:phasediagramsdefectlightandheavy}
\end{figure}

Consider first $b=1$. By symmetry,  we may restrict  in this case to  $\gamma \geq 1$. Figure \ref{fig:phasediagramsdefectlightandheavy} presents  the phase diagram in the $(\gamma, \tau)$ plane  for a   very light   ($\kappa = 0.03$) and a very heavy ($\kappa = 1.8$) domain wall.  For the light, nearly  tensionless,  wall the phase diagram approaches that of a homogeneous CFT. The low-$T$  solution is   single-center, and the Hawking-Page (HP)  transition occurs at $\tau \approx  1$. Light domain walls follow closely geodesic curves, and avoid  the horizon in a large region of parameter space.\,\footnote{One can  compute this phase  diagram analytically by expanding in  powers of  $\lambda$.}  

Comparing  the left with the right figure \ref{fig:phasediagramsdefectlightandheavy} shows that heavy walls facilitate the formation of the black hole and have   a harder time staying outside.  Indeed,  in the right figure the HP  transition occurs   at  lower $T$,   and the  warm phase   recedes to $L_1\gg L_2$. Furthermore, both the cold and the warm solutions  have  now an additional  AdS restpoint. This confirms  the   intuition that heavier walls repel  probe masses  more strongly, and can shield them from falling inside  the black hole.

\begin{figure}[h!] 
 \centering
\begin{subfigure}{.40\textwidth}
  \centering
  \includegraphics[width=1\linewidth]{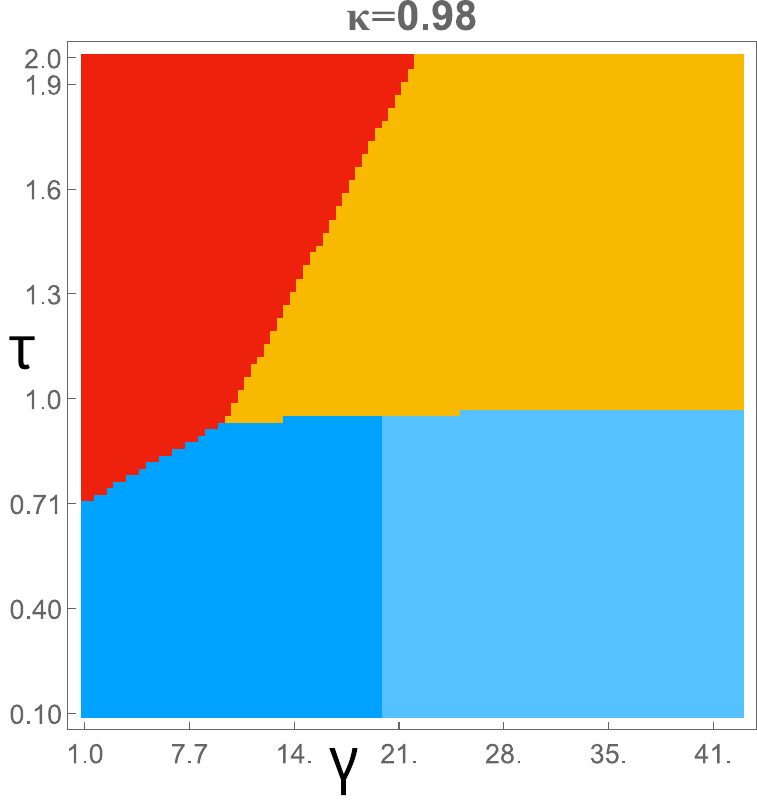}
 \end{subfigure}%
\hspace{19mm}
\begin{subfigure}{.4\textwidth}
  \centering
  \includegraphics[width=1\linewidth]{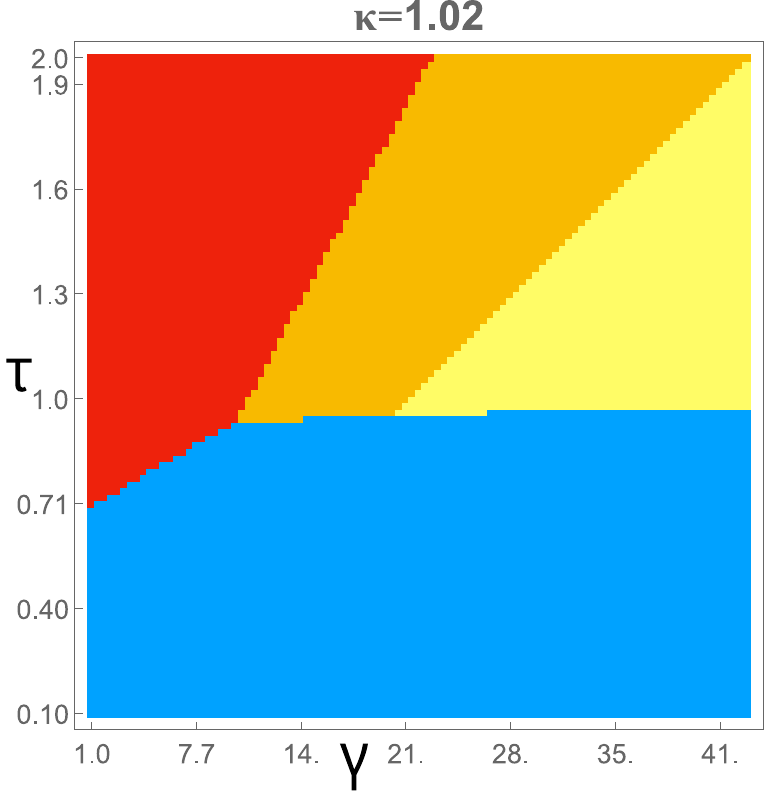}
 \end{subfigure}
 \caption{\small   Phase diagrams  for  intermediate-tension walls exhibiting sweeping transitions.  On the left   a restpoint of the vacuum solution  is swept away as  $\gamma$ increases beyond a critical value.  On the right   the same happens   in the warm solution but for decreasing  $\gamma$.  In these diagrams, the one-center warm solution is always {\small[H1,E2]}.}
\label{fig:phasediagramsdefectsweep}
\end{figure}

The  transition  that sweeps  away this AdS restpoint   is shown   explicitly  in the phase diagrams  of figure  \ref{fig:phasediagramsdefectsweep}.  Recall from the analysis of section \ref{sec:lowTphase} that in the low-$T$  phase such  transitions  happen  for   $\lambda < 1/\ell_1 \Longrightarrow \kappa < b = 1$.  Furthermore, the transitions  take  place at the critical mass ratios  $\mu_j^*\,$,  given by \eqref{mucritsweeping}. Since in cold  solutions the relation between $\mu$ and $\gamma$ is one-to-one, the dark-light blue critical lines are  lines  of constant $\gamma$. These  statements are in perfect agreement with the findings of   fig.\ref{fig:phasediagramsdefectsweep}. 

Warm  solutions of type {\small [H1,E1]}, respectively   {\small [E1,H1]}, exist for  tensions $\lambda >  1/\ell_1 \Longrightarrow \kappa > b$,   respectively $\lambda >  1/\ell_2 \Longrightarrow \kappa > 1$. In the case of  a defect,  these two ranges coincide. The stable black hole forms in the larger of the two slices, i.e.  for $\gamma >1$ in the $j=1$ slice. The sweeping transition occurs at  the critical mass ratio  $\mu_2^* = (b^2-\kappa^2)^{-1}$,  which through \eqref{tildeDir} and \eqref{tildemua} corresponds to a fixed value of $\tau_2$. Since $\tau = \tau_2(1+\gamma)$, the critical  orange-yellow line is  a straight line in the $(\gamma, \tau)$ plane, in accordance again with the findings  of  fig.\ref{fig:phasediagramsdefectsweep}.

 A noteworthy "empirical" fact is the rapidity of these  transitions as a function of $\kappa$. For $\kappa$ a little below or above the critical value the single-center cold, respectively warm phases almost disappear. Note also the cold-to-warm transitions are always near  $\tau \approx 1$. This is the critical value for Hawking-Page transitions in the homogeneous case,  as expected at large $\gamma$ when  the  $j=1$ slice  covers most of space.

The critical curves  for the cold-to-hot   and warm-to-hot transitions also look linear in the above figures, but this is an  illusion. Since the transitions are first order we must compare free energies. Equating  for example  the hot and cold free energies gives   after some   rearrangements (and with   $\ell_1=\ell_2:= \ell$)  
\begin{eqgroup}\label{criticalnotline}
 2\pi^2   \tau +  \frac{2}{\ell} \log g_I =   \frac{1}{2\tau_1}   \vert M_1\vert L_1^2 (1+ \frac{\mu}{\gamma})\ .
\end{eqgroup}

Now $\vert M_1\vert L_1^2$ can be expressed in terms of $\mu$ through (\ref{dirichletHorizonless}, \ref{mua}),and $\mu$ in the cold phase is a function of $\gamma$. Furthermore $ \log g_I/\ell= 4\pi {\rm tanh}^{-1} (\kappa/2)$  is  constant, see \eqref{gfactor}, and $\tau_1 = \tau/(1+\gamma^{-1})$. Thus   \eqref{criticalnotline} can be written as  a relation $\tau= \tau_{\rm hc}(\gamma)$, and we have verified with a careful fit that  $\tau_{\rm hc}$ is {not} a linear function of $\gamma$. 
 

\subsection{Non-degenerate vacua}    
\begin{figure}[h!] 
 \centering
\begin{subfigure}{.47\textwidth}
  \centering
  \includegraphics[width=1\linewidth]{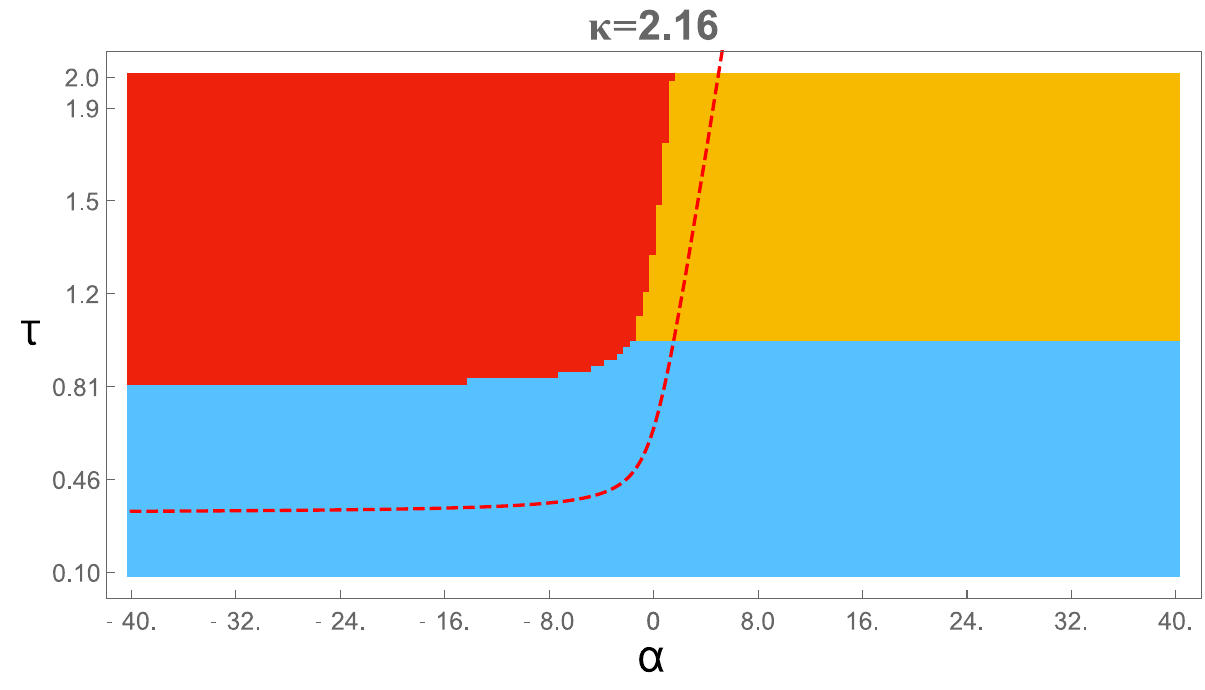}
 \end{subfigure}%
\hspace{4mm}
\begin{subfigure}{.47\textwidth}
  \centering
  \includegraphics[width=1\linewidth]{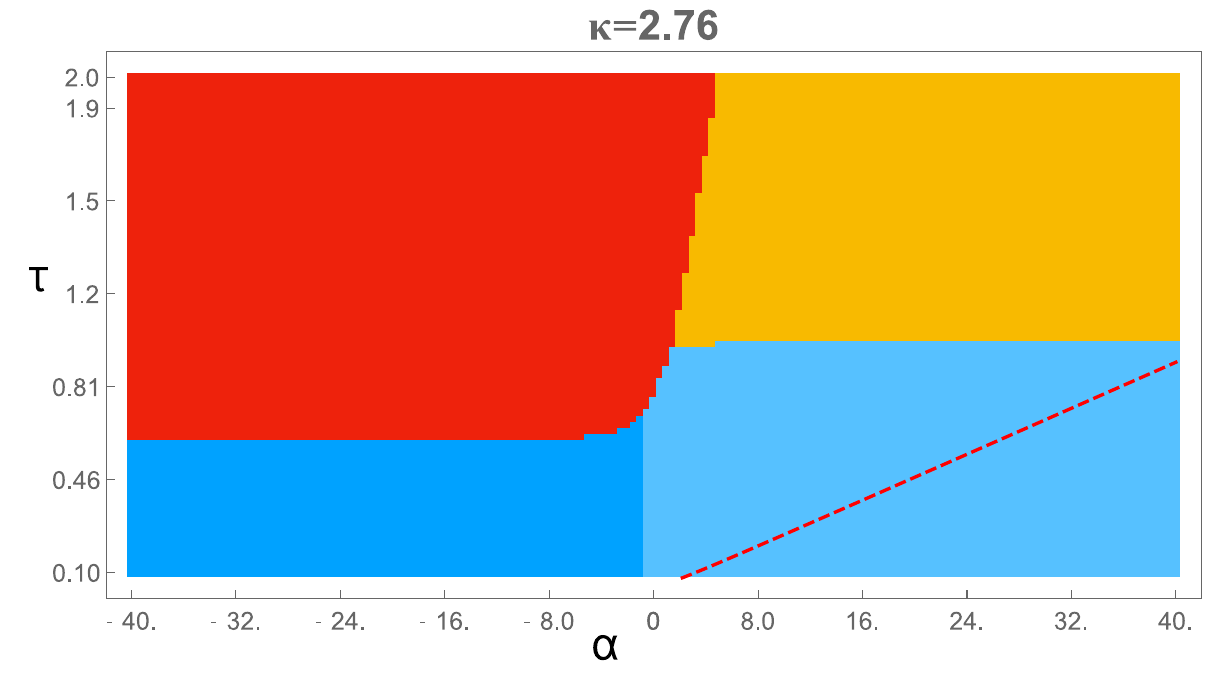}
 \end{subfigure}
\vskip 6mm
 \begin{subfigure}{.47\textwidth}
  \centering
  \includegraphics[width=1\linewidth]{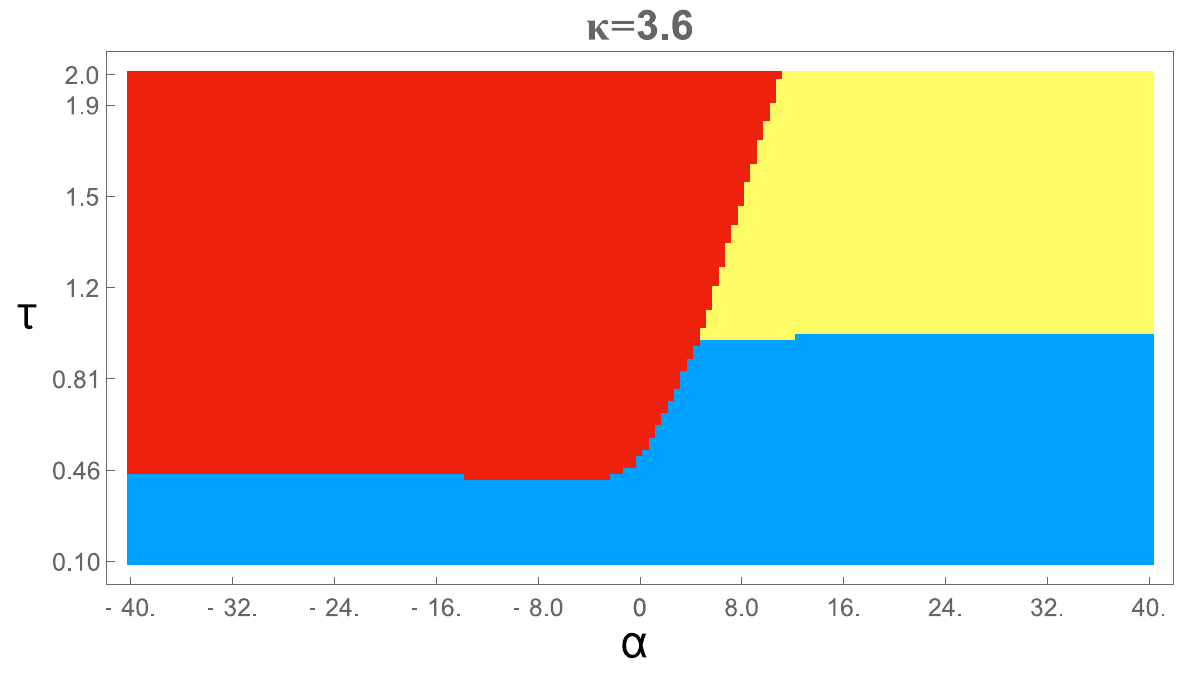}
 \end{subfigure}%
\hspace{4mm}
\begin{subfigure}{.47\textwidth}
  \centering
  \includegraphics[width=1\linewidth]{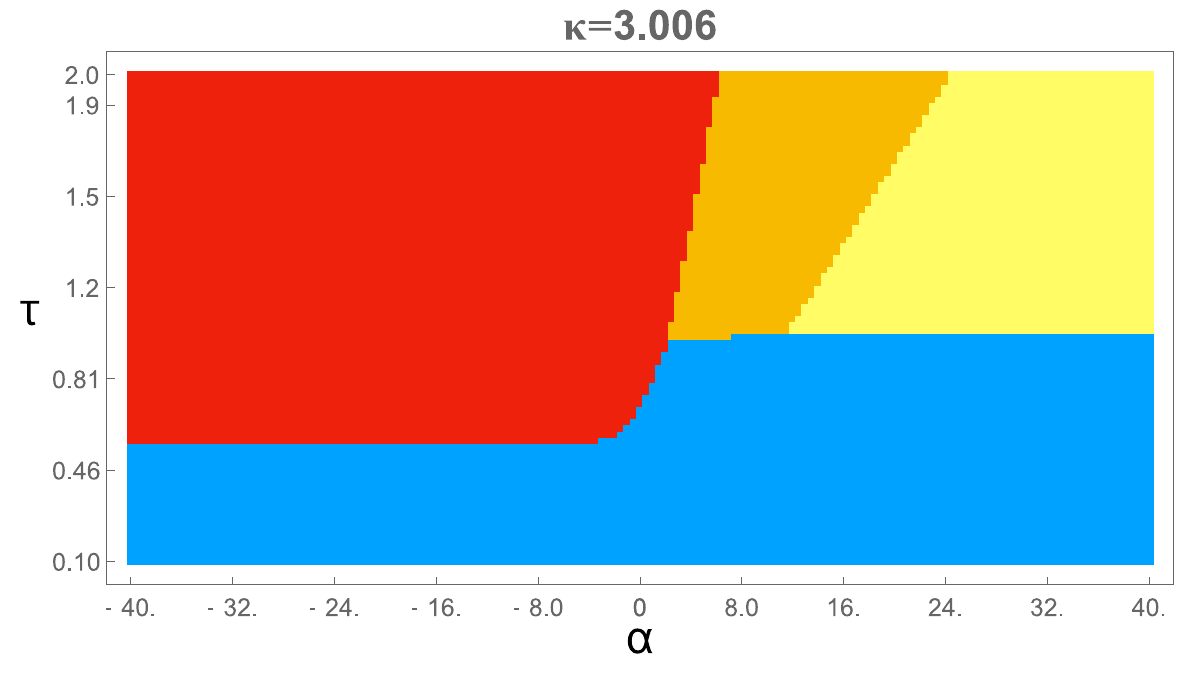}
 \end{subfigure}
 \caption{\small   Phase diagrams  for  $b=3$, and  values of the  tension that increase from the top-left figure clockwise. The  horizontal  and vertical axes are  $\alpha:= \gamma - \gamma^{-1}$ and $\tau$. The broken red curve is the bound $\tau = \tau_2^*(1+\gamma)$  below which the hot solution does not exist (there is no such bound  in the lower panels in which  the tension $\lambda > \lambda_0$). Note the absence of a warm phase in the left  ($\gamma <1$)  region of the diagrams. For the heaviest wall all non-hot solutions are double-center.}
\label{fig:allinterfacephasediagrams}
\end{figure}

Figure \ref{fig:allinterfacephasediagrams}   presents   the phase diagram  in the  case of non-degenerate AdS vacua,  $b=\ell_2/\ell_1 = c_2/c_1 = 3$,  and  for  different values of the tension in the allowed range,  $\kappa\in (2,4)$. Since there is no $\gamma \to  \gamma^{-1}$  symmetry, $\gamma$ here varies  between  0 to $\infty$. To avoid squeezing the $\gamma \in (0,1)$ region, we  use for  horizontal axis  $\alpha:= \gamma - \gamma^{-1}$.  This   is almost linear in the larger of $\gamma$ or $\gamma^{-1}$, when either  of these  is large,  but  the region $\gamma \approx  1$ is  distorted   compared to   \ref{fig:phasediagramsdefectlightandheavy} and \ref{fig:phasediagramsdefectsweep} of the previous section.
 
The most notable  new feature in  these phase diagrams is the absence of a warm phase in the region  $\gamma <1$.   This shows that it is impossible to keep the wall outside the black hole when the  latter forms on the false-vacuum side. From the perspective of  the dual  ICFT, see section \ref{puzzles}, the absence of  {\small [X,H1]}-type  solutions  means that no interfaces, however heavy,  can  keep  CFT$_1$ in the confined phase if  CFT$_2$ (the theory with larger central charge)  has already   deconfined. We suspect that this is a feature of  the thin-brane model which does not allow   interfaces  to be perfectly-reflecting \cite{Bachas:2020yxv}. 
 
Warm solutions  with the horizon in the pink slice appear to altogether disappear  above the  critical ratio of central charges $b_c=3$.\footnote{This critical value was also noticed in  ref.\,\cite{Simidzija:2020ukv}, who also note  that multiple branes can evade the bound confirming the intuition that it is a feature specific to  thin branes. As a matter of fact, although  {\small [X,H1]} solutions do exist for  $b <3$ as we show below, they have very large $\gamma$, outside the range of our  numerical plots, unless $b$ is very close to 1.} 

The boundary conditions corresponding to  topologies  of type {\small [X,H1]} are given by \eqref{hatDir}. We plotted the right-hand side of the second condition \eqref{hatDir} for different   values of $\lambda$ and  $\mu$ in their allowed range, and found no solution with positive $\tau_1$  for $b>3$. Analytic evidence for the existence of a strict $b_c=3$ bound can be found by considering the limit of a maximally isolating wall, $\lambda\approx \lambda_{\rm max}$, and  of a shrinking green slice   $\hat \mu\to -\infty$.  In this limit, the right-hand side of \eqref{hatDir} can be computed in closed form with the result

\begin{equation}
\tau_1(\hat \mu)  =  \frac{\pi}{\sqrt{-\hat\mu}} \biggl(2 - \sqrt{1+ \frac{\ell_2}{\ell_1}} \,\,\biggr)  + {\rm subleading}\ .
\end{equation}

We took  {\small X=E1} as dictated by the  analysis of sweeping transitions, see section \ref{sec:sweepingtransition} and in particular eq.\,\eqref{BHsweep}. This limiting   $\tau_1(\hat \mu)$ is negative for $b>3$,  and positive for $b<3$ where  warm  {\small [E1,H1]} solutions do exist, as claimed.

An interesting corollary is that end-of-the-world branes cannot  avoid the horizon of a black hole, since  the near-void limit,  $\ell_1 \ll \ell_2$, is in the range that has no {\small [X,H1]}  solutions. Indeed, in the models of doubly holographic evaporating black holes mentioned in sec.\ref{sec:IslandsBHparadox}, the EOW brane does indeed always falls in the horizon.


  \subsection{Unstable black holes} 
The   phase diagrams  in figs.\ref{fig:phasediagramsdefectlightandheavy}, \ref{fig:phasediagramsdefectsweep}, \ref{fig:allinterfacephasediagrams} show the   solution with the  lowest free energy in various  regions of parameter space. Typically, this dominant phase  coexists with  solutions that describe unstable or metastable black holes which   are ubiquitous in the  thin-wall model.\footnote{For a similar discussion of   deformed   JT gravity see   ref.\cite{Witten:2020ert}. Note that in  the  absence of a   domain wall,  the only static black hole solution  of pure Einstein gravity in  2+1 dimensions  is  the non-spinning BTZ black hole. 
}

\begin{figure}[h!] 
 \centering
   \includegraphics[width=1.04\linewidth]{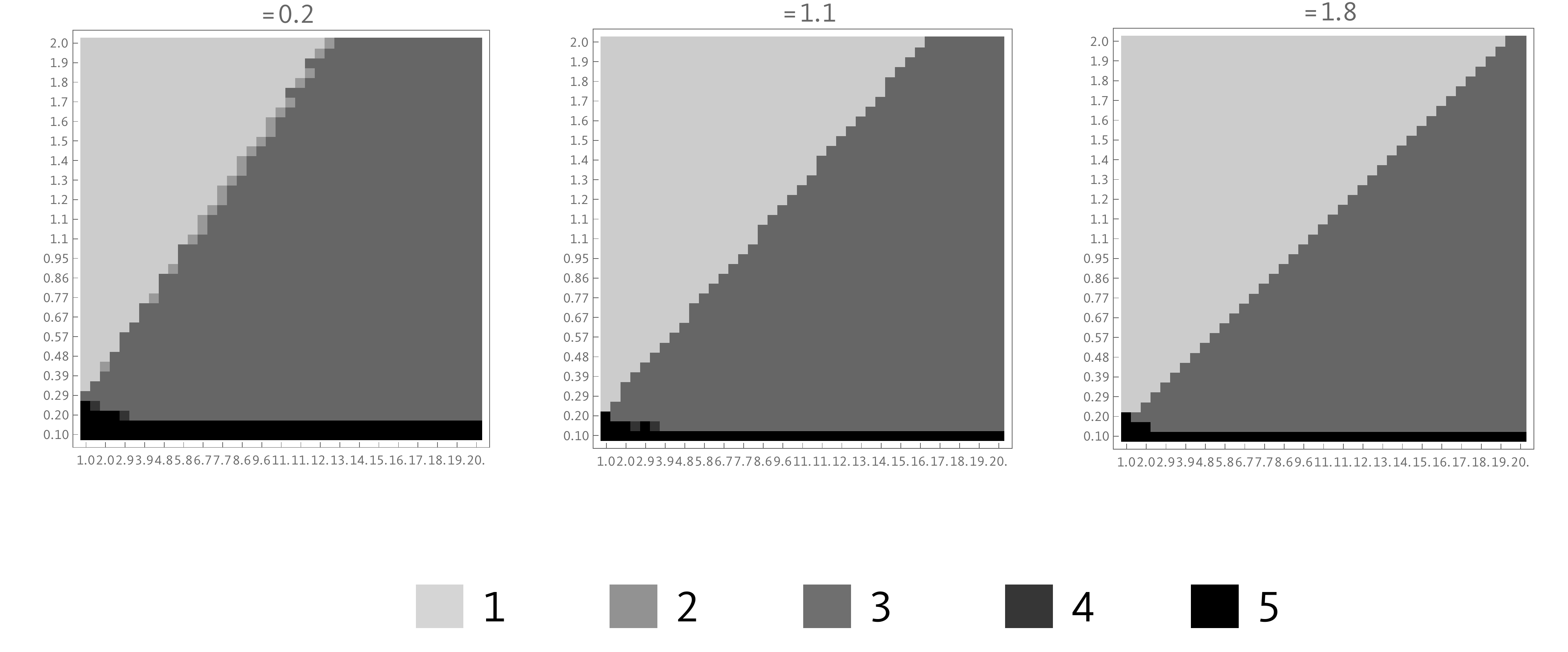}
   \caption{\small  The number of independent black hole solutions in the $(\gamma, \tau)$ parameter space  for $b=1$, and three values of the tension $(\kappa =  0.2;\ 1.1; {\rm and} \    1.8)$. The darker the shade the larger the number of black holes.
}
 \label{fig:13}
\end{figure}

Figure  \ref{fig:13} shows the number of  black hole solutions in the degenerate case,  $b=1$, for  small, intermediate and large wall tension, and in different regions  of the $(\tau, \gamma)$ parameter space. The axes are the same as in figs.\ref{fig:phasediagramsdefectlightandheavy} and \ref{fig:phasediagramsdefectsweep} but  the range of $\gamma$ is halved. 
At sufficiently high temperature the growing horizon captures the wall, and the only solution is the hot solution.  We see however that in a large region of intermediate temperatures  the hot solution coexists with two warm solutions. 
Finally, at  very  low temperature the hot solution coexists with  four  other black-hole solutions, two on either side of the wall. The dominant phase  in this region  is  vacuum, so the  black holes  play no role in the canonical ensemble.

The hot solution exists almost everywhere, except when $\lambda <\lambda_0$ and $\tau = \tau_2(1+\gamma) < \tau_2^*(1+\gamma)$  with $\tau_2^*$ given by \eqref{tau2star}. It has positive specific heat even when it is not the dominant phase. For warm black holes, on the other hand, the  specific heat can have  either sign. One can see  this  semi-analytically by focusing once again  on  our favorite   near-critical region $\lambda \approx \lambda_0$.  A simple inspection of fig.\ref{fig:continuouswarmgraph} shows that  in some range $ \tau_2^* < \tau_2 < \tau_2^{\rm max}$, so the hot solution coexists with two  nearby warm solutions. At the maximum $\tau_2^{\rm max}$,  where $d\tau_2/d\mu=0$, the warm  solutions merge and then disappear. Since the black hole is in the $j=1$ slice, $M_1= (2\pi T)^2$ and their energy reads
\begin{eqgroup}
 E_{\rm [warm]} =  \frac{1}{2}(\ell_1 M_1 L_1 + \ell_2 M_2 L_2)=
 2\pi^2 T^2 L_2  \bigl(  \ell_1  \gamma  + \ell_2 {\mu } \bigr) \ .
\end{eqgroup}

Taking a derivative with respect to $T$ with  $L_1, L_2$  kept fixed we  obtain 
\begin{eqgroup}\label{sheat}
\frac{d}{dT} E_{\rm [warm]} = \frac{2}{T}E_{\rm [warm]}
 + 
 2\pi^2 T^2   L_2^{\,2} \ell_2 \frac{d \mu}{d\tau_2}\ .
\end{eqgroup}

Near $\tau_2^{\rm max}$ the dominant contribution to this expression comes from the derivative ${d \mu/  d\tau_2}$   which jumps  from $-\infty$ to $+\infty$. It follows that  the warm black hole with the higher mass has negative specific heat, and should decay to its companion black hole either classically or in the quantum theory.\footnote{We have verified in several numerical examples that the black holes with negative specific heat are never the ones with  lowest free energy, a conclusion similar to the one reached in deformed JT gravity in \cite{Witten:2020ert}. 
}
 
It would be very interesting to calculate this decay process, but we leave this for future work. One last comment  concerns  transitions from the double-center vacuum geometries of type {\small [E1,E1]}, to warm solutions where the wall avoids the horizon.  One can ask what side of the wall the black hole chooses. A natural guess is that   it  forms in the deepest of the two AdS  wells. The relative depth is the ratio of blueshift factors at the two rest points, 
\begin{eqgroup}
   {\mathfrak  R} := \sqrt{\frac{g_{tt}\vert_{r_1=0}}{g_{tt}\vert_{r_2=0}}}\,  =   \,\frac{\ell_2}{\ell_1 \sqrt{\mu(\gamma)}} \,\ . 
\end{eqgroup}

One expects the black hole to form in the $j=1$ (green) slice if  ${\mathfrak  R}<1$ and in the $j=2$ (red) slice if  ${\mathfrak  R} > 1$. Our numerical plots confirmed in all cases this expectation. 

\section{Closing remarks}
While the thin brane model lends itself to an analytical analysis thanks to its simplicity, an important question is how much of this analysis will survive in top-down interface models, where the domain walls are typically thick. For instance, the order parameters of the Hawking-Page and sweeping transitions (the area of the horizon and the number of inertial observer rest-points), do not depend on the assumption of a thin wall, and should go through unscathed. The warm-to-hot transition, on the other hand, may be replaced by a smooth crossover, since there is no sharp criterion to decide whenever a thick wall enters the horizon or avoids it. From the field theory however, since this transition is related to the deconfinement transition of one of the CFTs, we have a well-defined order parameter in the shape of Polyakov loops. For this reason, we also expect such a phase transition to survive the upscale to the UV, although its interpretation from the bulk might be different.

Nonetheless, generating and examining a full UV complete example of this model would certainly be very interesting. Supergravity solutions dual to ICFT exist \cite{Bobev:2013yra,Chiodaroli:2011fn,Chiodaroli:2011nr}, and they usually involve a fibration of an AdS$_{D-1}$ space on internal manifold. As we move from one side of the internal manifold to the other, the geometry interpolates between two different AdS$_D$ vacua. It would be very interesting to study such solutions at finite temperature, but the main obstacle lies in extending those solutions to finite temperature in a way that emulates our minimal model. There is a way to do so by promoting the fibered AdS geometry to a black hole geometry\cite{Bak:2011ga,Uhlemann:2021nhu,Demulder:2022aij}; which remains a solution. However, this description is quite different from our minimal model, as it does not describe a situation where a black hole can be located on one side of the wall, rather the horizon is "fibered" along the internal manifold.

The question of computing the entanglement structure of the system in the various states is also of obvious interest, both in the possible connection to the sweeping transition, but also purely to understand the applicability of the RT prescription in the presence of membranes, as was done in \cite{Akal:2020twv,Deng:2020ent} for BCFT. We return to these questions in the last chapter.

Finally, another direction would be the extension of the minimal model, by addition of matter either on the bulk\cite{Zhao:2020qmn} or on the membrane. The choice of the matter should be motivated by some specific question one would like to study, this being particularly relevant in the context of strongly coupled condensed matter systems\cite{Zaanen:2015oix}.

\chapter{Steady states of holographic interfaces}
\label{chap:steadystatesofholo}
\epigraph{This chapter is based on 2107.00965}{}
In this chapter, we consider again the same minimal ICFT model, but we attempt to study states which are driven out of equilibrium. Indeed, the holographic duality has been mainly exploited at, or near thermal equilibrium, where a hydrodynamic description applies. For far-from-equilibrium processes our understanding is poorer (see \cite{Liu:2018crr} for a review). Although the semiclassical gravity description seems more tractable, highly-distorted horizons raise a host of unsolved technical and conceptual issues. To make progress, simple analytically accessible models such as the minimal one can prove to be valuable.

We will restrict here to the study of \textbf{N}on-\textbf{E}quilibrium \textbf{S}teady \textbf{S}tates (NESS) which are stationary states, characterized by persistent currents. They are the simplest example of a far-from-equilibrium state, which will allow for an analytical treatment. In (1+1) dimensional critical systems where energy transport is ballistic, they are particularly simple thanks to the stringent constraints imposed by conformal symmetry \cite{Bernard:2016nci}. In a pure CFT, these stationary states are equilibrium states in disguise; they can be obtained from thermal states by boosting them. By including a minimal interface in the system, we break this symmetry, and that will allow us to obtain "true" stationary states.

The salient feature of the gravity dual of such a state is a highly deformed, non-Killing event horizon, which we determine analytically. It lies behind the apparent horizon, seemingly in contradiction with well-known theorems \cite{hawking_ellis_1973} stating the opposite should be true. Other theorems also state that stationary non-Killing black holes should be excluded \cite{Hollands:2006rj,Moncrief:2008mr}. The way that these solutions manage to evade all these restrictions, is by non-compacity; both at the contact point with the interface, and by the infinite extent of the system, which is necessary to sustain the stationary nature of the state.

Another important insight brought about the shape of this horizon is about the entropy production at the interface. Due to the scattering of incoming excitations, there is naturally entanglement forming between the reflected and transmitted fluids. As we will argue, the gravity dual suggests that the interface is a perfect scrambler, namely that the quantum fluids exit the scattering thermalized. Although the picture is suggestive, a definitive proof of this fact should go through the computation of entanglement entropies, with the (H)RT prescription. This is surprisingly difficult in such spacetimes, and it is the topic of the next chapter.

This scrambling behaviour is reminiscent of flowing black funnels \cite{Fischetti:2012ps,Emparan:2013fha,Marolf:2013ioa,Santos:2020kmq} where a non-dynamical black hole acts as a source or sink of heat in the CFT. There are however important differences between the two setups. The non-back-reacting 1+1 dimensional black hole is a spacetime boundary that can absorb or emit arbitrary amounts of energy and entropy. Conformal interfaces, on the other hand, conserve energy and have a finite-dimensional Hilbert space. So even though one could mimic their energy and entropy flows by a two-sided boundary black hole whose (disconnected) horizon consists of two points with appropriately tuned temperatures, the rational, if any, behind such tuning is unclear. We will briefly comment on the gravity dual, which shows important differences, outlining that the two models are qualitatively different.

Throughout this chapter we use units in which $8\pi G =1$.

\section{The Boosted Black String}
Let us begin by the holographic dual of a simple stationary thermal state. As we mentioned, this state can simply be obtained by boosting an equilibrium thermal state, which introduces a steady current in the model. From the gravity point of view, this translates into a boost of the static BTZ geometry, introducing a spin $J$. The geometry that will be of interest here is then the uncompactified version of (\ref{banadosBTZmetric}) :
 \begin{equation}\label{metricspinningstring}
ds^2  =     \frac{\ell^2 dr^2}{({r^2 } - M\ell^2  + {J^2 \ell^2/ 4r^2} )}   - ({r^2 } - M\ell^2 )dt^2  + r^2 dx^2  - J \ell   dx dt \ .
\end{equation}
We denote by $x \in \mathbb{R}$ the uncompactified angle of (\ref{banadosBTZmetric}). This metric has a planar horizon which is "spinning" from left to right, namely it has a current of energy flowing in the direction of $J$.

It has an inner and outer horizon $r_{\pm}$ (see (\ref{banadosBTZmetric})), and as already explained we need $M\ell \geq |J|$ to avoid naked singularities. We will occasionally use the shorthand already introduced in (\ref{banadosBTZmetric}) :
\begin{eqgroup}
 h(r) = (r^2-M\ell^2+\frac{J^2\ell^2}{4r^2}) = \frac{(r^2-r_+^2)(r^2-r_-^2)}{r^2}\ .
 \label{hrformula}
\end{eqgroup}

Besides $r_{\pm}$, another special radius is $r_{\rm ergo}=\sqrt{M}\ell \geq r_+$. It delimits the ergoregion inside which no observer (powered by any engine) can stay at a fixed position $x$. To see this, consider any timelike vector field $v_t^\mu = (\dot{t},\dot{r},\dot{x})$ describing an observer's trajectory.  Then :
\begin{eqgroup}
     v_t^\mu v_{t\,\mu}<0 \Leftrightarrow  J \ell \dot{x}\dot{t}>r^2\dot{x}^2-(r^2-M\ell^2)\dot{t}^2+\frac{\ell^2 \dot{r}^2}{(r^2-M\ell^2+\frac{J^2\ell^2}{4r^2}}>0\ .
     \label{ergondition}
\end{eqgroup}

Crucially, we used $r^2-M\ell^2<0$ and $h(r)>0$ which hold only in the ergoregion, and outside the black hole. Together with $\dot{t}>0$ which holds for future-directed timelike vectors outside the horizon, we find that $\sign{J}\dot{x}>0$, so that the observer is indeed dragged along with the black hole. This is reminiscent of the Kerr black hole with which the metric (\ref{metricspinningstring}) shares several of its properties (see \cite{Carlip:1995qv} for a review). The outer horizon is a Killing horizon, while the inner one is a Cauchy horizon. The frame-dragging forces will force ingoing matter to cross the outer horizon at infinity, $x\approx \frac{J\ell t}{2r_+^2}\rightarrow\infty$. This is of course a pathology of the coordinates which are ill-adapted to the horizon. More appropriate coordinates are Eddigton-Finkelstein, which are essentially infalling with the observers and are defined through :
\begin{eqgroup}
dv  =  dt  +   \frac{\ell dr}{h(r)} \qquad {\rm and}\qquad  dy  = dx +  \frac{J\ell ^2 dr}{   2r^2  h(r)}  \label{finkelsteinchangeofcoord}\ .
\end{eqgroup}

In these coordinates the metric is non-singular at the (future) horizon :
\begin{eqgroup}
 ds^2= - h(r) dv^2 +2\ell dv dr+ r^2(dy-\frac{J\ell}{2r^2}dv)^2\ .
 \label{finkelsteinmetric}
\end{eqgroup}
Notice that by the change $(x,t)\rightarrow (-x,-t)$ in (\ref{finkelsteinchangeofcoord}), we can obtain the metric which is regular at the past horizon. However, since we consider geometries that are presumably formed by some physical process, we will consider the past horizon as unphysical.

\subsection{Dual CFT state}
\label{sec:steadystatenointerface}
In the context of holography, (\ref{metricspinningstring}) describes a NESS of the dual CFT. This has been discussed in many places, see \cite{Erdmenger:2017gdk,Craps:2020ahu,Pourhasan:2015bsa,Bhaseen:2013ypa,Chang:2013gba,Flory:2017xqr}. This can be confirmed explicitely by going to the Fefferman-Graham gauge (\ref{flatfeffermangraham}), through the following change of coordinates :
\begin{equation}
x^\pm = x\pm t \  , \qquad     r^2 =    \frac{\ell^2}{z^2} \Bigl(1  + z^2 \frac{\langle T_{--} \rangle}{\ell}  \Bigr) 
\Bigl(1  + z^2  \frac{\langle T_{++} \rangle}{\ell}  \Bigr)\ ,
\label{changeofcoordbtztofefferman}
 \end{equation}
 which brings (\ref{metricspinningstring}) to the form (\ref{flatfeffermangraham}). From it we read the dual CFT state :
 \begin{eqgroup}
    \frac{1}{2}  J
  = \langle T_{--} \rangle - \langle T_{++}\rangle   \qquad {\rm and} \qquad 
  \frac{1}{2}   M\ell  =   \langle T_{--}\rangle  +  \langle T_{++}\rangle \ .
  \label{CFTNESSstate}
 \end{eqgroup}
 
 It follows that  the dual state has constant fluxes   of energy in both directions, with a net flow    $ \langle T^{tx} \rangle =   J/2 $.  To abide by the standard notation for heat flow we will sometimes  write $J/2 = dQ/dt$. 
 
Generic  NESS  are  characterized by operators other than $T_{\alpha\beta}$, for instance by  persistent U(1) currents. To describe them  one must  switch  on non-trivial matter  fields, and the above simple analysis must  be modified. The vacuum solutions \eqref{metricspinningstring} describe, nevertheless, a  universal minimal class of NESS that exist in all holographic conformal theories.

There are many  ways of preparing these universal  NESS. For instance, one can couple  the two endpoints of the system to heat baths so that left- and right-moving excitations thermalize at different temperatures $\Theta_\pm$\,.\footnote{We use $\Theta$ for temperature to avoid confusion with the energy-momentum tensor. In gravity the heat baths can be replaced by non-dynamical boundary black holes, see below.} An alternative protocol (which avoids  the complications of reservoirs and leads) is  the  partitioning  protocol. Here one   prepares two semi-infinite systems at  temperatures $\Theta_\pm$, and joins  them at  some initial  time $t=0$. \footnote{To implement the partitioning protocol on the gravity side one should  replace  the constant $\langle T_{++} \rangle$ in \eqref{flatfeffermangraham} by   $\theta(x^-) \Theta_-^2 + \theta(-x^-) \Theta_+^2$, where $\theta(x)$ is the step function,  and similarly for $\langle T_{--} \rangle$. This only reproduces the flow of energy for $t>0$, while for the discontinuity at $t=0$ one would most likely need external sources. Analyzing such non-stationary geometries is beyond our scope here.
}
  
The steady state will  then  form  inside  a linearly-expanding interval in the  middle \cite{Bernard:2016nci}(see also \cite{Ecker:2021ukv} for an explicit construction in 4D SYM). In both cases, after transients have died out one expects 

\begin{equation}\label{CFTstatewithtemperatures}
\langle T_{\pm\pm} \rangle = \frac{\pi c}{12} \Theta_\pm^2 =     \pi^2  \ell   \Theta_\pm^2  \quad
\Longrightarrow \quad 
\langle T^{tx} \rangle = \frac{\pi c}{12} (\Theta_-^2  - \Theta_+^2)\ ,
\end{equation}
where   $c= 12\pi \ell$ is the   central charge of the CFT by the Brown-Henneaux formula (\ref{brownhenneauxcentralcharge}).

Equation \eqref{CFTstatewithtemperatures} for the flow of heat is  a (generalized)  Stefan-Boltzmann law  with Stefan-Boltzmann constant $\pi  c/12$. Comparing \eqref{CFTstatewithtemperatures}  to \eqref{CFTNESSstate}  relates the temperatures $\Theta_\pm$ to the parameters $M$ and $J$ of the black string. This idealized CFT  calculation is, of course,  only  relevant for systems in  which  the   transport of energy is predominantly ballistic. Eq.\,\eqref{CFTstatewithtemperatures} implies in particular the existence of a  quantum  of  thermal  conductance, see  the review \cite{Bernard:2016nci} and references therein. 

\begin{figure}[!h]
\centering
\includegraphics[width=0.8\linewidth]{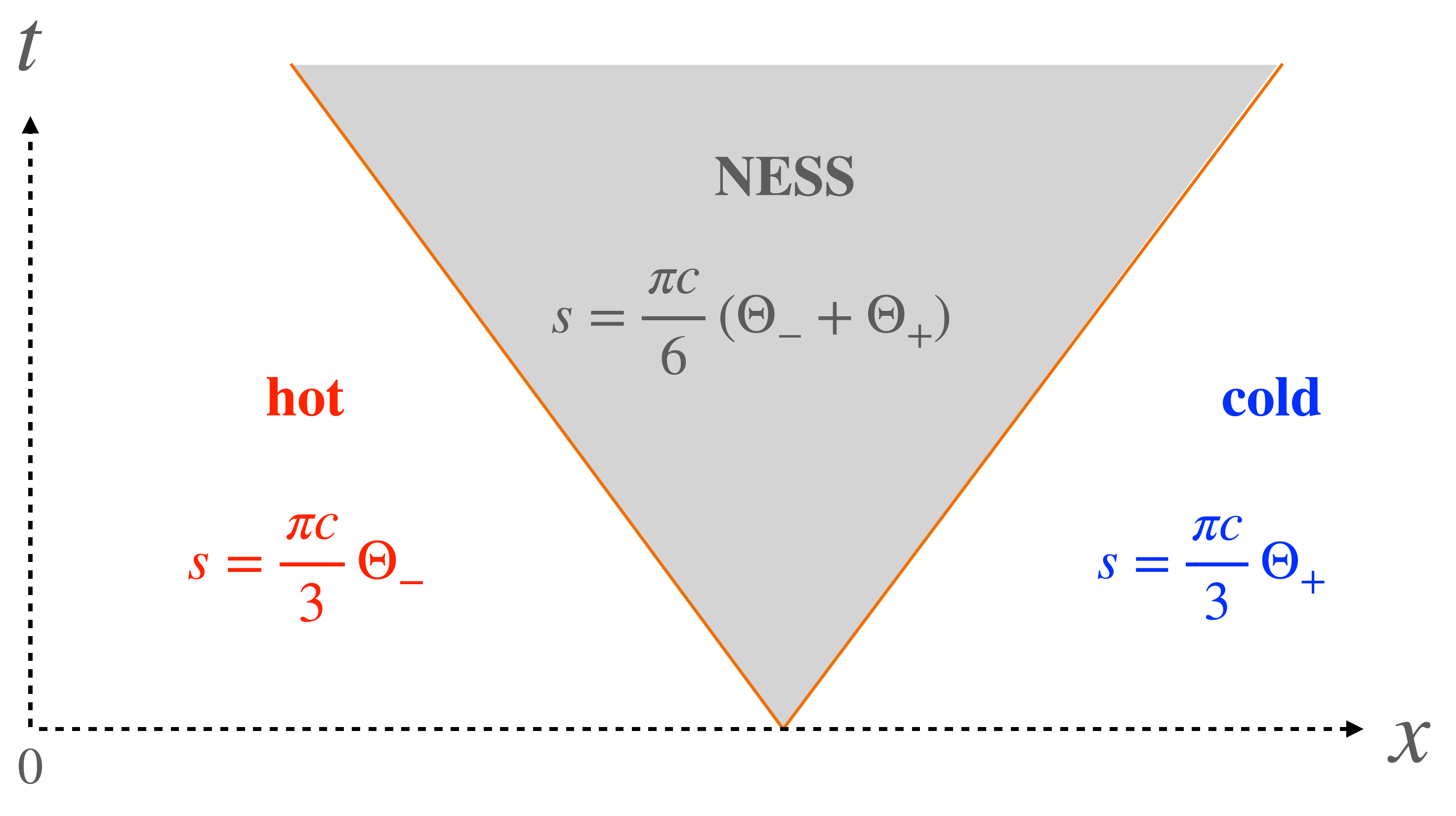}
     \caption{\small When two identical semi-infinite quantum  wires  at temperatures $\Theta_\pm$ are joined   at   $t=0$,  a  NESS  forms inside  an   interval that expands at constant speed  in both directions \cite{Bernard:2016nci}. The entropy density $s(t,x) $  is shown  in the three regions of the protocol. The  energy density  profile is identical, except  for   the replacement 
     $\Theta_\pm \to \frac{1}{2} \Theta_\pm^2$. 
    }
\label{fig:protocols}
 \end{figure}

It is  interesting  to also consider  the flow of entropy. This is illustrated   in figure  \ref{fig:protocols} which shows the  entropy density  $s \equiv   s^t$ in the three spacetime regions of   the partitioning   protocol. Inside the NESS region there is a constant flow of entropy from the hotter toward  the colder side
\begin{eqgroup}
  s_\pm    =  \pm \frac{\pi c }{6 } \Theta_\pm \ .
\end{eqgroup}
Here $s_\pm$ are the entropy densities of the chiral fluids defined through the first law $\delta \langle T_{\pm\pm} \rangle = \Theta_\pm \delta s_\pm$. The passage of the right-moving shockwave increases the local entropy at a rate $\pi c(\Theta_--\Theta_+)/6$, while the  left-moving wave reduces it at an equal  rate. Total entropy   is therefore  conserved, not surprisingly  since there are no interactions in this simple conformal 2D fluid. 

One can compute  the entropy holographically with the help of the Hubeny-Rangamani-Ryu-Takayanagi  formula\cite{Ryu:2006bv,Hubeny:2007xt}. For  a boundary region of size  $\Delta x$  the entanglement entropy reads (see sec. \ref{sec:simpleexamples} for a derivation).
\begin{equation}\label{entropyeasyNESS}
 S_{q\rm ent} = \frac{c}{6} \log \big[ \frac{\beta_+\beta_-}{\pi^2 \epsilon^2} \sinh (\frac{\pi \Delta x}{\beta_+} ) \sinh (\frac{\pi \Delta x}{\beta_-} )
 \big]  \ ,
\end{equation}
where $\beta_\pm = \Theta_\pm^{-1}$ and $\epsilon$ is a UV cutoff. From this one computes the entropy density in the steady state
\begin{eqgroup}\label{entropydensityNEsseasy}
s_{\rm NESS}  =\lim_{\Delta x\to \infty}   \frac{ S_{q\rm ent}}{\Delta x}  
  = \frac{\pi c }{ 6 } (\Theta_-+\Theta_+) = 2\pi r_+\ .
  \end{eqgroup}
The last equality, obtained with the help of   \eqref{CFTstatewithtemperatures}, \eqref{CFTNESSstate} and   \eqref{metricspinningstring},    recasts    $s_{\rm NESS}$ as the Bekenstein-Hawking entropy of the boosted black string (recall that our  units  are $8\pi G = \hbar =1$).  This agreement was one of the earliest tests \cite{Strominger:1997eq}  of the AdS/CFT correspondence.

\section{NESS of interfaces} \label{sec:NESSofinterface}
Although formally out-of-equilibrium, the  state of the previous section is a rather trivial example of a NESS. It can be obtained from the thermal state by a Lorentz boost, and  is therefore a Gibbs state  with chemical potential  for  the  (conserved) momentum in the $x$ direction. 

More interesting steady states  can be found when   left-  and right-moving  excitations  interact, for instance at impurities\cite{Bernard:2014qia,Sonner:2017jcf,Novak:2018pnv} or when the  CFT lives  in a non-trivial background  metric \cite{Hubeny:2009ru,Fischetti:2012ps,Marolf:2013ioa}.  Such interactions lead to long-range entanglement and decoherence,  giving  NESS that are not just thermal states in disguise.\footnote{\,Chiral separation also fails when  the CFT is deformed by  (ir)relevant     interactions. The  special case of the $T\bar T$   deformation  was studied, using both integrability and holography,  in \cite{Medenjak:2020bpe,Medenjak:2020ppv}. Interestingly,   the  persistent energy current  takes again  the  form \eqref{CFTstatewithtemperatures} with a deformation-dependent  Stefan-Boltzmann  constant. }
     
The case of  a conformal defect,  in particular,   has been  analyzed in ref.\cite{Bernard:2014qia}.  As explained in this reference the heat current is still given by  eq.\eqref{CFTstatewithtemperatures} but the Stefan-Boltzmann  constant is multiplied by ${\cal T}$, the energy-transmission coefficient of the defect (\ref{transmissionreflectioncoeffdefinition}). 

\begin{figure}[!h]
\centering
\includegraphics[width=0.5\linewidth]{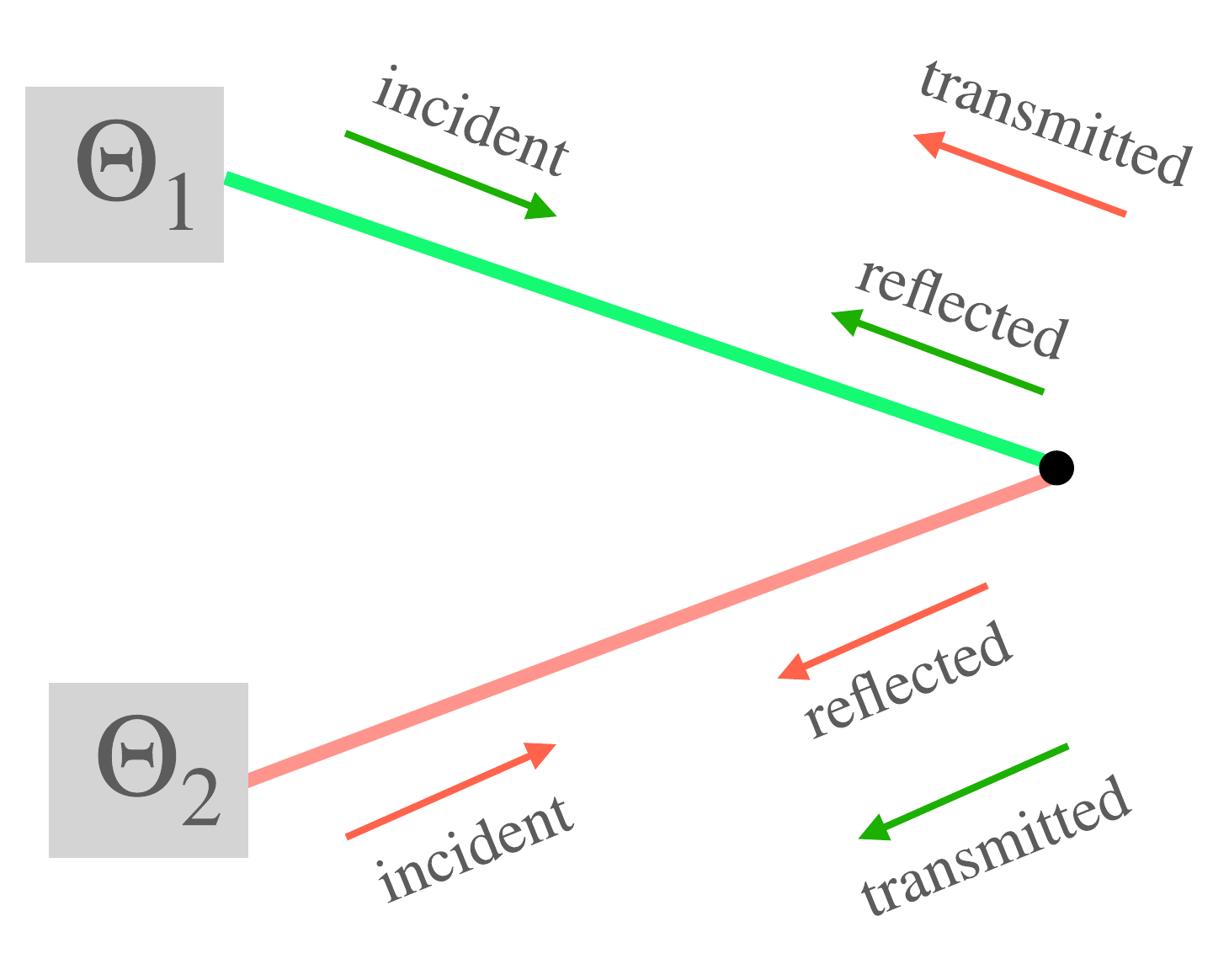}
\caption{\small The  energy fluxes given in \eqref{4flows}. The two half wires are colored red and green, and space is  folded at the interface (black dot). The incoming excitations are thermal while the state of the outgoing ones, consisting  of both reflected and transmitted fluids, depends on the nature of the junction as discussed in the main text.}
\label{fig:interfacefluxes}
 \end{figure}

The relevant setup is shown in figure \ref{fig:interfacefluxes}.  The fluids entering the NESS region from opposite directions are thermal at different temperatures   $\Theta_{1} \not= \Theta_{2}$. The difference, compared to the discussion of the previous section, is that the two half wires ($j=1,2$)  need not be identical, or   (even  when  they are) their junction is a scattering impurity where interactions between the currents can take place.
 
\subsection{Energy currents}\label{sec:nessenergycurrents}
  
Consider ${\cal R}_{j}$ and ${\cal T}_{j}$, the reflection and transmission coefficients for energy  incident on the interface  from the $j$th side (see (\ref{transmissionreflectioncoeffdefinition})). Then, the energy currents in the NESS read \footnote{The currents are given in  the folded picture in which the interface is a boundary of the tensor-product theory CFT$_1\otimes$CFT$_2$, and both incoming waves depend on $x^-$.} 
\begin{eqgroup}\label{4flows} 
& \langle T_{--}^{(1)} \rangle =  \frac{ \pi c_1}{12} \Theta_{1}^2\ , \qquad
 \langle T_{++}^{(1)} \rangle =  {\cal R}_{1} \frac{ \pi c_1}{12} \Theta_{1}^2  +  {\cal T}_{2} \frac{ \pi c_2}{12}\, \Theta_{2}^2\ ,
 \\ 
  & \langle T_{--}^{(2)} \rangle =  \frac{ \pi c_2}{12 } \Theta_{2}^2 \ , \qquad
  \langle T_{++}^{(2)} \rangle =  {\cal T}_{1} \frac{ \pi c_1}{ 12} \Theta_{1}^2  +  {\cal R}_{2} \frac{ \pi c_2}{12} \,\Theta_{2}^2 \ ,
 \end{eqgroup} 
where we simply considered the incident excitations to be thermalized (from the reservoirs at $\infty$) while the outgoing currents are a result of both transmission and reflection of the incoming currents. Crucially, we used the fact that the energy-transport coefficients across a conformal interface in 2d are universal, i.e independent  of  the nature of the incident excitations. As we have shown in sec.\ref{sec:reflectiontransmisstioncoefficients}, this assumes that the  Virasoro symmetry is not extended by extra spin-2 generators, which is true in our holographic model. We  have also used that the incoming and outgoing excitations do not interact away from the interface. 
 
Conservation of energy and the detailed-balance condition (which ensures that when $\Theta_{1}=\Theta_{2}$ the heat flow stops) imply the following relations among the reflection and transmission coefficients:
\begin{eqgroup}\label{detbal}
 {\cal R}_{j}+{\cal T}_{j} = 1\qquad {\rm and}  \qquad  c_1{\cal T}_{1}  =  c_2{\cal T}_{2} \ .
\end{eqgroup}

Hence, only one of the four transport coefficients is independent. Without  loss of  generality we assume that $c_2\geq  c_1$, i.e. that CFT$_2$  is the theory with more degrees of freedom. The average-null-energy  condition  requires $0\leq {\cal R}_{j},  {\cal T}_{j} \leq 1$, so from  \eqref{detbal} we conclude 
\begin{eqgroup}
 0\leq {\cal T}_{2} \leq  \frac{c_1}{c_2}  \qquad  {\rm or \ equivalently}    \qquad 1 \geq {\cal R}_{2}  \geq    1- \frac{ c_1}{c_2}\ .
\end{eqgroup}

As noticed in \cite{Meineri:2019ycm}, reflection positivity of  the Euclidean theory gives a weaker bound \cite{Billo:2016cpy} than this  Lorentzian bound. Note also that in the asymmetric case  ($c_2$ strictly bigger than $c_1$) part of the energy incident from side 2 is necessarily reflected.

Let $dQ/dt =  \langle T^{(1)\,tx}  \rangle  = - \langle T^{(2)\,tx}\rangle $ be the heat current across the interface. From eqs.\,\eqref{4flows} and \eqref{detbal} we find
\begin{eqgroup}\label{heatcurrentNESSinterface}
  \frac{dQ}{ dt}  =  \frac{\pi}{ 12} c_1{\cal T}_{1} \big(
  {  \Theta_{1}^2 } -  {  \Theta_{2}^2 } \big) \ .
\end{eqgroup}

Since in a unitary theory $c_1 {\cal T}_{1} $ is non-negative, heat flows as expected from the hotter to the colder side. The heat flow only stops for perfectly-reflecting interfaces (${\cal T}_{1} = {\cal T}_{2}= 0$),  or when  the two  baths are at equal  temperatures. For small  temperature differences, the heat conductance reads
\begin{eqgroup}
  \frac{d  Q}{dt }  =  \frac{\pi \Theta}{ 6} c_j{\cal T}_{j}    \delta\Theta  \ .
\end{eqgroup}

The  conductance per degree of freedom, ${\pi \Theta / 6}$, is thus multiplied by the transmission coefficient of the defect \cite{Bernard:2014qia}. Note finally that the interface is subject to a radiation force given by the discontinuity of pressure, 
\begin{eqgroup}
 F_{\rm rad} =  \langle T^{(1)\,xx}\rangle  -  \langle T^{(2)\,xx}  \rangle=  \frac{\pi}{6} \big( {c_1 {\cal R}_1   \Theta_1^2} - 
 {c_2 {\cal R}_2  \Theta_2^2}
 \big) \ ,
\end{eqgroup}
where we used \eqref{4flows} and \eqref{detbal}. The force is proportional to the reflection coefficients, as expected.

 
 \subsection{Entropy production}\label{sec:entropyproduction} 
There is a crucial  difference between the NESS of  section \ref{sec:steadystatenointerface}, and the NESS in the presence of the interface. In both cases the incoming fluids are in a thermal state. But while for a homogeneous wire they exit the system  intact, in the presence  of an interface they interact and become entangled. The  state of the outgoing excitations  depends therefore  on the nature of these interface   interactions. 
   
Let us  consider the  entropy density of the outgoing fluids, defined as the von Neumann entropy density for an interval   $[x, x+ \Delta x]$. We parametrize it by  effective temperatures, so that the entropy currents read
  \begin{eqgroup}\label{entropydensityNESSinterface} 
&  s_{-}^{(1)}  =  -\frac{ \pi c_1}{ 6} \,\Theta_{1}  \ , \qquad
  s_{+}^{(1)}   =  \frac{ \pi c_1}{ 6} \,\Theta_{1}^{{\rm eff}}
  \ ,
 \\[1ex] 
  &  s_{-}^{(2)}  =  -\frac{ \pi c_2}{6 }\, \Theta_{2}  \ , \qquad
     s_{+}^{(2)}   =  \frac{ \pi c_2}{ 6} \,\Theta_{2}^{{\rm eff}}\ .
 \end{eqgroup} 
We stress that  \eqref{entropydensityNESSinterface} is just a parametrization, the outgoing fluids need not be in a thermal state. 

\begin{figure}[!h]
\centering
\includegraphics[width=.8\linewidth]{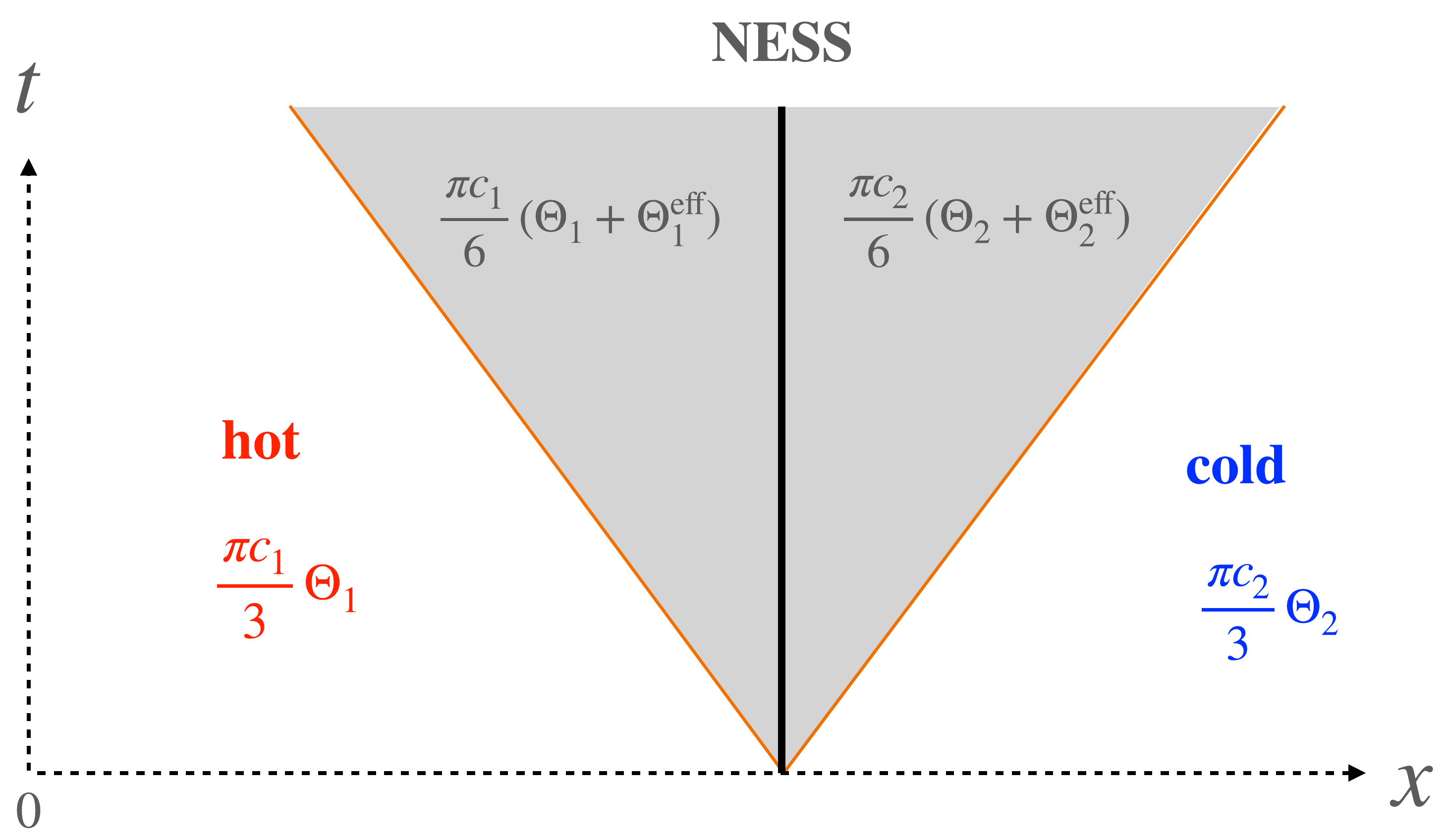}
     \caption{\small   The entropy densities in the four regions of the partitioning protocol discussed in the text (space  is here unfolded). The entropies of the outgoing fluids, which depend a priori on details of the scatterer, have been parametrized by two  effective temperatures. 
  }
\label{fig:protocols2}
 \end{figure}

In principle $\Theta_j^{\rm eff}$ may vary  as a function of $x$, but we expect them to approach constant values in the limit $t\gg \vert x\vert \gg \Delta x \to  \infty$. Figure \ref{fig:protocols2} is a cartoon of the entropy-density profile $s^t$ in various spacetime regions of the partitioning protocol. Entanglement at the interface produces thermodynamic  entropy that is  carried away by the two shock waves.  The total thermodynamic entropy on a full constant-time slice obeys
\begin{equation}\label{4Sflows} 
\frac{dS_{\rm tot}}{dt} = \frac{\pi c_1}{6}  (\Theta_1^{\rm eff}  - \Theta_1) +   \frac{\pi c_2}{6}  (\Theta_2^{\rm eff}  - \Theta_2)
    +   \frac{dS_{\rm def}}{dt} \ ,
\end{equation} 
where $S_{\rm def}$ denotes  the entropy of the interface.  Since this is bounded by the logarithm of the $g$-factor,  $S_{\rm def}$ cannot grow indefinitely and the last  term of  \eqref{4Sflows} can be neglected in a steady state.\footnote{ 
Defects  with  an infinite-dimensional Hilbert space  may evade this argument. But in the holographic model studied in this paper, $\log g \sim O(c_j)$ \cite{Azeyanagi:2007qj,Simidzija:2020ukv} and the last term in \eqref{4Sflows} can be safely  neglected at leading semiclassical order.}
 
The entanglement between outgoing excitations is encoded in a scattering matrix, which we may write schematically as
\begin{eqgroup}\label{Smatrix}
  \mathbb{S}(\psi_1^{\rm in}, \psi_2^{\rm in}, \psi_{\rm def}^{\rm in}\vert   \psi_1^{\rm out},  \psi_2^{\rm out}, \psi_{\rm def}^{\rm out}  )\ .
\end{eqgroup}
Here $\psi_j^{\rm in/out}$ are the incoming and outgoing excitations, and $\psi_{\rm def}^{\rm in/out}$  is the state of the defect before/after the scattering. Strictly speaking, there is no genuine S-matrix in conformal field theory. What describes the conformal interface is a formal operator ${\cal I}$, obtained by unfolding the associated boundary state  \cite{Bachas:2001vj,Bachas:2007td}. The above  $\mathbb{S}$ is an  appropriate Wick rotation of ${\cal I}$, as explained in ref.\cite{Bernard:2014qia}.

The density matrix  of the outgoing fluids depends a priori on the entire S-matrix,  not just on the  transport
 coefficients ${\cal T}_j$ and ${\cal R}_j$. In the large N limit we consider however, these amount to quantum corrections and our state is indeed described by $\langle T_{ij}\rangle$.

The second law of thermodynamics bounds the effective temperatures from below since the entropy production \eqref{4Sflows}  cannot be negative. The $\Theta_j^{\rm eff}$ are also bounded from above because the entropy density cannot exceed the  microcanonical one, $ s =  (\pi c\, u/3)^{1/2}$ with  $u$ the energy density of the chiral fluid. Using \eqref{4flows} and  the  detailed-balance condition this gives 
\begin{eqgroup}\label{boundsS}
 \Theta_{1}^{{\rm eff}} \leq \sqrt{ {\cal R}_1 \Theta_1^2 + {\cal T}_1 \Theta_2^2 } 
 \qquad {\rm and} \qquad
  \Theta_{2}^{{\rm eff}} \leq \sqrt{ {\cal R}_2 \Theta_2^2 + {\cal T}_2 \Theta_1^2 }  \ . 
\end{eqgroup}

The   bounds  are saturated   by perfectly-reflecting or transmitting interfaces,  i.e. when either ${\cal R}_j=1$ or  ${\cal T}_j=1$. This is trivial, because in such cases there is no entanglement between the outgoing fluids in this case.

Partially reflecting/transmitting interfaces that saturate the bounds \eqref{boundsS} act as perfect scramblers. Their existence  at weak coupling seems  unlikely,  but strongly-coupled holographic interfaces could be of this kind. We will later argue that the thin-brane holographic interfaces are perfect scramblers. This is supported by the fact (shown in section \ref{sec:eventversusapparent}) that far from the brane the event horizon approaches the equilibrium BTZ horizons, and hence the outgoing  chiral fluids are thermalized.   
 
Any domain-wall solution  interpolating between two BTZ geometries, with no other non-trivial asymptotic  backgrounds  should  be likewise  dual to a  NESS of a perfectly-scrambling interface. We suspect that many top-down  solutions  of this kind  exist, but they are hard to find. Indeed,  although many BPS domain walls are known in the supergravity literature, their  finite-temperature counterparts  are  rare. The one  example that we are aware of is the Janus AdS$_3$ black brane  \cite{Bak:2011ga}. But even for this computationally-friendly example, the far-from-equilibrium stationary solutions are not known. 

\section{Stationary branes}\label{sec:stationarybranes} 
To simplify the problem we will here resort to the more tractable thin-brane approximation, hoping that it captures some of  the essential physics of the stationary states. This model is the one described in sec. \ref{sec:MinimalICFT}, as well as the one studied in the related papers \cite{Bachas:2020yxv,Simidzija:2020ukv,Bachas:2021fqo}. 

\subsection{General setup}
Consider   two  BTZ metrics   \eqref{metricspinningstring}  glued  along a thin  brane whose worldvolume  is parametrized by  $\tau$ and  $\sigma$. Much like in the previous chapter, its embedding in the two coordinate patches  ($j=1,2$)  is given by six functions $\{ r_j(\tau, \sigma), \,t_j(\tau, \sigma), \,x_j(\tau, \sigma)\}$. Here, we must allow a little bit more generality than in the static case due to the loss of the time-inversion symmetry in the stationary case. The most general ansatz, such 
that the induced metric is $\tau$-independent,  is of the form :
\begin{equation}\label{ansatz} 
 x_j=x_j(\sigma), \ \ r_j=r_j(\sigma), \ \ t_j = \tau + f_j(\sigma)\ . 
\end{equation} 
We will denote by $h_{ab}$ the metric induced on the worldsheet of the membrane.

Compared to the static case, in (\ref{ansatz}) we allow for the addition of the function $f_j(\sigma)$ to the time coordinate of the membrane. We will see that extension of the ansatz will prove to be necessary to find solutions. In principle, one  can multiply $\tau$ on the right-hand side by  constants $a_j^{-1}$. But the metric \eqref{metricspinningstring} is invariant under rescaling  of the coordinates $r \to a r $,  $(t ,  x) \to a^{-1} (t, x) $, and of the parameters  $(M,  J) \to a^{2} (M, J) $, so we may  absorb the $a_j$ into  a redefinition of  the parameters $M_j, J_j$. Hence, without loss of generality,  we set $a_j = 1$.

Following the same convention as in chap.\ref{chap:phasesofinterfaces}, we choose the parameter $\sigma$ to be the redshift factor squared \footnote{This is a slight misnomer, since $\sigma$ becomes  negative in  the ergoregion. Nonetheless, we will adopt this jargon introduced in chap.\ref{chap:phasesofinterfaces}} for a stationary observer
\begin{eqgroup}\label{sigmarrelation}  
\sigma  =  r_1^2 -M_1\ell_1^2 = r_2^2 -M_2\ell_2^2 \ .
\end{eqgroup}

Eq. (\ref{sigmarrelation}) is one of the metric matching equations (\ref{metricmatching}). Of  the remaining embedding  functions,  the sum $f_1+f_2$ is pure gauge (it can be absorbed by a reparametrization of $\tau$) whereas the time  delay across the wall, $\Delta t (\sigma) \equiv f_2(\sigma) - f_1(\sigma)$, is a physical quantity. This and the two functions $x_j(\sigma)$  should be determined  by solving the  three remaining  
equations: (i) The rest of the metric matching conditions equating $h_{\tau\sigma}$ and $h_{\sigma\sigma}$, and (ii) one of the (trace-reversed) Israel-Lanczos  conditions \eqref{tracereverseIsrael}, which we remind here :
\begin{equation}
K^1_{ab}+K^2_{ab}  =   \lambda  h_{ab} \tag{\ref{tracereverseIsrael}}\ .
\end{equation}
The conventions used are the same as in the previous chapter, namely the normal vector to the wall is outward pointing and $\lambda$ is the brane tension. 

\subsection{Solution of  the  equations}\label{sec:equationsolutions}
In this section we solve the Israel-Lanczos equations. According to the convention chosen above, the solution is given in the `folded setup' where the interface is a conformal boundary for the product theory CFT$_1\otimes\,$CFT$_2$. Unfolding side $j$ amounts to sending $x_j \to -x_j$ and $J_j\to -J_j$.
\section{Solving the thin-brane equations}
\label{sec:solvingwallNESS}
From the form \eqref{metricspinningstring} of the bulk metric and the embedding ansatz \eqref{ansatz} of  a stationary brane,  we derive  the following  continuity equations   for the induced  metric. In this section we write $\hat{g}_{ab}$ for the induced metric instead of $h_{ab}$, to prevent any confusion with the function $h(r)$ : 
\begin{align}\label{A1} 
  \hat{g}_{\tau\tau} &=  M_1\ell_1^2 -  r_1^2 = M_2\ell_2^2 - r_2^2\ ,\\
  \label{A2} 
  \hat{g}_{\tau\sigma} & =(M_1\ell_1^2- r_1^2)f_1' -  \frac{J_1\ell_1}{2}  x_1' =  (M_2\ell_2^2 - r_2^2)f_2'  -  \frac{J_2\ell_2}{2}  x_2'\ ,\\
\hat{g}_{\sigma\sigma}  &= \frac{\ell_1^2r_1'^2}{h_1(r_1)}+r_1^2x_1'^2-J_1\ell_1 x_1'f_1'
      + (M_1\ell_1^2- r_1^2)f_1'^2\nonumber\ ,\\
        &= \frac{\ell_2^2r_2'^2}{h_2(r_2)}+r_2^2x_2'^2-J_2\ell_2 x_2'f_2'+(M_2\ell_2^2 - r_2^2)f_2'^2\label{A3}\ .
\end{align}
The primes denote  derivatives  with respect to $\sigma$, and  the function $h(r)$ has been  defined in \eqref{hrformula}, 
 \begin{equation}\tag{\ref{hrformula}}
 h(r) = {r^2 } - M\ell^2  + \frac{J^2 \ell^2}{4r^2}  = \frac{1}{r^2} (r^2 - r_+^2)(r^2-r_-^2)   \ .
 \end{equation}
 
Following sec.\ref{sec:solvingthewall}  we choose  the convenient parametrization $ \sigma = - \hat{g}_{\tau\tau}$,  so that $r_j^2 = \sigma + M_j\ell_j^2$ and $r_j' = 1/2r_j$. This parametrization need not be one-to-one, it will actually only cover half  of the   wall when  this latter  has a turning point.  With this choice the ergoplane is located at $r_j^2 = M_j \ell_j^2 \Longrightarrow \sigma =0$, and the functions $h_j$ can be written as 
\begin{equation}\label{hsigmaformula}
    h_j(\sigma) =    \frac{\sigma^2 + \sigma  M_j\ell_j^2 + J_j^2\ell_j^2/4 }{\sigma + M_j\ell_j^2} =
  \frac{ (\sigma -  \sigma_+^{{\rm H}j} )(\sigma - \sigma_-^{{\rm H}j} )  }{\sigma + M_j\ell_j^2}   \ ,
\end{equation}
where
\begin{eqgroup}\label{sHpm}
\sigma_\pm^{{\rm H}j} =  - \frac{M_j\ell_j^2}{2}   \pm  \frac{1 }{2}\sqrt{M_j^2 \ell_j^4 -J_j^2 \ell_j^2 } 
\end{eqgroup}  
are  the  locations of the  horizons in the $j$th chart.

From \eqref{A1}-\eqref{A3} one computes  the determinant of the induced metric 
\begin{equation}\label{det} 
  -  {\rm det}(\hat{g})  
   =    \frac{\sigma \ell_j^2}{4r_j^2 h_j} + {h_j r_j^2} x_j'^{\,2}\ .
\end{equation}
Note that it  does not depend on  the time-delay functions $f_j(\sigma)$, because these   can be absorbed by  the  unit-Jacobian reparametrization  
\begin{eqgroup}
\tilde\tau = \tau + f_j(\sigma), \quad \tilde \sigma = \sigma\ .
\end{eqgroup}
Eq.\eqref{det} can be used to express the $x_j^\prime$ (up to a sign)   in terms of  det $\hat{g}$. A combination of \eqref{A1} and \eqref{A2} expresses, in turn, the time delay across the wall in terms of the $x_j'$, 
\begin{eqgroup}\label{A7} 
 \sigma (f_2^\prime - f_1^\prime) = \frac{1}{2} (J_1\ell_1 x_1^\prime -  J_2\ell_2 x_2^\prime)\ .
\end{eqgroup}

To complete the calculation we need therefore  to  find  det$\hat g$ and then solve the equations \eqref{det} for $x_j'$. 

This is done with the help of the Israel-Lanczos conditions \cite{Lanczos,Israel:1966rt} (see sec.\ref{sec:thegluingequation} for more details) which express the discontinuity of the extrinsic curvature across the wall, (\ref{tracereverseIsrael}).

After Mathematica-aided computations, we arrive at the expressions for the extrinsic curvature (see app. \ref{app:appendixextrinsiccomputationNESS}) :
\begin{equation}
  K_{\tau\tau} =   \frac{h r^2x'}{ \ell  \sqrt{\vert \hat g \vert } } \qquad {\rm and} \qquad 
      K_{\tau\sigma} =    \frac{h r^2x'}{ \sigma \ell  \sqrt{\vert \hat g\vert }} \hat{g}_{\tau\sigma} +  \frac{ J \sqrt{\vert \hat g\vert } }{2\sigma}\ ,
\end{equation}
where $\hat{h}$  is  a shorthand notation for  ${\rm det}(\hat{g})$. The Israel-Lanczos  equations \eqref{tracereverseIsrael}  thus read
\begin{align}\label{Isr1} 
 &\frac{1}{\sqrt{\vert \hat g}\vert } \Bigl(  \frac{h_1r_1^2x_1'}{ \ell_1 }+\frac{h_2r_2^2x_2'}{ \ell_2 }\Bigr) 
 = - \lambda  \sigma\ , \\
\label{Isr2} 
  &\frac{1}{\sqrt{\vert \hat g}\vert } \Bigl(   \frac{h_1r_1^2  x_1'}{\ell_1 } + \frac{h_2r_2^2  x_2'}{\ell_2  }\Bigr)\hat g_{\tau\sigma}+  \frac{ \sqrt{\vert \hat g}\vert }{2 }(J_1+J_2)  =  - \lambda \sigma  \,\hat g_{\tau\sigma}\ .
\end{align}   
These  are compatible if and only if 
 \begin{eqgroup}\label{cons}
J_1+J_2 = 0 \ ,
\end{eqgroup}
which  translates to energy conservation in the boundary CFT. We have checked  that the third  equation, $[K_{\sigma\sigma}] = - \lambda \hat g_{\sigma\sigma}$,  is   automatically obeyed and thus redundant.

\subsection{The general solution} 
Squaring twice \eqref{Isr1}  and using \eqref{det} to eliminate the $x_j^{\prime\,2}$ leads to  a quadratic equation for the determinant. This has a singular solution $ {\rm det} (  \hat  g ) =0$, and a non-pathological  one 
\begin{equation}\label{B1}
 -{\rm det} (  \hat  g ) = \lambda^2 \sigma^3  \bigg[\frac{4h_1h_2r_1^2r_2^2}{ \ell_1^2\ell_2^2}
    - \big(\frac{h_1r_1^2}{ \ell_1^2} + \frac{h_2r_2^2}{ \ell_2^2} - \lambda^2\sigma^2  \big)^2 \bigg]^{-1}\ .
\end{equation}
Inserting  the   expressions for $r_j(\sigma)$ and $h_j(\sigma)$ leads  after some algebra to
  \begin{equation}\label{det1}
 -{\rm det} (  \hat  g ) = \frac{\lambda^2 \sigma}{A\sigma^2+2B\sigma+C} \ ,
\end{equation}
with coefficients :
\begin{eqgroup}
&    A = (\lambda_{\max}^2-\lambda^2) (\lambda^2-\lambda_{\min}^2)\ ,
\\
& B = \lambda^2(M_1+M_2) - \lambda_0^2(M_1-M_2)\ ,
\\
&    C = -(M_1-M_2)^2 + \lambda^2J_1^2\ .
\label{coeffs}
\end{eqgroup}
The critical tensions in these expressions are
 \begin{eqgroup}\label{3cr}
  \lambda_{\rm min}  =  \big\vert  \frac{1}{\ell_1} - \frac{1}{\ell_2}  \big\vert \ , \quad
 \lambda_{\rm max} = \frac{1}{\ell_1}+  \frac{1}{\ell_2}\ , \quad
 \lambda_0  = \sqrt{\lambda_{\rm max}\lambda_{\rm min}}\ . 
\end{eqgroup}
For a static wall, {\it i.e.} when $J_1=J_2=0$, the above formulae reduce, as they should,  to the ones obtained  ref.\cite{Bachas:2021fqo}.\,\footnote{When comparing with this reference  beware that it uses  the (slightly  confusing) notation $ \hat g_{\sigma\sigma} \equiv g(\sigma)$ so that,  since   the metric  is diagonal  in the static case,   det\,$\hat g = -\sigma g(\sigma)$. }  The only effect of the non-zero $J_j$  is actually  to shift  the  coefficient $C$ in \eqref{coeffs}.

The roots of the quadratic polynomial in the denominator of \eqref{det1}, 
\begin{eqgroup}\label{spmapp} 
\sigma_\pm =  \frac{-B \pm \sqrt{B^2 - AC} }{A} \ , 
\end{eqgroup}
determine the   behaviour of the solution. If $\sigma_+$ is   either complex or negative  (part of) the brane worldvolume  has ${\rm det}\, \hat g >0$ in  the ergoregion, so it is  spacelike and physically unacceptable. Acceptable solutions  have  $\sigma_+ >0$ or $\sigma_+=0$, and  describe walls that avoid, respectively enter the ergoregion as explained in the main text, see section \ref{sec:insideergoregion}.

The actual shape of the wall is found by inserting \eqref{det1} in \eqref{det} and solving for  $x_j'^{\,2} $. After some rearrangements this gives
\begin{align}
  \epsilon_1  \frac{x_1'}{\ell_1} &=  \frac{(\lambda^2+\lambda_0^2)\sigma + (M_1-M_2)} {2(\sigma+M_1\ell_1^2+J^2\ell_1^2/4\sigma) 
    \sqrt{A\sigma (\sigma - \sigma_+)(\sigma - \sigma_-) }}\ , \\
\epsilon_2  \frac{x_2'}{\ell_2} &=     \frac{(\lambda^2-\lambda_0^2)\sigma - (M_1-M_2)} {2(\sigma+M_2\ell_2^2+J^2\ell_2^2/4\sigma) 
    \sqrt{A\sigma (\sigma - \sigma_+)(\sigma - \sigma_-) }}\ , 
\end{align}
where $\epsilon_j=\pm$ are signs. They are fixed by  the linear equation \eqref{Isr1} with the result
\begin{eqgroup}
 \epsilon_j(\sigma)  =  -\frac{\sigma }{\vert\sigma\vert}\ . 
\end{eqgroup}

These signs agree with the known universal solution  \cite{Bachas:2001hpy,Bachas:2021fqo} near the AdS boundary, at $\sigma \to \infty$, and they ensure that walls entering the ergoregion have no kink. Expressing  the denominators   in terms of the horizon locations  \eqref{sigmaH} gives the equations \eqref{x1ness} and \eqref{x2ness} of the main text.

It is worth noting that the tensionless ($\lambda\to 0$)  limit of our solution  is singular. Indeed, on one hand,  extremising   the brane action and ignoring its back-reaction gives a  geodesic worldvolume,  but on the other hand for  $\lambda=0$ fluctuations of the string are unsuppressed. In fact,  when  $\lambda$ is small the wall starts as a geodesic near the AdS boundary but   always departs significantly in the interior.

The results of sec. \ref{sec:solvingwallNESS}  can be  summarised  as follows. First,   from \eqref{sigmarrelation} 
\begin{eqgroup}\label{rjsolutions}
  r_j(\sigma)    = \sqrt{\sigma + M_j \ell_j^{\,2}}\ .
\end{eqgroup}
Secondly, we find a sufficient and necessary condition on the parameters for the existence of solutions : 
\begin{eqgroup}
  J_1 = -J_2  
\end{eqgroup}
This ensures conservation of energy in the CFT, as seen from the holographic dictionary \eqref{CFTNESSstate}. Thirdly,   matching $h_{\tau\sigma}$ from  the two sides determines  the time delay in terms of the embedding functions $x_j$, 
\begin{eqgroup}\label{fj} 
\Delta t^\prime  \equiv f_2' - f_1'   =     \frac{ J_1 }{2 \sigma } ( \ell_1 x_1^\prime + \ell_2 x_2^\prime)\ ,
\end{eqgroup}
where primes denote derivatives with respect to $\sigma$. What remains is thus to find the functions $x_j(\sigma)$. 
 
To this end, we use the continuity of $h_{\sigma\sigma}$ and  the $\tau\tau$ component of \eqref{tracereverseIsrael}. It is useful and convenient to first solve these two equations for the determinant of the induced metric, with the result
\begin{eqgroup}\label{detg}
 -{\rm det} \hat g =  \frac{\lambda^2 \sigma}{{A \sigma^2 + 2B\sigma +C }}  = 
  \frac{\lambda^2 \sigma}{{A (\sigma - \sigma_+)(\sigma - \sigma_-)}}\ ,  
\end{eqgroup}
where 
\begin{eqgroup}\label{spm} 
 \sigma_\pm =  \frac{-B \pm \sqrt{B^2 - AC} }{A} \ ,
\end{eqgroup}
and the coefficients $A,B,C$ read
\begin{eqgroup}\label{NESSabc}
A = (\lambda_{\max}^2-\lambda^2) (\lambda^2-\lambda_{\min}^2)\ , & \quad 
B = \lambda^2(M_1+M_2) - \lambda_0^2(M_1-M_2)\ , 
\\
&    C = -(M_1-M_2)^2 + \lambda^2J_1^2\ .   
\end{eqgroup} 
 The three critical tensions entering  the above coefficients have been defined in the previous chapter, sec. (\ref{sec:criticaltensions}).
\begin{equation}\tag{\ref{criticaltensions}}
  \lambda_{\rm min}  =   \biggl\vert \frac{1}{\ell_1} - \frac{1}{\ell_2} \biggl\vert  \quad
 \lambda_{\rm max} = \frac{1}{\ell_1}+  \frac{1}{ \ell_2} \quad
 \lambda_0  = \sqrt{\lambda_{\rm max}\lambda_{\rm min}}  \ .
\end{equation}

Without loss of generality we  assume,  as earlier, that   $\ell_1\leq \ell_2$,  so the absolute value in   $ \lambda_{\rm min} $ is superfluous. Note that the expressions \eqref{detg} to \eqref{NESSabc} are the same as the ones for static branes, see (\ref{abc}), except  for the extra term $\lambda^2 J_1^2$ in the coefficient  $C$. 
  
The  determinant  of  the induced metric  can be expressed in terms of $x_j$ and $\sigma$ in each chart, $j=1$ and $j=2$. It does not depend on the time-shift functions $f_j$, which could be absorbed by a reparametrization of the metric with unit Jacobian. Having already extracted  ${\rm det}h$, one  can now invert these relations to find the $x_j'$, 
\begin{align}\label{x1ness} 
    \frac{x_1'}{\ell_1} &= -  \frac{{\rm sgn}(\sigma) \big[ (\lambda^2+\lambda_0^2)\sigma^2  + (M_1-M_2)\sigma \big] }
     {2(\sigma- \sigma_+^{\rm H1} )(\sigma- \sigma_-^{\rm H1}) 
    \sqrt{A\sigma  (\sigma - \sigma_+)(\sigma - \sigma_-) }}  \ ,\\
\label{x2ness} 
  \frac{x_2'}{\ell_2} &= -  \frac{{\rm sgn}(\sigma) \big[(\lambda^2-\lambda_0^2)\,\sigma^2 \, - (M_1-M_2) \sigma \big] }
  {2(\sigma- \sigma_{+}^{\rm H2} )(\sigma- \sigma_-^{\rm H2}) 
    \sqrt{A \sigma  (\sigma - \sigma_+)(\sigma - \sigma_-) }}\ , 
\end{align}  
where we denoted the horizons locations in the blueshift parametrization:
\begin{eqgroup}\label{sigmaH}
\sigma_{\pm}^{{\rm H}j} =  - \frac{M_j\ell_j^2}{2}   \pm  \frac{1 }{2}\sqrt{M_j^2 \ell_j^4 -J_j^2 \ell_j^2 }  \ .
\end{eqgroup}
These points are where the outer and inner horizons of the $j$th  BTZ metric  intersect the domain wall. 
 
Eqs. \eqref{rjsolutions} to  \eqref{sigmaH} give  the general stationary  solution of the thin-brane  equations  for any Lagrangian parameters $\ell_j$ and $\lambda$, and geometric parameters $M_j$ and $J_1=-J_2$.  The Lagrangian parameters are part of the basic data of the interface CFT, while the geometric parameters determine the CFT  state. When  $J_1=J_2=0$,  all  these expressions reduce  to the static  solutions described in chap.\ref{chap:phasesofinterfaces}.

\section{Inside  the ergoregion}\label{sec:insideergoregion} 
The qualitative behaviour of the domain wall is governed  by  the  singularities of  (\ref{x1ness}, \ref{x2ness}),  as one moves from the AdS boundary  at $\sigma \sim  \infty$ inwards. In addition to the BTZ horizons at $\sigma^{{\rm H}j}_\pm$, other potential singularities arise at $\sigma_\pm$ and at the entrance of the ergoregion $\sigma=0$. From \eqref{detg}  we see that the brane worldvolume would become spacelike beyond $\sigma=0$, if $\sigma_\pm$ are both either negative or complex. 

To avoid such pathological behaviour one of the following two conditions must  be met:
 
 \begin{itemize}
\item    \fbox{$\,\sigma_+>0 \,:$}  The singularity at $\sigma_+$ is in this case a turning point, and the   \\ wall does not extend  to lower values of $\sigma$. Indeed, as seen from \eqref{x1ness} and \eqref{x2ness}, the singularity as $\si_+$ is integrable.
\item   \fbox{$\,0= \sigma_+  > \sigma_-\,:$} In this case the worldvolume remains timelike as the wall enters  the ergoregion. The reader can verify from eqs.\,\eqref{detg}, \eqref{x1ness} and  \eqref{x2ness} that the embedding near  $\sigma=0$ is smooth. 
\end{itemize}

\begin{figure}[!h]
\centering
\includegraphics[width=.9\textwidth]{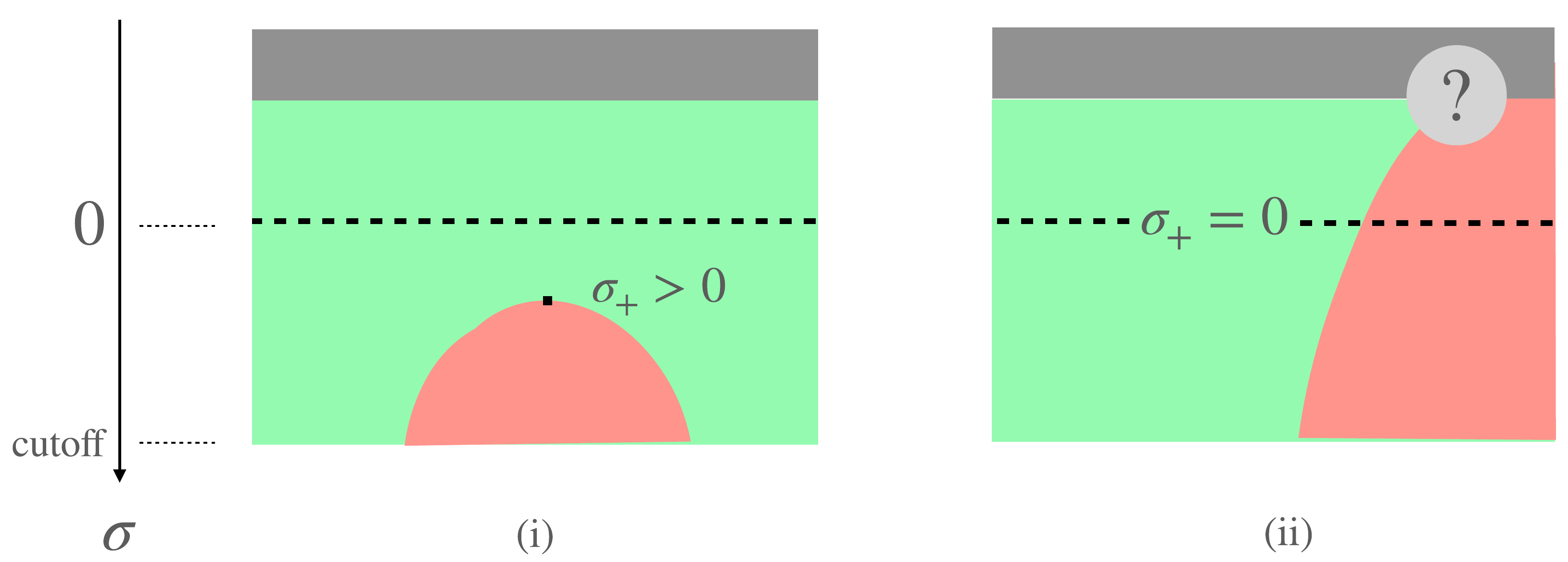}
     \caption{\small The  two kinds of   stationary-wall geometries:  (i) The wall avoids  the ergoregion, turns around   and intersects  the AdS boundary twice; or (ii) it enters the ergoregion and does not come out again. The broken line is the ergoplane, the two outer BTZ regions  are coloured in  green and pink, and the region behind the horizon in grey. The    horizon  in case (ii) will be described in detail in the coming section.}
\label{fig:wall_horizon}
 \end{figure}    
These  two   possibilities  are illustrated in figure \ref{fig:wall_horizon}. Branes entering the ergoregion are dual, as will become clear, to steady  states  of an isolated interface, while those that avoid  the ergoregion  are dual to  steady states of  an interface anti-interface pair. We will return to the second  case in section \ref{sec:pairofinterfaces}, here we focus on the isolated  interface. 

The condition  $\sigma_+=0$  implies  $C=0$ and $B\geq 0$. Using  \eqref{spm} and \eqref{NESSabc},  and the fact that the coefficient   $A$ is positive for  tensions in the allowable range ($\lambda_{\min} < \lambda < \lambda_{\max}$) we obtain
\begin{eqgroup}\label{sigma+0}
M_1-M_2 =  \pm \lambda  J_1 = \mp \lambda J_2  
\quad
  {\rm and} \quad 
  \lambda^2 (M_1+M_2)  \geq  \lambda_0^{\,2} (M_1-M_2) \ . 
\end{eqgroup}

Furthermore,  cosmic censorship requires that   $\ell_j M_j> \vert J_j\vert$ unless the bulk  singularity at  $r_j=0$  is  excised  (this is the case  in  the pink region  of the left  fig. \ref{fig:wall_horizon}. If none of the singularities is excised, the inequality in \eqref{sigma+0} is  automatically satisfied and hence redundant. 

With the help of the  holographic dictionary  \eqref{CFTNESSstate} one  can  translate the expression  \eqref{sigma+0} for $M_1-M_2$  to the   language of ICFT. Since the  incoming fluxes are thermal, $T_{--}^{(j)} = \pi^2 \ell_j \Theta_{j}^2$\, and \eqref{CFTNESSstate}  gives
\begin{eqgroup}
    M_j =   {4\pi^2 \Theta_{j}^2 }  - \frac{J_j}{\ell_j} \Longrightarrow  \ ,
    M_1 - M_2 =  4\pi^2 ( \Theta_{1}^2  -  \Theta_{2}^2 )   - J_1 (\frac{1}{\ell_1} + \frac{1}{\ell_2})   \ .
\end{eqgroup}
Combining with \eqref{sigma+0} gives the  heat-flow rate
\begin{eqgroup}
 J_1 =  2  \langle T^{(1)\,tx}  \rangle  =    {4\pi^2} \big[  \frac{1}{\ell_1} + \frac{1}{\ell_2} \pm \lambda \big]^{-1} 
( \Theta_{1}^2  -  \Theta_{2}^2   )\ .
\end{eqgroup}
This agrees with the ICFT  expression  \eqref{heatcurrentNESSinterface} if we identify the transmission coefficients  as follows (recall that $c_j=12\pi \ell_j$)
\begin{eqgroup}\label{transmissionfromentering}
{\cal T_j} = \frac{2}{ \ell_j}\big[ \frac{1}{ \ell_1} + \frac{1}{ \ell_2}  \pm \lambda \big]^{-1}\ .
\end{eqgroup}

It is gratifying to find that, for the choice of plus sign, \eqref{transmissionfromentering}  are precisely   the coefficients ${\cal T_j} $ computed in the  linearized approximation in ref.\cite{Bachas:2020yxv}. In essence, this is a non-perturbative derivation of the transmission coefficients of the interface in the minimal model. The correct choice of sign will be justified in a minute. 

Let us pause here to take stock of the situation. We found that  (i) the dual of an isolated interface {\it must} correspond  to a  brane that enters the ergoregion (otherwise it will turn back, and the dual will contain two interfaces), and (ii) that  the brane equations determine in this case the flow of heat in accordance with the CFT  result of  \cite{Bernard:2014qia,Meineri:2019ycm} and the transmission coefficients found in \cite{Bachas:2020yxv}.\footnote{The fact that our non-linear analysis agrees with the linearized-wave treatment of \cite{Bachas:2020yxv} is an indirect confirmation of the fact that the transport coefficients are universal.}

To complete the story, we must make sure that once inside the ergoregion the brane does not come out again. If it did, it would intersect the AdS boundary at a second point, so the solution would not be  dual to an isolated interface as claimed.

Inserting  $\sigma_+=0$ in the embedding functions  (\ref{x1ness},\,\ref{x2ness}) we find 
\begin{eqgroup}\label{xjentering} 
\frac{x_1'}{\ell_1}  & = -  \frac{(\lambda^2+\lambda_0^2)\sigma  + (M_1-M_2)} {2(\sigma- \sigma_+^{\rm H1} )(\sigma- \sigma_-^{\rm H1}) \sqrt{A (\sigma - \sigma_-) }}  \ ,
\\
\frac{x_2'}{\ell_2}  & = -  \frac{(\lambda^2-\lambda_0^2)\sigma   - (M_1-M_2) }{2(\sigma- \sigma_+^{\rm H2} )(\sigma- \sigma_-^{\rm H2}) \sqrt{A  (\sigma - \sigma_-) }}  \ ,
\end{eqgroup}
where  the   $\sigma_\pm^{{\rm H}j}$ are given by  \eqref{sHpm} and   
 \begin{eqgroup}\label{sigma-entering}
\sigma_- = -  \frac{2\lambda  }{A} \big[ \lambda (M_1+M_2) \pm  2  \lambda_0^2J_1 \big]   \ .
\end{eqgroup}
As already said, the  embedding is regular   at $\sigma=0$, i.e. the brane enters the ergoregion smoothly.  What it does next depends on which singularity it encounters first. If this were the square-root singularity at $\sigma_-$,  the  wall would    turn around (just like it does for  positive $\sigma_+$),  exit the ergoregion and intersect the AdS boundary at another  anchor point. This is the possibility that we want to exclude. 

Consider for starters  the simpler case  $\ell_1 = \ell_2 \equiv \ell$. In this  case $\lambda_0=0$ and $A = \lambda^2 ( 4/\ell^2 - \lambda^2)$, so  \eqref{sigma-entering}  reduces to  
\begin{eqgroup}\label{ineq22}
  \sigma_-  =  -  \frac{2\ell^2 (M_1+ M_2)  }{4- \lambda^2\ell^2}  \leq - {\rm min}(M_j) \ell^2  \ .
\end{eqgroup}
In the last step we used the fact that  both $M_j$ are positive, otherwise the conical singularity at $r_j=0  \Longleftrightarrow\sigma =  - M_j \ell^2$ would be naked. 

What \eqref{ineq22}  shows is that the putative turning point $\sigma_-$  lies behind the bulk  singularity in at least one of the two BTZ regions,  where our solution cannot be extended. Thus this turning point is never reached.

For general $\ell_1 \not=\ell_2$ a weaker  statement is true, namely that  $\sigma_-$ is shielded by an inner horizon for at least one $j$. For a proof, we maximize $\sigma_-$ with respect to the brane tension $\lambda$. We have performed this  calculation with Mathematica, but do not find it useful to reproduce details here. The key point for our purposes is that there are no solutions in which the brane enters the ergoregion, turns around before an inner horizon, and exits towards the AdS boundary. Since as argued by Penrose \cite{Penrose:1969pc},  Cauchy (inner) horizons are classically unstable,\footnote{For recent  discussions  of  strong cosmic censorship in the BTZ black hole see  \cite{Dias:2019ery,Papadodimas:2019msp,Balasubramanian:2019qwk,Emparan:2020rnp,Pandya:2020ejc}.}solutions in which the turning point lies behind one of them cannot be trusted. As such, the cauchy horizon should be viewed as the singularities of the black hole.

One last remark is in order concerning the induced brane metric $h_{ab}$. By redefining the worldvolume time,   $\tilde{\tau}=\tau+ {J\ell_1} \int  {x_1'(\si)}d\si /{2\si}$, we can bring this metric to the diagonal form :
\begin{equation}
    d\hat s^2 =-\si d{\tilde{\tau}}^2+
  \vert {\rm det}\, \hat g \vert \,   \frac{d\si^2}{\si}\qquad {\rm with} \qquad 
  {\rm det}\, \hat g =  \frac{\lambda^2   }{ {A  (\sigma_- - \sigma)}}\  . 
\end{equation}

The  worldvolume is   timelike for  all $\sigma >  \sigma_-$, as already advertised. More interestingly, the metric (felt by signals that propagate on the brane) is that of a two-dimensional black-hole  with  horizon at the ergoplane $\sigma=0$. This lies outside the bulk horizons $\sigma_+^{{\rm H}j}$, in agreement  with arguments showing   that the causal structure is always set by the Einstein metric \cite{Gibbons:2000xe}. Similar remarks in a closely-related context were made before  in ref.\cite{Frolov:1995qp}. 

The brane-horizon (bH) temperature,  
\begin{eqgroup}
4\pi  \Theta_{\rm bH} =    \big(-{\rm det}\, \hat g\vert_{\sigma=0}  \big)^{-1/2} \ ,  
\end{eqgroup}
is intermediate between   $\Theta_1$ and $\Theta_2$  as can be easily checked. For $\ell_1=\ell_2$ for example one finds $ 2\,\Theta_{\rm bH}^2 = \Theta_1^2+ \Theta_2^2$.  

\section{The non-Killing  horizon}\label{sec:nonkillinghorizon}
Since   $\sigma_-$ lies behind an inner  horizon,   the first singularities of  the embedding functions (\ref{x1ness},\ref{x2ness}) are at $\sigma_+^{{\rm H}j} $. A key  feature of the non-static solutions is that these outer  BTZ  horizons, which are apparent horizons as  will  become clear, do not meet  at the same point on the brane. 
For  $J_j\not=  0$ the following strict inequalities indeed hold 
\begin{equation}\label{orderH} 
  \sigma_+^{{\rm H}1} >  \sigma_+^{{\rm H}2}    \quad {\rm if} \ \ M_1 > M_2
\qquad  \sigma_+^{{\rm H}2} <  \sigma_+^{{\rm H}1}    \quad {\rm if} \ \  M_1< M_2   \ .
\end{equation}  
For small $J_j$ these inequalities are manifest  by Taylor expanding   \eqref{sigmaH},
\begin{equation}
 \sigma_+^{{\rm H}j}  = - \frac{J_j^2 }{M_j} + O(J_j^{\,4})\ . 
\end{equation}
We show  that they hold  for all $J_j$ in appendix \ref{app:horizonineq}.  

The  meaning of these inequalities  becomes clear if we use the holographic dictionary \eqref{CFTNESSstate}, the energy currents \eqref{4flows} and the detailed-balance condition \eqref{detbal} to write the $M_j$ as follows
\begin{eqgroup}\label{theMj} 
&M_1 = 2\pi^2 \left[ \Theta_1^2 (1+{\cal R}_1) + \Theta_2^2 (1- {\cal R}_1) \right]  \ ,
\\
&M_2 = 2\pi^2 \left[ \Theta_1^2 (1-{\cal R}_2) + \Theta_2^2 (1+{\cal R}_2) \right]\ .
\end{eqgroup}    
Assuming  $0\leq {\cal R}_j \leq 1$, we see that the hotter side of the interface has the larger $M_j$. What \eqref{orderH}  therefore says is  that the brane hits the BTZ horizon of the hotter side first. 
 
\subsection{The arrow of time}\label{sec:arrow}
Assume for  concreteness  $M_1> M_2$, the case $M_2>M_1$ being  similar.\footnote{Strictly speaking we also ask that the  brane hits both outer horizons   before the  inner (Cauchy) horizons, since we cannot trust our classical solutions beyond the  latter. As explained in  appendix \ref{app:horizonineq}, this condition is automatic when $M_2>M_1$, but not when $M_1>M_2$, where it is possible for  some range of parameters  to have $\sigma_+^{{\rm H}2} <\sigma_-^{{\rm H}1}$. Those specific solutions should be discarded, although we do not think that this difference has a deeper physical meaning.}

From eq.\eqref{sigma+0} we have  $M_1 = M_2 + \lambda \vert J_1\vert$. We do not commit yet  on the sign of $J_1$, nor on the   sign in \eqref{sigma+0}, but the product of the two should be  positive. From the holographic interpretation, we expect $J_1$ to be positive (heat flows from the hot to the cold side), so the correct sign in \eqref{sigma+0} is plus (to match the correct reflection/transmission coefficients), but we would like to understand how this condition arises from the gravity side. 

\begin{figure}[!h]
  \centering  
\includegraphics[width=0.97\linewidth]{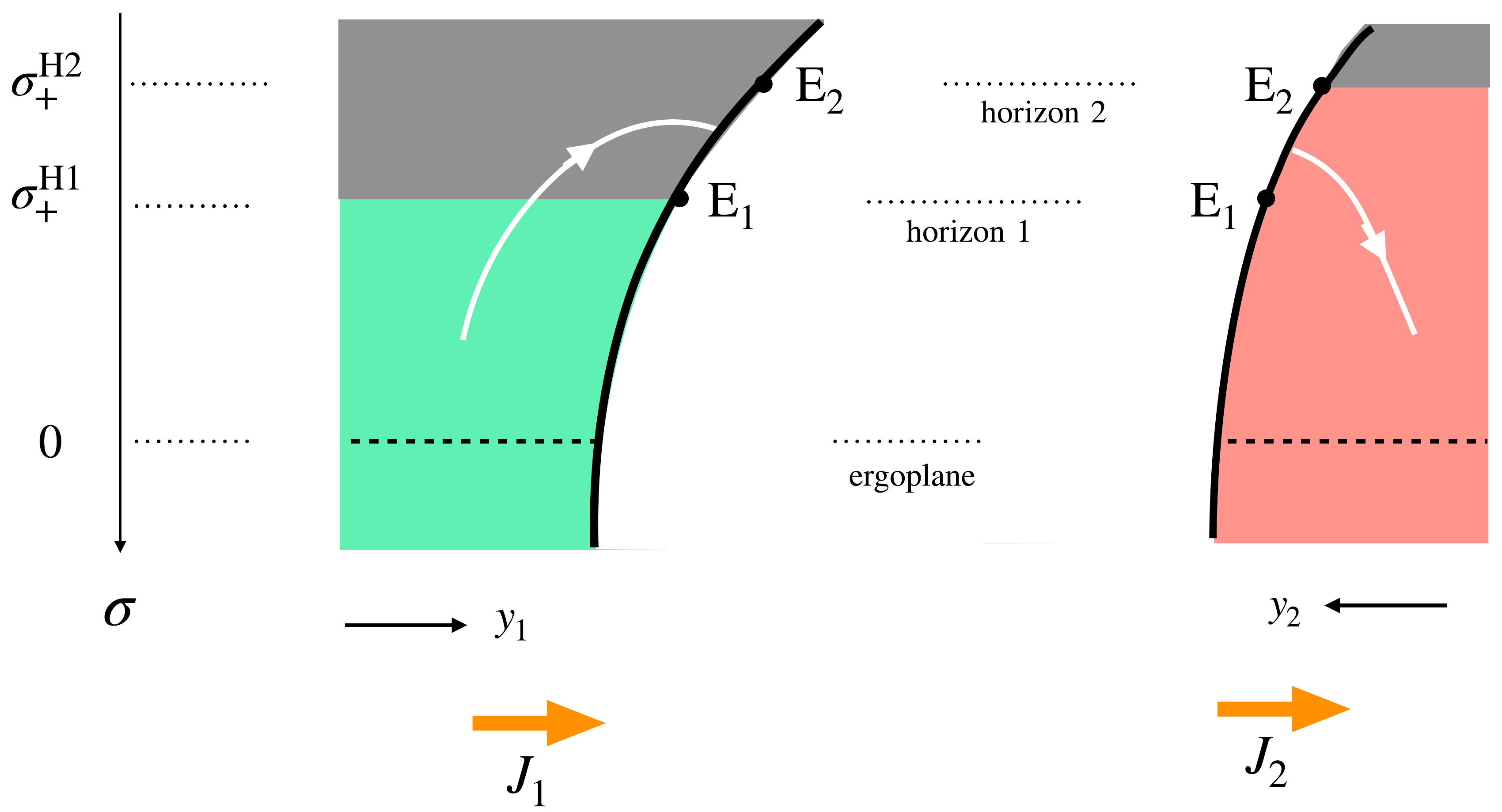}
     \caption{\small A brane (thick black curve) entering  the local outer  horizons  ${\cal H}1$ and ${\cal H}2$ (the boundaries of the grey regions in the figure) at two different points E1 and E2. The  piece [E1,E2] of the wall is behind the horizon of slice 1 but outside the horizon of slice 2. The thick orange arrows show the direction of heat flow. The  white curve is the worldline of an observer entering ${\cal H}1$, crossing the brane and emerging outside ${\cal H}2$. }
\label{fig:dischornew}
 \end{figure}

Figure  \ref{fig:dischornew} shows a sketch of the behaviour of the brane past the ergoplane. The vertical axis is  parameterised by $\sigma$ (increasing  downwards), and  the horizontal axes by the ingoing Eddington-Finkelstein  coordinates $y_j$ defined in \eqref{finkelsteinchangeofcoord}. These coordinates are regular at the future horizons, and reduce to the flat  ICFT coordinates $x_j$ at the AdS boundary. Therefore, they do not affect the CFT state, or in other words, they are pure gauge and act trivially on the asymptotic boundary.

Let us take a closer look  at the  wall embedding in Eddington-Finkelstein (EF) coordinates. From (\ref{x1ness},\ref{x2ness})   and  the identities $r_j' = 1/2r_j$  we get  
\begin{eqgroup}\label{yjness} 
 &y_1' =    \frac{\ell_1} {2(\sigma- \sigma_+^{\rm H1} )(\sigma- \sigma_-^{\rm H1})}
 \left[ 
     \frac{J_1\ell_1 }{2 \sqrt{\sigma + M_1\ell_1^2}}
  -  \frac{(\lambda^2+\lambda_0^2)\sigma  + \lambda \vert J_1\vert } { 
    \sqrt{A (\sigma - \sigma_-) }} 
 \right] \ ,
\\
&y_2' =     \frac{\ell_2} {2(\sigma- \sigma_+^{\rm H2} )(\sigma- \sigma_-^{\rm H2})}
 \left[ 
     \frac{J_2\ell_2 }{2 \sqrt{\sigma + M_2\ell_2^2}}
  -   \frac{(\lambda^2- \lambda_0^2)\sigma   -  \lambda \vert J_2\vert } { 
    \sqrt{A (\sigma - \sigma_-) }} 
 \right]  \ .
\end{eqgroup}
A little algebra shows that  the square brackets in the above expression vanish at the corresponding horizons $\sigma= \sigma_+^{{\rm H}j}$ if   $J_1=-J_2>0$. The  functions $y_j$ present no singularity at the horizon. By contrast, if  $J_1=-J_2 <0$ these functions are  singular:  $y_1 \to +\infty$ at $\sigma_+^{{\rm H}1}$, and $y_2 \to - \infty$ at $\sigma_+^{{\rm H}2}$. 

Consider first the case $J_1>0$. At first, it seems remarkable that in Eddigton-Finkelstein coordinates, the singularity at the horizon $\sigma= \sigma_+^{{\rm H}j}$ disappears and the brane enters smoothly in the black hole. After all, the membrane depends on $\lam$, $J_1$ and both $M_j$, while the Eddington-Finkelstein coordinates change depends only on one side's parameters, $M_j$, $J_j$. Indeed, there are seemingly "miraculous" cancellations at play at $\sigma_+^{{\rm H}j}$ so that the behavior of the membrane at this location is independent on the tension. This can somehow be explained by remembering that the Eddington-Finkelstein coordinates are adapted to infalling observers. It appears that close to the horizon, the shape of the membrane is independent of its tension, and its shape follows a falling observer's trajectory. 

In the case $J_1<0$, this cancellation of course does not happen, and the membrane does not enter the future horizon. However, if we go to outgoing E-F coordinates, we find that the membrane exits smoothly the past horizon of the white hole. 

Knowing this, we now understand why we find a pair of solutions dual to the stationary ICFT system. The first one, with $J_1>0$ generates a heat flow in the correct direction in the field theory; from hot to cold. In this case, $M_1=M_2+\lambda J_1$, and hence the sign in the expression (\ref{transmissionfromentering}) for the transmission coefficients is plus, in agreement with the result of ref.\cite{Bachas:2020yxv}.

The other solution with $J_1<0$ can be obtained by time-reversal, which leaves the $M_j$ unchanged. This gives an explanation for the interpretation of this companion solution on the field theory: it is the same stationary state, but time-inverted. Indeed, if one chooses the minus sign in the expressions \eqref{transmissionfromentering} (which is the "incorrect" transmission coefficients), and plugs them in (\ref{4flows}), while at the same time exchanging $T_{--}\leftrightarrow T_{++}$, one will recover the equations (\ref{4flows}), but now with the "correct" transmission and reflections coefficients. Indeed, in the time-reversed stationary states, the "incident" rays will be the left-movers. Thus, the minus sign of (\ref{transmissionfromentering}) appears when one chooses the incorrect time direction for the stationary states, and as such, misinterprets the scattering experiment.

In the end, both solutions describe exactly the same situation. By choosing the time direction to be forward, we will have to discard the time-reversed solution. In gravity, similarly to a white hole, which solves Einstein's equations but cannot be produced by gravitational collapse, we expect that no physical protocol can prepare the $J_1<0$ solution.


\subsection{Event versus apparent horizon}\label{sec:eventversusapparent} 
Denote by  ${\cal H}_1$  and ${\cal H}_2$  the horizons  of the two BTZ regions of the stationary geometry, and by  {\small E}$_1$ and  {\small E}$_2$ their  intersections with the brane worldvolume. We can foliate spacetime by Cauchy slices  $v_j = \bar v + \epsilon_j(r_j,x_j)$, where $\bar v$ is a uniform foliation parameter.\footnote{The non-trivial radial dependence in the definition of the Cauchy slice  is necessary  because constant  $v_j $ curves are   lightlike behind the $j$th horizon.} We use the same symbols for the  projections of  ${\cal H}_j$ and {\small E}$_j$ on a   Cauchy slice. Since simultaneous translations of $v_j$ are Killing isometries,  the  projections do not depend on $\bar v$.

Both  ${\cal H}_1$  and ${\cal H}_2$ are local (or apparent)  horizons, i.e. future-directed  light rays can only traverse them in one direction. But it  is clear  from  figure \ref{fig:dischornew}  that ${\cal H}_1$ cannot be part of the  event horizon of global spacetime. Indeed,  after  entering   ${\cal H}_1$ an observer moving to the right can traverse  the  [{\small E}$_1$, {\small E}$_2$] part of the wall,   emerge outside ${\cal H}_2$ in region  2, and from there   continue her journey to the boundary. Such  journeys are  only forbidden if  {\small E}$_1$= {\small E}$_2$,
i.e. for the  static equilibrium solutions. 

Before proceeding, let us briefly describe an alternative route that was considered before arriving at the correct resolution, since we believe it might be interesting to some readers. In our initial attempt, we searched for a way to prevent an observer from doing the trip depicted in white in (\ref{fig:dischornew}), keeping the apparent horizon as event horizons. A way to do that is to clip the membrane at {\small E1}. This is of course already not very satisfying on side 1, as the membrane terminates on the horizon, but we hoped there might be a way to justify this.

On side 2, this is even harder to justify, as we would have a "dangling" membrane, and it seems that an observer could go around the membrane and cross it from the "wrong side", which gives a picture of the bulk which is completely different. This new problem can be resolved thanks to the ergosphere. Indeed, no observer could make the trip to go behind $E_1$ from the pink
side, because the dragging forces would prevent him. There would thus be an artificial "end of the world", as no observer could pass behind the membrane. Incidentally, this also worked as an explanation for the correct sign choice of the currents $J_i$, as inverting them spoils this setup. Ultimately, this explanation was discarded, as there was no way to explain the abrupt ending of the membranes without introducing additional matter fields, and while for observers the geometry was consistent, the full geometry of the spacetime was quite strange, as with a spacelike trajectory one could go "around" the membrane. As we explain now, the real resolution is that the event horizon is deformed by the gluing of the spacetimes.

In order to analyze the problem systematically, we define an everywhere-timelike unit vector field that will define the arrow of time in the full spacetime. 
\begin{equation}\label{arrowoftime}
     t^\mu \partial_\mu  = \frac{\partial}{\partial v_j} +  \frac{h_j(r_j) -1}{2\ell_j}\frac{\partial}{\partial r_j}+ 
      \frac{J_j\ell_j}{2r_j^2} \frac{\partial}{\partial y_j}
      \qquad {\rm in \ the  \ {\it j}th\ region\,. } \ .
\end{equation}
Using the metric \eqref{finkelsteinmetric} the reader can check that $t^\mu t_\mu = -1$. To avoid charging the  formulae we  drop temporarily  the  index $j$. A future-directed null curve has a tangent vector
\begin{equation}\label{nullrays}
\dot x^\mu = ( \dot v , \dot r , \dot y )  \qquad {\rm where} \quad \dot x^\mu\dot x_\mu = 0 \ \ \ {\rm and} \ \ \ 
\dot x^\mu t_\mu   < 0 \ .
\end{equation}
The  dots denote  derivatives  with respect to a parameter on the curve. Solving the conditions \eqref{nullrays} gives 
\begin{equation}\label{nullray}
     \dot r = \frac{h}{2\ell}  \dot v   - \frac{r^2}{2\ell \dot v}  \bigl(\dot y -  \frac{J\ell}{2r^2} \dot v \bigr)^2\  \quad  {\rm and} \ \ \ \dot v   > 0  \ .
\end{equation}
We see that the arrow of time is defined by increasing $v$, and that behind the horizon, where $h(r)$ is negative, $r$ is   monotone decreasing with time, meaning a causal curve necessarily advances deeper into the black hole. This suffices to show that ${\cal H}_2$ {\it is} part of the event horizon  -- an observer crossing it  will never make it out to the boundary again, and will plummet to the Cauchy Horizon. Crossing the membrane does not help, as the observer finds himself behind ${\cal H}_1$, where he continues to plummet. 

As already shown, the story differs in region 1, as an observer may follow a curve similar to the one depicted in (\ref{fig:dischornew}). Here, the event horizon consists of a lightlike surface $\widetilde{\cal H}_1$  such that  no future-directed causal curve starting from a point behind it can reach the [{\small E}$_1$,{\small E}$_2$] part of the wall. Indeed, once a causal curve enters behind $\sigma^{\rm H1}_+$, it necessarily must sink in further because of (\ref{nullray}). Its only chance of escape is reaching the wall at [{\small E}$_1$,{\small E}$_2$], and thus if this is not possible, it is indeed located behind the event horizon.
  
\begin{figure}[!h]
 \centering  
\includegraphics[width=0.97\linewidth]{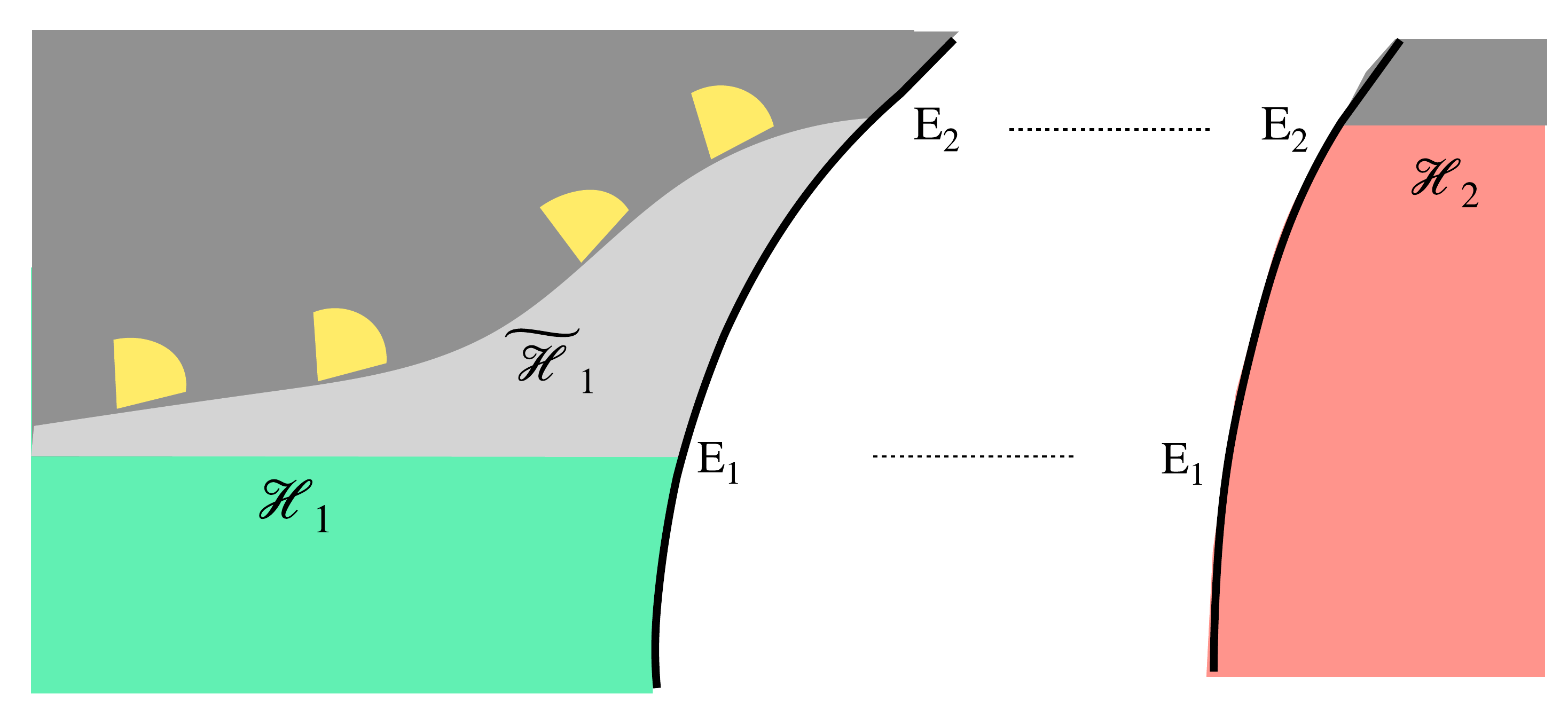}
     \caption{\small The event and apparent horizons, $\widetilde {\cal H}_1  \cup {\cal H}_2$ and $ {\cal H}_1  \cup {\cal H}_2$, as described in the text. The event horizon is connected but it is not Killing. Projections of the  local  light-cone on a Cauchy slice  are shown in yellow. The light grey region behind  ${\cal H}_1$ is outside the event horizon because signals can escape towards the right.}
\label{fig:eventvapparent}
 \end{figure}

Clearly, the full event horizon 
\begin{equation}\label{eventH}
  {\cal H}_{\rm event} =  \widetilde{\cal H}_1\, \cup\, {\cal H}_2
\end{equation}
must be  continuous and lie behind the apparent horizon ${\cal H}_1$ in region 1. This is illustrated in fig.\ref{fig:eventvapparent}. General theorems \cite{hawking_ellis_1973} actually show that a local horizon which is part of a  trapped compact surface cannot lie outside  the event horizon. But there is no clash with these theorems here because ${\cal H}_1$ fails to be  compact, both  at infinity and  at  {\small E}$_1$. 

To compute the projection of  $\widetilde{\cal H}_1$ on a  Cauchy slice, note first that it is a curve that passes through the point {\small E}$_2$. When a causal curve enters the apparent horizon $\si^{\rm H1}_+$, it will inexorably sink deeper into it. Its best chance at escaping the singularity it thus to maximize its speed toward the wall. The constraints on the tangent vector will be less stringent when the curve is lightlike. 

Time-reversing this thought process, it is easy to convince oneself that the projection of the horizon will be the curve that is everywhere tangent to the projection of the local light cone,  as shown  in fig.\ref{fig:eventvapparent}. Put differently, at every point on the curve we must minimize the angle between (the projection of) light-like vectors and  the positive-$y_1$ axis. This will guarantee that an observer starting behind $\widetilde{\cal H}_1$ will not  be able to move fast enough towards the right in order to hit the wall before the point {\small E}$_2$. 

Parametrising  the curve by $y_1$,  using \eqref{nullray}  and dropping again for simplicity the $j=1$ index we find 
\begin{eqgroup}\label{nullra}
  - \frac{dy}{dr}\,\bigg\vert_{\widetilde {\cal H}_1} \ = \   
  {\rm max}_{\, v_y >0}\ 
 \bigg[  \frac{r^2}{2\ell v_y}  \bigl(1  -  \frac{J\ell}{2r^2} {v_y }  \bigr)^2 - \frac{h}{2\ell}  {v_y}   \bigg]^{-1}    \ ,
\end{eqgroup}
where $v_y  \equiv dv/dy$. The extrema of this expression  are $v_y   = \pm r/\sqrt{M\ell^2 - r^2}$. Recall that we are interested in the region behind the BTZ  horizon  and in  future-directed light rays for which $v$ is monotone increasing  (whereas $r$ is monotone decreasing). For null rays moving to the right we should thus pick the positive $v_y$ extremum.
Inserting in \eqref{nullra} gives the differential equation obeyed by $\widetilde {\cal H}_1$, 

\begin{eqgroup}\label{nullr}
  \frac{dy}{dr}\bigg\vert_{\widetilde {\cal H}1 }  =   
  \frac{ 2 \ell}{ J\ell -  2 r  \sqrt{  M\ell^2 -r^2 } }\ .
\end{eqgroup}
The (projected) event horizon in region $j=1$ is the integral of \eqref{nullr} with  the  constant of integration fixed so that  the curve passes through   {\small E}$_2$. 
   
Here now comes the important point. The reader can check that near the BTZ  horizon, $r =  r_+^{\rm H1}(1+ \epsilon)$ with $\epsilon\ll 1$, the denominator in \eqref{nullr} vanishes like $\epsilon$. This is a non-integrable singularity, so $y(r)$ diverges at $r_+^{\rm H1}$ and hence ${\widetilde {\cal H}1}$ approaches   asymptotically   ${\cal H}1$ as  announced in section \ref{sec:entropyproduction}. Presumably, the holographic entropy will therefore asymptote to that of the equilibrium  BTZ horizon, given by \eqref{entropydensityNEsseasy}. This suggests that the chiral outgoing fluid is thermal, not only in the cold region 2 but also in the hotter region 1, at least far from the interface. But since the state of the outgoing fluid will be the same on the full slice, we conclude that it is thermal everywhere. Many questions remain about this deformed event horizon, particularly whether it is indeed related to the entropy of the black hole. Near the interface, the HRT surface measuring the entanglement entropy will certainly be deformed, but it is unclear if it will enter the apparent horizon or not. These questions will be explored in the next chapter.


\subsection{Remark on  flowing  funnels}
\label{sec:funnels}
\begin{figure}[!h]
 \centering  
\includegraphics[width=0.97\linewidth]{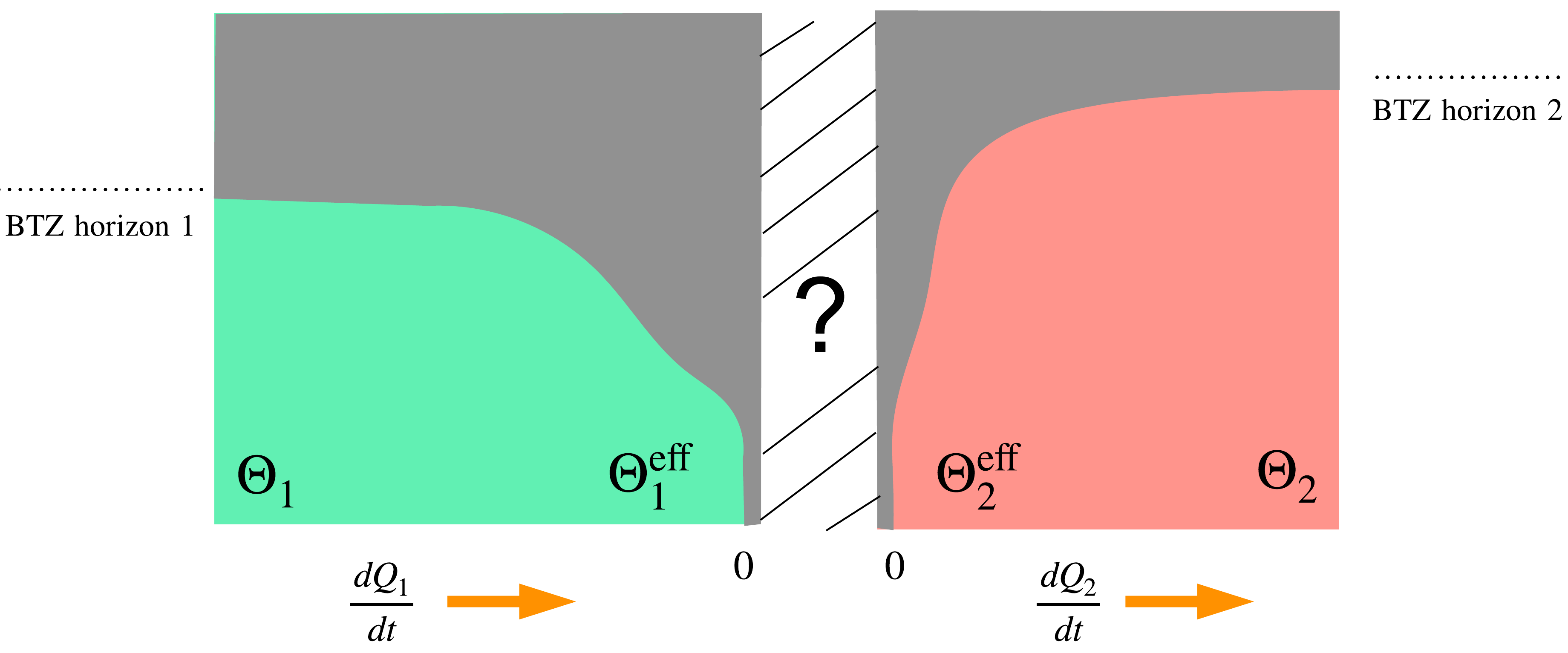}
     \caption{\small A two-sided flowing   funnel that can mimic  the  energy and entropy flows  of  the holographic interface. Tuning the horizon temperatures  so that  the  boundary black hole does not absorb any energy is, however, an adhoc condition.  }
\label{fig:bfun}
 \end{figure}

The fact that  outgoing fluxes are thermalized means that, in what concerns the  entropy and energy flows, the interface behaves like  a  black cavity. This latter can be modeled by a non-dynamical, two-sided boundary black hole whose (disconnected)  horizon consists of two points. To mimic the behaviour of the interface, the two horizon temperatures should be equal to the  $\Theta_j^{\rm eff}$ that saturate the bounds \eqref{boundsS}. This is illustrated in  figure \ref{fig:bfun}.                     
The precise shape  of the flowing horizon(s)  depends on the boundary  black hole(s) and is not important for our purposes here.  For completeness,  following ref.\cite{Fischetti:2012ps}, we outline how to derive it 
  in appendix \ref{app:flowingfunnel}.  Like the thin-brane horizon of figure \ref{fig:eventvapparent}, it approaches the BTZ horizons at infinity but differs in the central region (notably with a delta-function peak in the entropy density at $x= 0$, see app.\ref{app:flowingfunnel}). 
  
The key difference is however elsewhere. The two halves of the flowing funnel of figure \ref{fig:bfun} are a priori separate solutions, with the temperatures $\Theta_j$ and $\Theta_j^{\rm eff}$ chosen at will. To mimic the conformal interface one  must impose  continuity of the heat flow, 
\begin{equation}\label{heatfun}
\frac{dQ_1}{dt}  = \frac{\pi c_1}{ 12} (\Theta_1^2 -  (\Theta_1^{\rm eff})^2)  =  \frac{\pi c_2}{ 12} ( (\Theta_2^{\rm eff})^2 - \Theta_2^2) =  
\frac{dQ_2}{ dt}\ .
\end{equation}
This relates the  horizon  temperatures to each other and to those of the distant heat baths. It is however unclear whether any  local condition behind the event horizons can impose the  condition \eqref{heatfun}. 
  
\section{Pair of interfaces}\label{sec:pairofinterfaces}
In this last section we  consider  a pair of identical interfaces between two theories, CFT$_1$ and CFT$_2$.\footnote{Our branes are not oriented, so there is no difference between an interface and anti-interface. More general setups could include several different  CFTs and triple junctions of branes, but such systems are beyond the scope of the present work.}  

The interface  separation is  $\Delta x$. Let the theory that lives  in the finite interval be  CFT$_2$ and the theory outside  be CFT$_1$ (recall that we are assuming $\ell_2 \geq \ell_1$). At thermal equilibrium the system undergoes a first-order  phase  transition  at a critical temperature $\Theta_{\rm cr} = b/\Delta x $ where $b$ depends on the classical Lagrangian parameters $\lambda \ell_j$   \cite{Bachas:2021fqo, Simidzija:2020ukv}. 

Below $\Theta_{\rm cr}$ the brane avoids  the horizon and is  connected, while above $\Theta_{\rm cr}$ it breaks into two disjoint pieces that hit separately the singularity of the black hole. This   is  a variant of the  Hawking-Page phase transition  \cite{Hawking:1982dh} that  can  be interpreted  \cite{Witten:1998zw} as a deconfinement transition of  CFT$_2$.   

We would like to understand what happens when this  system is  coupled  to  reservoirs with  slightly different  temperatures $\Theta_\pm =\Theta \pm d\Theta$ at $x=\pm \infty$.  Because of the temperature gradient, the branes are now stationary, but  they conserve the topology  of their static ancestors. In the low-$\Theta$ phase the brane avoids  the ergoregion (which is displaced from the horizon infinitesimally) and stays connected,  while in the high-$\Theta$ phase it splits in two disjoint   branes that  enter the ergoregion and hit  separately a Cauchy horizon or a bulk singularity. 

\begin{figure}[!h]
 \centering  
\includegraphics[width=0.97\linewidth]{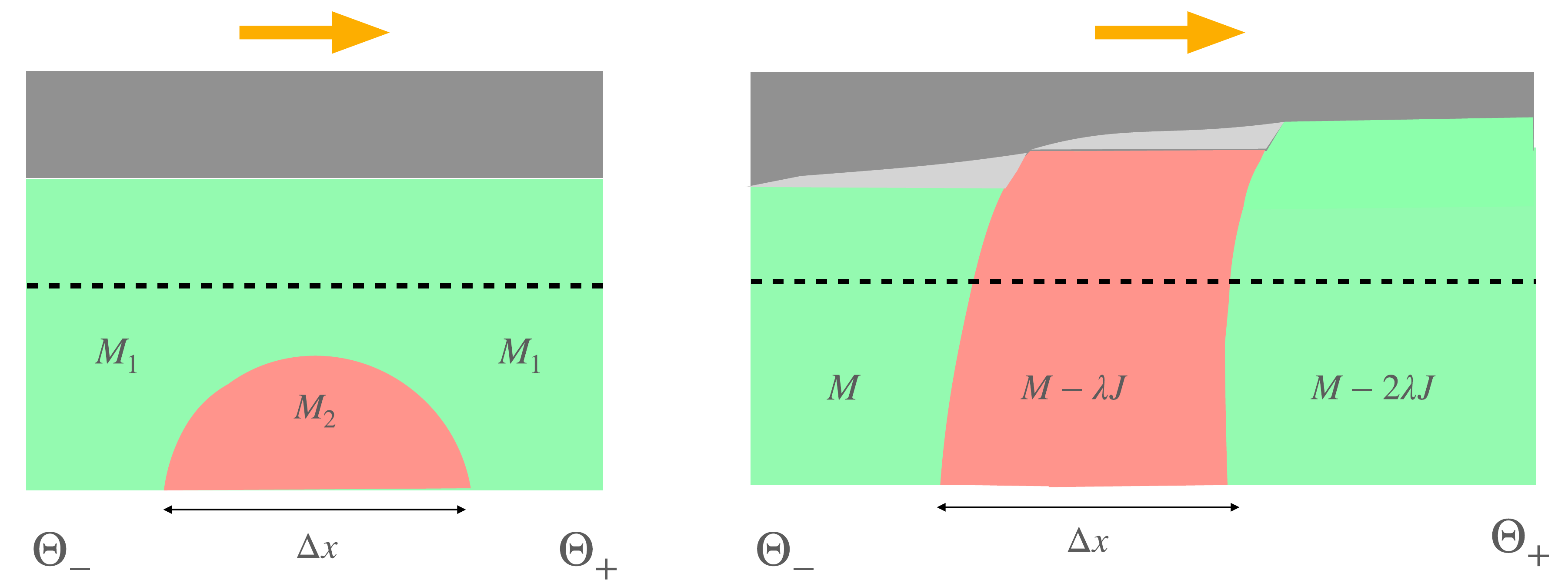}
     \caption{\small The  two types of NESS  for an interface pair. In the 'quantum phase' (left)heat conducts as if there was no scatterer, while in the 'classical phase' (right)   the  conductance is the same as for  an isolated CFT$_1$ defect dual to a  brane of  tension $2\lambda$. Also given  in the figure is the  BTZ mass parameter in  different  regions of the geometry. The yellow arrows show the direction of heat flow. }
\label{fig:pair}
 \end{figure}

The two phases  are  illustrated in figure \ref{fig:pair}. 

Consider the  high-$\Theta$ phase first. The    isolated-brane solution  of  sections \ref{sec:insideergoregion} and \ref{sec:nonkillinghorizon}  is  here juxtaposed to a solution  in which the roles of CFT$_1$ and CFT$_2$ are inverted.  The mass parameter  of the three BTZ regions decreases  in the direction of heat  flow, jumping by $ \lambda  J$ across each brane.
This is indeed the `ticket of entry'  to the ergoregion, as explained in  eq.\,\eqref{sigma+0}  and  section \ref{sec:arrow}.  

The total change of BTZ mass across the pair is the same as if the two branes had merged into a single one with twice  the tension.  Using  the holographic dictionary \eqref{CFTNESSstate}  and the fact that incoming fluxes at $x=\pm\infty$ are thermal with  temperatures  $\Theta_\pm$  one indeed  computes
  \begin{equation}\label{prob1}
  \underline{ {\rm high}\ \Theta} :  \qquad  \frac{dQ}{ dt}    = \frac{\pi^2 \ell_1}{1+ \lambda \ell_1} ( \Theta^2_- - \Theta^2_+) 
    \equiv  {\pi^2 \ell_1  }{\cal T}_{\rm pair}  ( \Theta^2_- - \Theta^2_+) \ ,
  \end{equation}
where the effective  transmission coefficient ${\cal T}_{\rm pair} $ is  that of  a  CFT$_1$ defect whose dual  brane has  tension  $2\lambda$. Note in passing that this  effective  brane tension can exceed the upper bound \eqref{criticaltensions} above which an individual brane inflates, and that an array of widely-spaced  branes can make the transmission coefficient arbitrarily small. 
   
The heat flow \eqref{prob1} is what one would obtain from classical scatterers.\footnote{The argument grew out  of  a conversation  with   Giuseppe Policastro who noticed that the  tensions of two juxtaposed branes effectively add up in  the calculation of ref.\cite{Bachas:2020yxv}} 

To understand why, think of   ${\cal T}_j$ and  ${\cal R}_j$ as classical transmission and reflection probabilities  for quasi-particles incident on the  interface from the side $j$. The probability of passing  through both interfaces is the sum of probabilities  of trajectories with any number of double reflections in between, 
\begin{equation}\label{prob} 
  {\cal T}_{\rm pair}=  {\cal T}_1 ( 1 +    {\cal R}_2^2 +  {\cal R}_2^4  + \cdots  )  {\cal T}_2 
  =  \frac{ {\cal T}_1 {\cal T}_2}{1 -   {\cal R}_2^2 } =  \frac{1}{1+ \ell_1 \lambda} \ ,
\end{equation}
where in the last step we used   the holographic relations \eqref{transmissioncoeffintermsoflambda}.  This gives precisely the result \eqref{prob1} as advertised. 
          
The low-$\Theta$ case  is drastically different.  The solution   is now obtained   by  gluing   a brane with  a turning point (i.e. $\sigma_+>0$, see  section \ref{sec:equationsolutions})  to its mirror image,  so that the brane has reflection symmetry. The bulk metric,  however, is not $\mathbb{Z}_2$ symmetric because in the mirror image we do not flip the  sign of   the BTZ `spin'   $J$.  This is required for  the continuity of the $dx/dt$ component of the bulk metric. While it may seem we are gluing brane with its time-reversed counterpart this is not the case here. The arguments that lead to this conclusion in sec.\ref{sec:eventversusapparent} relied on the regularity at the horizon, which is absent here. Similarly, the argument from the transmission coefficients in the field theory are not applicable here, since there might be interference effects on the energy fluxes in the middle of the two membranes. The  BTZ mass is  thus  the same at $x= \pm\infty$,  while its value in the CFT$_2$ region depends on the interface separation $\Delta x$. It follows from the holographic dictionary \eqref{CFTNESSstate} that the heat flow is  in this case  unobstructed,
 \begin{equation}\label{quan}
  \underline{ {\rm low} \Theta} : \qquad 
    \frac{dQ}{dt}    =  {\pi^2 \ell_1  } ( \Theta^2_- - \Theta^2_+)\ ,
\end{equation}
i.e. the effective transmission coefficient is ${\cal T}_{\rm pair} =1$. Superficially,  it looks  as if  two branes with equal and opposite tensions  have merged into a  tensionless one. 

In reality, however, the above phenomenon is deeply quantum. What the calculation says is that when a  characteristic thermal wavelength  becomes larger than  the interface separation, coherent scattering  results in   all incident energy being   transmitted. This is all the more  surprising since CFT$_2$ is in the confined phase, and one could  have expected that   fewer degrees of freedom  are available to conduct heat. The microscopic mechanism behind this surprising phenomenon deserves to be  studied further.

The above discussion stays valid for finite temperature difference $\Theta_+-\Theta_-$,   but the dominant phase cannot in this case be found  by comparing free energies. Nevertheless, as $\Delta x\to 0$ we expect from the dual ICFT that the interface-antiinterface pair fuses into the trivial (identity) defect \footnote{In some range of parameters ($\lam\geq\lam_0$, $\ell_2<3\ell_1$), the pair fuses into a non-trivial defect\cite{Bachas:2021fqo}, although this configuration is at best metastable.} \cite{Bachas:2007td}, whereas at very large $\Delta x$ the connected solution ceases
to exist. A transition is therefore bound to occur   between these  extreme separations. 

Let us comment finally on what happens if the interval theory is CFT$_1$, the theory with fewer degrees of freedom, and the outside theory is CFT$_2$.  Here the low-temperature phase only exists for sufficiently-large   tension if  $c_1< c_2<3 c_1$, and does not exist  if   $c_2> 3 c_1$ \cite{Bachas:2021fqo, Simidzija:2020ukv}.  The (sparse) degrees of freedom of the  interval  theory in this latter case are always in the high-temperature phase, and there can be no quantum-coherent conduction of heat. Reassuringly, this  includes the  limit $c_2/c_1 \to 0$ in which  the CFT$_1$ interval is effectively void. 

Note also that in the low-temperature phase the wire can be compactified to a circle and the heat current can be sustained without external reservoirs. This is not possible in the high-temperature phase. 

\section{Closing  remarks}  
The study of far-from-equilibrium  quantum systems is an exciting frontier both in condensed-matter physics and in quantum gravity. Holography is a   bridge between these two areas of research and  has led to  many new insights. Much remains however to be understood, and simple tractable models can help as testing grounds for new ideas. The holographic NESS  of this paper are   tractable thanks to several simplifying factors: 2d  conformal symmetry,  isolated impurities and the assumption of a thin brane.  If the first two can be justified in (very) pure ballistic systems, the thin-brane approximation is an ad-hoc assumption of convenience. Extending these results to  top-down  dual pairs is an important step to validate them.
  
Another obvious question concerns  the structure of entanglement and the  Hubeny-Rangamani-Ryu-Takayanagi  curves \cite{Ryu:2006bv,Hubeny:2007xt} in the above  steady states. While  it is known that geodesics  cannot probe  the region behind equilibrium  horizons \cite{Hubeny:2012ry}, they can reach behind both apparent and event horizons in  time-dependent backgrounds, see e.g.  \cite{AbajoArrastia:2010yt,Balasubramanian:2010ce,Hartman:2013qma}.  In the framework of the  fluid/gravity correspondence the   entropy current  associated with the event horizon is a local functional of the boundary data \cite{Bhattacharyya:2008xc}. It would be interesting to examine this question  in the present far-from-equilibrium context. In the next chapter, we provide some partial results in this direction.

Another interesting question  is  how  the deconfinement transition of the interval  CFT in  section \ref{sec:pairofinterfaces} relates to  the sudden jump in thermal conductivity of the system. Last but not least,  it would be nice to relate the production of entropy to the scattering matrix of microscopic interfaces, e.g.  for  the simplest free-field interfaces of \cite{Bachas:2001vj,Bachas:2007td,Bachas:2012bj}. Presumably, this is also a computation that could be completed holographically, with a slightly more complicated model.

\chapter{Entanglement entropy and holographic interfaces}
\label{chap:entanglemententropyandholoint}
\epigraph{Based on unpublished work }{}
This final chapter is dedicated to the study of the entanglement entropy structure in the models that were constructed in the previous two chapters. There are several motivations for why this is an interesting quantity to study.

First and foremost, it is in direct connection with the doubly holographic black hole evaporation models quickly described in sec.\ref{sec:IslandsBHparadox}. Thus if one wants to use this holographic model to test the Island formula, it is necessary to be able to compute RT surfaces. Indeed, this is what was studied in \cite{Anous:2022wqh}.

On a related note, the study of RT surfaces in such spacetimes is interesting in and of themselves because there generally are several competing RT surfaces. As we change the boundary interval, there will be discontinuous jumps in the RT surfaces, a sort of phase transition in the entanglement structure. In the context of the doubly-holographic models and the page curve, this is related to the Page time as is explained in sec.\ref{sec:IslandsBHparadox}.

From the point of view of the ICFT, such transitions might be related to some other property of the underlying state. For instance, one of our conjectures was that the presence (or not) of a center in one side of the ICFT dual (see \ref{sec:sweepingtransition}) might be related to the crossing (or not) of the RT surface computing the Von Neumann entropy of associated CFT. This turned out to be incorrect, but the connections of RT surfaces with bulk reconstruction \cite{Dong:2016eik} suggests there should be a way to relate the sweeping transition with the entanglement structure of the ICFT.

Another interesting direction was alluded in chap.\ref{chap:steadystatesofholo}, and it is the computation of RT surfaces in non-equilibrium solutions with deformed horizons. Indeed, it is still an open question how is the Bekenstein-Hawking formula modified whenever we consider out-of-equilibrium horizons, which are non-Killing. Is it the event horizon or the apparent horizon that must appear in the formula \cite{Engelhardt:2017aux}? In equilibrium situations, those two coincide, so the distinction cannot be made. Additionally, this may also allow us to understand and confirm the entropy production at the interface of the NESS state described in the previous chapter.

Finally, entanglement entropy computations allow one to compute the \textbf{Q}uantum \textbf{N}ull \textbf{E}nergy \textbf{C}ondition (QNEC) \cite{Bousso:2015mna}. As we will explain in the main text, the QNEC is an inequality (which should hold in any QFT) relating derivatives of the entanglement entropy and the one-point function of the Stress-Energy tensor. As such, it is a powerful tool to determine if simple holographic bottom-up models are at least consistent. For instance, the QNEC has been used to restrict quenches of CFTs \cite{PhysRevmukopad,Mezei:2019sla,Banerjee:2022dgv}, by using the RT prescription in the dual bulk. Such a quench geometry is obtained in much the same way as the models of chap.\ref{chap:phasesofinterfaces}, the difference being the membrane is now spacelike, instead of timelike. Whenever the QNEC is not obeyed in such a quench, it signals that it is probably unphysical. In the same vein, it would be interesting to compute the QNEC in the various geometries that were obtained in the previous chapters to check for consistency.

Since this chapter is composed of partial, incomplete and sometimes inconclusive results, it will necessarily seem much more scattered and less polished than the previous ones. While there isn't a clear-cut goal or thread, but rather several different directions, we believe that the work we present here is valuable, and hope that it will be completed and "packaged" in the more standard form of a paper sometime.

Throughout this chapter, we use units in which $8\pi G =1$.

\section{Geodesics in asymptotically Anti-de-Sitter spaces}
\label{sec:geodesicinasympads}
The main object of study in this chapter are RT and HRT surfaces in 3-dimensional asymptotically AdS spacetimes. In fact, we can restrict to the study of spacetimes which are locally AdS, since the only matter content of our models is the membrane, and everywhere else the vacuum equations of motion hold. In three dimensions, finding RT curves amounts to computing spatial geodesics anchored at the asymptotic boundary.

In this section, we compute and outline some properties of such geodesics, mainly in Poincaré space. Indeed, we solved the equations in one coordinate system, by exploiting diffeomorphisms between the different geometries, we can obtain the geodesics in any locally AdS spacetime.

Let us begin by recalling the Poincaré metric :
\begin{eqgroup}\label{poincaremetriclight}
ds^2 = \frac{\ell^2}{z^2}(dw_+ dw_- +dz^2)\ ,
\end{eqgroup}
where $w_-=x-t$, $w_+=x+t$ are the lightcone coordinates. Consider the most generic spacelike geodesic, of initial point $p_0=(w_+^0,w_+^0,z_0)$ and tangent vector $\dot{p}_0 = (\dot{w}_{+}^0,\dot{w}_{-}^{0},\dot{z}_0)$. We will consider WLOG affinely parametrized geodesics, namely :
\begin{eqgroup}\label{affinecondition}
 \dot{p}^\mu\dot{p}_{\mu}= \frac{\ell^2}{z^2}\left(\dot{w}_+ \dot{w}_- +\dot{z}^2\right)=1\ ,
\end{eqgroup}
where this condition will hold not only at the initial point, but everywhere along the geodesic, so we omit the "0" indices.

Denoting the affine parameter $\lam$, the action that we need to minimize is simply :
\begin{eqgroup}\label{actionofone}
L= \int \frac{\ell^2}{z^2}(\dot{w}_+\dot{w}_-+\dot{z}^2)d\lam\ .
\end{eqgroup}
Using cyclicity of $w_+$, $w_-$ we get two integration constants :
\begin{equation}\label{Kpmplusminus}
    K_{\pm} =\frac{\dot{w}_\pm \ell }{z^2}\ .
\end{equation}
Translations of the affine parameter give us the conservation of the tangent vector norm (\ref{affinecondition}) :
\begin{eqgroup}\label{conservationnormequation}
K_-K_+z^2+\frac{\ell ^2 \dot{z}^2}{z^2}=1\ .
\end{eqgroup}
Note that other symmetries like scale invariance ($z\rightarrow \omega z$, $w_{\pm}\rightarrow \omega w_{\pm}$) or boosts ($w_+\rightarrow \g w_+, w_-\rightarrow \g^{-1}w_-$) do not provide additional integration constants.

We can solve (\ref{conservationnormequation}) for $z(\lam)$. There are three cases which we must differentiate, $K_- K_+>0$, $K_+K_-<0$ and $K_+K_-=0$. These signs are simply determined by the initial vectors, since ${\rm sign}(K_- K_+)={\rm sign}(\dot{w}_-\dot{w}_+)$. Thus, in Poincaré coordinates, we will have three types of qualitatively different geodesics.
\subsection{Case $K_-K_+>0$}
\label{sec:casekpkmmorezero}
We begin by the $K_-K_+>0$ case. Upon integrating (\ref{conservationnormequation}) :
\begin{eqgroup}
\pm(\frac{\lam}{\ell}+C)={\rm Arctanh}\left(\sqrt{1-K_+K_- z^2}\right)\Leftrightarrow z(\lam)=\sqrt{\frac{1-\tanh^2(\lam/\ell+C)}{K_+K_-}}\label{zsolpositive}\ .
\end{eqgroup}
The $\pm$ signs denotes two branches of the solution which join at $z=\frac{1}{K_+K_-}$, for $\lam = -C\ell$. The $C$ integration constant can be absorbed into a translation of the affine parameter, we set $C=0$ so that the "turning point" of the geodesic is always at $\lam=0$. Then in (\ref{zsolpositive}) the plus sign must be chosen for $\lam>0$ and minus sign for $\lam<0$. For this choice, the initial $\lam_0$ is given as :

\begin{equation}\label{lam0positive}
    \lam_0=-\ell\;{\rm sign}(\dot{z}_0) {\rm Arctanh}\left(\sqrt{1-\frac{\ell^2}{z_0^2}\dot{w}_+^0\dot{w}_-^0}\right)\ .
\end{equation}
Plugging (\ref{zsolpositive}) into (\ref{Kpmplusminus}) and performing the integration yields :
\begin{equation}
    w_{\pm} = \frac{1}{K_\mp}\tanh(\lam/\ell)+G_\pm\ ,
\end{equation}
where $G_{\pm}$ are two integration constants. They are easily fixed using the initial conditions, and the full solution is given by :
\begin{equation}
    \begin{aligned}
            z(\lam)&=\frac{z_0^2}{\ell}\sqrt{\frac{1-\tanh^2(\lam/\ell)}{\dot{w}_+^0\dot{w}_-^0}} \Leftrightarrow \lam = \pm \ell \sqrt{1-\frac{z(\lam)^2}{z_0^2}\frac{\ell^2}{z_0^2} \dot{w}_+^0\dot{w}_-^0}\ ,\\
            w_{\pm}(\lam)&=\frac{\tanh(\lam/\ell)-\tanh(\lam_0/\ell)}{\ell \dot{w}_{\mp}^0}z_0^2+w_{\pm}^0\ ,\\
            \lam_0 &= -\ell\,{\rm sign}(\dot{z}_0) {\rm Arctanh}\left(\sqrt{1-\frac{\ell^2}{z_0^2}\dot{w}_+^0\dot{w}_-^0}\right)\ .
    \end{aligned}
    \label{eq:geodesicgeneralpositive} 
\end{equation}

These are the geodesics one usually thinks of in AdS space. At $\lam=\pm \infty$, they are anchored on the boundary of AdS. In the case where $\dot{w}^0_{+} =\dot{w}^0_{-}$, they are semi-circles centered on the asymptotic boundary. By boosting them one can obtain the general shape of the geodesic.

Note that while it would be nice to parametrize the geodesic by its initial point on the asymptotic boundary, this can't be done as they lie formally at infinity. Indeed, as we send $z_0=\eps\rightarrow 0$, to satisfy the affine condition we need $\dot{w}^0_{\pm}\propto \eps^2$ and $\dot{z}_0\propto \eps$, so on the boundary the tangent vector is degenerate.

\subsection{Case $K_-K_+<0$}
\label{sec:casekpkmlesszero}
We treat the case $K_-K_+<0$. The resolution of the equations is essentially the same, modulo the change ${\rm tanh}\rightarrow {\rm coth}$. Skipping the details, we have :
\begin{eqgroup}
    z(\lam)&=\frac{z_0^2}{\ell}\sqrt{\frac{1-\coth^2(\lam/\ell)}{\dot{w}^0_{+}\dot{w}^0_{-}}}\ ,\\
    w_{\pm}(\lam)&=\frac{\coth(\lam/\ell)-\coth(\lam_0/\ell)}{\ell \dot{w}^0_{\mp}}z_0^2+w^0_{\pm}\ ,\\
    \lam_0 &= -\ell\,{\rm sign}(\dot{z}_0) {\rm Arccoth}\left(\sqrt{1-\frac{\ell^2}{z_0^2}\dot{w}^0_{+}\dot{w}^0_{-}}\right)
    \label{eq:geodesicgeneralnegative} \ .
\end{eqgroup}

\begin{figure}[!h]
    \centering
    \includegraphics[width=0.4\linewidth]{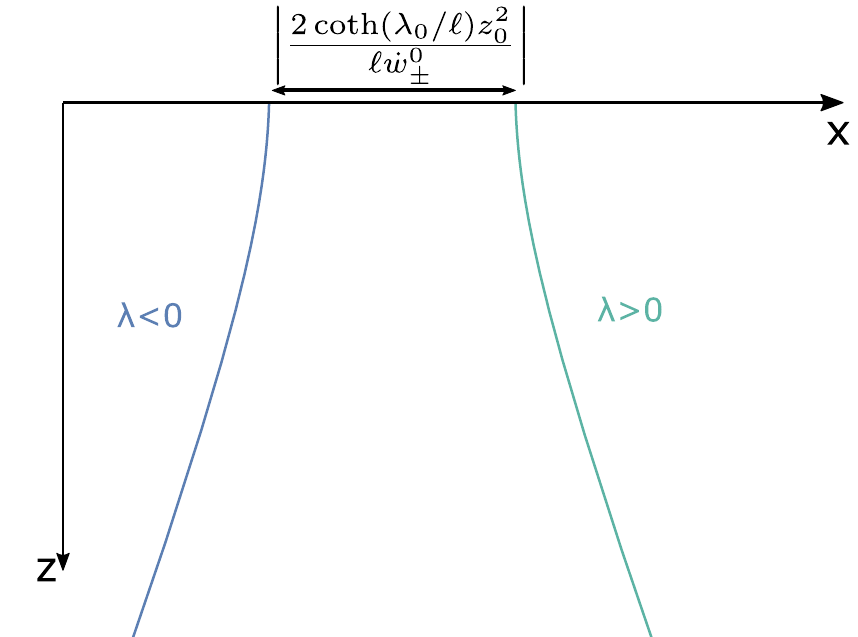}
    \caption{{\small Plot of the two disconnected branches of the geodesic of equation (\ref{eq:geodesicgeneralnegative}). Both branches escape out to the Poincaré horizon. If we maximally extend, we would find that they end up hitting the boundary again, as all spacelike geodesics have two anchor points on the boundary in global coordinates.}}
    \label{fig:strangegeodesic}
\end{figure}

Note that now, the point $\lam=0$ is not really a "turning point" anymore since as we approach it all the coordinates diverge as $\frac{1}{\lam}$. For the two signs of $\lam$, we get two disconnected solutions, with one end in the Poincaré horizon $z=\infty$ and the other on the boundary $z=0$. Naturally, the change $\lam\rightarrow -\lam$ is equivalent to a reversal of the initial tangent vector $\dot{w}_\pm^0\rightarrow -\dot{w}_\pm^0$,  along with a translation of the initial point of $-\frac{2\coth(\lam_0/\ell)z_0^2}{\ell \dot{w}_\mp^0}$. See fig.\ref{fig:strangegeodesic} for a sketch of the geodesic projected to the $t=cst$ plane (note that there are no geodesics of this type that stay in this plane, because $\dot{t}_0\neq0$ necessarily).
These geodesic seem very strange, but this is due to the coordinate system we chose. Indeed, the Poincaré horizon is a coordinate singularity, and if we were to map these geodesics to global coordinates, we would see that they are a portion of a geodesic that is doubly anchored at the boundary. Nonetheless, for the purposes of the RT prescription, the coordinate system in which we look for the geodesics will of course matter, since it will depend on the state of the CFT.
\subsection{Case $K_-K_+=0$}
The last case to consider is of "measure zero" in the set of initial conditions, the case $K_- K_+=0$. It includes in particular geodesics that are shot "straight" toward the bulk interior.

Solving (\ref{conservationnormequation}) in this case gives us 
\begin{equation}
    z=e^{{\rm sign}{\dot{z}_0} \lam/\ell}\ ,
\end{equation}
where we absorbed the integration constant into a translation of $\lam$.  The equation (\ref{Kpmplusminus}) is again readily solved, regrouping all the equations we obtain :
\begin{eqgroup}
    z &= {\rm exp}\left({\rm sign}(\dot{z}_0)\frac{\lam}{\ell}\right)\ ,\\
    w_{\pm} &= \frac{\dot{w}^0_{\pm}\ell}{2z_0^2}(e^{2{\rm sign}(\dot{z}_0)\lam/\ell}-e^{2\lam_0/\ell})+w^0_{\pm}\ ,\\
    \lam_0 &= \ell \; {\rm sign}(\dot{z}_0)\ln(z_0)\ .
    \label{eq:geodesicgeneralnull} 
\end{eqgroup}
Of course, in (\ref{eq:geodesicgeneralnull}), one of the $\dot{w}^0_\pm$ must be equal to zero if the equations are to be satisfied.

\section{Simple examples}
\label{sec:simpleexamples}
Let us apply the formulas we have derived above to some standard examples. Consider an interval of length $a$ on the boundary of Poincaré space. The dual CFT state is simply the vacuum on $\mathbb{R}^2$, so computing the entanglement entropy of the interval should reproduce the classic result by Calabrese and Cardy \cite{Calabrese_2009}.

Consider the initial point to be $(t=0, x=0, z=\eps)$ on the boundary, where we introduced the IR cutoff $\eps$. Since the geodesics exhibited in the previous section are parametrized in terms of the initial tangent vector $\dot{p}^0$ and not the final point, we must invert some relations to find the correct values for which the geodesic ends on $(t=0,x=a,z=0)$\footnote{In poincaré space, it is possible to invert the relation generally and to give the geodesics as parametrized by initial and final points. We don't find it useful to reproduce it here, as it is more cumbersome than the parametrization we exhibited and, while it can be convenient, we won't really need it for our purposes.}.

We already know $\dot{w}_+\dot{w}_->0$ in order to have a geodesic connecting two points on the boundary. The endpoint is located at $z=\eps$, at $\lam=-\lam_0$.  From the equation of $w_{\pm}(-\lam_0)$ (\ref{eq:geodesicgeneralpositive}), we find :
\begin{eqgroup}
 \dot{w}^0_+=\dot{w}^0_-\equiv \dot{w}_0=\frac{\eps}{\ell\sqrt{\frac{a^2}{4\eps^2}+1}}\label{solutionwdot}\ .
\end{eqgroup}
What remains is to compute the length of the geodesic. Thankfully, in the affine parametrization this is simply given by the difference in initial and final affine parameters:
\begin{eqgroup}
L = -2\lam_0 = 2\ell {\rm Arctanh}\left(\frac{\dot{w}_0\ell a}{2\eps^2}\right)=2\ell {\rm Arctanh}\left(\frac{a}{\sqrt{a^2+4\eps^2}}\right)\label{lengthgeodpoincaré}\ .
\end{eqgroup}
To obtain the entanglement entropy, all we have to do is divide by $4G=\frac{1}{2\pi}$ and take the limit $\eps\rightarrow 0$. We obtain at leading order :
\begin{eqgroup}\label{intervalentangentropypoincare}
S=2\pi L = 4\pi \ell \ln\left(\frac{a}{\eps}+\cal{O}(\eps^2)\right)=\frac{c}{3}\ln\left(\frac{a}{\eps}\right)+\cal{O}(\eps^2)\ ,
\end{eqgroup}
which correctly reproduces the expected result\cite{Calabrese_2009}.

Let us illustrate a slightly more complicated example, which will be crucial for the methods of computation we will use in the ICFT setups. We consider now the spinning string geometry, the metric (\ref{metricspinningstring}). In this case, spacetime is only stationary, but not static. The problem thus becomes immediately more complex; instead of studying curves restricted to a 2-dimensional Cauchy slice, we must allow for generic spatial geodesic in the full 3D geometry. Furthermore, one can try to solve the geodesic equations for the metric (\ref{metricspinningstring}), and it is feasible but a much harder problem.

Instead, we exploit the change of coordinate to bring us back to the Poincaré geometry. To find the diffeomorphism relating the two coordinate systems one can for instance equate the two parametrizations of the AdS hyperboloid described in sec.\ref{sec:generalitiesAdS}, and solve the resulting equations. We find ($\si = {\rm sign}(J)$):
\begin{eqgroup}
    w_+ &= \sqrt{\frac{r^2-r_+^2}{r^2-r_-^2}}\exp\left(\frac{(r_+-\si r_-)(x+t)}{\ell}\right)\ ,\\
    w_- &= \sqrt{\frac{r^2-r_+^2}{r^2-r_-^2}}\exp\left(\frac{(r_++\si r_-)(x-t)}{\ell}\right)\ ,\\
    z &= \sqrt{\frac{r_+^2-r_-^2}{r^2-r_-^2}} \exp\left(\frac{ x r_+- \si tr_-}{\ell}\right)\ .
    \label{nesstopoincare}
\end{eqgroup}

Of course, (\ref{nesstopoincare}) is far from unique, as we can apply any isometry of Poincaré and still obtain a valid change of coordinates. Note in passing that this coordinate change covers only a portion of Poincaré space, since $w_+>0\cap w_->0$, and only a portion of the spinning string, as the coordinate change becomes singular at the horizon $r=r_+$\footnote{The outer horizon is mapped to $w_-=0\cup w_+=0$. A more careful analysis would show that this surface is composed in fact of two horizons, the future, and past ones. If one extends to the rest of the Poincaré spacetime, we would obtain the two interior regions, as well as another exterior.}. There are other similar changes of coordinates that map the region $r_-<r<r_+$ and $0<r<r_-$, but we won't be needing them here.

Consider now the boundary region in stationary CFT delimited by two points $(t=0,r=1/\eps,x=0)$ and $(t=0,r=1/\eps,x=a)$. WLOG we take $\si=1$, namely $J>0$. Under the change of coordinates (\ref{nesstopoincare}), these are mapped to 
\begin{eqgroup}\label{pointsinpoincarefromNESS}
p_1=&\left(w_+=1,w_-=1, z= \eps \sqrt{r_+^2-r_-^2}\right)\ ,\\
p_2=&\left(w_+=\exp(\frac{r_+-r_-}{\ell}a),w_-=\exp(\frac{r_++r_-}{\ell}a), z= \eps \sqrt{r_+^2-r_-^2}\exp(\frac{xr_+}{\ell})\right)\ ,
\end{eqgroup}
where we kept only the leading order in $\eps$. We will write $p_i^b$ for the corresponding points on the 2D-boundary, with the $z$-coordinate excised.

Note that we do not care about the specific shape of the boundary interval we select between the two points. Indeed, any spacelike curve between $p_1$ and $p_2$ will have the same entanglement entropy, because they all lie in the same "causal diamond"\footnote{The causal diamond of a portion of a Cauchy slice $A$ is the region that is in causal contact uniquely with $A$. For instance, if $A$ is the interval between two points $x$, $y$, the causal diamond is determined by shooting lightrays from $x$ and $y$, and selecting the enclosed region. Because points in this region can only be affected by $A$, they can be expressed as a unitary transformation applied to $A$. See fig.\ref{fig:cauchysliceatx}. }. Any two of these curve houses states which are related by a unitary transformation, and thus have the same entanglement entropy, see sec.\ref{sec:entanglement}.

We can exploit this fact to simplify the problem even further. Indeed using the points (\ref{pointsinpoincarefromNESS}), we would still need to apply the HRT prescription, as they do not lie on a preferred Cauchy slice. Then, we use the isometries of Poincaré to bring these two points on a constant time slice. In this case, we can begin by a $x$-translation to bring $p_1$ to the origin. Then, we perform a boost to bring $p_2$ on the same $t=0$ Cauchy slice. Since these are isometries of Minkowski on the boundary, we do not need to perform them explicitly, as we now that the norm $|p^b_1-p^b_2|$ will be conserved. We find :
\begin{eqgroup}
|p_2^b-p_1^b|^2 = 4\exp(\frac{r_+ a}{\ell}) {\rm sinh}\left(\frac{r_+-r_-}{2\ell}a\right){\rm sinh}\left(\frac{r_++r_-}{2\ell}a\right)\equiv\ti{a}\ .
\label{intervallengthsquared}
\end{eqgroup}

In this way, we successfully reduced the problem to the same RT computation as previously with an interval of size $\ti{a}$. There is one difference, and that is in the UV cutoffs which are now different at the two endpoints. Recycling (\ref{solutionwdot}) and computing the correct $\lam_f$ given the different cutoffs, we obtain :
\begin{eqgroup}\label{entanglemententropyspinning}
S=2\pi L=2\pi(\lam_f-\lam_0)=\frac{c}{6} \log(\frac{ 4}{\eps^2 (r_+^2-r_-^2)}{\rm sinh}\left(\frac{r_+-r_-}{2\ell}a\right){\rm sinh}\left(\frac{r_++r_-}{2\ell}a\right))\ ,
\end{eqgroup}
which correctly recovers the expected entanglement entropy, see (\ref{entropyeasyNESS}). 

\begin{figure}[!h]
    \centering
    \includegraphics[width=0.6\linewidth]{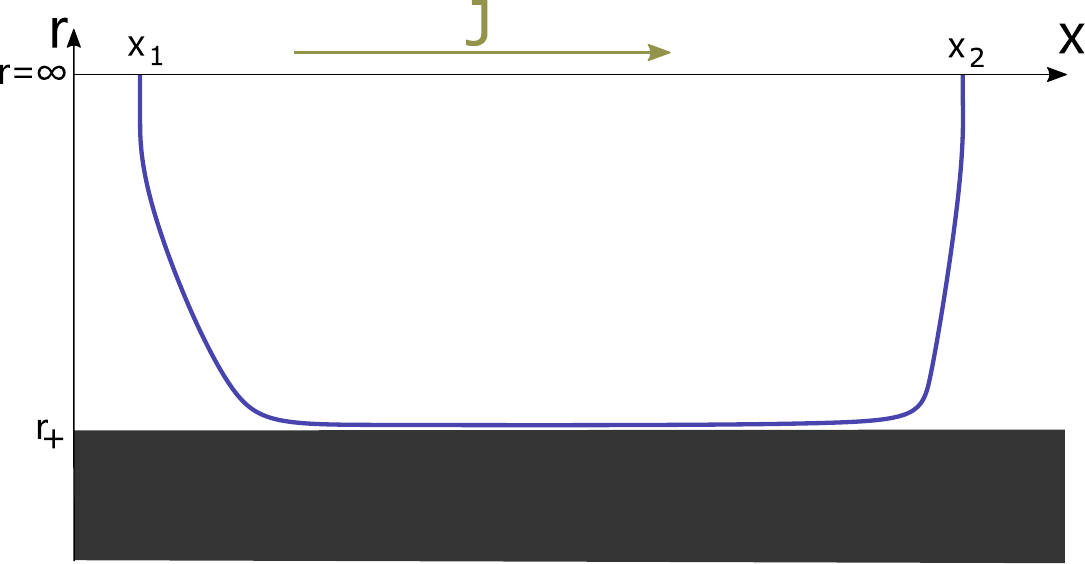}
    \caption{\small Sketch of an HRT surface between two equal time points on the boundary, $x_1$ and $x_2$. The HRT surface is not confined to the constant time Cauchy slice depicted here, rather it is its projection onto this plane. We can see that the HRT surface drops very quickly close to the horizon, where it can reduce its length. Spacelike geodesics are not forbidden from entering the horizons, but there are no spacelike geodesics entering the horizon with two anchor points on the same boundary, see app.\ref{app:geodesicsinstringu}}
    \label{fig:sketchRTNess}
\end{figure}

With this method, inverting (\ref{nesstopoincare}), it is also possible to obtain the equation of the HRT surface in the original coordinate system. In particular, the spacelike geodesic that we obtain does not remain in a constant time plane, as we should expect for a non-static geometry. A sketch of the HRT surface, projected on the $t=0$ plane, can be found on fig. \ref{fig:sketchRTNess}. Notice that the HRT surface cannot breach the horizon, and as $a\rightarrow \infty$, it will approach it more and more, which gives the Bekenstein-Hawking formula for the coarse-grained entropy (see (\ref{entropydensityNEsseasy})).

\section{Entanglement structure of static ICFT states}
In this section, we apply the lessons we have just learned to compute the entanglement structure of static ICFT states on $\mathbb{R}^2$. We will also mention possible applications to the compact geometries of chap.\ref{chap:phasesofinterfaces}.

\subsection{RT surfaces with interfaces}
In this section, we wish to apply the RT prescription to spacetimes containing a gravitating domain wall. As the wall connects two separate bulks, we need a prescription to continue the RT surface across the wall. For instance, in the case of end-of-the-world branes which appear in Island computations \cite{Chen:2020hmv,Takayanagi:2011zk}, the prescription states that the RT surface should end perpendicularly on the End of the World brane\footnote{This is only in the case where the EOW brane is described by the simplest Lagrangian, only involving its tension. If additional fields are living on it, the prescription will change}. We will show that this can be recovered this as a special limit of the more general interface prescription.

Consider again the action determining a spacelike geodesic, but on a spacetime which is composed of two pieces. Each piece will contain a domain wall which we parametrise as $x^\mu_{mi}(\xi^a)$, where the $i$ denotes its embedding on side $i$. The geodesic is parametrised by $\lam$, and crosses the membrane at $\lam^*$ :
\begin{eqgroup}
L=\int_{\lam_0}^{\lam^*}\sqrt{g^1_{\mu\nu} \dot{x}^\mu \dot{x}^\nu}d\lam+\int_{\lam^*}^{\lam_0}\sqrt{g^2_{\mu\nu} \dot{x}^\mu \dot{x}^\nu}d\lam = \int_{\lam_0}^{\lam^*}\cal{L}_1(x,\dot{x})d\lam+\int_{\lam^*}^{\lam_0}\cal{L}_2(x,\dot{x})d\lam\ .
\label{twofaceactionspacelike}
\end{eqgroup}

In what follows, we omit the second half of the action as it is treated similarly. In the variation of (\ref{twofaceactionspacelike}), the initial and final points are fixed, $\delta x(\lam_0)=\delta x(\lam_f)=0$. On the other hand, the variation at $\lam^*$ receives a softer constraint; it must simply remain on the membrane. Using the parametrization of the membrane, we can write is as :
\begin{eqgroup}\label{membranevariation}
\delta x^\rho(\lam^*) = \frac{\pa x^\rho_m}{\pa\xi^a}\delta \xi^a\ ,
\end{eqgroup}
where the $\delta \xi^a$ are arbitrary variations. The variation of (\ref{twofaceactionspacelike}) then reads :
\begin{eqgroup}
\delta L = \int_{\lam_0}^{\lam^*}\frac{\delta \cal{L}_1}{\delta x^\si(\lam)}\delta x^\si(\lam) + \frac{\pa\cal{L}_1}{\pa \dot{x}^\rho}\delta x^\rho(\lam^*)+1\leftrightarrow 2\ ,
\label{variationofL}
\end{eqgroup}
where we used the shorthand $\frac{\delta \cal{L}_1}{\delta x^\si(\lam)}=\frac{\pa \cal{L}_1}{\pa x^\si}-\frac{d}{d\lam}\frac{\pa \cal{L}_1}{\pa \dot{x}^\si}$, which is the term which will give the Euler-Lagrange equations. The second term in (\ref{variationofL}) appears from an integration by parts.

Thus by the arbitrariness of the variation $\delta x^\si(\lam)$, we obtain the geodesic equations which must be obeyed by the geodesic on both sides. The "crossing constraints" are obtained by requiring that the boundary term vanishes. They are :
\begin{eqgroup}
&\frac{\pa\cal{L}_1}{\pa \dot{x}^\rho}\frac{\pa x^\rho_m}{\pa\xi^a}=\frac{\pa\cal{L}_2}{\pa \dot{x}^\rho}\frac{\pa x^\rho_m}{\pa\xi^a}\ ,\\
\Leftrightarrow &\frac{t^{1\,\rho}_a \dot{x}^\si_1 g_{1\,\rho\si}}{\sqrt{g_{1\mu\nu}\dot{x}_1^\mu \dot{x}_1^\nu}}=\frac{t^{2\,\rho}_a \dot{x}^\si_2 g_{2\,\rho\si}}{\sqrt{g_{2\mu\nu}\dot{x}_2^\mu \dot{x}_2^\nu}}\ ,
\label{crossingconstraint}
\end{eqgroup}
where we labeled by $i$ the curves lying on side $i$. We introduced the notation $t^{i\,\rho}_a =  \frac{\pa x^\rho_m}{\pa\xi^a}$ which are the tangent vectors to the membrane.

The constraints (\ref{crossingconstraint}) tell us that the projection of membrane tangent vectors on the geodesic should be conserved while crossing. 

One can get a bit more intuition on the meaning of this constraint by considering the membrane to contain only spacelike directions. In this case, the tangent vectors $t^{i\,\mu}_a$ are all spacelike, and using the metric matching condition (\ref{metricmatching}), we have that the scalar products $t^{i\,\mu}_at^{i\,\nu}_b g_{i\mu\nu}=h_{ab}$ are equal on both sides, since they are the membrane's induced metric. This allows us to divide on both sides of (\ref{crossingconstraint}) by the norm of the tangent vectors. The resulting equations then equate the angles of the geodesic with the membrane :
\begin{eqgroup}\label{anglematching}
\sphericalangle \left(t^1_a,\dot{x}_1\right)=\sphericalangle \left(t^2_a,\dot{x}_2\right)\ .
\end{eqgroup}

Note that the condition (\ref{crossingconstraint}) can only be interpreted as (\ref{anglematching}) in the case where the metric matching is satisfied. Indeed, there is an intuitive way of understanding why is this the case. The metric matching condition is essentially a continuity equation on the metric. This should imply that the Christoffel symbols are at most step-wise discontinuous. Then by the geodesic equation :
\begin{eqgroup}
\ddot{x}^\mu = \Gamma^\mu_{\nu\rho}\dot{x}^\nu \dot{x}^\rho\ .
\end{eqgroup}
By integrating we find that $\dot{x}^\mu$ should be continuous across the interface. The precise way this is realized is simply (\ref{anglematching}).

In the case where some directions of the membrane are timelike, the norm squared of the tangent vectors will be negative. We could still divide on both sides by the absolute value of the norm, but the expression that we obtain in this way is difficult to visualize as an "angle".

\subsection{Vacuum ICFT}
\label{sec:vacuumICFT}
We are now ready to compute RT surfaces in setups containing a gravitating wall. We consider in this section the joining of two CFTs in their vacuum state, on an infinite interval. The solution is described in sec.\ref{sec:nearboundarysol}, we rewrite here the crucial formulas to set slightly different conventions. The dual metrics can be given in Poincaré coordinates :
\begin{eqgroup}\label{ICFTpoincaremetric}
ds^2_j = \frac{\ell_j^2}{z_j^2}\left(-dt^2+dx_j^2+dz_j^2) \right)\ ,
\end{eqgroup}
and the membrane equation is very simple in these coordinates :
\begin{eqgroup}
\label{membraneequationvacuum}
x_j(z_j) = -\frac{\tan(\psi_j)}{z_j}\ ,
\end{eqgroup}
where the solution is given in the folded picture, and the conventions are to keep the part of spacetime that lies to the "increasing x" side of the membrane (for convenience we inverted here the convention with respect to previous chapters). The identification is done by setting $\ell_1/z_1=\ell_2/z_2\Leftrightarrow \cos(\psi_1)/z_1=\cos(\psi_2)/z_2$ along the membrane.

\begin{figure}[!h]
    \centering
    \includegraphics[width=0.6\linewidth]{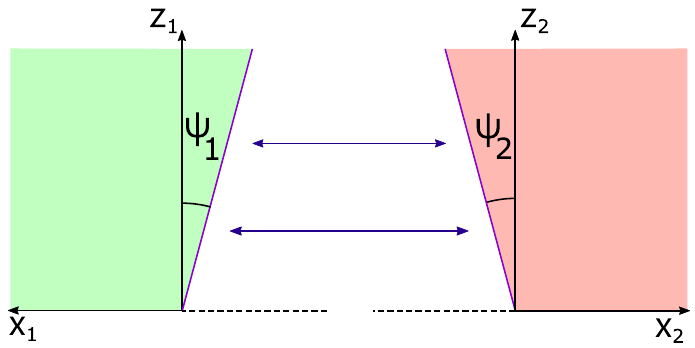}
    \caption{\small Example of a glued geometry dual to a vacuum ICFT state. The depicted angles $\psi_i$ have positive values in the picture. The arrows denote that the two membranes should be identified.}
    \label{fig:conventionsvacuummembrane}
\end{figure}

Remember that in the minimal model, the only parameters available are the AdS radii $\ell_i$ and the tension $\lam$ of the membrane, as here, the state is fixed to the vacuum. From the gluing equations (\ref{constraint on angles}), we can trade $\ell_2/\ell_1$ and $\lam$ for the angles $\psi_i$. The WLOG assumptions $\ell_1\leq \ell_2$ as well as the Israel equations then impose :
\begin{eqgroup}
0< \psi_1\leq \pi/2\;\;;-\psi_1<\psi_2<\psi_1\ .
\label{anglesconstraints}
\end{eqgroup}
So, for any two angles in the range (\ref{anglesconstraints}), we have an ICFT dual for definite $\frac{c_2}{c_1}$ and membrane tension $\lam$.

\begin{figure}[!h]
    \centering
    \includegraphics[width=0.5\linewidth]{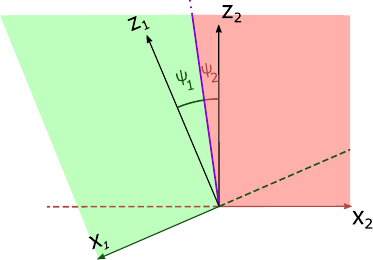}
    \caption{\small The same solutions depicted in fig.\ref{fig:conventionsvacuummembrane}, but presented in an unified coordinate chart. One can verify that the way we glued the two spacetimes, they respect the gluing condition $cos(\psi_1)/z_1=\cos(\psi_2)/z_2$.}
    \label{fig:vacuumembranefullgeometry}
\end{figure}

The simplicity of the solution in this case allows us to explicitly construct a coordinate system unifying the two spacetime pieces\cite{Bachas:2001hpy}. Then, the full solution can be expressed in a single coordinate chart, see fig.\ref{fig:vacuumembranefullgeometry}.

We will see that in this picture, crossing RT surfaces can be expressed by a beautiful geometric construction, which is due to \cite{Sonner:2017jcf}. In the following section, we reproduce the results already obtained there, and we push the analysis of the RT surfaces in slightly more detail than in the aforementioned paper.

\subsection{Intervals containing the interface}
\label{sec:intervalcontaininterface}
We consider first the application of the RT prescription to boundary intervals containing the interface. Consider first two points on the boundary at equal times, located on each side of the interface, $p_1 =(\tau_1=0,x_1=\sigma_1)$ and $p_2=(\tau_2=0,x_2=\sigma_2)$ (in the folded picture).  Because the geometry is static, we can restrict ourselves to a constant time Cauchy slice to search for the RT surface. As mentioned in sec.\ref{sec:casekpkmmorezero}, spacelike geodesics confined to the Cauchy slice are simply semi-circles whose center lies on the AdS boundary.

\begin{figure}[!h]
    \centering
    \includegraphics[width=0.6\linewidth]{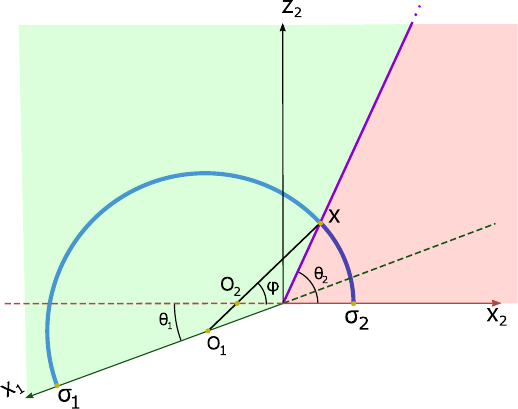}
    \caption{{\small Example of a crossing geodesic, for $\psi_2<0$. The angle $\varphi$ together with $O_2=(\lam,0)$ determines $\si_2$ as well as the membrane intersection $x$. By extending the line $xO_2$ to intersect the boundary of side 1 (green), we find $O_1$ which is the center of the second semi-circle geodesic. The angles $\th_1=\psi_1+\psi_2$ and $\th_2=\frac{\pi}{2}+\psi_2$ are simply introduced for convenience for the computations of app.\ref{app:Derivationboundaryintervalgeodesic}}.}
    \label{fig:geodesicoftraversinginterval}
\end{figure}

The sought-out RT surface will be composed of circle arcs meeting on the interface. To satisfy the crossing constraints (\ref{crossingconstraint}), we need to appropriately choose the radii of the two geodesics so that the circles are tangent on the membrane (such that their incident angle is continuous). As first noticed in \cite{Sonner:2017jcf}, this problem can be solved by a simple Euclidean geometry construction, outlined in fig.\ref{fig:geodesicoftraversinginterval}

The only caveat of the construction of fig.\ref{fig:geodesicoftraversinginterval} is that the RT surface is parametrized in terms of $O_1=(\lam,0)$ and $\varphi$, instead of the boundary points $\si_1$ and $\si_2$. 

By repeated use of the law of sines, we obtain the following formulas (see app.\ref{app:Derivationboundaryintervalgeodesic} for the detailed derivation) :
\begin{eqgroup}
    \si_2 &= (-\lam)\left(\frac{\cos(\psi_2)}{\cos(\al)}-1\right)\ ,\\
    \si_1 &= \frac{(-\lam)\sin(\al+\psi_2)}{\sin(\al-\psi_1)}\left(1+\frac{\cos(\psi_1)}{\cos(\al)}\right)\ ,\\
    \mu &= \frac{\si_2}{\si_1}=\frac{\sin(\al-\psi_1)\left(\frac{\cos(\psi_2)}{\cos(\al)}-1\right)}{\sin(\al+\psi_2)\left(\frac{\cos(\psi_1)}{\cos(\al)}+1\right)}\ ,
    \label{eq:si1si2alphboundarywithinterface}
\end{eqgroup}
where we defined $\alpha=\varphi-\psi_2$ simply because it simplifies the look of the expressions. There are some constraints on $\lam$ and $\alpha$ to generate a well-defined RT surface (for instance, to avoid values for which the construction yields an RT surface traversing on the wrong side of the asymptotic boundary). The conditions read :
\begin{eqgroup}
 \lam<0&\rightarrow \psi_1\leq \al \leq\pi/2\Leftrightarrow 0<\frac{\si_2}{\si_1}<1\ ,\\
 \lam>0&\rightarrow \pi/2 \leq \al \leq \pi-\psi_1 \Leftrightarrow \frac{\si_2}{\si_1}>1\ .
 \label{conditionlamalphawithinterface}
\end{eqgroup}
Thus, for $\psi_1<\al<\pi-\psi_1$ we cover all possible boundary intervals.
Notice that the absolute value of $\lam$ does not influence the ratio $\mu=\si_2/\si_1$, which is a consequence of scale invariance. By a lengthy computation, one can show that $\mu(\alpha)$ is a monotonically increasing function, which shows that there is a unique RT surface given $\si_1$ and $\si_2$.

What remains is to compute the length of the geodesic to obtain the entanglement entropy. Parametrizing the semicircle as ($x=R\cos(\th)$, $z=R \sin(\th)$) the length of the geodesic depends only on the initial and final opening angles :
\begin{eqgroup}\label{geodlengthwithangle}
L=\ell \ln(\frac{\sin\th_f(1+\cos\th_0)}{\sin\th_0(1+\cos\th_f)})\ .
\end{eqgroup}
Using (\ref{geodlengthwithangle}), and denoting the $z_i$-cutoffs $\eps_i$, we find the length of the full geodesic (see app.\ref{app:Derivationboundaryintervalgeodesic}):
\begin{eqgroup}
\label{lengthgeodboundaryinterfacecase}
L=&\ell_1 \ln(\frac{\si_1}{\eps_1}) + \ell_2\ln(\frac{\si_2}{\eps_2})\\
&\underbrace{+\ell_1 \log\left(\frac{2}{\tan\left(\frac{\al-\psi_1}{2}\right)\left(1+\frac{\cos(\al)}{\cos(\psi_1)}\right)}\right)+\ell_2 \log\left(\frac{2\tan\left(\frac{\al+\psi_2}{2}\right)}{1-\frac{\cos(\al)}{\cos(\psi_2)}}\right)}_{g(\xi)}\ .
\end{eqgroup}

The quantity on the second line is independent of $\lam$, and thus depends only on the ratio $\mu$. The fact that (\ref{lengthgeodboundaryinterfacecase}) takes the form $L=\ell_1 \ln(\frac{\si_1}{\eps_1}) + \ell_2\ln(\frac{\si_2}{\eps})+g(\mu)$ is no coincidence. To understand why, consider the more general case in which $p_1 =(\tau_1=0,x_1=1)$, $p_2=(\tau_2\neq 0, x_2=\sigma_2)$ (where we used a scale transformation and $\tau$-translation to choose the $p_1$ coordinates). As we have shown in \ref{sec:interfaces}, there is one Virasoro that survives the introduction of the interface. Constraining ourselves to the global conformal transformations to remain in the vacuum state, we can use those to bring $\tau_2$ to zero (more specifically, we use a special conformal transformation centered on $p_1$). These CFT transformations translate to isometries of the bulk so it follows that the geodesic length between the two points is also unchanged. Using these isometries, it is possible to restrict the general form of the geodesic distance between two points.

However, we can identify several quantities which are invariant under all the isometries such as $\xi=\frac{-(t_1-t_2)^2+(x_1-x_2)^2}{4x_1x_2}$, $\frac{x_i}{z_i}$ or $\frac{-(t_1-t_2)^2+(x_1-x_2)^2+(z_1-z_2)^2}{4z_1z_2}$. Thus, this isn't sufficient to constrain it to the form identified above. We must take into account the additional fact that the geodesics are anchored to the boundary (or more precisely on the IR cutoff surface). A more elegant way to restrict the generic form, is to exploit the conformal symmetry of the dual theory. Indeed, as described in \cite{Sonner:2017jcf}(see also \cite{McAvity:1993ue}) another interpretation of boundary-anchored geodesic is that they compute two-point functions of operators of large scaling dimension. Then, we can transfer the form of the generic two-point function into the form of the geodesic length between two points $p_i = (t=\tau_i,x=\si_i,z=\eps_i)$  :
\begin{eqgroup}\label{lengthgeneralform}
L &= \ell_1 \ln(\frac{\si_1}{\eps_1}) + \ell_2\ln(\frac{\si_2}{\eps_2}) +g(\xi)\ ,\\
\xi &=\frac{-(\tau_1-\tau_2)^2+(x_1-x_2)^2}{4x_1x_2}\ .
\end{eqgroup}
This formula embodies the procedure we used to obtain (\ref{entanglemententropyspinning}), by a succession of coordinate changes. We can use (\ref{lengthgeneralform}) together with (\ref{lengthgeodboundaryinterfacecase}) to compute the entanglement entropy of any boundary interval for which the two points lie on either side of the interface.

The only missing piece is that we must invert an equation to extract the function $\alpha(\mu)$. This can in fact obtained analytically, although the expression is not very elegant. The defining equation is obtained with (\ref{eq:si1si2alphboundarywithinterface}), and after some massaging :
\begin{eqgroup}\label{calphaequationsimplecrossing}
\mu = \frac{c_2\sqrt{1-c_\al^2}+c_\al s_2}{c_1\sqrt{1-c_\al^2}-c_\al s_1}\times\frac{c_\al+c_1}{c_2-c_\al}\ ,
\end{eqgroup}
where we use the shorthands $c_i = \cos(\psi_i)$, $c_\al = \cos(\al)$. Eq. (\ref{calphaequationsimplecrossing}) can be reduced to a 4th-order polynomial equation after squaring. Solving yields 4 candidate solutions for $c_\al(\mu)$, but after plugging them back in (\ref{calphaequationsimplecrossing}), only two remain. Imposing the further condition $|c_\al|<1$, only one remains, as we expected from the monotonicity of $\mu(\al)$ :
\begin{eqgroup}\label{mualphasimplecrossing}
\cos(\al) = \frac{(-1+\mu^2)(c_1+c_2)-\sqrt{2}(\mu-1)\sin(\frac{\psi_1+\psi_2}{2})\sqrt{(1+\mu^2)-(\mu-1)^2\cos(\psi_1-\psi_2)+4\mu \cos(\psi_1+\psi_2)}}{2(1+\mu^2+2\mu \cos(\psi_1+\psi_2))}\ .
\end{eqgroup}

The equation is quite unpalatable, but by plugging it into (\ref{lengthgeodboundaryinterfacecase}), together with the RT prescription we obtain an analytic expression of the entanglement entropy of any interval in the boundary CFT, which is remarkable.

Let us conclude by considering the "End of the World" membrane limit, where side 1 disappears. This is realized by taking the limit $\frac{\ell_1}{\ell_2}\rightarrow 0$, which is $\psi_1\rightarrow \pi/2$ in terms of the angles. In this limit, since side 1 effectively disappears, the geodesic in fig.\ref{fig:geodesicoftraversinginterval} becomes anchored at the membrane. Because the green boundary and the membrane are co-linear in this limit, $O_1$ moves to the origin, and the geodesic becomes perpendicular to the wall. We recover in this way the BCFT prescription \cite{Takayanagi:2011zk} for RT surfaces.

\subsection{Same-side intervals}
We consider now the case where the two endpoints of the boundary interval are located on the same side. Let us begin with the case where the two points are located on the true vacuum, i.e. the CFT with lower central charge. In this case, it is easy to see that the RT surface will be the same as in the trivial case with no boundary. From now on, we will call "trivial" the geodesics that do not cross the membrane.

From the constraints (\ref{anglesconstraints}) it is clear that a trivial geodesic will always be allowed, as the semicircle will never intersect the membrane. Let us consider a putative non-trivial geodesic, that crosses into the false vacuum. This curve will be necessarily longer than the trivial geodesic. Indeed, from (\ref{geodlengthwithangle}), the portion of the geodesic in the false vacuum will obtain a multiplicative factor of $\ell_2/\ell_1>1$, with respect to the length it would have had in the true vacuum. Adding this to the fact we are deviating from the trivial geodesic path for this excursion into the other side, it is clear that its length must be higher than the trivial one.

Although this argument alone does not exclude the existence of non-minimal extremal curves of this type, we will confirm they do not exist through the geometric construction that will follow.  The entanglement structure of intervals lying on the true vacuum is thus unchanged from the "pure" case given by (\ref{intervalentangentropypoincare}).

Things get more interesting as we move the boundary segment to the false vacuum. Both arguments given in the previous paragraphs fail here. For once, $\psi_2$ can be negative, and so given big enough intervals the trivial geodesic will intersect the membrane, such that the true geodesic must necessarily be non-trivial. On the other hand, even when $\psi_2>0$, taking some extra distance to cross onto the other side might be worth it. Indeed, this can be seen as taking a "shortcut" through the false vacuum, where distances are shrunk by a factor of $\ell_1/\ell_2$.

\begin{figure}[!h]
    \centering
    \includegraphics[width=.6\linewidth]{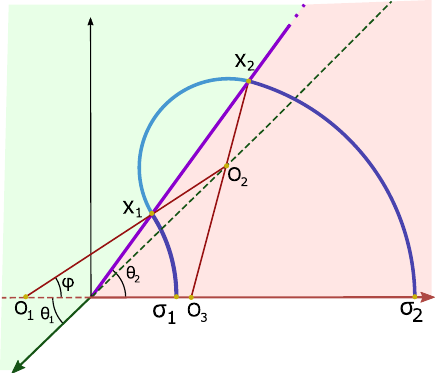}
    \caption{\small Example of a double crossing geodesic, for $\psi_2<0$. The full geodesic is composed of three semicircles of centers $O_1$, $O_2$ and $O_3$. The construction goes similarly as the single crossing case. We begin with $O_1=(\lam,0)$ and $\varphi$. By extending the line $O_1x_1$, we intersect the green boundary at $O_2$. We find $x_2$ as the other intersection with the semi-circle of center $O_2$, and repeat by extending $x_2O_2$, constructing $O_3$. The semi-circle of center $O_3$ intersects the boundary at $\si_2$.}
\label{fig:doublecrossinggeodesic}
\end{figure}

The typical non-trivial geodesic is depicted in fig.\ref{fig:doublecrossinggeodesic}.
The Euclidean construction is again parametrized by the angle $\varphi$ and center $O_1$. The full geodesic is composed of three circle arcs that are tangent to each other on the membrane. Again, with repeated applications of the sine law one can obtain expressions for the final and initial points (see app. \ref{app:doublecrosscomputation} for details) :
\begin{eqgroup}
    \si_1 &= -\lam \left(\frac{\cos(\psi_2)}{\cos(\alpha)}-1\right)\ ,\\
    \si_2 &= -\frac{\lam\sin(\al+\psi_2)}{\sin(\psi_1-\al)}\frac{\sin(\al+\psi_1)}{\sin(\al-\psi_2)}\left(1+\frac{\cos(\psi_2)}{\cos(\al)}\right)\ ,\\
    \mu &= \frac{\si_2}{\si_1} = \frac{\sin(\al+\psi_2)}{\sin(\psi_1-\al)}\frac{\sin(\al+\psi_1)}{\sin(\al-\psi_2)}\frac{\left(\cos(\alpha)+\cos(\psi_2)\right)}{\left(\cos(\psi_2)-\cos(\alpha)\right)}\ .
    \label{si1si2doublecrossinginterface}
\end{eqgroup}

While the formulas seem only slightly more complicated than (\ref{eq:si1si2alphboundarywithinterface}), the situation is quite a bit more involved in this case. Let us first look at the allowed range for $\alpha=\varphi-\psi_2$. To find it, we enforce $\si_1>0$ and $\si_2>0$, as well as that the intersection on the membrane takes place at $z>0$.  By symmetry we will later restrict further to $\si_2>\si_1$. After tedious but straightforward computations, we find :
\begin{eqgroup}
 \lam<0&\rightarrow |\psi_2|\leq \al \leq \psi_1\Rightarrow \frac{\si_2}{\si_1}>1\ ,\\
 \lam>0&\rightarrow \pi-\psi_1 \leq \al \leq \pi-{\rm Max}(\psi_2,0) \Rightarrow \frac{\si_2}{\si_1}<1\ .
 \label{conditionlamalphadoublecrossing}
\end{eqgroup}

From (\ref{si1si2doublecrossinginterface}), $\al\rightarrow \pi-\al$ sends $\frac{\si_1}{\si_2}\rightarrow \frac{\si_2}{\si_1}$. Thus, it should suffice to consider only one of the two cases of (\ref{conditionlamalphadoublecrossing}). But as we can see, the ranges of the two options for ${\rm sign}(\lam)$ don't seem to coincide under this mapping. In fact, this is simply an artifact of our Euclidean construction that introduced a dissymmetry between $\si_1$ and $\si_2$, because we considered the intersections $x_1$ to lie in the "positive" side of the half-line $O_1x_1$. By considering intersections on the other half-line, the range ${\rm Max}(\psi_2,0)<\alpha<0$ (with $\lam>0$) is indeed allowed, restoring the symmetry $\si_2\leftrightarrow \si_1$.

For what follows we will usually consider ${\rm Max}(\psi_2,0)<\alpha<\psi_1$ and thus $\mu>1$, since it is the situation depicted in fig.\ref{fig:doublecrossinggeodesic}. Unlike in the previous section, the function $\mu(\alpha)$ is not monotonic. Indeed, it has a unique minimum at $\alpha_{\rm crit}$ :
\begin{eqgroup}\label{critalphaminimum}
\sin(\alpha_{\rm crit}) = \frac{1}{4}\left(-\frac{\sin(2\psi_1)}{\cos(\psi_2)}+\sqrt{\frac{\sin^2(2\psi_1)}{\cos^2(\psi_2)}+8\left(1-\frac{\cos(2\psi_1+\psi_2)}{\cos(\psi_2)}\right)}\right) \ ,
\end{eqgroup}
which we verify to always lie in the allowed $\al$ range. This implies that at least in some range, there are two distinct non-trivial geodesics. 

Looking at the boundary limits, we have to distinguish the two cases $\psi_2<0$ and $\psi_2>0$ :
\begin{eqgroup}\label{mulimits}
\lim_{\al\rightarrow \psi_1}\mu(\al) &= \infty\ ,\\
\lim_{\al\rightarrow {\rm Max}(\psi_2,0)}\mu(\al) &=\begin{cases}
          \infty \quad &\text{if} \, \psi_2\geq0 \\
          \frac{1-\cos(\psi_2)}{1+\cos(\psi_2)} \quad &\text{if} \, \psi_2< 0
     \end{cases}\ .
\end{eqgroup}

\begin{figure}[!h]
\centering
\begin{subfigure}[b]{0.7\textwidth}
   \includegraphics[width=1\linewidth]{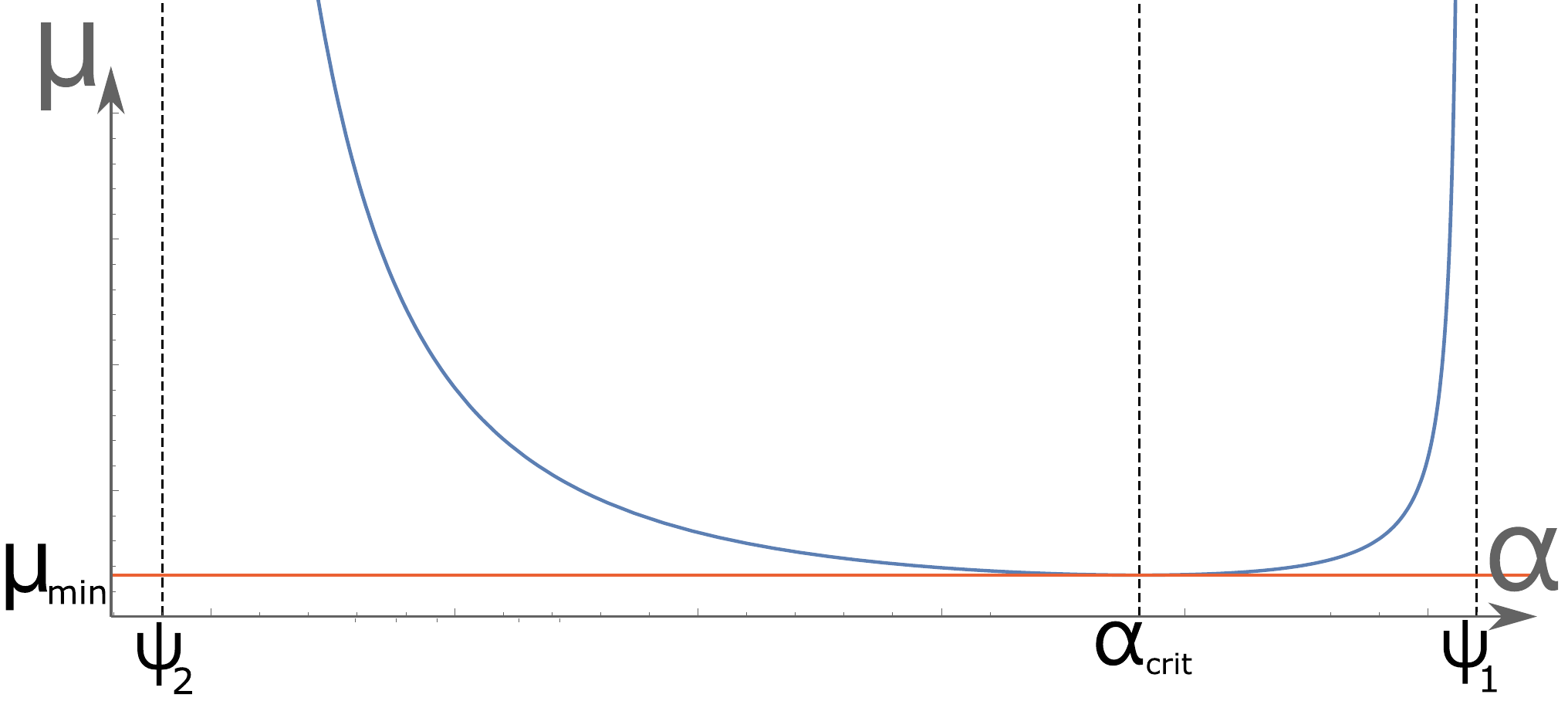}
   \caption{{\small Example plot of $\mu(\al)$ in the case $\psi_2>0$.}}
\end{subfigure}

\begin{subfigure}[b]{0.7\textwidth}
   \includegraphics[width=1\linewidth]{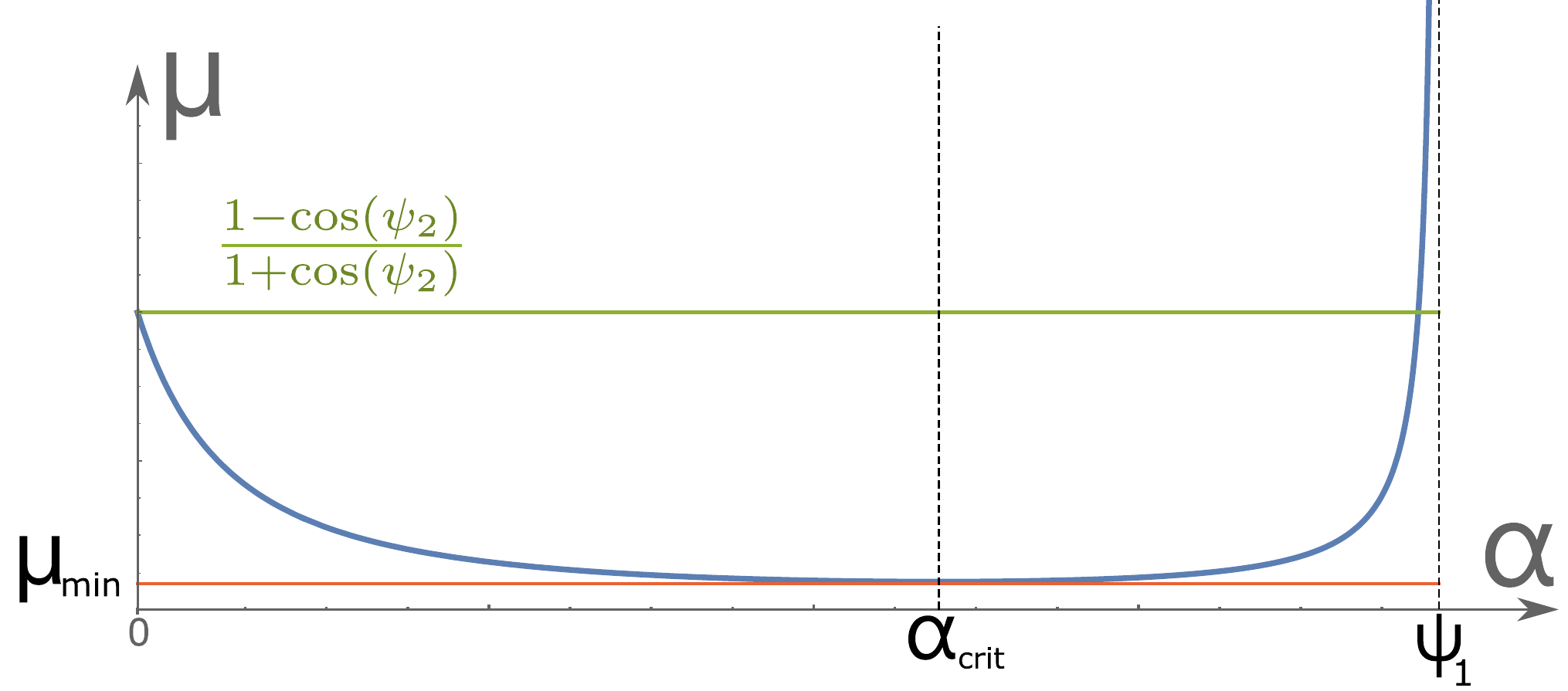}
   \caption{{\small Example plot of $\mu(\al)$ in the case $\psi_2<0$. The green line denotes the value of $\mu$ above which the trivial solution ceases to exist.}}
\end{subfigure}

\caption{\small Plots of $\mu(\al)$ for both signs of $\psi_2$. The red line indicates the value of $\mu$ under which no non-trivial solutions exist, attained at $\al_{crit}$. For $\psi_2>0$, above $\mu_{min}$ two distinct non-trivial solutions always exist, while it is not the case when $\psi_2<0$. The geodesics corresponding to values of $\al<\al_{crit}$ are never the shortest.}
\label{fig:plotofmubothsignspsi2}
\end{figure}

In fig. \ref{fig:plotofmubothsignspsi2} we plot the curve $\mu(\al)$ for both signs of $\psi_2$.

An interesting remark arising from this analysis is yet again the appearance of the mysterious "critical tension" $\lam_0$ (see \ref{sec:criticaltensions}), which is the tension corresponding to $\psi_2=0$. For $\psi_2>0$ or overcritical tensions, there are always two candidate non-trivial geodesics, while for $\psi_2<0$ or subcritical tensions, this is only the case for small enough $\mu$. In this case, this will have no bearing on the RT surface as we will see that the geodesics for $\al<\al_{crit}$ are never dominant. However, in the interpretation of \cite{Sonner:2017jcf}, where the geodesics compute two-point functions, the subdominant ones do contribute to corrections, so in this case, the crossing of the critical tension does produce a measurable field theory effect. In spite of all these hints, a deeper understanding of the meaning of this critical tension is still lacking.

A last remark concerns the disappearance of trivial geodesics. It is easy to see that when $\psi_2<0$ and $\mu>\frac{1-\cos(\psi_2)}{1+\cos(\psi_2)}=\mu(\al=0)\equiv \mu_{min}$, they cross the membrane and thus cease to exist. As we approach this limit from below, the crossing geodesics with $\al<\al_{crit}$ approaches the trivial one, matching at $\mu=\mu_{min}$. Perhaps there is a deeper explanation for the simultaneous disappearance of these two geodesic paths from the field theory POV, but we have not been able to elucidate it.

As in the previous case, we can obtain the geodesic length analytically (see \ref{app:doublecrosscomputation} for more details) :
\begin{eqgroup}
L = \ell_2 \ln(\frac{\si_1}{\eps_1})+\ell_2\ln(\frac{\si_2}{\eps_2})+\underbrace{\ell_2\ln(\frac{4 \tan(\frac{\al+\psi_2}{2})}{\tan(\frac{\al-\psi_2}{2})\left(1-\frac{\cos^2(\al)}{\cos^2(\psi_2)}\right)})+\ell_1 \ln(\frac{ \tan(\frac{\al+\psi_1}{2})}{\tan(\frac{\psi_1-\al}{2})})}_{g(\xi)}\label{fulllengthdoublecross}\ .
\end{eqgroup}
As expected, the length is invariant under $\al\rightarrow \pi-\al$ which is the inversion $\si_2\leftrightarrow \si_1$, and we have arranged it in the form (\ref{lengthgeneralform}). This length is to be compared to the trivial one which can be written as:
\begin{eqgroup}
L_{\rm triv}&=\ell_2 \ln(\frac{4 R^2}{\eps_1\eps_2})\quad R=\frac{\si_2-\si_1}{2}\ .
\end{eqgroup}

To compare the two lengths we should use (\ref{si1si2doublecrossinginterface}) to express $L_{\rm triv}$ in terms of $\al$. Let us call $\Delta L(\al) = L(\al)-L_{\rm triv}(\al)$ the difference between trivial and non-trivial geodesic.

Unfortunately, the expression is much too complicated to hope to have a closed expression for $\al_{\rm trans}$, the value where dominating geodesic switches. However, we can again inspect the different limits to distill some properties. For instance, we can show that no matter what, when $\ell_1<\ell_2$, there will always be non-trivial geodesics dominating for large enough intervals. Indeed, in the limit $\al\rightarrow \psi_1$, where $\si_2/\si_1\rightarrow \infty$ :
\begin{eqgroup}\label{deltaLlimitpsi1}
L-L_{\rm triv} &= (\ell_2-\ell_1)\ln(\psi_1-\al)+\cal{O}(1)\ .
\end{eqgroup}

Thus in this limit, the non-trivial geodesic with $\al>\al_{\rm crit}$ always dominates. In doing this analysis, we expected to find a "critical angle" for the brane, over which the non-trivial geodesic stopped existing, as found in \cite{Geng:2021mic}, in the context of Boundary CFT. However, as they mention in the paper, this critical angle appears only in larger than three dimensions for the bulk, so the absence of a critical angle in our model is consistent with their BCFT findings (the BCFT can be obtained in the limit $\ell_1\rightarrow 0$, where CFT$_1$ disappears and the interface becomes a boundary). It would be interesting to extend the results of \cite{Geng:2021mic} in higher dimensions to the more general case of Interface CFTs.

Similarly, we can inspect the lower limit $\al\rightarrow {\rm Max}(\psi_2,0)$. The interesting case is when $\psi_2>0$, otherwise the non-trivial and trivial geodesics coincide as $\al\rightarrow 0$. For $\psi_2>0$ :
\begin{eqgroup}\label{deltaLlimitpsi2}
L-L_{\rm triv}(\al=\psi_2) &= \ell_2 \ln(\frac{\sin(\psi_1-\psi_2)}{\cos^2(\psi_2)\sin(\psi_1+\psi_2)})+\ell_1\ln(\frac{\tan(\frac{\psi_1+\psi_2}{2})}{\tan(\frac{\psi_1-\psi_2}{2})})\ ,
\end{eqgroup}
which is always positive, and therefore in this limit the trivial geodesic is always preferred. 

\begin{figure}[!h]
\centering
\includegraphics[width=0.7\linewidth]{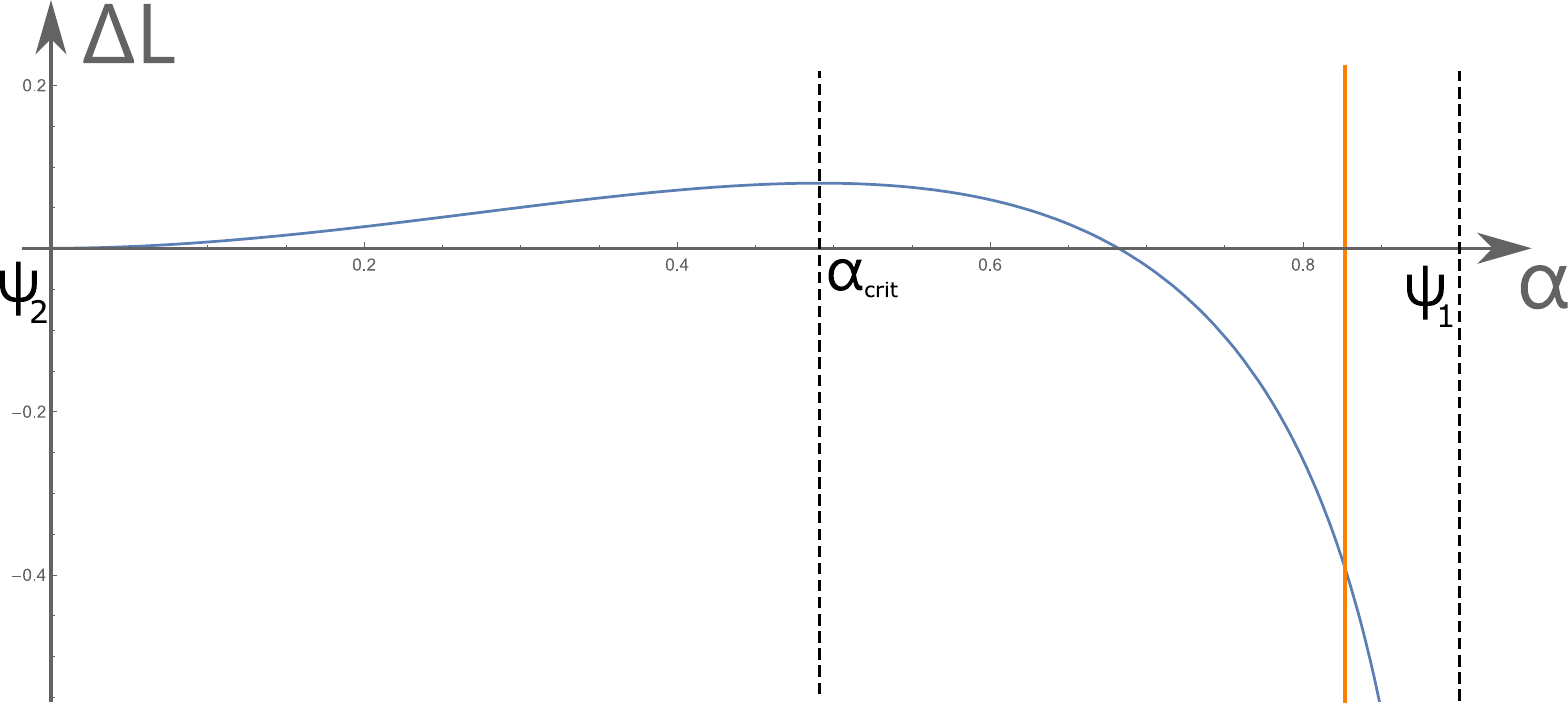}
\caption{{\small Plot of $\Delta L=L-L_{\rm triv}$ as a function of $\alpha$, in the case $\psi_2<0$. The local maximum is located at $\al_{crit}$. The orange line denotes the value of $\al$ above which the trivial geodesics cease existing (in particular, values of $\Delta L$ after this point are meaningless). As we can see, this happens after the non-trivial geodesic becomes dominant. We verified numerically that this is always the case; namely the geodesic switch always happens because the non-trivial geodesic becomes shorter, not because the trivial one ceases existing.}}
\label{fig:deltaLtypical}
\end{figure}

By plotting the curve and playing with the parameters, we can make further statements. The typical curve $L-L_{\rm triv}$ is depicted in fig.\ref{fig:deltaLtypical}.

As we can see, and as we verified numerically by swiping over the parameter range for $\psi_i$, it has a single maximum. Incidentally, we verified that this maximum is attained precisely for $\al_{\rm crit}$ (\ref{critalphaminimum}), which is when $\mu(\alpha)$ is minimal. There is probably a way to prove this analytically, but we haven't been able to do so yet. It can however be understood intuitively, because, at $\al=\al_{crit}$, the non-trivial geodesic is tangent at the membrane which is the worst-case scenario since we don't benefit of the shortcut through side 1. As we grow $\mu$, either by increasing or decreasing $\al$, the geodesic can pass deeper into the true vacuum, thus shortening its path. Combining this with the limit (\ref{deltaLlimitpsi2}), this shows that the non-trivial geodesics which lie at $\al<\al_{\rm crit}$ are always subleading when they exist.

To complete our analysis we would like to invert (\ref{si1si2doublecrossinginterface}) to find $\al(\mu)$. This appears to be a straightforward exercise much like in the previous section, and it is in theory, but in practice, we have encountered hurdles that have prevented us from obtaining a closed analytical expression in this case. Indeed, consider the equation that is to be solved :
\begin{eqgroup}\label{calphaequationmudoublecross}
\mu = \frac{c_\al+c_2}{c_2-c_\al}\times\frac{\left(\sqrt{1-c_\al^2}c_2+s_2c_\al\right)\left(\sqrt{1-c_\al^2}c_1+s_1 c_\al\right)}{\left(s_1c_\al-\sqrt{1-c_\al^2}c_1\right)\left(\sqrt{1-c_\al^2}c_2-s_2 c_\al\right)}\ .
\end{eqgroup}

This can be solved analytically by squaring away the square roots which yields a 4th order polynomial in $c_\al$\footnote{Strictly speaking we obtain a 6th order polynomial, but there are 2 roots that are evident, and aren't valid solutions to (\ref{calphaequationmudoublecross}).}. We do not include the full expressions for the roots as they are two cumbersome, but they take the form :
\begin{eqgroup}\label{doublecrossinversionform}
c_\al &= A+\sqrt{q}\pm\sqrt{p_2-p_1}\ ,\\
c_\al &= A-\sqrt{q}\pm\sqrt{p_2+p_1}\ ,
\end{eqgroup}
where $A,\,q$ and $p_i$ depend on $\psi_i$ and $\mu$. The complicated step lies in choosing which of the four candidate solutions are actually solutions of (\ref{calphaequationmudoublecross}). This is done by replacing (\ref{doublecrossinversionform}) into (\ref{calphaequationmudoublecross}) and verifying it is a solution, as well as checking that the $c_\al$ lies in the range (\ref{conditionlamalphadoublecrossing}). We have attempted this with the aid of Mathematica, but the resulting expressions were humongous, and their simplification relied on several assumptions on the range of parameters, which made it computationally impossible to simplify analytically.

\begin{figure}[!h]
    \centering
    \includegraphics[width=0.5\linewidth]{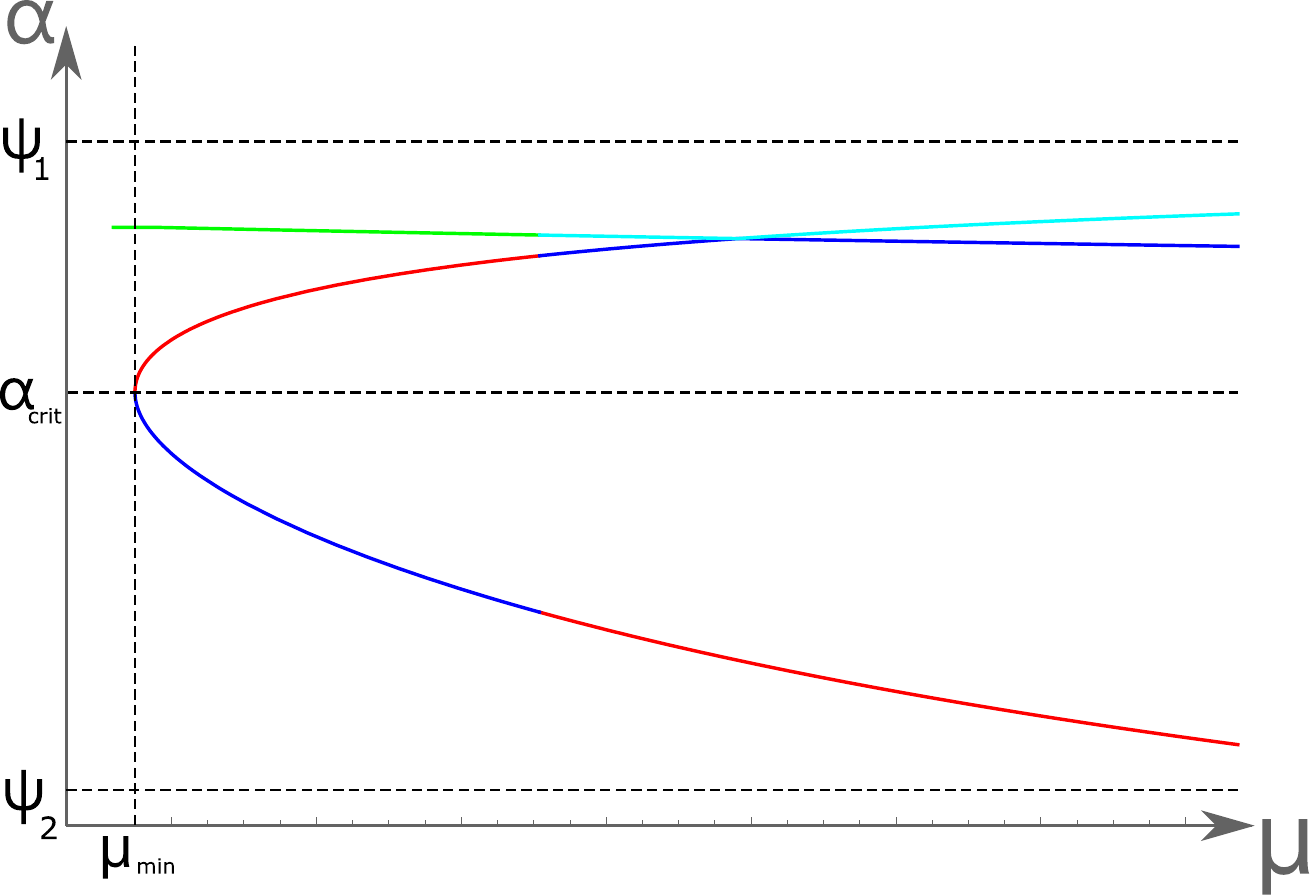}
    \caption{{\small The four candidate solutions of (\ref{doublecrossinversionform}) are plotted in different colors. One can verify that the true solutions of (\ref{calphaequationmudoublecross}) are the ones constituting the paraboloïd shape in the picture. We see on the plot both a discontinuous jump in the solutions, as well as a kink. By mixing and matching the different candidates, we obtain a smooth curve. }}
    \label{fig:exampleofswitcheroo}
\end{figure}

Thus, we turn to numerical verifications. This allows us to find the correct solutions among (\ref{calphaequationmudoublecross}), but only for a given value of $\mu$, $\psi_i$. One could hope that the correct solutions do not change as we move the parameters, but this is not the case. See fig.\ref{fig:exampleofswitcheroo} for an example

Looking at fig. \ref{fig:exampleofswitcheroo}, we need to do some "cut and patch" of different solutions to actually obtain the inverse of the curve displayed in fig.\ref{fig:plotofmubothsignspsi2}. This can be understood by looking at the form (\ref{doublecrossinversionform}). Say that $q$ has a double root for some $\mu=\mu^*$, at fixed $\psi_i$. Then locally around that point, $\sqrt{q}\approx \sqrt{q''(\mu^*)(\mu-\mu^*)^2} = \sqrt{q''(\mu^*)}|\mu-\mu^*|$. Because of the absolute value, the solutions (\ref{doublecrossinversionform}) will have a kink at $\mu^*$. Then, to obtain smooth solutions we must combine the $\pm\sqrt{q}$ solutions for $\mu<\mu^*$, with the $\mp\sqrt{q}$ solutions for $\mu>\mu^*$. Another situation is of the type $f(\mu)/\sqrt{q(\mu)}$, where $f(\mu)$ has a simple root at $\mu^*$, while $q(\mu)$ has a double root. In that case, this will induce a discontinuity of $2f'(\mu^*)/\sqrt{q''(\mu*)}$ in the function. Both cases are displayed in fig.\ref{fig:exampleofswitcheroo}.

As it turns out, we found that those "inversions" of sign can happen for $\sqrt{q}$, $\sqrt{p_2-p_1}$ and $p_1$ as well as it contains a term $\propto 1/\sqrt{q}$. To be able to obtain an inverse of (\ref{si1si2doublecrossinginterface}) for any $\psi_i$, one would need to analytically determine the roots of the aforementioned quantities. Unfortunately, this turns out to be impossible as the equations involved are much too complicated, involving high powers and various trigonometric functions.

Thus, in the double-crossing case a closed form like (\ref{mualphasimplecrossing}) is lacking. Of course, (\ref{doublecrossinversionform}) combined with a quick numerical check to select the correct solutions still allow us to compute entanglement entropies relatively quickly, as locally we have even an analytic expression.

\section{Application to more general geometries}
The extended analysis of the previous sections was in the hopes of being able to bootstrap these computations to compute the entanglement structure of the solution chap.\ref{chap:steadystatesofholo} and possibly also for the solutions of chap.\ref{chap:phasesofinterfaces}. Unfortunately, for many cases we will see that this method is not applicable.

\subsection{The thermal ICFT state}
Let us begin analysing the non-compact thermal static solution with an interface, which is the limit $J\rightarrow 0$ of the NESS state in chap.\ref{chap:steadystatesofholo}. We remind the bulk metric :
\begin{eqgroup}\label{reminderBTZ}
ds^2=-(r_j^2-M\ell_j^2)dt^2+\frac{\ell_j^2 dr_j^2}{r_j^2-M\ell_j^2}+r_j^2 dx^2\ .
\end{eqgroup}
The solution is given by (\ref{xjentering}) for $J=0$, $M_1=M_2$. One can verify that in this case, the wall's induced metric is AdS$_2$, and we can integrate (\ref{xjentering}) to obtain the explicit form for the wall :
\begin{eqgroup}\label{blackstringbrane}
x_j(r) = -\frac{{\rm sign}(\lam^2\pm\lam_0^2)}{\sqrt{M}}{\rm arcoth}\left(\frac{\sqrt{b_\pm^2\ell_j^2+r^2}}{r}\right)\ ,
\end{eqgroup}
where the $\pm$ is $+$($-$) for $j=1$($j=2$), while $b_\pm^2=\frac{(\lam^2-\lam_{\rm min}^2)(\lam_{\rm max}^2-\lam^2)}{M(\lam^2\pm\lam_0^2)^2}$. The gluing is made by identifying $r_j^2-M\ell_j^2$ along the membrane.

The goal is now to find a coordinate transformation that brings this solution to the vacuum one above. The additional difficulty to what was done in sec.\ref{sec:simpleexamples} is that now we have the membrane shape to keep track of. Since the membrane breaks some of the isometries of Poincaré, it will not be sufficient to find the coordinate change mapping (\ref{blackstringbrane}) to (\ref{poincaremetriclight}), but the shape of the membrane will also be important. The procedure we employ to identify the correct coordinate change is as follows. 

First, we find an arbitrary coordinate change which brings us to the correct bulk metric, namely Poincaré in this case. After that, we restrict ourselves only to isometries of Poincaré space. Generally, some of these isometries will act non-trivially on the membrane, but they will leave the bulk metric unchanged. Using those, we attempt to bring the membrane to the form (\ref{membraneequationvacuum}). For this step, one can use the dual field theory to greatly facilitate the problem. Indeed, according to sec.\ref{sec:MinimalICFT}, given the boundary state and the interface shape, the bulk dual is completely fixed. Thus, instead of trying to match the 2-dimensional membranes in the bulk, it is sufficient to match the 1-dimensional interfaces in the dual theory, by using only global conformal transformations.

Following this procedure, we begin by with the following change of coordinates :
\begin{eqgroup}\label{btztopoinc}
x_p'\pm t_p'&=\frac{\sqrt{r^2-M\ell^2}}{r} e^{\sqrt{M}(x\pm t)}\ ,\\
z_p' &= \frac{\sqrt{M\ell^2}}{r} e^{x\sqrt{M}}\ ,
\end{eqgroup}
which brings (\ref{reminderBTZ}) to the Poincaré metric. By taking the limit $r\rightarrow \infty$, we see that on the boundary the interface is mapped from $x=0$ to :
\begin{eqgroup}\label{firststepboundary}
-t_p'{}^2+x_p'{}^2=1 \ .
\end{eqgroup}

This initially seems strange as (\ref{firststepboundary}) describes two disconnected interfaces. However, from the coordinate change (\ref{btztopoinc}) we reach only the upper right quadrant of the plane $(x_p'-t_p',x_p'+t_p')$, which ensures that we reach only one of the two disconnected pieces. The other 3 quadrants would be reached by considering the maximally extended black hole string (which, as it contains another boundary, it also contains another membrane), whereas our change of coordinates covers only the exterior of the solution (\ref{reminderBTZ}). In any case, that is all we need to compute RT surfaces, since for the static situation they will not enter the horizon.

What remains to be done is to find a global conformal transformation that maps (\ref{firststepboundary}) to the $x=0$ interface. This case is fairly simple, as (\ref{firststepboundary}) is a hyperbola, which becomes a circle if we wick rotate to Euclidean coordinates. As is well known, any circle can be conformally mapped to a line by a special conformal transformation combined with a translation. In this case, the combination of the two transformations read :
\begin{eqgroup}\label{specialconformalcircletoline}
x_p &=\frac{2 (x'^2-t'^2-1)}{1+x'^2-t'^2+2 x'}\ ,\\
t_p &=\frac{4  t'}{1+x'^2-t'^2+2 x'}\ ,
\end{eqgroup}
which indeed maps (\ref{firststepboundary}) to the interface $x=0$. Notice that from the allowed quadrant mentioned before, we reach only the portion $-2<t_p<2$ of the interface. At the risk of repeating ourselves, the rest of the geometry corresponds to portions of the maximally extended black string solutions, which are of no interest to us when computing entanglement entropies of intervals lying on a single asymptotic boundary. 

From (\ref{specialconformalcircletoline}) we can compute the associated coordinate transformation in the bulk (see sec.\ref{sec:symmetriesofAdS3}), which is simply a special conformal transformation of the three coordinates $t,x,z$ and it is an isometry of Poincaré since the $\frac{1}{z^2}$ prefactor will cancel the overall scale factor :
\begin{eqgroup}
x_p &=\frac{2 (x'_p{}^2-t'_p{}^2+z'_p{}^2-1)}{1+x'_p{}^2-t'_p{}^2+z'_p{}^2+2 x'_p)}\ ,\\
t_p &=\frac{4  t'_p}{1+x'_p{}^2-t'_p{}^2+z'_p{}^2+2 x'_p}\ ,\\
z_p &=\frac{4z'_p}{1+x'_p{}^2-t'_p{}^2+z'_p{}^2+2 x'_p}\ .
\label{specialconformalcircletolinebulk}
\end{eqgroup}
Combining everything, we have the coordinate change that brings the static ICFT black string solution to the static ICFT vacuum solution :
\begin{eqgroup}\label{fullcoordinatechange}
z_p &= \frac{2 \sqrt{M\ell^2} }{\cosh(\sqrt{M}t)\sqrt{r^2-M\ell^2}+r\cosh(x\sqrt{M})}\ ,\\
t_p&= \frac{2 \sqrt{r^2-M \ell^2}\sinh(\sqrt{M}t)}{\cosh(\sqrt{M}t)\sqrt{r^2-M\ell^2}+r\cosh(x\sqrt{M})}\ ,\\
x_p&=\frac{2r\sinh(\sqrt{M}x)}{\cosh(\sqrt{M}t)\sqrt{r^2-M\ell^2}+r\cosh(x\sqrt{M})}\ .
\end{eqgroup}

We can then leverage the equations of the previous section to compute, in principle analytically, the entanglement structure of the theory in the thermal equilibrium state. Analyzing the resulting formulas for the entanglement entropy is unfeasible because of their complexity, so we must content ourselves with plots. 

Consider boundary intervals at constant time, $t=0$, and $x_i>0$ in the folded picture. Notice that under (\ref{fullcoordinatechange}), these are mapped to constant time intervals with $0<x_p^i=\si_i<2$. However, since the difference $L-L_{triv}$ depends only on the ratio $\si_2/\si_1$, we do not expect a qualitative change in behavior for the transitions between trivial and non-trivial geodesics, in the case where the two points lie on the same side. Since this transition generally takes place for $\si_2/\si_1\gg1$ (or $\si_2/\si_1\ll1$), and $\si_2\approx 2$ if $x_2\gg1$, we estimate that the switching to the double-crossing geodesic happens for $x_1<<\frac{2}{\sqrt{M}}$. 

Indeed, this can be understood intuitively because the horizon of the BH puts a cap to how deep can the geodesic go. Thus, crossing to the true vacuum side is not as enticing as in the vacuum solution, since the geodesic will have to curve back to the false vacuum side as soon as it reaches the horizon. Therefore for this excursion to be worthwhile, the geodesic should start ever closer to the interface as $M$ increases (and the horizon approaches the boundary), which is confirmed by our estimation $x_1\ll\frac{2}{\sqrt{M}}$.

Note that one difference is that in the BTZ case, the inversion symmetry $x_i\rightarrow 1/x_i$ no longer holds, as the cross-ratio is expressed differently in this coordinate system. In particular, non-trivial geodesics will exist only if one of the points satisfies $x_i\ll \frac{1}{\sqrt{M}}$. This is in contrast to the vacuum case, where we have non-trivial geodesics as soon as $\si_1/\si_2 \gg 1$ (or $\ll 1$), regardless of the value of the individual $\si_i$.

\begin{figure}[!h]
\centering
\begin{subfigure}[t]{0.49\textwidth}
   \includegraphics[width=1\linewidth]{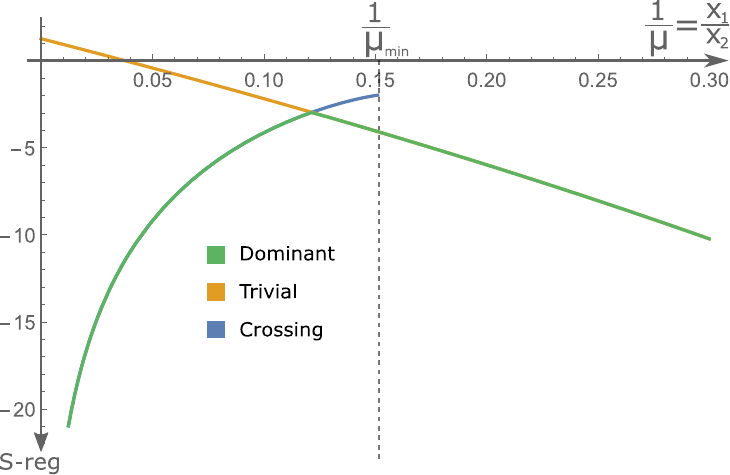}
   \caption{{\small Entanglement entropy plot, zoomed in for small values of $x_1$, to see the transition between the trivial and non-trivial RT surface. At $1/\mu_{min}$, the non-trivial geodesic ceases existing }}
\end{subfigure}
\begin{subfigure}[t]{0.49\textwidth}
   \includegraphics[width=1\linewidth]{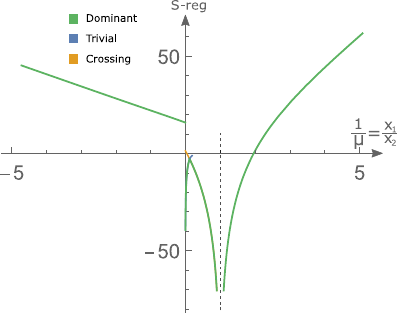}
   \caption{{\small Entanglement entropy plot, zoomed out to see the linear increase in entropy which is characteristic of thermal states. Notice that the slope is different for the positive and negative values; this is expected as the horizons are located at ${\scriptscriptstyle r^i=\sqrt{M\ell_i^2}}$, thus yielding different proportionality constants. We obtain a singularity at $x_1/x_2=1$, where the points coincide. The geodesic length tends to 0, but as we subtracted the cutoff value, it becomes singular.}}
\end{subfigure}
\caption{{\small Two plots of the entanglement entropy in the BTZ solution. We fix $x_2=1$ in side 2, and vary $x_1$ (negative values mean $x_1$ is on side 1). To be precise, we plot the entanglement entropy regularised by subtracting the divergent parts $\ell_i \ln(\frac{1}{\eps})$, where $\eps$ is the IR cutoff. This explains why we obtain negative values. The plots are made for $\psi_2<0$, but they are qualitatively the same for $\psi_2>0$.}}
\label{fig:blackstringentropy}
\end{figure}

Beyond that, there is not much to say about the entanglement structure in the black string case, save for linear growth of the entropy as the interval grows larger, which is the tell-tale of a thermal state. We plot two curves depicting the entanglement entropy for intervals containing the interface, and intervals contained in the false vacuum in fig.\ref{fig:blackstringentropy}.

The study of RT surfaces in the thermal state of the ICFT has very interesting applications in itself, especially in Page-curve computations in eternal black hole setups \cite{Sonner:2017jcf,Almheiri:2019yqk}. However from the point of view of the entanglement structure, we did not expect any surprises, and the difference between vacuum and thermal ICFT is similar to the case without interface.

\subsection{The NESS ICFT state}
\begin{figure}[!h]
    \centering
    \includegraphics[width=0.7\linewidth]{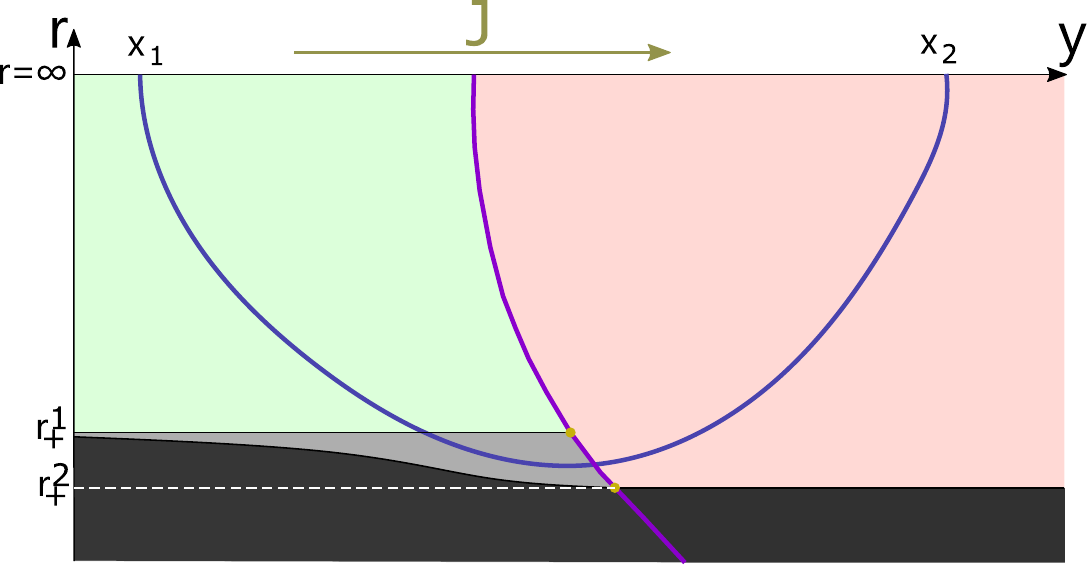}
    \caption{{\small A depiction of an HRT surface (in blue) dipping inside the apparent horizon of the hotter side. The blackened region is the inside of the event horizon, while the grey one sits inside the apparent horizon. While there do not exist spacelike geodesics which cross the apparent horizon and have two anchors at the boundary, single-anchored geodesic can cross the apparent horizon. By clipping such a geodesic at the membrane, and crossing outside the apparent horizon on side 2, we could obtain the depicted HRT surface. Here, we depicted the geodesic as avoiding the event horizon, but we do not know if this is always true.}}
    \label{fig:RTinsidehorizonsketch}
\end{figure}

The computation of the previous section was done as a preparation for the real goal, which is computing HRT surfaces in the NESS state of chap.\ref{chap:steadystatesofholo}. Indeed, in this state we hope to find exotic behavior for the HRT surfaces, since the apparent horizons do not match at the membrane. While spacelike geodesics with two anchor points on the boundary cannot cross these, it would be conceivable for an HRT surface to cross the membrane outside one horizon, and emerge inside the other, see fig.\ref{fig:RTinsidehorizonsketch}. Indeed, while in the spinning string, a geodesic with two anchor points on the boundary cannot cross the apparent horizon, there are geodesics with a single anchor point on the boundary that do (see app.\ref{app:geodesicsinstringu}).

In fact, such an occurence seems almost inevitable as we consider a growing boundary interval containing the interface. As shown in fig. \ref{fig:sketchRTNess}, spacelike geodesics tend to approach the (apparent) horizon, and thus as they hit the membrane from the colder side, we would expect them to emerge behind the apparent horizon of the hotter side.

The other compelling reason to look at the entanglement structure of this state is to definitively prove that the interface indeed acts as a perfect scrambler, and to understand the entropy production from the gravity perspective. It is not immediately clear if such a thing can be computed within the stationary NESS state. Indeed, to compute the entropy production at the interface, one would naturally consider the entanglement entropy for an interval containing it. In the state preparation depicted in fig.\ref{fig:protocols2}, the shockwave due to the quench will alter the dual geometry, and the RT surface will be modified, resulting in a time-varying entanglement entropy for the interval. Then, taking the derivative of this quantity would yield the entropy production at the interface.

However, from the geometries computed in chap.\ref{chap:steadystatesofholo}, we have access only to the gray area of fig.\ref{fig:protocols2}, i.e. only the stationary state. Then, the entropy of an interval containing the interface will naturally be constant simply because the HRT surface computing its value will not have time-dependence. This is unavoidable by the very fact we are considering a stationary state. Still, in the microscopic model, even in the stationary state, the knowledge of the "S-matrix" of the interface should allow us to compute the entropy production. The question of whether this information is accessible from the minimal model under consideration is still open. 

One possible direction would be to consider the entropy (left and right-moving) currents which will naturally be constant in time. While the incoming currents are certainly thermal(they are prepared to be such) the outgoing currents from the interface can have a non-trivial \emph{spatial} profile, which tells us how different wave modes get scrambled after scattering\cite{Sonner:2017jcf,Novak:2018pnv,Ecker:2021ukv}. Therefore, looking at the behavior of the intervals entropy under a spatial variation of the endpoints might be a way to probe the entropy production at the interface. Note that at infinity these currents become also thermal, as we have verified by applying the HRT prescription for intervals far removed from the interface, and confirming the trivial surface does not intersect the membrane (thus recovering \ref{entanglemententropyspinning}, which shows that the dual state in the considered region is indeed thermal).

At any rate, all of that rests on our ability to compute said HRT surfaces. We will make a first attempt by mirroring the methods of the previous subsection. We will be considering the metric in Eddington-Finkelstein coordinates, as we expect the interiors of the black strings to play an important role here, unlike the static case. We remind the form of the relevant metric :
\begin{eqgroup}\label{finkelsteinmetricreminder}
ds^2=-\frac{(r^2-r_+^2)(r^2-r_-^2)}{r^2}dv^2+r^2(dy-\frac{{\rm sign}(J)r_-r_+}{r^2}dv)^2+2\ell dvdr \ ,
\end{eqgroup}
where we expressed it in terms of the location of the horizons $r_{\pm}$, see (\ref{hrformula}) for their definitions in terms of $M$, $J$.

Composing (\ref{nesstopoincare}) with (\ref{finkelsteinchangeofcoord}) we can find a change of coordinates mapping to Poincaré :
\begin{eqgroup}
    x_p+t_p=w_+&=\frac{r+r_+}{r+r_-}\exp(\frac{(r_+-r_-)(s y+v)}{\ell})\ ,\\
    x_p-t_p=w_-&=\frac{r-r_+}{r+r_-}\exp(\frac{(r_++r_-)(s y-v)}{\ell})\ ,\\
    z&=\frac{\sqrt{r_+{}^2-r_-{}^2}}{r+r_-}\exp(\frac{s y r_+-v r_-}{\ell})\ ,
    \label{finktopoinc}
\end{eqgroup}
where $s={\rm sign}(J)$, which we will take to be positive from now on. In (\ref{finktopoinc}), we chose the change of coordinates such that the geometry is mapped to the quadrants $w_+>0$, where the exterior is $w_->0$, and the interior $w_-<0$.

Like before, let us look what is the shape of the interface in these coordinates. By plugging $y=0$, $r=\infty$ we find that the interface in the new coordinates satisfies the equation :
\begin{eqgroup}\label{branepoincfromNESS}
    w_+^{r_++r_-}w_-^{r_+-r_-}=1 \ .
\end{eqgroup}

Of course we recover (\ref{firststepboundary}) when $J=0$ and $r_-=0$.

The next step in the procedure would be to find a global conformal transformation that maps (\ref{branepoincfromNESS}) to the interface $x=0$. Unfortunately, we are out of luck here; it can be shown that no such transformation exists. In a way, this outcome should have been expected from the get-go. Unlike the "pure" case with no membrane, here the non-equilibrium NESS state is not an equilibrium state in disguise. In hindsight, it would have been stranger to be able to map it to the static vacuum case, as that would imply that the NESS state we had found was merely an equilibrium situation expressed in peculiar coordinates.

That is however a serious setback in our hopes of computing RT surfaces in this setup. Indeed, without access to the beautiful Euclidean geometry description of sec.\ref{sec:vacuumICFT}, the problem at hand seems much harder to tackle even numerically, let alone analytically. In the next section, we outline some attempts at semi-numerical algorithms, which are still a work in progress.

\subsection{Other static geometries}
Let us finally briefly mention the possibility to map the vacuum ICFT case to the static geometries of chap.\ref{chap:phasesofinterfaces}. Here again, the question is disappointingly quickly answered by looking at the induced metric on the membrane. Indeed, as remarked before, the membrane metric in the vacuum case of sec.\ref{sec:vacuumICFT} is AdS$_2$; in particular, its Ricci scalar is a constant. For the solutions of chap.\ref{chap:phasesofinterfaces} (eq.(\ref{fullsolutionstatic})), this is not the case except when $M_1=M_2$. So  the analytical results of the vacuum ICFT case are applicable only to this very narrow subset of solutions.

For the $M<0$ case, the relevant mapping linking Poincaré (\ref{poincaremetric}) to global coordinates (\ref{EquilibriumAdSmetric}) :
\begin{eqgroup}\label{globaltopoincmap}
    &z_p = \frac{\ell^2\sqrt{-M}}{\sqrt{r^2-M\ell^2}\cos(t\sqrt{-M})+r \cos(x\sqrt{-M})}\ ,\\
    &t_p = \frac{\ell\sqrt{r^2-M\ell^2}\sin(\sqrt{-M} t)}{\sqrt{r^2-M\ell^2}\cos(\sqrt{-M}t)+r\cos(\sqrt{-M}x)}\ ,\\
    &x_p = \frac{\ell r\sin(\sqrt{-M}x)}{\sqrt{r^2-M\ell^2}\cos(\sqrt{-M}t)+r\cos(\sqrt{-M} x)}\ .
\end{eqgroup}

Note that we must constrain $z_p>0$, so that only a portion of the global space is mapped by (\ref{globaltopoincmap}). This change of coordinates is specifically adapted for the interface located at $x=0$ (the other interface is located at $x=\frac{\pi}{2\sqrt{-M}}$, as can be seen by integrating the solutions (\ref{fullsolutionstatic}) in the case $M_1=M_2=M<0$). This also explains the "disappearance" of one of the two interfaces under (\ref{globaltopoincmap}). Consider the Cauchy-Slice $t=0$; under (\ref{globaltopoincmap}), the full slice is mapped to Poincaré coordinates except for a single point, which is $x=\frac{\pi}{2\sqrt{-M}}$, the position of the second interface. The same transformation (\ref{globaltopoincmap}), but with translated $x$, would have mapped the two interfaces although they would not have the simple shape $x=cst$, which is needed if we want to apply the results of sec.\ref{sec:vacuumICFT}.

Since we can map the full Cauchy slice $t=0$ (except one point), the findings of sec.\ref{sec:vacuumICFT} are almost directly applicable to the global solutions. Of particular interest are remarks made about the connections of various RT surfaces with the critical tension $\lam_0$ (see the paragraph after fig.\ref{fig:plotofmubothsignspsi2}). Indeed, in the case where $M_1=M_2<0$, the sweeping transition happens precisely at $\lam=\lam_0$ (which does not hold when $M_1\neq M_2$). With speculation, this can lead us to two possible conjectures concerning the sweeping transition. In the following, "candidate RT surface" is an allowable spacelike geodesic in the RT prescription, which is not necessarily the minimal one.
\begin{itemize}
    \item Centerless slices admit no trivial candidate RT surfaces for sufficiently big boundary intervals (while centerful always do)
    \item Centerless slices admit only one crossing candidate RT surface for sufficiently big boundary intervals (while centerful admits two)
\end{itemize}
In the simpler case $M_1=M_2$, both conditions are satisfied. Unfortunately, the first conjecture can be immediately falsified. Indeed, according to the conditions (\ref{mucritsweeping}), we can have centerful solutions with tension $\lam<\lam_0$, given $\ell_1<\sqrt{2}\ell_2$. For these solutions, the near-boundary membrane has $\psi_2<0$. So for big enough boundary intervals, the trivial geodesic will intersect the membrane and cease existing, invalidating the first conjecture. One could amend the conjecture to link the disappearance of the trivial geodesic to the critical tension $\lam_0$, instead of the sweeping transition, but it is also unclear if this statement holds for $M_1\neq M_2$.

The second conjecture cannot be dismissed as easily, and one must again turn to numerical methods. This is a work in progress, as we are still in search of a satisfying all-encompassing algorithm as explained in the following section. 

\section{Numerical attempts}
We describe briefly in this section some attempts at obtaining HRT surfaces in these more complicated geometries with the use of numerical methods. The algorithms we will present were developed mainly for the use in the stationary NESS geometry, but with the scope of applying them also to the static geometries of chap.\ref{chap:phasesofinterfaces}. Ironically, the simplicity of the thin-brane model is actually the main difficulty in implementing numerical methods. Indeed, for a smooth metric standard shooting algorithms would presumably work to find the sought-out HRT surfaces.

Let us restate the problem to set notation. The spacetime in consideration is composed of two halves of locally AdS$_3$ spaces (labeled $1$ and $2$), which are joined through a codimension two membrane, whose embedding equation in each side $x^\mu_{m,i}(\xi^a)$ is known. Given any two spacelike separated points $p_1$ and $p_2$, located on an IR cutoff surface that is arbitrarily close to the boundary, we seek to find all possible spacelike geodesics connecting them. We have identified two different possible directions for the algorithms, which have both their pro and cons.

\subsection{Semi-Numerical algorithms}
In the case of a homogeneous spacetime, we have shown in sec.\ref{sec:geodesicinasympads} that we are able to find analytically the geodesics connecting two given points in the bulk. We can use that to greatly reduce the amount of numerics needed, hence the name "semi-numerical". When we need to connect two points in the same slice, we can consider that we have the geodesic path analytically (with some caveats we will mention), as long as we have an analytic change of coordinates mapping our metric to Poincaré space.

In what follows, we consider only possible non-trivial geodesics, i.e. geodesics that cross the interface. 

\subsubsection{The minimization method}
An allowable non-trivial geodesic will cross the membrane in such a way that the crossing constraints (\ref{crossingconstraint}) are satisfied. Consider first the case where the $p_i$ lie on different sides of the interface. In this algorithm, we begin by picking an arbitrary point on the interface parametrised by $\xi^a_*$, such that the corresponding points on both sides ($x^\mu_{m,i}(\xi^a_*)$) are spacelike separated from the boundary points $p_i$. Given that, we obtain a curve composed of two geodesic pieces joined at the interface. This path will in general not satisfy the crossing constraints, and that is where the numerics kick in.

The crossing equations (\ref{crossingconstraint}) will determine the correct $\xi^a_*$. Although those equations can be expressed completely analytically in our setups, it is in general not possible to solve them so. Thus, we use standard numerical methods, such as Newton-Raphson algorithm which needs the equation to be differentiable, or secant methods which do not, both of them being natively implemented in Mathematica through the "FindRoot" function.

In the case where $p_1$ and $p_2$ are on the same side, the algorithm is similar. The difference is that now our problem is parametrised by two arbitrary points on the membrane, and we obtain two sets of crossing equations. In this case, we need to solve twice as many algebraic equations, with twice as many free parameters. Note that we have additional conditions that force all chosen points to be spacelike separated from one another.

On paper, this method looks great; after all, it requires numerics only in the resolution of a set of algebraic equations. However, it has several important complications in practice. The first big problem is the "spacelike separated" condition. In poincaré coordinates, this condition is trivially verified by checking $-\Delta t ^2 +\Delta x^2+\Delta z^2>0$. In other coordinates, the conditions are usually non-trivial, and one has to employ the coordinate change to Poincaré to verify them. This is an important issue because it is difficult to implement these constraints into the numerical algorithms that solve the crossing equations.

Generically what tends to happen is that the algorithm attempts to evaluate the equations at a point $\xi^a_*$ which does not satisfy the constraints, crashing the search as we transition to timelike geodesics. One can mend this problem somewhat by "sprinkling" the interface with several initial points $\xi^a_*$, in the hopes of one falling close enough to the solution such that the numerical solver converges. This is of course not ideal, first of all because it increases significantly the computational cost, but mainly because one can never be certain if a solution was not found due to insufficient sprinkling, or due to it not existing at all. Additionally, selecting an efficient sprinkling is also non-trivial as the allowable regions are non-compact.

This problem disappears completely in the case where we consider static geometries. In this case we can restrict ourselves from the get-go to a specific Cauchy slice, which reduces the dimensionality of the problem, but more importantly removes the timelike direction. In this case, any two points chosen in this slice will be spacelike separated. Thus, this method seems more adapted to computation of RT surfaces in the geometries of chap.\ref{chap:phasesofinterfaces}.

\begin{figure}[!h]
    \centering
    \includegraphics[width=0.37\linewidth]{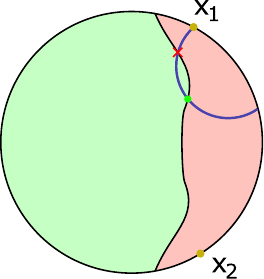}
    \caption{{\small Sketch of a problem that can occur when implementing numerical algorithms. In the figure, we aim to find a double crossing geodesics anchored at $x_1$ and $x_2$. Say in the minimization method we are looking for a geodesic traversing at the green point. The geodesic path crosses the membrane at the red cross, before reaching the green point, and thus there are no geodesics crossing at the green dot. Verifying that such edge cases do not happen generically is not an easy task.}}
    \label{fig:accidentalcrossing}
\end{figure}

Another, less prevalent but still important problem will be unwanted intersections. Because the spacetime is excised at the interface, there will be situations in which the geodesic crosses the interface in unintended places, see fig.\ref{fig:accidentalcrossing}. To avoid such unwanted solutions, we should check for spurious intersections with the membrane. This is actually a difficult problem, as the resulting equations are again not analytically solvable, requiring another numerical resolution. These constraints are difficult to reconcile with the numerical solving of the crossing equation, for the same reasons explained for the spacelike condition.

We have attempted to use this method in the static geometries of chap.\ref{chap:phasesofinterfaces}, with reasonable success in the case of symmetric intervals. It remains a work in progress however to iron out the "unwanted crossings" issue, as well as to analyse the resulting numerical curves. For this reason, we don't find it useful to present the very preliminary results that we have at hand.

\subsubsection{The shooting method}
In this second approach, we parametrise the candidate HRT surface by its initial conditions. We start on $p_1$, where we pick arbitrarily a spacelike unit vector, $\dot{p}_1$. This is again best done in Poincaré coordinates, where the unit condition (\ref{affinecondition}) is explicitly solvable. This yields an expression for the geodesic in terms of $(p_1,\dot{p}_1)$. The step here which (generally) requires a numerical resolution is the determination of the intersection with the membrane. One has two options to express this condition. This equation can be written down either in Poincaré coordinates (where presumably, the membrane equation is more complicated, but the geodesics are simpler), or in the coordinate system naturally adapted to the membrane solution. If the geodesic is parametrised with $\lam$, this equation yields $\lam^*$, the intersection point with the membrane, if it exists. If no solution is found, one draws another initial vector $\dot{p}_1$ and repeat the procedure.

Then, the crossing conditions (\ref{crossingconstraint}) determine new initial conditions $(p_m,\dot{p}_m)$ for the geodesic on the other side, which we can follow until we either reach the boundary, escape to infinity, or hit the membrane again and repeat the process. In principle, we obtain in this way a parametrisation of the final point $p_f$, in terms of the initial conditions $p_f(p_1,\dot{p}_1)$. What remains to be done is solve the equation $p_f(p_1,\dot{p}_1)=p_2$ for $\dot{p}_1$, which again can in general only be done numerically given the complexity of the function $p_f$.

The pros of this approach is that we do not have to worry about the "spacelike separated condition", as it is automatically satisfied by the choice of a spacelike tangent vector $\dot{p}_1$. Otherwise, it is a trade-off; instead of having to solve the crossing equations, we need to solve for the final point of geodesic. Whether this is advantageous or not depends on the specifics of the geometry at hand. 

Generally, we run in much of the same problems which are difficult to avoid with numerical solution-finding. First of all, we may miss the intersection of the geodesic with the membrane with an inappropriate initialisation value for $\lam$, which again would force us repeat the numerical resolution by sprinkling possible initial $\lam_0$. The other problem lies in the inversion of $p_f(p_1,\dot{p}_1)$. While it should be a continuous function, in attempts we made we found it to be extremely sensitive on the initial condition. Coupled to the fact that there are geodesics that escape to infinity (fig.\ref{fig:strangegeodesic}), this makes the numerical inversion of $p_f$ unreliable at best...

\subsection{Application to the interface NESS state}
In this small section we presents the attempts to apply the shooting method to the stationary states obtained in chap.\ref{chap:steadystatesofholo}. The reason we chose the shooting method over the minimization method is because in the first place, we wanted to search for geodesics like those depicted in fig. \ref{fig:RTinsidehorizonsketch}. As in this case the specific final point $p_f$ was not the first priority, the shooting method seemed the more appropriate.

Recall the membrane embedding (\ref{ansatz}):
\begin{equation}\tag{\ref{ansatz}} 
 x_j=x_{m,j}(\sigma), \ \ r_j=r_{j}(\sigma), \ \ t_j = \tau + f_j(\sigma)\ .
\end{equation} 
The derivative of the horizon entering membrane was exhibited in sec.\ref{sec:insideergoregion}:
\begin{eqgroup}\label{repeatNESSsolutionder}
x_{m,j}'(\si)  &= -  \ell_j\frac{(\lambda^2\pm\lambda_0^2)\sigma  + \lam J_i} {2(\sigma- \sigma_+^{\rm Hj} )(\sigma- \sigma_-^{\rm Hj}) \sqrt{A (\sigma - \sigma_-) }}\ ,\\
f_1'=f_2'&= \frac{1}{4\si}\left(J_1\ell_1 x_{m,1}'-J_2\ell_2x_{m,2}' \right)\ ,
\end{eqgroup}
the various constants are defined in sec.\ref{sec:insideergoregion}. Most importantly, recall that $\si=r_j^2-M_j\ell_j^2$.

We will be working in Eddington-Finkelstein coordinates (\ref{finkelsteinchangeofcoord}) :
\begin{equation}
dv  =  dt  +   \frac{\ell dr}{h(r)} \qquad {\rm and}\qquad  dy  = dx +  \frac{J\ell ^2 dr}{   2r^2  h(r)}  \tag{\ref{finkelsteinchangeofcoord}}\ .
\end{equation}

In the shooting method the main difficulty is the determination of the intersection with the membrane, so it is important to express it as simply as possible. Thus we will explicitely perform the integration of (\ref{finkelsteinchangeofcoord}) and (\ref{repeatNESSsolutionder}). In the specific case (\ref{repeatNESSsolutionder}) where the membrane enters the horizon, its world-volume becomes AdS$_2$\footnote{It is as of now unclear what is the condition that determines whether the world-volume of the membrane is AdS$_2$. It curiously appears to be the case whenever the membrane is dual to a \emph{single} static interface(i.e. of equation $x=cst$), as is the case for horizon-entering membranes or the ones appearing in the vacuum ICFT case. It would be interesting to explore whether this could be explained by the application of a Fefferman-Graham-like prescription on the worldvolume of the membrane.}, and $x_j'$ can be integrated in terms of elementary functions, instead of incomplete elliptic ones, which would add an additional numerical step to the computation. 

Integrating the coordinate change (with the convention that at the boundary it is vanishing), we find :
\begin{eqgroup}\label{integratedFinkelstein}
y &= x+\frac{J\ell^2}{4 (r_+^{2}-r_-^{2})}\left(\frac{\ln(\frac{|r-r_+|}{r+r_+})}{r_+}-\frac{\ln(\frac{|r-r_-|}{r+r_-})}{r_-}\right)\ ,\\
v &= t+\frac{\ell}{2 (r_+^{2}-r_-^{2})}\left(r_+\ln(\frac{|r-r_+|}{r+r_+})-r_-\ln(\frac{|r-r_-|}{r+r_-})\right)\ ,
\end{eqgroup}
where $r_\pm$ are the horizons. 

For the membrane, the equation is a little bit more involved :
\begin{eqgroup}\label{integratedNESSmembrane}
x_m(r) &= -\frac{|J|\ell^2}{2(r_+^2-r_-^2)}\left[\sign{J}\frac{g(r,\si^H_+)}{r_+}-\sign{(\lam^2\pm \lam_0^2)\si^H_-+J \lam}\frac{g(r,\si^H_-)}{r_-}\right]\ ,\\
f_1(r) &=-\frac{J^2}{4}\left(\frac{\ell_1^3}{2(r_{1+}^2-r_{1-}^2)}\left[\sign{J_1}\frac{g(r,\si^{H1}_+)}{r_{1+}\si_+^{H1}}-\sign{(\lam^2+ \lam_0^2)\si^{H1}_-+J_1 \lam}\frac{g(r,\si^{H1}_-)}{r_{1-}\si^{H1}_-}\right]+1\leftrightarrow 2\right)\ ,\\
g(r,a) &= \sqrt{a}\int_{\sqrt{-\si_-}}^{\sqrt{\si(r)-\si_-}}\frac{dt}{t^2-a}\\ &=-\frac{1}{2}\left(\ln\left[\frac{\sqrt{\si(r)-\si_-}+\sqrt{a-\si_-}}{\left|\sqrt{\si(r)-\si_-}-\sqrt{a-\si_-}\right|}\right]-\ln\left[\frac{\sqrt{-\si_-}+\sqrt{a-\si_-}}{\left|\sqrt{-\si_-}-\sqrt{a-\si_-}\right|}\right]\right)\ ,
\end{eqgroup}
where we use the shortcut $\si(r)=\sqrt{r^2-M\ell^2}$ when useful, and we omitted the indices denoting the side when possible. The convention here is that the membrane crosses the ergosphere at $x=0$ in the spinning string coordinates.

The membrane equation $(\tau+v_m(\si),r(\si),y_m(\si))$ in Eddington-Finkelstein coordinates is obtained by plugging (\ref{integratedNESSmembrane}) into (\ref{integratedFinkelstein}).
It becomes obvious from (\ref{integratedNESSmembrane}) that there is no hope of solving the membrane intersection equation analytically. Note that because the membrane embedding is invariant under $v$-translations we only need to solve one equation to find the intersection with the geodesic. Namely :
\begin{eqgroup}\label{intersectequation}
y_m(r_g(\lam^*))=y_g(\lam^*)\ ,
\end{eqgroup}
where $y_m$ denotes the interface embedding, and $r_g$, $y_g$ are the coordinates of the spacelike geodesic whose equation can be obtained by composing (\ref{eq:geodesicgeneralpositive}-\ref{eq:geodesicgeneralnull}) with (\ref{finktopoinc}). 

For the crossing equation, we define the normal and tangent vectors (omitting the index $j$ denoting the side):
\begin{eqgroup}\label{tangentandnormal}
t_\tau^\mu &= (1,0,0)\ ,\\
t_\si^\mu &= (v'_{m,j}(\si),\frac{1}{2\sqrt{\si+M\ell^2}},y_{m,j}'(\si))\ ,\\
n_\mu &= \frac{1}{\cal{N}}(0,-y_{m,j}'(\si),\frac{1}{2\sqrt{\si+M\ell^2}})\ ,
\end{eqgroup}
where $\cal{N}^2=n_\mu n^\mu$, and $n_\mu t^\mu_a=0$.

We can trivialize the crossing equations by expressing the geodesic tangent vectors in the basis (\ref{tangentandnormal}) :
\begin{eqgroup}\label{geodintangentbasis}
\dot{y}^\nu_{gi}(\lam^*) = a^\tau_i t_{i\tau}^\mu + a^\si_i t_{i\si}^\mu + b_i n^\mu_i\ .
\end{eqgroup}

With this parametrization, and for an affinely parametrized geodesic, the crossing equations reduce to :
\begin{eqgroup}
a_1^a a_1^b h^1_{ab}=a_2^a a_2^b h^2_{ab}\ ,
\end{eqgroup}
where $h^i_{ab}=t^\nu_{ia}t^\mu_{ib}g^i_{\mu\nu}$ is the induced metric on the interface. By the matching equations, $h^1_{ab}=h^2_{ab}$, and since the metric is non-degenerate we deduce that the crossing conditions simply force $a_1^a=a_2^a$, $b_1=b_2$ ! In other words, expressed in the appropriate basis (\ref{tangentandnormal}), the geodesic tangent vector is unchanged upon crossing.

With these equation in mind we applied the shooting algorithm described in the previous section which yields the numerical function $p_f(p_1,\dot{p}_1)$. The last remaining step, which is the (numerical) inversion of this function is until now unsuccessful, as most of the time the numerical search diverges, for the reasons mentioned in the previous section. By randomizing the initial condition, we were not able to find geodesics of the type depicted in fig. \ref{fig:RTinsidehorizonsketch}, although we found crossing geodesics that were outside both apparent horizons. Nonetheless, this does not yet show that these peculiar geodesics do not exist. In fact, we noticed that as we move $y_2$ far from the interface boundary, only very precise initial conditions yield geodesics that cross the membrane close to $r^+_2$ (refer to fig.  \ref{fig:RTinsidehorizonsketch}). Thus a random search would have a hard time finding those curves, which could explain why we haven't stumbled upon one.

The best way to search for such geodesics is actually to implement a hybrid of the shooting and minimization algorithm. One picks a point on the boundary, as well as a point on the membrane (which lies below one of the apparent horizons), and computes the geodesic connecting those two points analytically. Then, with the crossing condition, we obtain a shooting problem on the other side, and we repeat until that yields a geodesic with two boundary endpoints. This is currently being worked out.

\section{QNEC in ICFT setups}
The Quantum Null Energy Condition (QNEC) is an attempt to generalize the Null Energy Condition (NEC) which states that the stress-energy tensor of classical matter should obey $T_{\mu\nu}(x)v^\mu v^\nu\geq 0$, for any null vector $v^\mu$, and at any point $x$. Such conditions are introduced ad-hoc in General Relativity in an attempt to restrict the allowable stress-energy tensors to "realistic" ones, to prevent the appearance of unphysical spacetime geometries (but they also are an essential ingredient in many general theorems \cite{Bousso:2015mna}), and thus such conditions should hold for types of matter that we consider to be physical. As long as we consider only (reasonable) classical fields, this condition is satisfied.

Unfortunately, it is violated quantum mechanically when we apply it to $\langle T_{\mu\nu}\rangle v^\mu v^\nu$\cite{Epstein:1965zza}, as quantum effects such as the Casimir energy can produce negative energy densities locally. To restore the validity of these inequalities, what one usually does is to consider an averaged version, obtaining, in this case, the \textbf{A}veraged NEC (ANEC) \cite{Hartmananec:2016lgu,Faulkneranec:2016mzt} :
\begin{eqgroup}
\label{ANEC}
\int_C \langle T_{\mu\nu}\rangle v^\mu v^\nu d\lam \geq 0\ ,
\end{eqgroup}
where the integration is done over an integral curve of the null vector field $v^\mu$. The statement is that the inequality should hold for any null-vector field and any of its associated integral curves $C$. The ANEC then has a chance of surviving quantum effects, as they can produce local negative energy densities, but we expect the energy to remain positive when averaging them out for stable systems. The big disadvantage is that the condition is no longer local, and thus becomes much less powerful.

For this reason, there have been efforts to formulate a QNEC, namely a quantum version of the NEC which retains the locality property. The bound should of course accommodate for the fact that locally the energy could be negative. Such an inequality was conjectured by Bousso et. al\cite{Bousso:2015mna}, and proven using holographic techniques \cite{Koeller:2015qmn,Akers:2016ugt}, and later also directly from field theory techniques \cite{Balakrishnan:2017bjg,Ceyhan:2018zfg}. The QNEC is particularly interesting as it relates the energy density to variations of the entanglement entropy on an interval, which are not usually thought to be correlated quantities in non-gravitational systems. 

\begin{figure}[!h]
    \centering
    \includegraphics[width=0.4\linewidth]{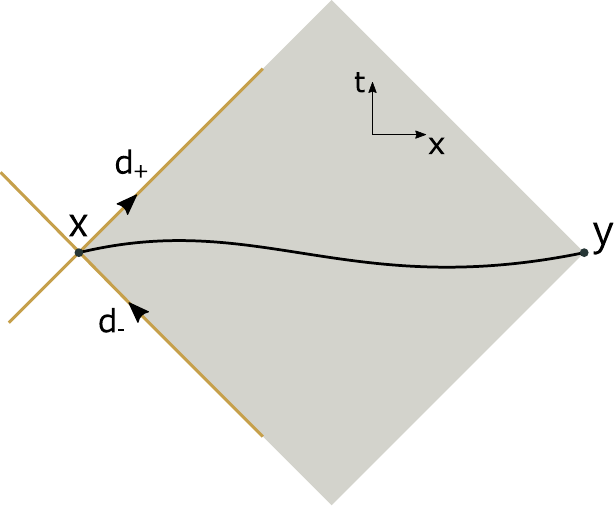}
    \caption{{\small The points $x$ and $y$ delimit a portion of a Cauchy slice, for which we can compute the entanglement entropy. The shaded region is the causal diamond, and any spacelike curve lying inside it and anchored at $x$ and $y$ will have the same entanglement entropy, as the state on these slices are related by unitary transformations. In yellow, we denote the lightlike direction under which we move $x$ do compute the derivatives used in the QNEC.}}
    \label{fig:cauchysliceatx}
\end{figure}

We will be interested in the case of a $2$-dimensional CFT on flat space, in which the QNEC takes a more restrictive form\cite{Wall:2011kb}. Consider a generic point $x$ in the geometry, for which we want to write the QNEC. We begin by picking a spatial slice ending at $x$, and with possibly another boundary at $y$ (see fig. \ref{fig:cauchysliceatx}). By embedding this slice inside a Cauchy slice, we can associate with it an entanglement entropy, which we denote $S(\rho,x,y)$, where $\rho$ explicits the dependence on the state of the CFT. We can then compute the change in $S(\rho,x,y)$ under variations of $x$ along the lightlike directions. The QNEC is the inequality :

\begin{eqgroup}\label{QNECinCFT2D}
2\pi \langle T_{\pm\pm}(x)\rangle \geq \pa_{\pm}^2 S(\rho,x,y)+\frac{6}{c}(\pa_{\pm} S(\rho,x,y))^2\ .
\end{eqgroup}

In particular, notice that the term involving the central charge $c$ is exclusive to CFTs, and makes the inequality stricter. For a QFT, this term is missing. The derivative $\pa_{\pm}$ refer variations of $x$, keeping $y$ fixed.

A consistency check is the verification that under a conformal transformation (which will change the energy density (\ref{TTransform})), the inequality conserves the same expression. Indeed, one can verify that the quantity on the RHS of (\ref{QNECinCFT2D}) has the same transformation properties as the stress-energy tensor\cite{Wall:2011kb}. Note also that (\ref{QNECinCFT2D}) is really a family of inequalities for $\langle T_{\pm\pm}(x)\rangle$, and the strictest one is obtained by maximizing the RHS w.r.t. $y$.

Consider as an example the case of the vacuum state, $\langle T_{\pm\pm}\rangle =0$. Consider any two spacelike separated points on the boundary, denoted $x = (x^+,x^-)$, $y=(y^+,y^-)$ in lightcone coordinates. They define a spacelike interval for which we compute the entropy. With a few boosts, we can generalize the formula (\ref{intervalentangentropypoincare}) to :
\begin{eqgroup}\label{entropyvacuumgeneral}
S({\rm vac},x,y)=\frac{c}{6}\log(\frac{|x^+-y^+||x^--y^-|}{\eps^2})\ .
\end{eqgroup}

One can easily verify that the bound of the QNEC is identically vanishing for (\ref{entropyvacuumgeneral}), for any choice of $y$. Thus, for the vacuum state, the QNEC is saturated and becomes an equality. If one could determine whether the QNEC is saturated without an explicit check, then one could determine the state given the entanglement structure (or vice-versa) and obtain a sort of "energy-entanglement" equivalence. This is reminiscent of bulk reconstruction, where in some sense the bulk geometry (and hence the bulk Stress-energy tensor) emerges from the entanglement structure of the dual CFT\cite{Dong:2016eik}. 

From the saturation of the QNEC in the vacuum state, one can then easily show that it will also saturate in any state that can be obtained by conformally mapping the vacuum. This follows directly from the fact that (\ref{QNECinCFT2D}) changes covariantly with conformal transformations. For instance, one can verify directly with (\ref{entanglemententropyspinning}) that the steady state, dual to the metric (\ref{finkelsteinmetricreminder}) saturates the QNEC everywhere. With the same logic, one can push the reasoning much further. 

The QNEC will be saturated for any state, as long as the dual RT surfaces do not cross bulk matter. Indeed, without matter, the bulk metric is locally AdS$_3$, so we can find a change of coordinates which brings us to the Poincaré metric, at least for the portion of spacetime that the RT surface traverses. Then the computation follows as in the vacuum case, and the QNEC is saturated. In fact, one can quantify the deviation from saturation in the case of perturbative matter in the bulk\cite{Khandker:2018xls,Ecker:2019ocp}. To be more precise, the aforementioned references look at the non-saturation of the inequality (\ref{QNECinCFT2D}) for some fixed $y$. However, we believe that the more interesting quantity to inspect would actually be the best possible QNEC bound, namely :
\begin{eqgroup}\label{thetrueQNEC}
    2\pi \langle T_{\pm\pm}(x)\rangle \geq {\rm Max}_{y}\left(\pa_{\pm}^2 S(\rho,x,y)+\frac{6}{c}(\pa_{\pm} S(\rho,x,y))^2\right)\ .
\end{eqgroup}

When considering the best bound as in (\ref{thetrueQNEC}), it is not clear anymore that bulk matter necessarily invalidates the saturation. In that context, the geometries that we have built-in chap. \ref{chap:phasesofinterfaces} and \ref{chap:steadystatesofholo} are the perfect playground to test such questions, as they provide a very simple form of matter in the bulk, the thin membrane. Computing the QNEC in these geometries is also an important check for the thin-brane approximation. Indeed, for intervals for which the associated RT surfaces cross the membrane, it is not obvious that the QNEC should hold generically. Were we to find cases where it is violated, it would indicate some breakdown of the holographic bottom-up model we considered.

\subsection{QNEC for static geometries}
As we have learned throughout this chapter, computing (H)RT surfaces on our stitched geometries is generically not an easy task. Thus it is not surprising that computing the QNEC bound offers the same challenges. Once one has a robust algorithm to compute HRT surfaces anchored at arbitrary boundary points, the QNEC follows easily, but as of now, it is still lacking. We will thus focus on geometries that can be brought back to the vacuum Poincaré case, described in detail in sec.\ref{sec:vacuumICFT}. 

Recall that for these geometries we have determined that the entanglement entropy S associated with a crossing RT surface on an interval bounded by $x=(\tau_x,\si_x)$ and $y=(\tau_y,\si_y)$ has the following structure :
\begin{eqgroup}\label{newentropy}
S &= \frac{c_i}{6} \ln(\frac{\si_x}{\eps})+\frac{c_{i}}{6}\ln(\frac{\si_y}{\eps})+g(\xi)\ ,\\
\xi &= \frac{-(\tau_x-\tau_y)^2+(\si_x-\si_y)^2}{4\si_x\si_y}\ ,
\end{eqgroup}
where the $c_i$ and $g(\xi)$, should be chosen accordingly with the location of the two points, see the formulas (\ref{fulllengthdoublecross}) and (\ref{lengthgeodboundaryinterfacecase}) for the specifics.

We exploit the form (\ref{newentropy}) to compute its lightlike derivatives without having to compute any more RT surfaces. We specify to the case "+" version of the QNEC, as the staticity of the state guarantees that both directions will be equivalent.

Consider first the case where $\si_x$ is on side 1, and $\si_y$ is on side 2. Expressing $\xi$ and $\si_x$ in terms of the lightlike coordinates $w_\pm^x$, we find :
\begin{eqgroup}\label{rewrittenQNECforpointone}
\frac{\pa^2 S}{(\pa w_+^x)^2}+\frac{6}{c}\frac{\pa S}{\pa w_+^x}= \frac{\pi(w_-^x-w_-^y)^2(w_+^x+w_+^y)^2}{32 \si_x^4 \si_y^2}\left[ g''(\xi)+\frac{12\pi}{c_1}g'(\xi)^2\right]\ .
\end{eqgroup}

In the case where $\si_x$ is on side 2, one simply has to change $c_1\rightarrow c_2$. Notice that the bound is independent of the cutoff $\eps$, and thus we can safely take the limit.

By scale invariance and time-translation, let us begin by fixing $\si_x=1$, $\tau_x=0$ on side 1, and consider (to reduce dimensionality) $y =(\tau_y=0,\si_y)$ on side 2. In the notation of sec. \ref{sec:intervalcontaininterface}, we have $\mu=\frac{\si_y}{\si_x}$. In our Euclidean construction, the natural variable is the angle $\al$, and thus properly speaking, we have $g(\xi)=g(\al(\xi))$. The formula (\ref{mualphasimplecrossing}) gives us $\al(\mu)$, and we can express this for the cross-ratio by the equality $\xi=\mu-2+\frac{1}{\mu}$. This gives :
\begin{eqgroup}\label{muofxi}
\mu(\xi)=1+2\xi\pm 2\sqrt{\xi(1+\xi)}\ .
\end{eqgroup}

The two possible solutions will give the same conclusion, as they amount to the inversion $\mu\rightarrow 1/\mu$ which is a symmetry of the problem. Thus in the end, we have $g(\xi)=g(\al(\mu(\xi)))$ where each of the functions is analytical, although extremely cumbersome. 

\begin{figure}[!h]
\centering
\begin{subfigure}[b]{0.68\textwidth}
\centering
   \includegraphics[width=0.8\linewidth]{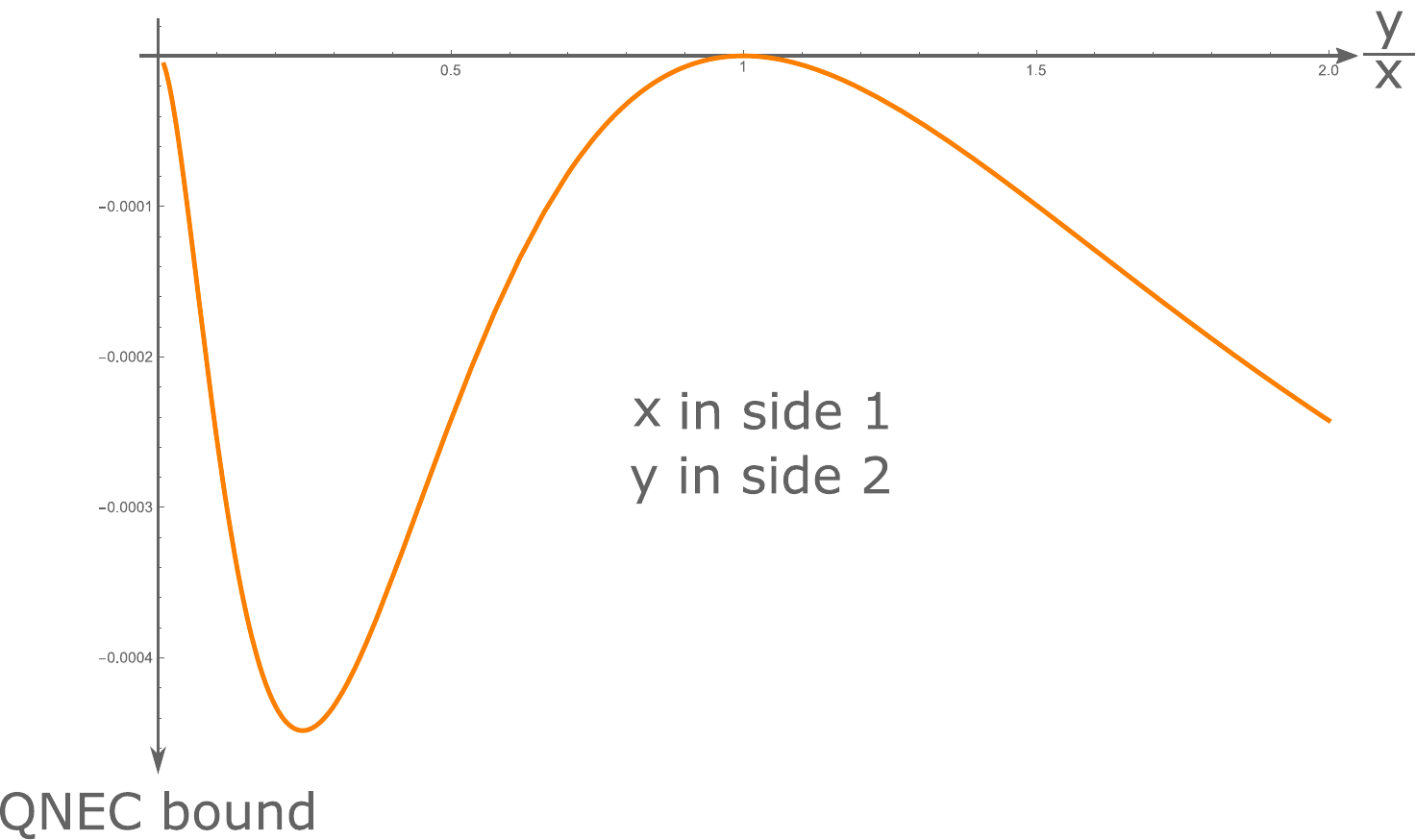}
   \caption{{\small QNEC bound plot for $x$ lying on side 1. The bound is saturated for the symmetric position, $y=x$.}}
\end{subfigure}

\begin{subfigure}[b]{0.68\textwidth}
\centering
   \includegraphics[width=.8\linewidth]{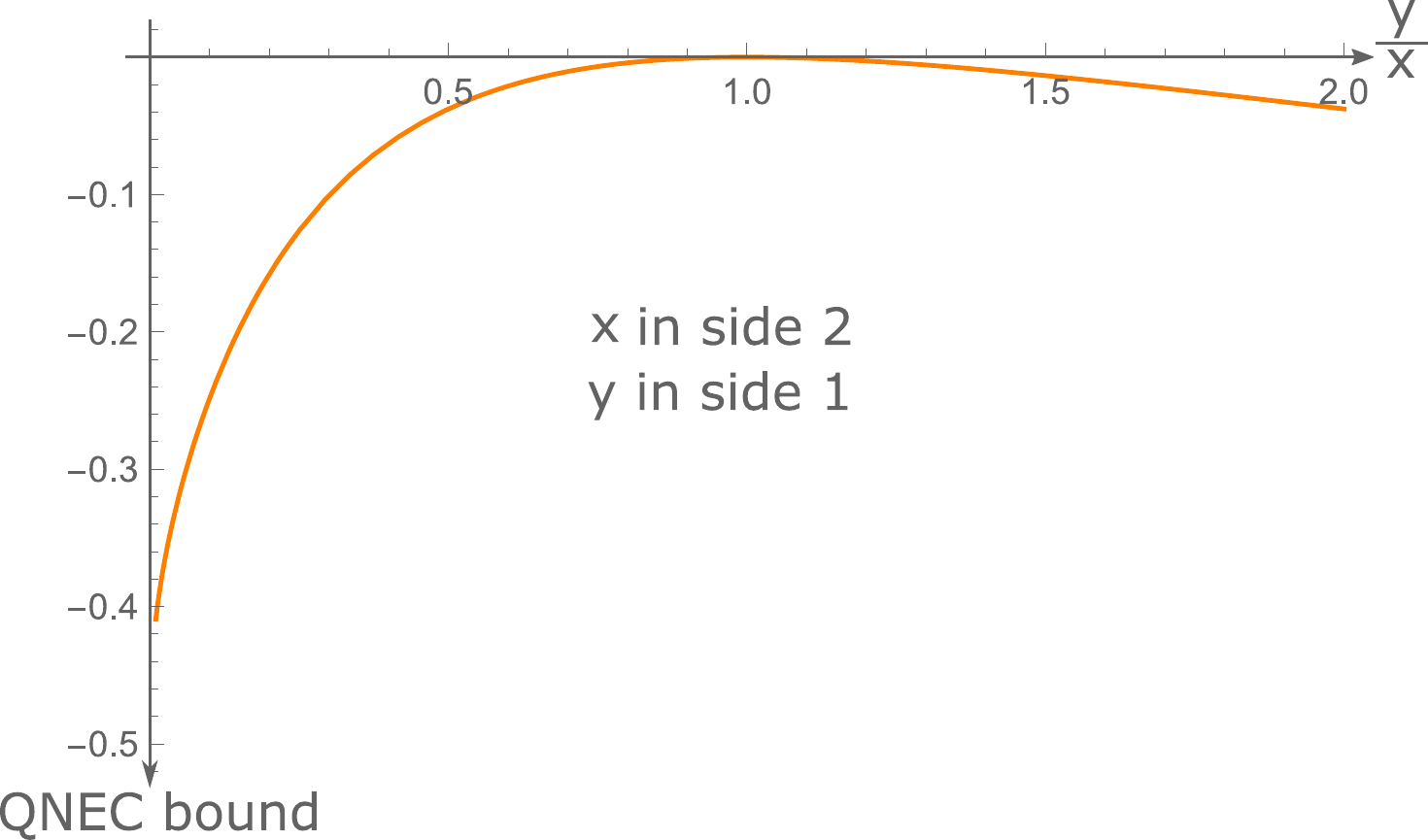}
   \caption{{\small QNEC bound plot for $x$ lying on side 2. The bound is saturated for the symmetric position, $y=x$.}}
\end{subfigure}

\caption{{\small Two plots of the QNEC bound, computed for boundary intervals that contain the interface of the ICFT, in the vacuum state. They are plotted as as function of varying $y$, having fixed $x$ (one can assume $x=1$ WLOG).}}
\label{fig:QNECboundssinglecrossing}
\end{figure}

In fig. \ref{fig:QNECboundssinglecrossing} we plot the QNEC bound for the crossing geodesic, as we vary the endpoint $y$ over the equal time slice.

Re-assuringly, the bounds obtained while varying the endpoint $y$ are all negative, which means that in the vacuum state the ICFT model is at least consistent with the QNEC. The interesting point is that we find a specific $y$ for which the bound is 0, meaning (\ref{thetrueQNEC}) saturates and this is despite the fact that the corresponding RT surface does cross bulk  matter. The $y$ in question is however rather special, as it is the point that lies symmetrically to $x$ with respect to the interface (equivalently, the point for which $\xi=0$).

The simple fact that (\ref{thetrueQNEC}) is saturated is actually unsurprising; indeed, in the interface models we consider one could always take $y\rightarrow x$, for which case the saturation is evident as we can always avoid the membrane. Nonetheless, it is noteworthy that we managed to find a saturating point even in the case where the RT surface crosses bulk matter. Whether this saturation holds more generally, or it is just due to the increased symmetry of the location of $y$, is yet to be determined.

Consider now the limits as the endpoint $y$ approaches the interface, from side $2$ (x in side $1$). We see the QNEC approaching $0$, and this can be understood assuming the smoothness of the entanglement entropy in this limit. After $y$ crosses the interface, the two points lie on side $1$ and the RT surface is the trivial one, giving the saturating bound. As $y$ approaches the boundary, we see the QNEC reaching this bound.

The story is different when $y$ lies on side $1$ ($x$ on side $2$). As $y$ approaches the interface, we see the QNEC reaches a non-saturating bound. Indeed, as $y$ crosses the interface, the dominating RT surface will not be the trivial one, but rather the double-crossing one depicted in fig. \ref{fig:doublecrossinggeodesic}. Such RT surfaces will not in general saturate the QNEC bound, and as such neither is the limit of the bound as $y$ tends to the interface.

To continue the analysis, the natural next step is to actually compute the QNEC for the two points $x$, $y$ lying on side 2. The computation goes similarly to what we have just shown, the only tricky step being the problems with the analytic form of $\al(\mu)$, see fig. \ref{fig:exampleofswitcheroo}. Again, we unfortunately haven't finished the work needed to present clear results so this is added to the list of works in progress. This also goes for the computation of the QNEC in more complicated states, as it cannot be done without first having a firm grasp on the RT construction.

\section{Closing remarks}
This section was dedicated to presenting partial results regarding the entanglement structure of the various ICFT states that were obtained in the previous chapters. The initial goal of exploiting the Euclidean geometry construction of sec.\ref{sec:vacuumICFT} and reducing all the entropy computations to this case was unsuccessful, which forced us to consider numerical methods. Nonetheless, we showed that this does allow full analytic control for static geometries, in which the induced metric of the membrane is AdS$_2$. 

Efforts in the numerical direction were centered around the main idea to exploit the AdS$_3$-locality of the considered spacetimes. By locally mapping them to Poincaré coordinates, we limit the necessity for numerical resolution only at a few points. For now, the results obtained are still lacking, and the priority for future work will be to iron out the details that prevent the successful application of the algorithms. We believe that understanding the entanglement structure of the ICFT NESS will not only offer insights into the curious entropy production at the interface, but also regarding the properties of non-killing, out-of-equilibrium horizons.

Finally, the recovery of the QNEC bound in these systems follows almost automatically from the entanglement structure, so it is a natural quantity to consider. Besides being a consistency check, of particular interest is the question of its saturation.  Indeed, some authors have claimed that for CFT in $D\geq 3$, the QNEC is always saturated \cite{Leichenauer:2018obf,Casini:2022rlv}. This fails in $2$D as we have noticed in our models, but it is a tantalizing question whether the saturation still holds for the best bound (\ref{thetrueQNEC}). This question deserves further study.

\chapter*{Conclusion and outlook}
The main focus of the work presented in this thesis was the study of a simple model describing a minimal Interface CFT and its holographic dual, which in the large N and 't Hooft coupling limit reduces to Einstein gravity with a gravitating membrane. The initial motivation to consider such a model was their intimate connection with the (then) recent progress on the black hole information paradox\cite{Penington:2019npb,Almheiri:2019yqk}. Indeed, the examples in which the Island formula can be applied usually involve Boundary CFT duals, which can be obtained as a limiting case from ICFT.

The simpler case of the $3$-dimensional bulk was initially introduced as a stepping stone in preparation for the study of full Supergravity solutions. However, the model proved to be much richer than expected, such that it became the main focus of this work. Furthermore, studying minimal models of this type has the advantage of yielding more generally applicable results. This broadens the scope of the results we obtain here to fields like condensed matter theory\cite{Zaanen:2015oix,Hartnoll:2009sz,Adams:2012th,Erdmenger:2020fqe}, where minimal holographic models are often used to allow the computations at strong coupling, not easily accessible from the field theory. Finally, having access to non-trivial holographic models which allow for analytic results provides a nice playground to test and further our understanding of AdS/CFT.

Aiming at applications in the black hole information problem, in chap. \ref{chap:phasesofinterfaces} we considered the ICFT model at finite temperature whose gravity duals include black holes. Surprisingly, the sole presence of the gravitating membrane produced a very rich phase diagram for this system. While our analysis was thorough within this minimal model, it identified several interesting directions that deserve future study. 

One key issue is the validity of the thin-brane approximation and how our results would fare in full top-down ICFT/gravitating wall pairs. Another approximation of our treatment is that we discarded bulk solutions involving membrane fusion (fig. \ref{fig:nonsmoothjoining}) even though such configurations sometimes appeared naturally in cases where the membrane intersected itself. While they should not alter the conclusions on the phase space drastically, this more complete model could exhibit some interesting properties of the pair of interfaces. 

A related question concerns the "exotic fusion" geometries in which, as we bring the two interfaces together, they do not fuse into the trivial defect (fig. \ref{fig:bubblesolution}). Their mere existence is quite surprising\cite{Brunner:2007td,Brunnerother:2013ora}, and it might be related to the "membrane fusion" geometries mentioned above. This is because deeper regions of the AdS bulk are supposed to probe the IR region of the dual field theory. As we go to the IR the two interfaces approach one another and eventually appear as one, "fusing". Thus, an allowable non-trivial membrane fusion in the bulk might signal the presence of a potential exotic interface fusion in the field theory.

One last problem that deserves future work is the elucidation of the meaning of the sweeping transition (sec. \ref{sec:sweepingtransition}) from the boundary theory. Our strong feeling is that they must be connected to the entanglement structure of the field theory, as it is the quantity that is related to the bulk reconstruction. We have arrived at some conjectures in chap. \ref{chap:entanglemententropyandholoint}, but they are still very tentative. Related to this problem is the role of the critical tension $\lam_0$. It is still unclear if this is simply a quantity that appears in the computations, or whether it carries a deeper meaning. 

In chap.\ref{chap:steadystatesofholo} we looked at an extension of the static states of chap. \ref{chap:phasesofinterfaces} to more general non-equilibrium stationary states. This is of interest both to confirm the work that was already done from the field theory perspective \cite{Liu:2018crr,Bernard:2014qia,Meineri:2019ycm} and extend it, but also to generate tractable non-equilibrium black hole solutions, which are still poorly understood. These goals were met. First, our calculation confirmed the universality of the energy-transport coefficients shown in ICFT in \cite{Meineri:2019ycm}. Secondly, it offers a much simpler way to compute these coefficients on the gravity side, when compared to the original scattering calculations of \cite{Chapman:2018bqj}. Lastly, perhaps the least expected result was the discovery of the maximal entropy production at the interface and the analytical recovery of the deformed, non-killing event horizon on the gravity side.

These results naturally lead to the fascinating question of the relation of the deformed horizon with entanglement entropy. We touched on this issue in chap.\ref{chap:entanglemententropyandholoint} and hope to complete our partial results in the near future. The well-known Bekenstein-Hawking formula, which has been re-derived with the RT prescription in AdS/CFT, has been mainly applied to black holes with static killing horizons. In this model, we have an example of an (analytically controlled) event horizon that is neither Killing, nor coinciding with its "apparent" counterpart. It is the perfect setup to elucidate the role played by these various horizons in holography. 

One other question that needs elucidation concerns the thermal transmission of the interface pairs. For a single interface, we recovered results previously obtained perturbatively\cite{Bachas:2020yxv}, but in the case of more than one interface, there appeared two distinct phases. When the membrane avoids the black hole and curves back to the boundary, the conductance suddenly becomes perfect. This behavior certainly pertains to some sort of coherent scattering, which allows all the thermal flux through as if there were no interfaces. The precise way in which this happens in the field theory is however unclear and deserves further study.

The final chapter \ref{chap:entanglemententropyandholoint} ties the project together as it brings us back to the initial goal of connecting our models to the Island computations. We have already mentioned that the computations of RT surfaces in the obtained geometries was the obvious next step in their study, and that is what is undertaken in this chapter. Based on the vacuum computations of \cite{Anous:2022wqh} which we present, we outline different angles (analytical and numerical) to tackle computations of the sought-out RT surfaces.

While some interesting results were obtained, we haven't yet managed to reach the initial goal: showing the existence of horizon probing RT surfaces in the non-equilibrium case, and understanding the entropy production at the interface. The main focus of future work will be the completion of these goals. Finally, another tangent direction that was mentioned is the computation of the QNEC bound for these same geometries. Its application to the ICFT model will cement its consistency, while potentially providing interesting insights on the necessary conditions for its saturation. The QNEC becomes much more interesting when one considers the Wick-rotated version of our models \cite{Simidzija:2020ukv,PhysRevmukopad}, in which the interface now plays the role of a quench. It could be interesting to see which restrictions it imposes if any, and whether it tells us something about the non Wick-rotated version.

Although much was accomplished, much still remains to be worked out, as is always the case in physics.

\appendix
\chapter{Misc}
\label{chap:misc}
\section{Extrinsic curvature}
\label{sec:ExtrinsicAppendix}
Consider a surface of codimension one embedded in a manifold of dimension $D$. It is parametrized by (D-1) parameters denoted $s^a$. In a Lorentzian manifold, one should differentiate according to the nature of the hypersurface (lightlike, spacelike or timelike). Unless otherwise stated, we assume here that the surface is timelike. Generally, its equation can be written as 
\begin{eqgroup}
x^\mu = x^\mu(s^a)\ ,
\label{surfaceequation}
\end{eqgroup}
where $x^\mu(s^a)$ are $D$ functions specifying the surface's shape. We can then naturally define a set of $D-1$ tangent vectors to the surface :
\begin{eqgroup}
t_a^\mu = \frac{\pa x^\mu}{\pa s^a}\ ,
\end{eqgroup}
which allows us to define the induced metric :
\begin{eqgroup}
h_{ab}=g_{\mu\nu} t_a^\mu t_b^\nu\ ,
\label{tangentmetric}
\end{eqgroup}
where $g_{\mu\nu}$ is the metric of the ambient spacetime. To define the normal covector $n_\mu$, one can use the generalized cross-product :
\begin{eqgroup}
n_\mu = \frac{1}{2}\eps_{\mu \nu_1\ldots \nu_{D-1}}\eps^{a_1\ldots a_{D-1}}t^{\nu_1}_{a_1} \ldots t^{\nu_{D-1}}_{a_{D-1}}\ ,
\label{normalvectorcrossproduct}
\end{eqgroup}
where the $\eps$ symbols are the levi-civita tensors corresponding to the metric $g$ and $h$ (with the appropriate normalisation with $\sqrt{|g|}$). Using (\ref{normalvectorcrossproduct}), the normal vector is normalized, $n^\mu n_\mu = 1$. (For a lightlike surface, $n^\mu n_\mu = 0$, and for a spacelike surface, $n^\mu n_\mu = -1$).

In the case where the surface can be defined in terms of an equation $f(x^\mu)=0$, there is a much simpler way to obtain the normal co-vector :
\begin{eqgroup}
 n_\mu = \frac{\pa_\nu f}{\sqrt{\pa_\nu f \pa^\nu f}}\ ,
 \label{easynormal}
\end{eqgroup}
which is obtained by varying the defining equation along tangent directions. The normal vector of course satisfies :
\begin{eqgroup}
 n_\mu t^\mu_a =0\ .
\end{eqgroup}

Using it, we can define the projection operator on the surface :
\begin{eqgroup}
 \Pi^\mu_\nu = \delta^\mu_\nu - n_\nu n^\mu
 \label{projectionop}
\end{eqgroup}

Note that formally, $n_\mu$ and hence $\Pi^\mu_\nu$ are defined only on the surface $x^\mu(s^a)$. Thus they are not vector fields on the ambient manifold. One can however arbitrarily extend them to all of the spacetime; this allows us to more easily define operations involving them, and in the end, it will not matter since we will evaluate everything on the surface.

The projection operator can be applied to any tensor to isolate its components tangent to the surface. For instance, applied on the metric :
\begin{equation}
    h_{\mu\nu}=\Pi^\al_\mu \Pi^\be_\nu g_{\al\be}=\Pi_{\mu\nu}\ .
    \label{metricisprojection}
\end{equation}
This is another way to see the metric (\ref{tangentmetric}), as expressed in the embedding coordinates. One can verify we have of course $h_{ab}=t_a^\mu t_b^\nu h_{\mu\nu}$. Thus, the projection tensor and the surface metric are essentially interchangeable.

We have all we need to define Extrinsic curvature :
\begin{eqgroup}
 K_{\mu\nu}= \Pi^\al_\mu \Pi^\be_\nu \co_\al n_\be(=\Pi^\al_\mu \co_\al n_\nu) \label{Extrinsic1}\ .
\end{eqgroup}

Although it is not immediately obvious, one can show that $K_{\mu\nu}$ is symmetric. This comes from the fact that locally, we can always write (\ref{easynormal}) by the implicit function theorem. In this form, $n_\mu$ is locally hypersurface orthogonal, namely it satisfies $n_{[\rho} \co_\mu n_{\nu]}=0$. Contracting this with $n^\rho$ yields $K_{[\mu\nu]}=0$ as required. 

Since $K_{\mu\nu}$ is a tensor defined only on the surface, it is easier to work in the $s^a$ coordinates :
\begin{eqgroup}
 K_{ab}=t^\mu_a t^\nu_b K_{\mu\nu} = t^\mu_a t^\nu_b \co_\mu n_\nu\ .
 \label{Extrinsic2}
\end{eqgroup}

To compute $\co_\mu n_\nu$, one has to extend the vector field $n_\nu$ to the full embedding space. Alternatively, expanding (\ref{Extrinsic2}) :
\begin{eqgroup}
 K_{ab}=t^\mu_a t^\nu_b (\pa_\mu n_\nu + \Gamma^\rho_{\mu\nu} n_\rho)=t^\nu_b \pa_a n_\nu + t^\mu_a t^\nu_b  \Gamma^\rho_{\mu\nu} n_\rho\ .
 \label{goodwaytocomputeK}
\end{eqgroup}

With the notation (\ref{goodwaytocomputeK}), one does not need to go through the trouble of extending $n_\nu$; indeed, the embedding space derivative has been "absorbed" by $t_a^\mu$, so that we need only take derivatives in the tangent directions, in which $n_\mu$ is well-defined.

The trace of the extrinsic curvature that will enter the Gibbons-Hawking action is then simply $K = K_{ab}h^{ab}= K_{\mu\nu}h^{\mu\nu}$.

\chapter{Thermal holographic interfaces}
\section{Extrinsic curvature formulas}
\label{app:appendixextrinsiccomputationstatic}
In this section we collect some formulas for the extrinsic curvature of the membrane, in the parametrization:
\begin{eqgroup}\label{staticmembraneparametrisation}
x^\mu_m(\si,\tau)=(\tau,\sqrt{\si-M\ell^2},x_m(\si))\ ,
\end{eqgroup}
where the order of coordinates is $(t,r,x)$, and the metric (\ref{labeledmetricsstatic}). The tangent and normal (co)vector then are :
\begin{eqgroup}
t_\tau^\mu &= (1,0,0)\ ,\\
t_\si^\mu &=(0,\frac{1}{2\sqrt{\si-M\ell^2}},x_m'(\si))\ ,\\
n_\mu &= \frac{1}{\cal{N}}(0,-x_m'(\si),\frac{1}{2\sqrt{\si-M\ell^2}})\ ,
\end{eqgroup}
where $\cal{N}=\frac{1}{\ell}\sqrt{\frac{(r')^2}{r^2} + (-M\ell^2 + r^2) (x')^2}$ is the normalisation of the normal vector. Note that we write $\sqrt{\si-M\ell^2}=r(\si)\equiv r$ where useful.

Formula (\ref{goodwaytocomputeK}) gives us the Extrinsic curvature :
\begin{eqgroup}\label{extrinsiccurvaturestatic}
K_{\tau\tau}&= \frac{r (r^2-M\ell^2)x'(\si)}{\ell\sqrt{\frac{r'{}^2\ell^2}{r^2}+(r^2-M\ell^2)x'^2}}=\frac{r^2x'}{\ell}\times\sqrt{\frac{\si}{g(\si)}}\ ,\\
K_{\si\tau} &= 0\ ,\\
K_{\si\si} &= \frac{x' r'{}^2\ell^2(3\si+M\ell^2)+r\si x'(r\si x'{}^2+\ell^2 r'')+\frac{1}{2}\ell^2 \si  x''}{\ell \si\sqrt{\si g(\si)}}\ ,
\end{eqgroup}
where all $r$ should be seen as function of $\si$. Note we used the shorthand $g(\si)$ defined in (\ref{matching2}). Then writing down the traced-reversed Israel condition (\ref{tracereverseIsrael}), for the $\tau \tau$ component, we find :
\begin{eqgroup}
\frac{r^2_1x'_1}{\ell_1}+\frac{r^2_2x'_2}{\ell_2}=-\lam \sqrt{\si g(\si)}\ ,
\end{eqgroup}
which is indeed the equation (\ref{extrinsicIsraelequationstatic}) used in the main text. The $\si\si$ Israel condition is much more complex as seen from the lengthy formula (\ref{extrinsiccurvaturestatic}) for $K_{\si\si}$. Nonetheless, one can write it down and plug in the solution (\ref{fullsolutionstatic}), and after a computer-aided simplification, it indeed evaluates to $0=0$, such that it is redundant with the other equation.

\section{Renormalized on-shell action}
\label{app:renormonshellactoin}
We reproduce here the Euclidean action of the holographic-interface  model, in units $8\pi G=1$. It is the sum of  bulk, brane, boundary and corner contributions, (see \cite{Takayanagi:2019tvn} for an explanation of the corner term)
\begin{eqgroup}\label{fullEuclideanAction}
S_{\rm gr} =    -\frac{1}{2}& \int_{{\mathbb  S}_1}d^3x 
\sqrt{g_1}(R_1+\frac{2}{\ell_1^2})  -\frac{1}{2}
 \int_{{\mathbb  S}_2}d^3x\sqrt{g_2}(R_2+\frac{2}{\ell_2^2}) +   \lam \int_{\mathbb   W} d^2s
     \sqrt{h_m} \nonumber  \\
    &  -\int_{\partial {\mathbb  S}_1} d^2s \sqrt{h_1} K_1 
     -\int_{\partial {\mathbb  S}_2} d^2s \sqrt{h_2} K_2 
     +  \int_{\rm C} (\theta - \pi)  \sqrt{h_c} + {\rm  c.t. }\ ,
\end{eqgroup}
 where the counterterms,  abbreviated  above by c.t.,  read \cite{Balasubramanian:1999re}
\begin{eqgroup}
  {\rm c.t.} =   \frac{1}{\ell_1}\int_{\mathbb{B}_1} \sqrt{h_1}  
  +  \frac{1}{\ell_2}\int_{\mathbb{B}_2} \sqrt{h_2}  - 
  \int_{\mathbb{B}_1\cap \mathbb{B}_2} (\theta_1+\theta_2)  \sqrt{h_c} \ .
\end{eqgroup}

We remind that $\mathbb  S_j$   are the  spacetime  slices whose boundary is the sum of the asymptotic boundary  $\mathbb{B}_j$ and  of the string worldsheet $\mathbb{M}$, i.e. $\partial \mathbb  S_j = \mathbb{B}_j \cup\mathbb{M}$. 

The induced metrics are denoted by the letter $h$. The $K_j$  are the traces  of the extrinsic curvatures  on  each  slice computed with the outward-pointing normal vector. Finally, in addition to the standard  Gibbons-Hawking-York boundary terms, one must  add the Hayward  term  \cite{Hayward:1993my, Takayanagi:2019tvn} at   corners of   $\partial \mathbb  S_j$  denoted by C.\,\footnote{These play no role  here, but they can be important in the case of string junctions.}

There is at least one such  corner at  the cutoff surface,  $\mathbb{B}_1\cap\mathbb{B}_2$, where $\theta -\pi$ is the sum of the angles $\theta_j$ defined in figure \ref{fig:nearboundarymem}.

Let us break the action into an interior and a  conformal boundary  term, $S_{\rm gr} = S_{\rm int} + S_{\mathbb{B}}$, with  the former including contributions from the worldsheet $\mathbb{M}$.  Using  the field equations $R_j = - {6/\ell_j^2}$ and $K_1\vert_{\mathbb{M}}+K_2\vert_{\mathbb{M}}  = 2\lam$, and  the volume elements that follow from  \eqref{labeledmetricsstatic}    and (\ref{matching1}, \ref{matching2}), 
\begin{equation}\sqrt{g_j}d^3x   =\ell_j r_j dr_j dx_j dt   \quad    {\rm and} \quad 
  \sqrt{h_m}d^2s =  \sqrt{fg}  d\sigma dt\ ,
\end{equation}

We can write the interior on-shell action as follows\,: 
\begin{equation}\label{interioraction}
 S_{\rm int}  =     \frac{2}{\ell_1}  \int_{{\Omega}_1} r_1\,   dr_1  dx_1 dt 
  + \frac{2}{\ell_2}  \int_{{\Omega}_2} r_2   dr_2  dx_2 dt   -
\lambda \int_{\mathbb{M}} \sqrt{fg} d \sigma dt \ .
 \end{equation}
We have been  careful to distinguish the spacetime slice S$_j$ from the   coordinate chart $\Omega_j$, because we will now use Stoke's theorem  treating $\Omega_j$ as part of flat Euclidean space, 
\begin{eqgroup}\label{Stokes}
   \sum_{j=1,2}  \frac{2}{\ell_j}  \int_{{\Omega}_j} r_j   dr_j  dx_j   =  \sum_{j=1,2} 
 \frac{1}{\ell_j} \oint_{{\partial \Omega}_j} r_j^2 (\hat r_j\cdot d\hat n_j) \ ,
 \end{eqgroup}
with $d\hat n_j dt$ the surface element on the boundary  $\partial \Omega_j$, and $\hat{r}_j$ the unit vector in direction of increasing $r_j$. Crucially, the boundary of $\Omega_j$ may  include a horizon which is a  regular interior submanifold  of the Euclidean spacetime and is not therefore part of $\partial \mathbb  S_j$. In particular, there is no Gibbons-Hawking-York  contribution there. 
 
The boundary integral in \eqref{Stokes} receives contributions from the three  pieces of ${\partial \Omega}_{1,2}$ : the asymptotic cutoff surface  $\mathbb{B}_1\cup\mathbb{B}_2$, the horizon if there is one, and the worldsheet $\mathbb{M}$.  

Consider first the contributions from the asymptotic cutoff and horizon. The normal vector $\hat{n}_j$, in this case, is simply $\pm\hat{r}_j$, so the contribution is proportional to the size $\Delta x$ of the boundary. For the membrane, the surface element in the Euclidean space $\Omega_j$ is :
\begin{eqgroup}\label{surfaceelementmembrane}
  d\hat{n}_j = (-x_j'(\si),r_j'(\si))\ ,
\end{eqgroup}
where here this is the outgoing normal vector for one of the two membrane pieces. Both of them will contribute equally to the integral, we integrate on them in opposing directions, while also having opposing normal vectors. Combining this together yields :
\begin{eqgroup}
 \frac{1}{\ell_j} \oint_{{\partial \Omega}_j} r_j^2 (\hat r_j\cdot d\hat n_j)=\frac{r_j^2 L_j}{\ell_j}-\frac{r_{h\,j}^2\Delta x_j\Bigl\vert_{Hor}}{\ell_j}-\int_{\si\in\Sigma}\frac{r^2_j(\si)x_j'(\si)}{\ell_j}\ .
\end{eqgroup}
In the last term, $\Sigma$ denotes the two "half-membranes", i.e. twice the interval $\si \in [\si_+,\infty]$.

Very conveniently, this last term precisely cancels the third  term in  \eqref{interioraction} by virtue of the Israel-Lanczos equation \eqref{extrinsicIsraelequationstatic}. In this way, we fortunately do not have to bother with integrals on the brane worldvolume.

Thus, after all the dust has settled,  the  action can be written as  the sum of terms evaluated either at   the black-hole horizon or  at the cutoff. After integrating over periodic time (which simply contributes to a prefactor of $\frac{1}{T}$ in front of the expressions) the interior part of the action, \eqref{interioraction},  reads
\begin{eqgroup}\label{interioractioncomputed}
S_{\rm int}  =  \frac{r_{1}^2 L_1}{\ell_1T}-\frac{M_1\ell_1\Delta x_1\vert_{\rm Hor}}{T}+ \frac{r_{2}^2 L_2}{\ell_2T}-\frac{M_2\ell_2\Delta x_2\vert_{\rm Hor}}{T}\ .
\end{eqgroup}
If the slice $\mathbb  S_j$ does not contain a horizon the corresponding contribution is absent, $\Delta x_1\vert_{\rm Hor}=0$. We wrote $r_{j}$ for the cutoff radius, which is to be sent to infinity at the end. Note that we replaced $r_{hj}=M_j\ell_j^2$, the horizon radius.

We now   turn to the conformal-boundary contributions from the lower line in the action \eqref{fullEuclideanAction}. For a fixed-$r_j$  surface, the   outward-pointing unit normal expressed as a 1-form   is {\bf n}$_j =   dr_j /\sqrt{r_j^2 - M_j \ell_j^2}$. Dropping the index $j$ for simplicity, one finds after a little algebra (virtually identical to the computations of sec.\ref{sec:HawkingPage})  :
\begin{eqgroup}
   K_{xx} = K_{tt} = - \frac{r}{\ell}  \sqrt{r^2 - M\ell^2}  \Longrightarrow
   \sqrt{\hat g} K =  - \frac{1}{\ell} (2r^2 - M\ell^2) \ .
\end{eqgroup}

Combining the Gibbons-Hawking-York terms and the counterterms gives

\begin{eqgroup}
S_{\mathbb{B}} =  \frac{1}{\ell_1 T} ( r_1 \sqrt{r_1^2 - M_1\ell_1^2} - 2r_1^2 +  M_1\ell_1^2)\Delta x_1\Bigl\vert_{\mathbb{B}_1}+(1 \leftrightarrow 2) \ .
\end{eqgroup}

Expanding  for large  cutoff radius, $r_j\vert_{\mathbb{B}_j} \to \infty$,   and dropping the terms that vanish in the limit we obtain

\begin{eqgroup}\label{boundaryactionresult}
  S_{\mathbb{B}} =  \frac{1}{\ell_1 T} (  - r_1^2 +  \frac{1}{2} M_1\ell_1^2) 
  L_1+(1 \rightarrow 2) \ .
\end{eqgroup}
Upon adding up    \eqref{interioractioncomputed} and \eqref{boundaryactionresult} the leading divergent term cancels, giving the following result for the renormalized on-shell action: 
\begin{eqgroup}\label{finalI}
 S_{\rm gr}  =   \frac{M_1\ell_1}{2 T}  \bigl( 
L_1   -  2  \Delta x_1 \bigl\vert_{{\rm Hor}} \bigr) +\frac{M_2\ell_2}{2 T}  \bigl(
L_2   -  2  \Delta x_2 \bigl\vert_{{\rm Hor}} \bigr) \ .
\end{eqgroup}
We used here the fact that  $\Delta x_j\vert_{{\mathbb{B}}_j} = L_j$,  and that $r_j^2 = M_j\ell_j^2$ at the horizon when  one exists. We also used implicitly  the fact that for smooth strings the Hayward term receives  no contribution  from  the interior and is removed by the counterterm at the boundary. 
   
 
As a check of this on-shell action  let us compute the  entropy. Using our   formula for the internal energy $\langle E\rangle  = \frac{1}{2}  (M_1 \ell_1 L_1 + M_2 \ell_2 L_2)$, and  $ S_{\rm gr} = F/T=\langle E\rangle/T  - S$  we find
\begin{eqgroup}
 S &=  \frac{1}{T}\bigl(M_1 \ell_1  \Delta x_1 \bigl\vert_{{\rm Hor}} + 
 M_2 \ell_2  \Delta x_2 \bigl\vert_{{\rm Hor}}\bigr) \ , \\
&= 4\pi^2 T \bigl(  \ell_1  \Delta x_1 \bigl\vert_{{\rm Hor}} +  
 \ell_2  \Delta x_2 \bigl\vert_{{\rm Hor}}\bigr)
 = \frac{\cal{A}({\rm horizon})}{4G} \ .
\end{eqgroup}
In the lower line we used the fact that  $M_j = (2\pi T)^2$ and $r_j^{\rm H} = 2\pi T \ell_j$ for slices  with horizon, plus our choice of  units $8\pi G=1$. The  calculation thus reproduces correctly the  Bekenstein-Hawking entropy.


\section{Opening arcs as  elliptic integrals}
 \label{app:2}
 In this appendix we   express the opening arcs, (\ref{Dira}-\ref{Dirc}), in terms of complete  elliptic integrals of  the first, second  and third kind, 
\begin{align}
  {\rm\bf K}(\nu) &= \int_0^1 \frac{dy}{\sqrt{(1-y^2)(1-\nu y^2)}}  \ ,\\
  {\bf E}(\nu)&= \int_0^1\frac{  \sqrt{1-\nu y^2}\,dy }{ \sqrt{1-y^2}} \ ,\\
  {\bf \Pi}(u,\nu)&= \int_0^1\frac{dy}{(1- uy^2)\sqrt{(1-y^2)(1-\nu y^2)}}\ . \\
\end{align}
 Consider the boundary conditions\eqref{Dira}. The other conditions (\ref{Dirb},\ref{Dirc}) differ only  by the  constant periods or horizon arcs, $P_j$ or $\Delta x_j\vert_{\rm hor}$. 

Inserting  the expression  \eqref{fullsolutionstatic} for $x_1^\prime$   gives 
\begin{eqgroup}
   L_1 = - \int_{\sigma_+}^\infty \frac{\ell_1 \,d\sigma }{(\sigma+M_1\ell_1^2)}\, 
    \frac{(\lambda^2+ \lambda_{0}^2)\,\sigma    +M_1-M_2}
    {\sqrt{ A \sigma  (\sigma-\sigma_+)(\sigma-\sigma_-)}} \ ,  
\end{eqgroup} 
and likewise  for  $L_2$.  The roots $\sigma_\pm$ are given by (\ref{abc},\ref{spm}). We assume that we are not in the  case  {\small [H2, H2]} where $M_1=M_2>0$, nor in the fringe  case $\sigma_+ = - M_j \ell_j^2$ when  the string goes through an AdS center. These cases will be treated separately. 

Separating  the integral in two parts,  and trading  the integration variable $\sigma$ for  $y$,  with $y^2 := \sigma_+/\sigma$, we obtain 
\begin{eqgroup}
    L_1 =  - \frac{2\ell_1}{\sqrt{A\,\sigma_+}}\bigg[ \frac{M_1 - M_2}{M_1\ell_1^2} 
   \int_0^1 \frac{dy}{\sqrt{(1-y^2)(1-  \nu
     y^2  )}}&\ ,\\
    +\Bigl(  (\lambda^2+ \lambda_{0}^2)  -   \frac{M_1 - M_2}{M_1\ell_1^2}  \Bigr) \int_0^1 \frac{y^2 dy}{
    (1-  u_1y^2) \sqrt{(1-y^2)(1-  
     \nu y^2  )}}  \bigg]&\ ,
\end{eqgroup}
where $\nu =   \sigma_-/\sigma _+ $ and $u_1 = - M_1\ell_1^2/\sigma_+$. Identifying the elliptic integrals finally gives
\begin{eqgroup}\label{ellipticnormal} 
 L_1 =  - \frac{2\ell_1}{\sqrt{A\sigma_+}}\bigg[ \frac{M_1-M_2}{M_1\ell_1^{\,2}}
 \Bigl( {\rm\bf K}(\nu) - {\bf \Pi}(u_1, \nu)\Bigr)  + (\lambda^2+ \lambda_{0}^2) {\bf \Pi}(u_1, \nu)\bigg]\ ,
\end{eqgroup}  
 and a corresponding expression for $L_2$ 
\begin{eqgroup}
 L_2 =  - \frac{2\ell_2}{\sqrt{A\,\sigma_+}}\bigg[ \frac{M_2-M_1}{M_2\ell_2^{\,2}}\, 
 \Bigl( {\rm\bf K}(\nu) - {\bf \Pi}(u_2, \nu)\Bigr)  +
  (\lambda^2 - \lambda_{0}^2)\, {\bf \Pi}(u_2, \nu)\bigg]\ ,
\end{eqgroup}             
with $u_2 = - M_2\ell_2^2/\sigma_+$.  The prefactors in \eqref{ellipticnormal} diverge when $M_1\to 0$ but the singularity is removed by expanding ${\bf \Pi}(u_1, \nu)$ around $u_1=0$. In this limit

\begin{eqgroup}
   L_1(M_1=0) & = - \frac{ 2\ell_1}{\sqrt{A\,\sigma_+}}
   \bigg[ \frac{M_2}{\sigma_-}({\rm\bf E}(\nu)-{\rm\bf K}(\nu))
   + (\lambda^2 +\lambda_0^2) {\rm\bf K}(\nu) \bigg]\ ,\\
   L_2(M_2=0) & = - \frac{ 2\ell_2}{\sqrt{A\,\sigma_+}}
   \bigg[ \frac{M_1}{\sigma_-}({\rm\bf E}(\nu)-{\rm\bf K}(\nu))
   + (\lambda^2 - \lambda_0^2) {\rm\bf K}(\nu) \bigg]\ ,
\end{eqgroup}                  
with  ${\rm\bf E}(\nu)$   the complete elliptic integral of the second kind. 
    
The  $M_1=M_2>0$  geometries  correspond  to the high-temperature phase  where    $M_j  = (2\pi T)^2$,   $\sigma_+=0$ and  $\sigma_- =  - (4\pi T \lambda)^2/A$. The integrals \eqref{Dirc}  simplify to elementary functions in this case. This seems to be related to the fact that in this case, the membrane metric becomes AdS$_2$.: 
\begin{eqgroup}
  L_1 - \Delta_1^{\rm Hor}=  - \frac{\ell_1 (\lambda^2+ \lambda_{0}^2) }{\sqrt{A  \vert \sigma_- \vert} }
  \int_0^\infty \frac{ds}{(s + a)\sqrt{s+1}}=- \frac{\ell_1 (\lambda^2+ \lambda_{0}^2) }{\sqrt{A  \vert \sigma_- \vert} }\frac{2}{\sqrt{1-a}}{\rm arctanh}(\sqrt{1-a})\ ,
\end{eqgroup}
with $a = A\ell_1^2/4\lambda^2$. Using the expression \eqref{abc} for $A$, and going through the samesteps for $j=2$,  the expression greatly simplifies :
\begin{eqgroup}\label{arctanh}
  L_1 - \Delta_1^{\rm Hor}  &=   - \frac{1}{\pi T}\, {\rm tanh}^{-1} \left( \frac{\ell_1  (\lambda^2 + \lambda^2_0)}{2\lambda}\right)\ , \\
  L_2 - \Delta_2^{\rm Hor} &=   - \frac{1}{\pi T}\, {\rm tanh}^{-1}  \left( \frac{\ell_2 (\lambda^2 -  \lambda^2_0)}{2\lambda}\right) \ .
\end{eqgroup}  

Interestingly,  since $\Delta_2^{\rm Hor}$  must be  positive,  $T L_2$ is bounded from below in the range  $\lambda < \lambda_0$ as discussed in section \ref{sec:highTphase}. 
  
In the high-temperature  phase the on-shell action,  \eqref{finalI},   reads
 \begin{eqgroup}
 I_{\rm gr}^{\rm (high-T)} =  4 \pi^2 T\Bigl[ -\frac{1}{2} 
  (\ell_1 L_1 + \ell_2 L_2) +  \ell_1 (L_1-  \Delta_1^{\rm Hor})+  \ell_2 (L_2-  \Delta_2^{\rm Hor})\Bigr]\ .
 \end{eqgroup}
 
 Using  the expressions \eqref{arctanh} and rearranging the arc-tangent functions gives
 \begin{eqgroup}
  I_{\rm gr}^{\rm (high-T)} :=   \frac{E}{T}  - S  =  - 2\pi^2 T(\ell_1 L_1 + \ell_2 L_2)  - \log\, g_I\ ,
 \end{eqgroup}
 where  the interface entropy  $S=\log\, g_I$ is given by \eqref{gfactor}. By this re-arrangement we are thus able to recover the interface entropy in a somewhat roundabout way, which nonetheless confirms the correctness of our solutions.
 
 
\section{Sweeping is continuous}
\label{app:3}
In this appendix, we show that sweeping  transitions are continuous. We focus for definiteness on the  sweeping of the $j=2$ AdS  center  at zero temperature (all other cases work out   the same). The transition  takes place when $\mu$  crosses   the critical value $\mu_2^*$  given by eq.\,\eqref{mucritsweeping}. Setting $\mu = \mu_2^*(1 - \delta)$ in expression \eqref{mub} gives

\begin{eqgroup}
f_2(\mu) = \frac{\ell_2}{ \sqrt{A}} 
\int_{s_+}^\infty ds 
\frac{(\lam^2-\lam_0^2)(s-\mu \ell_2^2)\,+ \delta }
{(s - \mu \ell_2^{ 2})\sqrt{A s ( s- s_+)(s-s_-)} }\ ,
\end{eqgroup}

\begin{eqgroup}\label{Japp}
= \frac{2\ell_2(\lam^2-\lam_0^2)}{\sqrt{As_+}}\,{\bf K} \bigl(\frac{s_-}{s_+}\bigr)\,+\,\frac{\ell_2 \,\delta}{\sqrt{A}}\,
\underbrace{\int_{s_+}^\infty
 \frac{ds }{(s-\mu \ell_2^2)\sqrt{s (s- s_+)(s-s_-)}}}_{J}\ . 
\end{eqgroup}
The first term is continuous at $\delta=0$, but the second requires some care because the integral $J$  diverges. This is because for  small $\delta$ 
\begin{eqgroup}
 s_+  -   \mu \ell_2^2 =\  \frac{\delta^2}{4 \lam^2 \mu_2^*  }+\cal{O}(\delta^3)\ , 
\end{eqgroup}
as one finds by explicit computation of  the expression \eqref{s+-cold}. we set $\delta=0$,  $J$ diverges near the lower integration limit. To bring the singular behavior to $0$ we perform the change of variable $u^2 =s -s_+$, so that 
\begin{eqgroup}
 J = \int_0^\infty \frac{2 du}{(u^2 +  \delta^2/4\lam^2 \mu_2^* )\sqrt{ (u^2+s_+^*)(u^2 +s_+^*-s_-^*)}}\ , 
\end{eqgroup}
where we kept only the leading order in $\delta$,  and $s_\pm^*$ are the roots at $\mu=\mu_2^*$. Since $s_+^*$ and $s_+^* -s_-^*$ are positive and finite, the small-$\delta$ behavior of the integral is (after rescaling appropriately $u$)
\begin{eqgroup}
J  =  \frac{4 \lambda \vert \mu_2^* \vert }{\vert\delta\vert \sqrt{s_+^*(s_+^*-s_-^*)}}
 \underbrace{\int_0^\infty \frac{du}{u^2+1}}_{\pi/2} 
+ {\rm finite}      \ .
\end{eqgroup}
Inserting in expression \eqref{Japp} and doing  some  tedious algebra leads finally to a discontinuity  of  the function   $f_2(\mu)$ equal to   sign$ (\delta) \pi/\sqrt{\mu_2^*}$.  This is precisely what is required  for $L_2$, \eqref{dirichletHorizonless}, to be  continuous when  the red ($j=2$) slice goes from type {\small E1} at negative $\delta$ to type {\small E2} at positive $\delta$. 

This, however, does not tell us anything about the continuity of higher derivatives of $L_2(\mu)$. The expansions that one needs to make become exponentially bigger, and it becomes nigh impossible to keep track of all the terms, as even Mathematica struggles to perform the simplifications. However, we made a numerical analysis with several numerical parameters. What this analysis revealed is that the continuity of the Free energy seems to extend at least to the third derivative (going deeper produced numerical instabilities which made it hard to conclude). For this reason, it is safe to say that the transition is completely smooth in terms of Free energy and its derivatives. It is thus a phase transition of another nature.

\section{Bubbles exist}
 \label{app:bubbles}
We show here that the bubble phenomenon of section \ref{sec:Faraday} is indeed realized in a region of the parameter space of the holographic model.

This is the region of non-degenerate gravitational vacua ($\ell_2$ strictly bigger than $\ell_1$) and a sufficiently light domain wall. Specifically,  we  will show that for  $\lambda$ close to its minimal value,  $\lambda_{\rm min}$,  the arc $L_1(\mu=0)$ is negative, so the wall self-intersects and $\mu_0$ is necessarily finite. 

Let  $\lambda =  \lambda_{\rm min}(1+  \delta)$  with $\delta \ll 1$. Setting $\mu=0$ and expanding eqs.\,\eqref{s+-cold} for small  $\delta$ gives
\begin{eqgroup}
A = \frac{8 \lambda_{\rm min}^2\, \delta  }{ \ell_1\ell_2} +\cal{O}(\delta^2)  , \quad 
 s_+ =  \frac{\ell_2}{4 \lambda_{\rm min}}  +\cal{O}(\delta)  , 
\quad 
 s_-  = - \frac{ \ell_1 }{ \ 2 \lambda_{\rm min} \delta } +\cal{O}(1)\ .
\end{eqgroup}
Plugging into (\ref{ellipticnormal})  with $ M_2 = \mu M_1\approx 0$ we find :
\begin{equation}
\sqrt{\vert M_1\vert}  
L_1 =  -\frac{2 }{\ell_1 \sqrt{A s_+ }}\left[ {\bf K}\left(\frac{s_-}{s_+}\right) +  (1-\frac{2\ell_1}{\ell_2})
{\bf \Pi}\left({\ell_1^2}{s_+},\, \frac{s_-}{s_+}\right)\right]\ ,
\end{equation}
where we have only kept leading orders in $\delta$. Now we need the asymptotic form  of the elliptic integrals when their argument diverges
\begin{equation}
        {\bf K} \left[-\frac{a}{\delta}\right] \approx   {\bf \Pi}\left[u ,-\frac{a}{\delta}\right] \approx 
 -\frac{\ln(\delta)\sqrt{\delta}}{2\sqrt{a}}+\cal{O}(\sqrt{\delta}) \ ,
\end{equation}
for $\delta\to 0_+$ with $u, a$ fixed. Using $a = 2\ell_1/\ell_2$ finally gives
\begin{eqgroup}
  \sqrt{\vert M_1\vert} L_1 \approx \bigl(\frac{\ell_2}{\ell_1}-1\bigr)^{1/2} \ln(\delta) +
{\rm subleading} \ .
 \end{eqgroup}
 
For  $\delta \ll 1 $ this is negative, proving our claim. Note that we took the green slice to be of type {\small E2}, as follows from our analysis of the sweeping transitions for light domain walls -- see  section \ref{sec:sweepingtransition}.

\chapter{Steady holographic interfaces}
\section{Extrinsic curvature formulas}
\label{app:appendixextrinsiccomputationNESS}
In this section we collect some formulas for the extrinsic curvature of the membrane, in the parametrisation :
\begin{eqgroup}\label{NESSmembraneparametrisation}
x^\mu_m(\si,\tau)=(\tau+f(\si),\sqrt{\si-M\ell^2},x_m(\si))\ ,
\end{eqgroup}
where the order of coordinates is $(t,r,x)$, and the metric (\ref{metricspinningstring}). The tangent and normal (co)vector then are :
\begin{eqgroup}
t_\tau^\mu &= (1,0,0)\ ,\\
t_\si^\mu &=(f'(\si),\frac{1}{2\sqrt{\si-M\ell^2}},x_m'(\si))\ ,\\
n_\mu &= \frac{1}{\cal{N}}(0,-x_m'(\si),\frac{1}{2\sqrt{\si-M\ell^2}})\ ,
\end{eqgroup}
where $\cal{N}=\sqrt{\frac{(r')^2}{r^2}\left(\frac{\si}{h(r)}\right) + h(r) \frac{(x')^2}{\ell^2}}$ is the normalisation of the normal vector. Note that we write $\si-M\ell^2=r^2(\si)\equiv r^2$ where useful, and we used the definition (\ref{hrformula}) for $h(r)$.

Note that the tangent vector is independent from $f(\si)$, this is expected as it does not change the geometry of the membrane embedding. It will however have an impact on the matching conditions.

Formula (\ref{goodwaytocomputeK}) gives us the Extrinsic curvature :
\begin{eqgroup}\label{extrinsiccurvatureNESS}
K_{\tau\tau}&= \frac{r x'(\si) h(r)}{\ell^2\sqrt{\frac{r'{}^2\si}{r^2 h(r)}+\frac{h(r)x'^2}{\ell^2}}}=\frac{r^2 x'(\si) h(r)}{\ell\sqrt{\si}\sqrt{|\hat{g}|}}\ ,\\
K_{\si\tau} &= \frac{\frac{r f'(\si)x'(\si)h(r)}{\ell^2}-\frac{J\ell x'(\si)}{2rh(r)}}{\sqrt{\frac{r'{}^2\si}{r^2 h(r)}+\frac{h(r)x'^2}{\ell^2}}}=\frac{\frac{r^2 x' h(r)f'}{\ell}-\frac{J\ell^2 x'}{2h(r)}}{\sqrt{\si}\sqrt{|\hat{g}|}}\ ,
\end{eqgroup}
where all $r$ should be seen as function of $\si$. Note we used the shorthand $\hat{g}={\rm det}(\hat{g})$ defined in (\ref{det}). We did not include the equation for $K_{\si\si}$ as it is very cumbersome, and is not used anywhere in the main-text.

Then writing down the traced-reversed Israel condition (\ref{tracereverseIsrael}). We indeed find (\ref{Isr1}) and (\ref{Isr2}) as in the main-text. As for the $(\si \si)$ equation, it can be shown with alot of handwork that it is indeed satisfied when the other two are.
\section{Horizon inequalities}
\label{app:horizonineq}
In section \ref{sec:nonkillinghorizon} we  asserted  that  BTZ  geometries  whose ergoregions can be  glued together by  a thin brane obey the inequalities 
\begin{eqgroup}\label{orderapp} 
  \sigma_+^{{\rm H}1} &>  \sigma_+^{{\rm H}2}    \quad {\rm if} M_1 > M_2\ ,\\ 
  \sigma_+^{{\rm H}2} &<  \sigma_+^{{\rm H}1}    \quad {\rm if}  M_1< M_2    \ ,
 \end{eqgroup}  
where the horizon locations are  
\begin{eqgroup}\label{sHapp}
\sigma_{\pm}^{{\rm H}j} =  - \frac{M_j\ell_j^2}{2}   \pm  \frac{1}{2}\sqrt{M_j^2 \ell_j^4 -J^2 \ell_j^2 }  \ ,
\end{eqgroup}  
and $J  \equiv  \vert J_1\vert = \vert J_2\vert >0$. This ordering of the outer horizons is  manifest if one expands at the  leading order for  small $J$. We  want to show  that it is  valid  for all  values of   $J$. 

If as $J$ is cranked up  the ordering was at some point reversed, then we would have $\sigma_+^{{\rm H}1} =   \sigma_+^{{\rm H}2} $, or equivalently  
\begin{eqgroup}\label{equalityhorizons1}
 M_2\ell_2^2 - M_1\ell_1^2 =  \sqrt{ M_2^2 \ell_2^4 - J^2\ell_2^2} - \sqrt{ M_1^2 \ell_1^4 - J^2\ell_1^2} \ . 
\end{eqgroup}
Squaring twice to eliminate the square roots gives
\begin{equation} \label{equalityhorizons2} 
    J^{2} = \frac{4\ell_1^2\ell_2^2(M_1-M_2)(M_2\ell_2^2-M_1\ell_1^2)}{(\ell_2^2-\ell_1^2)^2} \ .
\end{equation}
Without loss of generality we  assume, as elsewhere in the text, that $\ell_1 \leq \ell_2$. If $M_2>M_1$, then automatically $M_2\ell_2^2 > M_1\ell_1^2$ and   \eqref{equalityhorizons2} has no solution for real $J$. In this case the ordering  \eqref{orderapp} cannot be reversed. 

If on the other hand $M_1>M_2$ and $M_2\ell_2^2-M_1 \ell_1^2 >  0$  we need to  work harder. Inserting $J^2$  from  (\ref{equalityhorizons2})  back in the original equation \eqref{equalityhorizons1} gives after rearrangements
\begin{eqgroup}
    (\ell_2^2-\ell_1^2) (M_2\ell_2^2-M_1 \ell_1^2)  =  &\ell_2^2\bigl|
  (M_2\ell_2^2-M_1\ell_1^2) -  \ell_1^2(M_1-M_2)
    \bigr|\ , \\
    &- \ell_1^2\left| (M_2\ell_2^2-M_1\ell_1^2) -\ell_2^2(M_1-M_2) \right| \ ,
\end{eqgroup}
where the absolute values come from the square roots. This equation is not automatically obeyed whenever its doubly-squared version is. A solution  only exists if 
\begin{equation}
    M_1-M_2\leq \frac{\ell_2^2M_2-\ell_1^2M_1}{\ell_2^2}
    \label{eq:existenceineq}  \Leftrightarrow \frac{M_1}{M_2}\leq \frac{2\ell_2^2}{\ell_2^2+\ell_1^2} \ .
\end{equation}
Remember now  that we only  care about solutions with walls in the ergoregion, for which $M_1-M_2= \lam J$, see eq.\eqref{sigma+0}. Plugging  in \eqref{equalityhorizons2} this gives
\begin{equation}
     M_2 =   \bigl[ 1 - \frac{4\lam_0^2\\lam^2  }{\lam_0^{4} +4 {\lam^2}/{\ell_1^2}} \bigr] M_1
    \label{eq:JsolT}\ ,
\end{equation}
with $\lambda_0^2 = (\ell_2^2 - \ell_1^2)/\ell_1^2\ell_2^2$, see eq.\eqref{criticaltensions}. Consistency  with the bound \eqref{eq:existenceineq} for a brane tension in the  allowed range then requires
\begin{equation}
  \lambda_{\rm min} <   \lam\leq  \frac{\ell_1\lam_0^2}{2}\ ,
\end{equation}
where $ \lambda_{\rm min} = (\ell_2-\ell_1)/\ell_1\ell_2$. As one  can easily check,  this implies  $\ell_1 > \ell_2$ which  contradicts our initial assumption. We conclude that \eqref{equalityhorizons1} has no solution, and   the ordering \eqref{orderapp} holds for all $J$. 

For completeness, let us also consider the ordering of the inner horizons. Clearly $\sigma_+^{{\rm H}j} >    \sigma_-^{{\rm H}j}$ always, and for small $J$  also  $\sigma_+^{{\rm H}1} >    \sigma_-^{{\rm H}2}$ and $\sigma_+^{{\rm H}2} >    \sigma_-^{{\rm H}1}$. To violate  these last inequalities we need  $ \sigma_+^{{\rm H}1} =   \sigma_-^{{\rm H}2}$ or $\sigma_+^{{\rm H}2} = \sigma_-^{{\rm H}1}$ for some finite $J$, or equivalently 
\begin{eqgroup}\label{equalityinnerhorizons1} 
 M_2\ell_2^2 - M_1\ell_1^2 =  \mp \bigl( \sqrt{ M_2^2 \ell_2^4 -  J^2\ell_2^2} + \sqrt{ M_1^2 \ell_1^4 - J^2\ell_1^2} \,\bigr)  \ . 
\end{eqgroup}
 Squaring twice gives back  eq.\eqref{equalityhorizons2} which has no solution if $M_2>M_1$.  But if  $M_1>M_2$ and $M_2\ell_2^2-M_1 \ell_1^2 >  0$,  solutions to $ \sigma_+^{{\rm H}2} =   \sigma_-^{{\rm H}1}$ cannot be ruled out. Indeed, inserting $J$ from \eqref{equalityhorizons2} in \eqref{equalityinnerhorizons1} with the $+$ sign gives 
\begin{eqgroup}
    (\ell_2^2-\ell_1^2) (M_2\ell_2^2-M_1 \ell_1^2)  =  &\ell_2^2 \bigl|
  (M_2\ell_2^2-M_1\ell_1^2) -  \ell_1^2(M_1-M_2)
    \bigr|\ , \\
    & + \ell_1^2\left| (M_2\ell_2^2-M_1\ell_1^2) -\ell_2^2(M_1-M_2) \right| \ ,
\end{eqgroup}
which requires that 
\begin{eqgroup}
  \frac{\ell_2^2M_2-\ell_1^2M_1}{\ell_2^2} \leq   M_1-M_2\leq \frac{\ell_2^2M_2-\ell_1^2M_1}{\ell_1^2} \ .
\end{eqgroup}

These conditions are  compatible with  $M_1-M_2= \lam J$ and $\lam$ in the allowed range, so the outer horizon of slice 2 need not  always come before  the Cauchy horizon of slice 1. 

Finally one may  ask if the   inner (Cauchy)  horizons can join  continuously, i.e. if $\sigma_-^{{\rm H}1} =  \sigma_-^{{\rm H}2}$ is allowed.  A simple calculation shows that this is indeed possible for $\ell_2/\ell_1 <3$, a critical ratio of central charges that also  arose  in  references \cite{Simidzija:2020ukv,Bachas:2021fqo}. We don't know if this is a coincidence, or if some deeper reason lurks below.

  
\section{Background on flowing  funnels}\label{app:flowingfunnel}
In this appendix, we collect some  formulae on the  flowing funnels discussed  in section \ref{sec:funnels}. We start with the most general asymptotically-locally-AdS$_3$  solution in  Fefferman-Graham coordinates,  generalizing  the Banados geometries to arbitrary boundary metric, see \ref{flatfeffermangraham}, but here $g_{(0)}$ will be arbitrary instead of flat :
\begin{eqgroup}\label{FGapp}
 ds^2 =   \frac{\ell^2  dz^2}{z^2} + \frac{1}{z^2} g_{\alpha\beta}(x, z)  dx^\alpha dx^\beta\ ,
 \end{eqgroup}
 where $g_{\alpha\beta}$ is a quartic polynomial in $z$ (written here as a matrix) 
\begin{eqgroup}\label{FG1app}
g(x,z)  =g_{(0)} + z^2 g_{(2)} +  \frac{z^4}{4} g_{(2)}g_{(0)}^{-1}g_{(2)} \ .
\end{eqgroup}
In this equation  $g_{(0)}$ is the boundary metric and $g_{(2)}$ is  given by (see sec.\ref{sec:bottomupapproach} for more details :
 \begin{eqgroup}\label{FG2app}
  g_{(2)\;\alpha\beta}  =   \frac{\ell^2}{2 }  R_{(0)} g_{(0)\alpha\beta} +  {\ell  }  \langle T_{\alpha\beta}\rangle   \ ,
  \end{eqgroup}
where  $R_{(0)}$ is the  Ricci scalar  of $g_{(0)}$, and $\langle T_{\alpha\beta}\rangle$ the expectation value of the energy-momentum tensor. 
This must   be  conserved, $\nabla^a_{(0)}  \langle T_{ab}\rangle = 0$, and should obey the  trace anomaly  equation $g_{(0)}^{ab} \langle T_{ab}\rangle = -(c /24\pi)     R_{(0)}$.

 We  may take  the   boundary metric to be  that of the Schwarzschild  black hole
 (this differs from the metric in \cite{Fischetti:2012ps}, but since it is not dynamical we are free to  choose our preferred 
 boundary metric), 
 \begin{eqgroup}
  ds^2_{(0)} =  -f(x)\, dt^2 + \frac{dx^2}{f(x)} \qquad {\rm with} \quad f(x) = \frac{x }{x + a }\ . 
 \end{eqgroup}
The horizon at $x= 0 $  has  temperature $\Theta_{S} = (4\pi a)^{-1}$.  Using the familiar tortoise coordinates we can write
 \begin{eqgroup}
    ds^2_{(0)} =   f(x) (-dt^2 + dx_*^2)  \qquad {\rm where} \quad x_*  =  x + a \log x \ .
 \end{eqgroup}
 Let  $w^\pm = x_*  \pm t$.  
 The expectation value 
  of the energy-momentum tensor in the black-hole  metric can be  expressed in terms of   $ \phi = \log f(x)$
 as follows
\begin{eqgroup}\label{C5} 
 \langle T_{\pm\pm}\rangle = \frac{\ell}{2}\, \big[ \partial^2_\pm\phi - \frac{1}{2} (\partial_\pm \phi)^2 \big]   + k_\pm(w^\pm)\ ,  \quad
  \langle T_{+-}\rangle =   -\frac{\ell}{2}\,  \partial_+\partial_-\phi\ , 
\end{eqgroup}
 with  $k_\pm$  arbitrary functions of $w^\pm$ that depend on the choice of state. At $x\gg a$  where the metric is flat,    $k_\pm$ determines   the incoming and outgoing fluxes of energy. In a stationary solution,  these must be constant. If a heat bath at temperature $\Theta_+$ is placed at infinity,   $k_+ = \pi^2\ell  \Theta_+^2$. The function $k_-$,  on the other hand,  is fixed by  requiring that there is no outgoing flux at the Schwarzschild horizon.  

From 
\begin{eqgroup}\label{C7}
\frac{\ell}{2} \big[ \partial^2_\pm\phi - \frac{1}{2} (\partial_\pm \phi)^2 \big] =  - \frac{\ell  (a^2 + 4ax)}{16 (x+a)^4}  \ ,
\end{eqgroup}
we deduce   $ \langle T_{--}\rangle\vert_{x=0} = 0  \Longrightarrow k_- = \ell/16 a^2 = \pi^2\ell  \Theta_{S}^2$. The outgoing flux at infinity  is thermalized at the black hole temperature,  as expected. 
 
Inserting the expressions  (\ref{FG2app}-\ref{C7})  in   \eqref{FGapp} and \eqref{FG1app} gives the flowing-funnel metric in Fefferman-Graham  coordinates. These are however singular coordinates, not well adapted for calculating the event horizon as shown in \cite{Fischetti:2012ps}.  Following this reference, one can compute  the horizon by going  to BTZ coordinates -- this is possible because all solutions are  locally equivalent in three dimensions. The change   from any metric   \eqref{FGapp}\,-\,\eqref{FG1app}  to  local BTZ coordinates has been worked out in ref.\,\cite{Rooman:2000ei} (see also \cite{Krasnov:2001cu}) and can be used to compute the black-funnel shapes.  A noteworthy feature   is that the funnels  start  vertically inwards at $x=0$ \cite{Fischetti:2012ps} making  a delta-function contribution to the area density. Note that figure \ref{fig:bfun} shows two independent flowing funnels with Schwarzschild  temperatures $\Theta_S = \Theta_1^{\rm eff}$
and $\Theta_2^{\rm eff}$. 

\chapter{RT surfaces and entanglement entropy}
\section{Euclidean construction for single-crossing geodesic}
\label{app:Derivationboundaryintervalgeodesic}
We use fig. \ref{fig:geodesicoftraversinginterval} and aim to express the depicted points as a function of $\lam$ (where $O_2=(\lam,0)$), $\varphi$ and $\th_i$. The basic equation we will use is the law of sines inside various triangles. We will work with fig. \ref{fig:geodesicoftraversinginterval} as a reference, but if one is careful the formulas are applicable also to situations not depicted by the figure, for instance, when $\lam>0$. When writing something like $O_1O_2$, we mean the length between the segment $[O_1,O_2]$. This in particular could become negative in the formulas we will give, signaling that $O_1$ and $O_2$ cross. This is always accompanied by a sign change in the opposing angle so that the law of sines still works.

By the triangle $(O_2,O,X)$, we have :
\begin{eqgroup}\label{o2xsin}
\frac{O_2X}{\sin(\th_2)}=\frac{OO_2}{\sin(\th_2-\phi)}\Leftrightarrow O_2X= \frac{(-\lam)\sin(\th_2)}{\sin(\th_2-\phi)}\ ,
\end{eqgroup}
where we use $OO_2=-\lam$.

Then we find $OO_1$ using the triangle $(O,O_2,O_1)$ :
\begin{eqgroup}\label{oo1sin}
OO_1=\frac{OO_2\sin(\pi-\phi)}{\sin(\phi-\th_1)}=\frac{(-\lam)\sin(\phi)}{\sin(\phi-\th_1)}\ .
\end{eqgroup}

The last piece we need is $O_1X$, using the triangle $(O_1,X,O)$ :
\begin{eqgroup}\label{o1xsin}
O_1X=\frac{OO_1 \sin(\pi-\th_2+\th_1)}{\sin(\th_2-\phi)}\ .
\end{eqgroup}

With that in mind we can compute the anchor points $\si_1$, $\si_2$, as well as $Ox$ as we should require it to be positive :
\begin{eqgroup}
O\si_2 &= O_2X-OO_2 = (-\lam)\left(\frac{\sin(\th_2)}{\sin(\th_2-\phi)}-1\right)\ ,\\
O\si_1 &= OO_1+O_1X = \frac{(-\lam)\sin(\phi)}{\sin(\phi-\th_1)}\left(1+\frac{\sin(\th_2-\th_1)}{\sin(\th_2-\phi)}\right)\ ,\\
OX&=\frac{(-\lam)\sin(\phi)}{\cos(\th_2-\phi)}\ .
\end{eqgroup}

We can re-express these in terms of the initial angles with the identities $\th_2=\frac{\pi}{2}+\psi_2$ and $\th_1=\psi_1+\psi_2$.

\begin{eqgroup}
    O\si_2 &= (-\lam)\left(\frac{\cos(\psi_2)}{\cos(\psi_2-\phi)}-1\right)\ ,\\
    O\si_1 &= \frac{(-\lam)\sin(\phi)}{\sin(\phi-\psi_2-\psi_1)}\left(1+\frac{\cos(\psi_1)}{\cos(\psi_2-\phi)}\right)\ ,\\
    OX &= \frac{(-\lam) \sin(\phi)}{\cos(\phi-\psi_2)}\ .
    \label{si1si2xsinphi}
\end{eqgroup}

Then, by replacing $\phi-\psi_2\equiv\al$, we obtain the equation (\ref{eq:si1si2alphboundarywithinterface}) used in the main-text.

We would like now to compute the length of this geodesic. We will apply the formula (\ref{lengthgeneralform}) to the two pieces composing the geodesic. When the geodesic is anchored at the boundary, we cut it off at $z_i=\eps_i$. Denoting $R$ the radius of the semi-circle, the initial angle $\th_0$ is given by :
\begin{eqgroup}
    \frac{\eps_0}{R}=\sin(\al_0)\approx \al_0\ .
\end{eqgroup}

Thus because of the presence of the cutoffs, the length of the geodesic does depend on its radius, unlike what was suggested by the formula (\ref{lengthgeneralform}).

The semi-circle going from $\si_2\rightarrow x$ has radius $R_2$ given by :
\begin{eqgroup}
    R_2=O_2\si_2=\frac{(-\lam)\cos(\psi_2)}{\cos(\al)}\ .
\end{eqgroup}

The starting angle for this portion is given by the cutoff $\eps_2$, while $\th_f=\phi$. This gives the length $L_2$ of this portion :
\begin{eqgroup}
    L_2 = \ell_2 \ln(\frac{2R_2\sin(\al+\psi_2)}{\eps_2(1+\cos(\al+\psi_2))})= \ell_2 \ln(\frac{2R_2}{\eps_2}\tan(\frac{\al+\psi_2}{2}))\ ,
\end{eqgroup}
keeping only the leading order in $\eps_2$.

The other semi-circle from $X\rightarrow \si_1$ has radius $R_1$ :
\begin{eqgroup}
    R_1 = O_1X=\frac{(-\lam)\cos(\psi_1)\sin(\al+\psi_2)}{\sin(\al-\psi_1)\cos(\al)}\ ,
\end{eqgroup}
while the initial and final angles are $\th_0=\al-\psi_1$ and $\th_f=\frac{\eps_1}{R_1}$. Then its length $L_1$ is :
\begin{eqgroup}
    L_1=\ell_1 \ln(\frac{2R_1(1+\cos(\al-\psi_1))}{\eps_1\sin(\al-\psi_1)})=\ell_1\ln(\frac{2R_1}{\eps_1}\frac{1}{\tan(\frac{\al-\psi_1}{2})})\ .
\end{eqgroup}

The full length of the geodesic is obtained simply by summing the two contributions, $L=L_1+L_2$. After some algebra and re-arranging, we obtain the formula (\ref{lengthgeodboundaryinterfacecase}) used in the main-text.

\section{Euclidean construction for double-crossing geodesic}
\label{app:doublecrosscomputation}
We use fig. \ref{fig:doublecrossinggeodesic} to express the depicted points as a function of $\lam$ (where $O_2=(\lam,0)$), $\varphi$ and the $\th_i$. As in the previous section, this is done by repeated use of the law of sines.
 
Beginning with the triangle $(O,O_1,O_3)$, we have :
\begin{eqgroup}
    OO_2=\frac{(-\lam)\sin(\phi)}{\sin(\th_1-\phi)}\ .
\end{eqgroup}

With the triangle $(O,O_2,O_3)$ :
\begin{eqgroup}\label{OO3doubl}
    OO_3 = OO_2\frac{\sin(2\th_2-\th_1-\phi)}{\sin(2\th_2-\phi)}=-\frac{\lam\sin(\phi)\sin(2\th_2-\th_1-\phi)}{\sin(\th_1-\phi)\sin(2\th_2-\phi)}\ .
\end{eqgroup}

To find $\si_2$, we need the radius of the semi-circle centered at $O_3$. We use the triangle $(O,O_3,X_2)$ :
\begin{eqgroup}\label{O3X2double}
    O_3X_2=OO_3\frac{\sin(\th_2)}{\sin(\th_2-\phi)}\ ,
\end{eqgroup}
where $OO_3$ is given by (\ref{O3X2double}). Then $O\si_2=OO_3+O_3X_2$ :
\begin{eqgroup}\label{si2prelidoub}
    O\si_2 &=OO_3(1+\frac{\sin(\th_2)}{\sin(\th_2-\phi)})\ .
\end{eqgroup}

Similarly, $O\si_1=O_1\si_1-OO_1=O_1X_1-OO_1$. We use triangle $(O,O_1,X_1)$ :
\begin{eqgroup}
O_1X_1&=(-\lam)\left(1-\frac{\sin(\th_2)}{\sin(\th_2-\phi)}\right)\ ,\\
O\si_1 &= (-\lam)\left(\frac{\sin(\th_2)}{\sin(\th_2-\phi)}\right)\ .
\end{eqgroup}

Collecting the expression and replacing with the angles $\psi_i$ :
\begin{eqgroup}
O\si_1 &= (-\lam) \left(\frac{\cos(\psi_2)}{\cos(\psi_2 - \phi)}-1\right)\ ,\\
    O\si_2 &= \frac{(-\lam)\sin(\phi)}{\sin(\psi_1+\psi_2-\phi)}\frac{\sin(\phi+\psi_1-\psi_2)}{\sin(\phi-2\psi_2)}\left(1+\frac{\cos(\psi_2)}{\cos(\psi_2 - \phi)}\right)\ ,
\end{eqgroup}
which yield the formulas (\ref{si1si2doublecrossinginterface}) when doing the change of variables $\al=\phi-\psi_2$. In what follows, we express everythin in terms of $\al$ instead of $\phi$.

For the Length of the geodesics, we must now compute the length of three separate segments, using the formula (\ref{lengthgeneralform}). Some extra angles need to be determined on fig. \ref{fig:doublecrossinggeodesic}, but we don't details that here.

Consider first the segment connecting $\si_1\rightarrow X_1$. The radius is given by $R_1$ :
\begin{eqgroup}
 R_1=O_1X_1=\frac{(-\lam) \cos(\psi_2)}{\cos(al)}\ .
\end{eqgroup}

The starting angle $\th_0$ is given by the cutoff, $\eps_1$, while the ending $\th_f=\al+\psi_2$.

This gives a contribution to the length :
\begin{eqgroup}
 L_1 = \ell_2 \ln\left(\frac{2R_1}{\eps_1}\tan\left(\frac{\al+\psi_2}{2}\right)\right)\ .
\end{eqgroup}

For the second segment $X_1\rightarrow X_2$, the Radius of the geodesic is irrelevant. The initial angle is $\th_0=\pi-(\al+\psi_1)$, final $\th_f=\psi-(\psi_1-\al)$. This gives the contribution $L_3$ :
\begin{eqgroup}
    L_3 = \ell_1 \ln\left(\frac{\tan(\frac{\al+\psi_1}{2})}{\tan(\frac{\psi_1-\al}{2})}\right)\ .
\end{eqgroup}

For the last segment $L_2$, the radius $R_2$ is :
\begin{eqgroup}
    R_3=O_3X_2=OO_3 \frac{\sin(\th_2)}{\sin(\th_2-\phi)} = \frac{-\lam \sin(\al+\psi_2)\sin(\al+\psi_1)}{\sin(\psi_1-\al)\sin(\al-\psi_2)}\frac{\cos(\psi_2)}{\cos(\al)}\ .
\end{eqgroup}

The corresponding length $L_2$ is then :
\begin{eqgroup}
    L_3=\ell_2\ln(\frac{2R_3}{\eps_2\tan(\frac{\al-\psi_2}{2})})\ .
\end{eqgroup}

Adding the contributions of the $L_i$ gives the length of the full geodesic. After re-arranging, we obtain (\ref{fulllengthdoublecross}) used in the main-text.

\section{Peculiar geodesics in the spinning black string geometry}
\label{app:geodesicsinstringu}
In this section we briefly look at the shape of the different spacelike geodesics in the spinning string geometry. In particular, we highlight the fact that there are horizon entering geodesics, but that they cannot have two anchors on the boundary.

The geodesics are obtained from the equations of sec.\ref{sec:geodesicinasympads}, composed with the change of coordinates (\ref{finktopoinc}). Consider first the type $K_+K_->0$, described in sec.\ref{sec:casekpkmmorezero}. In poincaré space, this geodesic has two anchor points on the boundary. However, this does not necessarily translate to Finkelstein coordinates. To see this, it is useful to consider the inverse of (\ref{finktopoinc}) (for $J>0$) :
\begin{eqgroup}\label{inversefinktopoinc}
    r^2 &= r_+^2 + \frac{(w_+ w_-) (r_+^2 - r_-^2)}{z_p^2}\ ,\\
    v &= \frac{\ell}{2}\left(\frac{1}{r_+-r_-}\ln(\frac{w_+(r+r_-)}{r+r_+})-\frac{1}{r_++r_-}\ln(\frac{w_-(r+r_-)}{r-r_+})\right)\ ,\\
    y &= \frac{\ell}{2}\left(\frac{1}{r_+-r_-}\ln(\frac{w_+(r+r_-)}{r+r_+})+\frac{1}{r_++r_-}\ln(\frac{w_-(r+r_-)}{r-r_+})\right)\ ,\\
\end{eqgroup}
where the $r$ appearing in the RHS should be thought as depending on $w_\pm$, $z_p$.

When using Poincaré coordinates to compute geodesics, we must keep in mind the region of Poincaré which describes Finkelstein space, namely $w_+>0$. Thus, when a geodesics is of the type $K_+K_->0$, either it remains in the region $w_+>0$, and thus it is doubly anchored in Finkelstein space (see fig.\ref{fig:sketchRTNess}), or it reaches the $w_+=0$ boundary. In this case, by (\ref{inversefinktopoinc}), the geodesics escapes at $(v\rightarrow -\infty,r=r_+, y\rightarrow -\infty)$. As such, it follows closely the apparent horizon towards the "coordinate" horizon, both in time and space. Note that the divergence is such that $\frac{v}{y}\approx \frac{r_+}{r_-}$. In the context of a pure CFT, these geodesics do not contribute to the RT prescription, as they have only one anchor point. In ICFT, they might be "stopped" by the membrane, and upon crossing to the other side, head back to the boundary, so it is plausible that they play a role in this case.

By (\ref{finktopoinc}), the outer horizon of the black hole is mapped to $w_-=0$. Thus, one might argue that it should be possible to enter the horizon of the black hole and head back to the boundary: a geodesic in Poincaré that dips in the $w_-<0$ region and heads back to the boundary should generate such a horizon probing geodesic in Finkelstein coordinates. However, from (\ref{eq:geodesicgeneralpositive}), the curves $w_{\pm}(\lam)$ are obviously monotonic, barring such a possibility. Thus if a geodesic crosses the region $w_-<0$, it will remain in it and thus be unable to leave the black hole. Its precise fate depends on the geodesic, it could again escape at infinity before reaching the singularity if it reaches $w_+=0$, otherwise, it will reach $r=0$ as $z_p\rightarrow 0$, as seen from (\ref{inversefinktopoinc}). Again, these geodesics are irrelevant in the context of a pure CFT, but might be very important in the context of ICFT. If they intersect the membrane on their way to the singularity, that could give rise to geodesics of the type depicted in fig.\ref{fig:RTinsidehorizonsketch}.

Consider now the case $K_+K_-<0$, see sec.\ref{sec:casekpkmlesszero}. Now, as explained in the main text, even in Poincaré we are in presence of two disconnected branches each with one anchor point on the boundary. Considering them as separate geodesics, we look at the mapping of one of the two branches to Finkelstein space. The discussion is very similar, namely, we must pay attention to the important surfaces $w_\pm=0$ in Poincaré space. However, in this case as the affine parameter $\lam$ approaches $0$ (we assume we begin with $\lam_0<0$) we will have $w_{\pm}\rightarrow -{\rm Sign}(w_{\mp})\infty$. As such, we are assured to either enter the horizon at $w_-=0$, or escape at infinity at $w_+=0$. Indeed, as in this case the signs of $\dot{w}_\pm$ are opposing, one of the two situations will happen.

We do not comment on the last case $K_+K_-=0$, as it is irrelevant to the numerical algorithms. Nothing special happens with respect to the two other cases, the same reasoning applies.
\footnotesize{\bibliography{bibo}}

\end{document}